\documentclass[twocolumn,traditabstract,longauth]{aa}
\usepackage{graphicx}
\usepackage{amsmath,amsfonts,amssymb}
\usepackage{txfonts}
\usepackage{color}
\usepackage{fixltx2e}
\usepackage{natbib}
\usepackage{epsfig}
\usepackage{afterpage}
\usepackage{ifthen}
\usepackage[breaklinks,colorlinks,citecolor=blue]{hyperref}
\bibpunct{(}{)}{;}{a}{}{,}

\def\setsymbol#1#2{\expandafter\def\csname #1\endcsname{#2}}
\def\getsymbol#1{\csname #1\endcsname}

\def\Planck{\textit{Planck}}

\def\alltwentythirteenresultspapers{\nocite{planck2013-p01, planck2013-p02, planck2013-p02a, planck2013-p02d, planck2013-p02b, planck2013-p03, planck2013-p03c, planck2013-p03f, planck2013-p03d, planck2013-p03e, planck2013-p01a, planck2013-p06, planck2013-p03a, planck2013-pip88, planck2013-p08, planck2013-p11, planck2013-p12, planck2013-p13, planck2013-p14, planck2013-p15, planck2013-p05b, planck2013-p17, planck2013-p09, planck2013-p09a, planck2013-p20, planck2013-p19, planck2013-pipaberration, planck2013-p05, planck2013-p05a, planck2013-pip56, planck2013-p06b, planck2013-p01a}}

\def\alltwentyfifteenresultspapers{\nocite{planck2014-a01, planck2014-a03, planck2014-a04, planck2014-a05, planck2014-a06, planck2014-a07, planck2014-a08, planck2014-a09, planck2014-a11, planck2014-a12, planck2014-a13, planck2014-a14, planck2014-a15, planck2014-a16, planck2014-a17, planck2014-a18, planck2014-a19, planck2014-a20, planck2014-a22, planck2014-a24, planck2014-a26, planck2014-a28, planck2014-a29, planck2014-a30, planck2014-a31, planck2014-a35, planck2014-a36, planck2014-a37, planck2014-ES}}

\newbox\tablebox    \newdimen\tablewidth
\def\leaderfil{\leaders\hbox to 5pt{\hss.\hss}\hfil}
\def\endPlancktable{\tablewidth=\columnwidth 
    $$\hss\copy\tablebox\hss$$
    \vskip-\lastskip\vskip -2pt}
\def\endPlancktablewide{\tablewidth=\textwidth 
    $$\hss\copy\tablebox\hss$$
    \vskip-\lastskip\vskip -2pt}
\def\tablenote#1 #2\par{\begingroup \parindent=0.8em
    \abovedisplayshortskip=0pt\belowdisplayshortskip=0pt
    \noindent
    $$\hss\vbox{\hsize\tablewidth \hangindent=\parindent \hangafter=1 \noindent
    \hbox to \parindent{$^#1$\hss}\strut#2\strut\par}\hss$$
    \endgroup}
\def\doubleline{\vskip 3pt\hrule \vskip 1.5pt \hrule \vskip 5pt}

\def\L2{\ifmmode L_2\else $L_2$\fi}

\def\DeltaT{\ifmmode \Delta T\else $\Delta T$\fi}
\def\deltat{\ifmmode \Delta t\else $\Delta t$\fi}
\def\fknee{\ifmmode f_{\rm knee}\else $f_{\rm knee}$\fi}
\def\Fmax{\ifmmode F_{\rm max}\else $F_{\rm max}$\fi}
\def\solar{\ifmmode{\rm M}_{\mathord\odot}\else${\rm M}_{\mathord\odot}$\fi}
\def\Msolar{\ifmmode{\rm M}_{\mathord\odot}\else${\rm M}_{\mathord\odot}$\fi}
\def\Lsolar{\ifmmode{\rm L}_{\mathord\odot}\else${\rm L}_{\mathord\odot}$\fi}
\def\inv{\ifmmode^{-1}\else$^{-1}$\fi}
\def\mo{\ifmmode^{-1}\else$^{-1}$\fi}
\def\sup#1{\ifmmode ^{\rm #1}\else $^{\rm #1}$\fi}
\def\expo#1{\ifmmode \times 10^{#1}\else $\times 10^{#1}$\fi}
\def\,{\thinspace}
\def\lsim{\mathrel{\raise .4ex\hbox{\rlap{$<$}\lower 1.2ex\hbox{$\sim$}}}}
\def\gsim{\mathrel{\raise .4ex\hbox{\rlap{$>$}\lower 1.2ex\hbox{$\sim$}}}}

\def\simprop{\mathrel{\raise .4ex\hbox{\rlap{$\propto$}\lower 1.2ex\hbox{$\sim$}}}}
\def\deg{\ifmmode^\circ\else$^\circ$\fi}
\def\pdeg{\ifmmode $\setbox0=\hbox{$^{\circ}$}\rlap{\hskip.11\wd0 .}$^{\circ}
          \else \setbox0=\hbox{$^{\circ}$}\rlap{\hskip.11\wd0 .}$^{\circ}$\fi}
\def\arcs{\ifmmode {^{\scriptstyle\prime\prime}}
          \else $^{\scriptstyle\prime\prime}$\fi}
\def\arcm{\ifmmode {^{\scriptstyle\prime}}
          \else $^{\scriptstyle\prime}$\fi}
\newdimen\sa  \newdimen\sb
\def\parcs{\sa=.07em \sb=.03em
     \ifmmode \hbox{\rlap{.}}^{\scriptstyle\prime\kern -\sb\prime}\hbox{\kern -\sa}
     \else \rlap{.}$^{\scriptstyle\prime\kern -\sb\prime}$\kern -\sa\fi}
\def\parcm{\sa=.08em \sb=.03em
     \ifmmode \hbox{\rlap{.}\kern\sa}^{\scriptstyle\prime}\hbox{\kern-\sb}
     \else \rlap{.}\kern\sa$^{\scriptstyle\prime}$\kern-\sb\fi}
\def\ra[#1 #2 #3.#4]{#1\sup{h}#2\sup{m}#3\sup{s}\llap.#4}
\def\dec[#1 #2 #3.#4]{#1\deg#2\arcm#3\arcs\llap.#4}
\def\deco[#1 #2 #3]{#1\deg#2\arcm#3\arcs}
\def\rra[#1 #2]{#1\sup{h}#2\sup{m}}

\def\dots{\relax\ifmmode \ldots\else $\ldots$\fi}
\def\WHzsr{\ifmmode $W\,Hz\mo\,sr\mo$\else W\,Hz\mo\,sr\mo\fi}
\def\mHz{\ifmmode $\,mHz$\else \,mHz\fi}
\def\GHz{\ifmmode $\,GHz$\else \,GHz\fi}
\def\mKs{\ifmmode $\,mK\,s$^{1/2}\else \,mK\,s$^{1/2}$\fi}
\def\muKs{\ifmmode \,\mu$K\,s$^{1/2}\else \,$\mu$K\,s$^{1/2}$\fi}
\def\muKRJs{\ifmmode \,\mu$K$_{\rm RJ}$\,s$^{1/2}\else \,$\mu$K$_{\rm RJ}$\,s$^{1/2}$\fi}
\def\muKHz{\ifmmode \,\mu$K\,Hz$^{-1/2}\else \,$\mu$K\,Hz$^{-1/2}$\fi}
\def\MJysr{\ifmmode \,$MJy\,sr\mo$\else \,MJy\,sr\mo\fi}
\def\MJysrmK{\ifmmode \,$MJy\,sr\mo$\,mK$_{\rm CMB}\mo\else \,MJy\,sr\mo\,mK$_{\rm CMB}\mo$\fi}
\def\microns{\ifmmode \,\mu$m$\else \,$\mu$m\fi}

\def\muK{\ifmmode \,\mu$K$\else \,$\mu$\hbox{K}\fi}
\def\microK{\ifmmode \,\mu$K$\else \,$\mu$\hbox{K}\fi}
\def\muW{\ifmmode \,\mu$W$\else \,$\mu$\hbox{W}\fi}
\def\kms{\ifmmode $\,km\,s$^{-1}\else \,km\,s$^{-1}$\fi}
\def\kmsMpc{\ifmmode $\,\kms\,Mpc\mo$\else \,\kms\,Mpc\mo\fi}
\providecommand{\sorthelp}[1]{}

\def\WMAP{WMAP}
\def\COBE{\textit{COBE}}

\def\healpix{\texttt{HEALPix}}
\def\commander{\texttt{Commander}}

\renewcommand{\d}[0]{\vec{d}}

\newcommand{\n}[0]{\vec{n}}

\newcommand{\s}[0]{\vec{s}}
\renewcommand{\a}[0]{\vec{a}}
\newcommand{\m}[0]{\vec{m}}

\newcommand{\F}[0]{\tens{F}}

\newcommand{\T}[0]{\tens{T}}

\renewcommand{\L}[0]{\tens{L}}
\newcommand{\g}[0]{\vec{g}}
\newcommand{\N}[0]{\tens{N}}

\renewcommand{\S}[0]{\tens{S}}

\newcommand{\Te}[0]{T_{\rm e}}
\newcommand{\EM}[0]{\rm EM}

\def\tp{^{\rm T}}
\def\inv{^{-1}}

\title{\Planck\ 2015 results. X. \\Diffuse component separation: Foreground maps}
\author{\small
Planck Collaboration: R.~Adam\inst{79}
\and
P.~A.~R.~Ade\inst{91}
\and
N.~Aghanim\inst{64}
\and
M.~I.~R.~Alves\inst{64}
\and
M.~Arnaud\inst{77}
\and
M.~Ashdown\inst{73, 6}
\and
J.~Aumont\inst{64}
\and
C.~Baccigalupi\inst{90}
\and
A.~J.~Banday\inst{100, 10}
\and
R.~B.~Barreiro\inst{69}
% ...
}
\institute{\small
APC, AstroParticule et Cosmologie, Universit\'{e} Paris Diderot, CNRS/IN2P3, CEA/lrfu, Observatoire de Paris, Sorbonne Paris Cit\'{e}, 10, rue Alice Domon et L\'{e}onie Duquet, 75205 Paris Cedex 13, France\goodbreak
\and
Aalto University Mets\"{a}hovi Radio Observatory and Dept of Radio Science and Engineering, P.O. Box 13000, FI-00076 AALTO, Finland\goodbreak
\and
African Institute for Mathematical Sciences, 6-8 Melrose Road, Muizenberg, Cape Town, South Africa\goodbreak
\and
Agenzia Spaziale Italiana Science Data Center, Via del Politecnico snc, 00133, Roma, Italy\goodbreak
\and
Aix Marseille Universit\'{e}, CNRS, LAM (Laboratoire d'Astrophysique de Marseille) UMR 7326, 13388, Marseille, France\goodbreak
\and
Astrophysics Group, Cavendish Laboratory, University of Cambridge, J J Thomson Avenue, Cambridge CB3 0HE, U.K.\goodbreak
\and
Astrophysics \& Cosmology Research Unit, School of Mathematics, Statistics \& Computer Science, University of KwaZulu-Natal, Westville Campus, Private Bag X54001, Durban 4000, South Africa\goodbreak
\and
Atacama Large Millimeter/submillimeter Array, ALMA Santiago Central Offices, Alonso de Cordova 3107, Vitacura, Casilla 763 0355, Santiago, Chile\goodbreak
\and
CITA, University of Toronto, 60 St. George St., Toronto, ON M5S 3H8, Canada\goodbreak
\and
CNRS, IRAP, 9 Av. colonel Roche, BP 44346, F-31028 Toulouse cedex 4, France\goodbreak
\and
CRANN, Trinity College, Dublin, Ireland\goodbreak
\and
California Institute of Technology, Pasadena, California, U.S.A.\goodbreak
\and
Centre for Theoretical Cosmology, DAMTP, University of Cambridge, Wilberforce Road, Cambridge CB3 0WA, U.K.\goodbreak
\and
Centro de Estudios de F\'{i}sica del Cosmos de Arag\'{o}n (CEFCA), Plaza San Juan, 1, planta 2, E-44001, Teruel, Spain\goodbreak
\and
Computational Cosmology Center, Lawrence Berkeley National Laboratory, Berkeley, California, U.S.A.\goodbreak
\and
Consejo Superior de Investigaciones Cient\'{\i}ficas (CSIC), Madrid, Spain\goodbreak
\and
DSM/Irfu/SPP, CEA-Saclay, F-91191 Gif-sur-Yvette Cedex, France\goodbreak
\and
DTU Space, National Space Institute, Technical University of Denmark, Elektrovej 327, DK-2800 Kgs. Lyngby, Denmark\goodbreak
\and
D\'{e}partement de Physique Th\'{e}orique, Universit\'{e} de Gen\`{e}ve, 24, Quai E. Ansermet,1211 Gen\`{e}ve 4, Switzerland\goodbreak
\and
Departamento de F\'{\i}sica, Universidad de Oviedo, Avda. Calvo Sotelo s/n, Oviedo, Spain\goodbreak
\and
Department of Astronomy and Astrophysics, University of Toronto, 50 Saint George Street, Toronto, Ontario, Canada\goodbreak
\and
Department of Astrophysics/IMAPP, Radboud University Nijmegen, P.O. Box 9010, 6500 GL Nijmegen, The Netherlands\goodbreak
\and
Department of Physics \& Astronomy, University of British Columbia, 6224 Agricultural Road, Vancouver, British Columbia, Canada\goodbreak
\and
Department of Physics and Astronomy, Dana and David Dornsife College of Letter, Arts and Sciences, University of Southern California, Los Angeles, CA 90089, U.S.A.\goodbreak
\and
Department of Physics and Astronomy, University College London, London WC1E 6BT, U.K.\goodbreak
\and
Department of Physics, Florida State University, Keen Physics Building, 77 Chieftan Way, Tallahassee, Florida, U.S.A.\goodbreak
\and
Department of Physics, Gustaf H\"{a}llstr\"{o}min katu 2a, University of Helsinki, Helsinki, Finland\goodbreak
\and
Department of Physics, Princeton University, Princeton, New Jersey, U.S.A.\goodbreak
\and
Department of Physics, University of California, Santa Barbara, California, U.S.A.\goodbreak
\and
Department of Physics, University of Illinois at Urbana-Champaign, 1110 West Green Street, Urbana, Illinois, U.S.A.\goodbreak
\and
Dipartimento di Fisica e Astronomia G. Galilei, Universit\`{a} degli Studi di Padova, via Marzolo 8, 35131 Padova, Italy\goodbreak
\and
Dipartimento di Fisica e Scienze della Terra, Universit\`{a} di Ferrara, Via Saragat 1, 44122 Ferrara, Italy\goodbreak
\and
Dipartimento di Fisica, Universit\`{a} La Sapienza, P. le A. Moro 2, Roma, Italy\goodbreak
\and
Dipartimento di Fisica, Universit\`{a} degli Studi di Milano, Via Celoria, 16, Milano, Italy\goodbreak
\and
Dipartimento di Fisica, Universit\`{a} degli Studi di Trieste, via A. Valerio 2, Trieste, Italy\goodbreak
\and
Dipartimento di Fisica, Universit\`{a} di Roma Tor Vergata, Via della Ricerca Scientifica, 1, Roma, Italy\goodbreak
\and
Dipartimento di Matematica, Universit\`{a} di Roma Tor Vergata, Via della Ricerca Scientifica, 1, Roma, Italy\goodbreak
\and
Discovery Center, Niels Bohr Institute, Blegdamsvej 17, Copenhagen, Denmark\goodbreak
\and
Dpto. Astrof\'{i}sica, Universidad de La Laguna (ULL), E-38206 La Laguna, Tenerife, Spain\goodbreak
\and
European Southern Observatory, ESO Vitacura, Alonso de Cordova 3107, Vitacura, Casilla 19001, Santiago, Chile\goodbreak
\and
European Space Agency, ESAC, Planck Science Office, Camino bajo del Castillo, s/n, Urbanizaci\'{o}n Villafranca del Castillo, Villanueva de la Ca\~{n}ada, Madrid, Spain\goodbreak
\and
European Space Agency, ESTEC, Keplerlaan 1, 2201 AZ Noordwijk, The Netherlands\goodbreak
\and
Facolt\`{a} di Ingegneria, Universit\`{a} degli Studi e-Campus, Via Isimbardi 10, Novedrate (CO), 22060, Italy\goodbreak
\and
Gran Sasso Science Institute, INFN, viale F. Crispi 7, 67100 L'Aquila, Italy\goodbreak
\and
HGSFP and University of Heidelberg, Theoretical Physics Department, Philosophenweg 16, 69120, Heidelberg, Germany\goodbreak
\and
Haverford College Astronomy Department, 370 Lancaster Avenue, Haverford, Pennsylvania, U.S.A.\goodbreak
\and
Helsinki Institute of Physics, Gustaf H\"{a}llstr\"{o}min katu 2, University of Helsinki, Helsinki, Finland\goodbreak
\and
INAF - Osservatorio Astrofisico di Catania, Via S. Sofia 78, Catania, Italy\goodbreak
\and
INAF - Osservatorio Astronomico di Padova, Vicolo dell'Osservatorio 5, Padova, Italy\goodbreak
\and
INAF - Osservatorio Astronomico di Roma, via di Frascati 33, Monte Porzio Catone, Italy\goodbreak
\and
INAF - Osservatorio Astronomico di Trieste, Via G.B. Tiepolo 11, Trieste, Italy\goodbreak
\and
INAF/IASF Bologna, Via Gobetti 101, Bologna, Italy\goodbreak
\and
INAF/IASF Milano, Via E. Bassini 15, Milano, Italy\goodbreak
\and
INFN, Sezione di Bologna, Via Irnerio 46, I-40126, Bologna, Italy\goodbreak
\and
INFN, Sezione di Roma 1, Universit\`{a} di Roma Sapienza, Piazzale Aldo Moro 2, 00185, Roma, Italy\goodbreak
\and
INFN, Sezione di Roma 2, Universit\`{a} di Roma Tor Vergata, Via della Ricerca Scientifica, 1, Roma, Italy\goodbreak
\and
INFN/National Institute for Nuclear Physics, Via Valerio 2, I-34127 Trieste, Italy\goodbreak
\and
IPAG: Institut de Plan\'{e}tologie et d'Astrophysique de Grenoble, Universit\'{e} Grenoble Alpes, IPAG, F-38000 Grenoble, France, CNRS, IPAG, F-38000 Grenoble, France\goodbreak
\and
IUCAA, Post Bag 4, Ganeshkhind, Pune University Campus, Pune 411 007, India\goodbreak
\and
Imperial College London, Astrophysics group, Blackett Laboratory, Prince Consort Road, London, SW7 2AZ, U.K.\goodbreak
\and
Infrared Processing and Analysis Center, California Institute of Technology, Pasadena, CA 91125, U.S.A.\goodbreak
\and
Institut N\'{e}el, CNRS, Universit\'{e} Joseph Fourier Grenoble I, 25 rue des Martyrs, Grenoble, France\goodbreak
\and
Institut Universitaire de France, 103, bd Saint-Michel, 75005, Paris, France\goodbreak
\and
Institut d'Astrophysique Spatiale, CNRS (UMR8617) Universit\'{e} Paris-Sud 11, B\^{a}timent 121, Orsay, France\goodbreak
\and
Institut d'Astrophysique de Paris, CNRS (UMR7095), 98 bis Boulevard Arago, F-75014, Paris, France\goodbreak
\and
Institute of Astronomy, University of Cambridge, Madingley Road, Cambridge CB3 0HA, U.K.\goodbreak
\and
Institute of Theoretical Astrophysics, University of Oslo, Blindern, Oslo, Norway\goodbreak
\and
Instituto de Astrof\'{\i}sica de Canarias, C/V\'{\i}a L\'{a}ctea s/n, La Laguna, Tenerife, Spain\goodbreak
\and
Instituto de F\'{\i}sica de Cantabria (CSIC-Universidad de Cantabria), Avda. de los Castros s/n, Santander, Spain\goodbreak
\and
Istituto Nazionale di Fisica Nucleare, Sezione di Padova, via Marzolo 8, I-35131 Padova, Italy\goodbreak
\and
Jet Propulsion Laboratory, California Institute of Technology, 4800 Oak Grove Drive, Pasadena, California, U.S.A.\goodbreak
\and
Jodrell Bank Centre for Astrophysics, Alan Turing Building, School of Physics and Astronomy, The University of Manchester, Oxford Road, Manchester, M13 9PL, U.K.\goodbreak
\and
Kavli Institute for Cosmology Cambridge, Madingley Road, Cambridge, CB3 0HA, U.K.\goodbreak
\and
LAL, Universit\'{e} Paris-Sud, CNRS/IN2P3, Orsay, France\goodbreak
\and
LAPTh, Univ. de Savoie, CNRS, B.P.110, Annecy-le-Vieux F-74941, France\goodbreak
\and
LERMA, CNRS, Observatoire de Paris, 61 Avenue de l'Observatoire, Paris, France\goodbreak
\and
Laboratoire AIM, IRFU/Service d'Astrophysique - CEA/DSM - CNRS - Universit\'{e} Paris Diderot, B\^{a}t. 709, CEA-Saclay, F-91191 Gif-sur-Yvette Cedex, France\goodbreak
\and
Laboratoire Traitement et Communication de l'Information, CNRS (UMR 5141) and T\'{e}l\'{e}com ParisTech, 46 rue Barrault F-75634 Paris Cedex 13, France\goodbreak
\and
Laboratoire de Physique Subatomique et Cosmologie, Universit\'{e} Grenoble-Alpes, CNRS/IN2P3, 53, rue des Martyrs, 38026 Grenoble Cedex, France\goodbreak
\and
Laboratoire de Physique Th\'{e}orique, Universit\'{e} Paris-Sud 11 \& CNRS, B\^{a}timent 210, 91405 Orsay, France\goodbreak
\and
Lawrence Berkeley National Laboratory, Berkeley, California, U.S.A.\goodbreak
\and
Lebedev Physical Institute of the Russian Academy of Sciences, Astro Space Centre, 84/32 Profsoyuznaya st., Moscow, GSP-7, 117997, Russia\goodbreak
\and
Max-Planck-Institut f\"{u}r Astrophysik, Karl-Schwarzschild-Str. 1, 85741 Garching, Germany\goodbreak
\and
Max-Planck-Institut f\"{u}r Extraterrestrische Physik, Giessenbachstra{\ss}e, 85748 Garching, Germany\goodbreak
\and
McGill Physics, Ernest Rutherford Physics Building, McGill University, 3600 rue University, Montr\'{e}al, QC, H3A 2T8, Canada\goodbreak
\and
National University of Ireland, Department of Experimental Physics, Maynooth, Co. Kildare, Ireland\goodbreak
\and
Niels Bohr Institute, Blegdamsvej 17, Copenhagen, Denmark\goodbreak
\and
Optical Science Laboratory, University College London, Gower Street, London, U.K.\goodbreak
\and
SB-ITP-LPPC, EPFL, CH-1015, Lausanne, Switzerland\goodbreak
\and
SISSA, Astrophysics Sector, via Bonomea 265, 34136, Trieste, Italy\goodbreak
\and
School of Physics and Astronomy, Cardiff University, Queens Buildings, The Parade, Cardiff, CF24 3AA, U.K.\goodbreak
\and
School of Physics and Astronomy, University of Nottingham, Nottingham NG7 2RD, U.K.\goodbreak
\and
Sorbonne Universit\'{e}-UPMC, UMR7095, Institut d'Astrophysique de Paris, 98 bis Boulevard Arago, F-75014, Paris, France\goodbreak
\and
Space Research Institute (IKI), Russian Academy of Sciences, Profsoyuznaya Str, 84/32, Moscow, 117997, Russia\goodbreak
\and
Space Sciences Laboratory, University of California, Berkeley, California, U.S.A.\goodbreak
\and
Special Astrophysical Observatory, Russian Academy of Sciences, Nizhnij Arkhyz, Zelenchukskiy region, Karachai-Cherkessian Republic, 369167, Russia\goodbreak
\and
Sub-Department of Astrophysics, University of Oxford, Keble Road, Oxford OX1 3RH, U.K.\goodbreak
\and
Theory Division, PH-TH, CERN, CH-1211, Geneva 23, Switzerland\goodbreak
\and
UPMC Univ Paris 06, UMR7095, 98 bis Boulevard Arago, F-75014, Paris, France\goodbreak
\and
Universit\'{e} de Toulouse, UPS-OMP, IRAP, F-31028 Toulouse cedex 4, France\goodbreak
% ...
}

\begin{document}

\abstract{\Planck\ has mapped the microwave sky in temperature over
  nine frequency bands between 30 and 857\,GHz and in polarization
  over seven frequency bands between 30 and 353\,GHz in
  polarization. In this paper we consider the problem of diffuse
  astrophysical component separation, and process these maps within a
  Bayesian framework to derive an internally consistent set of
  full-sky astrophysical component maps.  Component separation
  dedicated to cosmic microwave background (CMB) reconstruction is
  described in a companion paper. For the temperature analysis, we
  combine the \Planck\ observations with the 9-year \WMAP\ sky maps
  and the Haslam et al.\ 408\,MHz map to derive a joint model of CMB,
  synchrotron, free-free, spinning dust, CO, line emission in the 94
  and 100\,GHz channels, and thermal dust emission. Full-sky maps are
  provided for each component, with an angular resolution varying
  between 7\parcm5 and 1\deg.  Global parameters (monopoles, dipoles,
  relative calibration, and bandpass errors) are fitted jointly with
  the sky model, and best-fit values are tabulated. For polarization,
  the model includes CMB, synchrotron, and thermal dust
  emission. These models provide excellent fits to the observed data,
  with rms temperature residuals smaller than 4\muK\ over 93\,\% of
  the sky for all \Planck\ frequencies up to 353\,GHz, and fractional
  errors smaller than 1\,\% in the remaining 7\,\% of the sky. The
  main limitations of the temperature model at the lower frequencies
  are internal degeneracies among the spinning dust, free-free, and
  synchrotron components; additional observations from external
  low-frequency experiments will be essential to break these
  degeneracies. The main limitations of the temperature model at the
  higher frequencies are uncertainties in the 545 and 857\,GHz
  calibration and zero-points.  For polarization, the main outstanding
  issues are instrumental systematics in the 100--353\,GHz bands on
  large angular scales in the form of temperature-to-polarization
  leakage, uncertainties in the analogue-to-digital conversion, and
  corrections for the very long time constant of the bolometer
  detectors, all of which are expected to improve in the near future.
}


\authorrunning{Planck Collaboration}
\titlerunning{Diffuse component separation: Foreground maps}

\maketitle

\alltwentyfifteenresultspapers
\alltwentythirteenresultspapers

\section{Introduction}
\label{sec:introduction}

This paper, one of a set associated with the 2015 release of data from
the \Planck\footnote{\Planck\ (\url{http://www.esa.int/Planck}) is a
  project of the European Space Agency (ESA) with instruments provided
  by two scientific consortia funded by ESA member states (in
  particular the lead countries France and Italy), with contributions
  from NASA (USA), and telescope reflectors provided by a collaboration
  between ESA and a scientific consortium led and funded by Denmark.}
mission \citep{planck2014-a01}, presents a coherent astrophysical
model of the microwave sky in both temperature and polarization, as
derived from the most recent \Planck\ observations. For temperature,
the analysis incorporates also the 9-year \WMAP\ observations and a
408\,MHz survey \citep{haslam1982}, allowing separation of
synchrotron, free-free, and spinning dust emission.

In March 2013, the \Planck\ Consortium released its first temperature
measurements of the microwave sky, summarized in terms of nine
frequency maps between 30 and 857\,GHz \citep{planck2013-p01}. The
richness of these data has enabled great progress in our understanding
of the astrophysical composition of the microwave sky. The current
\Planck\ data release presents additionally high-sensitivity, full-sky
maps of the polarized microwave sky, offering a fresh view on both
cosmological and astrophysical phenomena.

With increased data volume and quality comes both greater scientific
potential and more stringent requirements on model complexity and
sophistication. The current \Planck\ data release is more ambitious
than the 2013 release in terms of component separation efforts,
accounting for more astrophysical effects and components. In this
round, three related papers summarize the \Planck\ 2015 component
separation products and approaches. First, cosmic microwave background
(CMB) reconstruction and extraction are discussed in
\citet{planck2014-a11}. Second, this paper presents the diffuse
astrophysical foreground products derived from the 2015
\Planck\ observations, both in temperature and polarization. Third,
\citet{planck2014-a31} discusses the scientific interpretation of the
new low-frequency \Planck\ foreground products.

In this paper we establish a single parametric model of the microwave
sky, accounting simultaneously for all significant diffuse
astrophysical components and relevant instrumental effects using the
Bayesian \texttt{Commander} analysis framework
\citep{eriksen2004,eriksen2006,eriksen2008}. In the 2013 data release,
the same framework was applied to the \Planck\ temperature
measurements for frequencies between 30 and 353\,GHz, considering only
angular scales larger than 40\arcm\ FWHM.  This resulted in
low-resolution CMB, CO, and thermal dust emission maps, as well as
a single low-frequency foreground component combining contributions
from synchrotron, free-free, and spinning dust emission
\citep{planck2013-p06}.  Here we extend that analysis in multiple
directions. First, instead of 15.5~months of temperature data, the new
analysis includes the full \Planck\ mission data, 50\,months of LFI
and 29\,months of HFI data, in both temperature and
polarization. Second, we now also include the 9-year
\WMAP\ observations between 23 and 94\,GHz \citep{bennett2012} and a
408\,MHz survey map \citep{haslam1982}, providing enough frequency
constraints to decompose the low-frequency foregrounds into separate
synchrotron, free-free, and spinning dust components.  Third, we now
include the \Planck\ 545 and 857\,GHz frequency bands, allowing us to
constrain the thermal dust temperature and emissivity index with
greater precision, thereby reducing degeneracies between CMB, CO, and
free-free emission.  At the same time, we find that the calibration
and bandpass measurements of these two channels represent two of the
most important sources of systematic uncertainty in the analysis.
Fourth, the present analysis implements a multi-resolution strategy to
provide component maps at high angular resolution. Specifically, the
CMB is recovered with angular resolution 5\arcm\ FWHM
\citep{planck2014-a11}, thermal dust emission and CO $J$=2$\rightarrow$1
lines are recovered at 7\parcm5 FWHM, and synchrotron, free-free, and
spinning dust are recovered at 1\deg\ FWHM.

As in the 2013 data release, the CMB solutions derived using this
Bayesian approach form the basis of the \Planck\ 2015 CMB temperature
likelihood on large angular scales. This is described in detail in
\citet{planck2014-a13}, which also presents a detailed
characterization of the low-multipole CMB angular power spectrum. The
low-frequency astrophysical model presented here is used
as input for the temperature-to-polarization bandpass mismatch
corrections for the LFI polarization maps \citep{planck2014-a03}.

The paper is organized as follows. Section~\ref{sec:method} gives an
overview of the computational framework implemented in the
\texttt{Commander} code.  Section~\ref{sec:data} describes the data
selection and processing.  Section~\ref{sec:model}, gives an overview
of the relevant astrophysical components and systematic effects.
Sections~\ref{sec:temperature} and \ref{sec:polarization} give the
main temperature and polarization products. We summarize in Section
~\ref{sec:conclusions}.

\section{Algorithms}
\label{sec:method}

\subsection{Data, posterior distribution and priors}
\label{sec:posterior}

Most of the results derived in this paper are established within a
standard Bayesian analysis framework, as implemented in the
\texttt{Commander} code, in which an explicit parametric model,
$\s(\theta)$, is fitted to a set of observations, $\d$, either by
maximizing or mapping out the corresponding posterior distribution,
\begin{equation}
P(\theta|\d) = \frac{P(\d|\theta) P(\theta)}{P(\d)} \propto
\mathcal{L}(\theta) P(\theta).
\end{equation}
Here $\theta$ denotes some general set of free parameters in the
model, $\mathcal{L}(\theta)=P(\d|\theta)$ is the likelihood, and
$P(\theta)$ denotes a set of priors on $\theta$. The evidence,
$P(\d)$, is a constant with respect to the parameter set, and is
neglected in the following.

The data are defined by a set of pixelized frequency-channel sky maps,
$\d = \{\d_{\nu}\}$, comprising the three Stokes parameters $I$, $Q$ and
$U$. In this paper, however, we analyse temperature and polarization 
separately; therefore the data vector comprises either $I$ or $\{Q,U\}$.

We start by assuming that the data at a given frequency $\nu$ may be
described as a linear sum of signal $\s_{\nu}$ and noise $\n_\nu$,
\begin{equation}
\mathbf{d}_\nu = \s_\nu + \n_\nu,
\end{equation}
where $\n_{\nu}$ is assumed to be Gaussian-distributed with a known
covariance matrix $\N_\nu$. For the signal, we adopt the following
parametric expression:
\begin{align}
  \s_{\nu}(\theta) &= s_{\nu}(\a_i, \beta_i, g_{\nu}, \m_{\nu},
  \Delta_{\nu}) \\ & = g_{\nu} \sum_{i=1}^{N_{\textrm{comp}}}
  \F_\nu^i(\beta_i, \Delta_\nu)\a_{i} + \T_{\nu}\m_{\nu},
\end{align}
where $\a_i$ is an amplitude map for component $i$ at a given
reference frequency, $\beta_i$ is a general set of spectral
parameters for the same component, $g_\nu$ is a multiplicative
calibration factor for frequency $\nu$, $\Delta_\nu$ is a linear
shift in the bandpass central frequency, and $\m_{\nu}$ is a set
of template correction amplitudes, such as monopole, dipole, or
zodiacal light corrections for temperature, or
calibration leakage templates for polarization. The corresponding
spatially fixed templates are organized column-wise in a template
matrix $\T_{\nu}$. The mixing matrix, $\F_\nu^i(\beta_i, \Delta_\nu)$,
accounts for the effect of spectral changes as a function of frequency
for component $i$, parametrized by $\beta_i$, as well as bandpass
integration effects and unit conversions. For numerical stability, all
internal calculations are performed in units of brightness
temperature, and $\a$ is therefore naturally defined in the same
units at some specified reference frequency.

The posterior distribution takes the usual form,
\begin{align}
P(\theta|\d_\nu) &= P(\d_\nu|\a_i, \beta_i, g_{\nu}, \m_{\nu},
\Delta_{\nu}, C_{\ell}) P(\a_i, \beta_i, g_{\nu}, \m_{\nu},
\Delta_{\nu}, C_{\ell}) \\ &= \mathcal{L}(\a_i, \beta_i,
g_{\nu}, \m_{\nu}, \Delta_{\nu})
P(\a_i)P(\beta_i)P(\a_{\textrm{cmb}}|C_{\ell})\nonumber,
\end{align}
where we have included the CMB power spectrum, $C_{\ell}$, and also
implicitly adopted uniform priors on $\g_\nu$, $\Delta_\nu$, $\m_\nu$,
and $C_{\ell}$. Because the noise is assumed to be Gaussian and
independent between frequency channels, the likelihood reads
\begin{equation}
\mathcal{L}(\a_i, \beta_i, g_{\nu}, \m_{\nu}, \Delta_{\nu}) \propto
\exp{\left(-\frac{1}{2} \sum_{\nu}[\d_\nu-\s_\nu(\theta)]\tp\N^{-1}[\d_\nu-\s_\nu(\theta)]\right)}
\end{equation} 
Likewise, we further assume the CMB signal to be Gaussian distributed
with a covariance matrix, $\S(C_{\ell})$, given by the power spectrum,
and the corresponding CMB prior factor therefore reads
\begin{equation}
P(\a_{\textrm{cmb}}|C_{\ell}) =
\frac{e^{-\frac{1}{2}\a_{\textrm{cmb}}\tp \S^{-1}(C_{\ell})
    \a_{\textrm{cmb}}}}{\sqrt{|\S(C_{\ell})|}}.
\end{equation}

The only undefined factors in the posterior are the amplitude and
spectral parameter priors, $P(\a^i)$ and $P(\beta^i)$.  These
represent the most difficult problem to handle from a conceptual point
of view, since the prior is to some extent a matter of personal
preference. However, we adopt the following general practices in this
paper. First, for low-resolution analyses that include fitting of
template amplitudes (e.g., monopoles and dipoles), we always impose a
strict positivity prior, i.e., $\a_i > 0$, on all signal amplitudes
except the CMB. Without such a prior, there are large degeneracies
between the zero-points of the amplitude maps and the individual
template amplitudes. Second, for the high angular resolution analysis,
we fix the template amplitudes at the low-resolution values and
disable the positivity prior, in order to avoid noise bias. Third, to
further break degeneracies, we adopt fiducial values for the monopole,
dipole, and calibration factors for a few selected channels,
effectively imposing a set of external priors from CMB dipole
measurements and \ion{H}{i} cross-correlation \citep{planck2013-p03f}
to anchor the full solution. Fourth, for the spectral parameters, we
adopt Gaussian priors with means and variances informed by the high
signal-to-noise values observed in the Galactic plane, which for all
practical purposes are independent of the adopted priors.
Intuitively, we demand that a map of the spectral parameter in
question should not be much different in the data-dominated and the
prior-dominated regions of the sky. Fourth, one of the components in
the temperature model is free-free emission, which has two free
parameters, namely the effective emission measure, $\EM$, and the
electron temperature, $\Te$. The latter of these is very poorly
constrained with the current data set except in the central Galactic
plane, and we therefore adopt a smoothness prior on this paper to
increase the effective signal-to-noise ratio, demanding that is must
be smooth on $2\deg$ FWHM scales. This in turn has a large
computational cost by making the overall foreground parameter
estimation process non-local, and $T_e$ is therefore only varied in
fast maximum-likelihood searches, not in expensive sampling
analyses. Its effect on other parameters is, however, minimal,
precisely because of its low signal-to-noise ratio.  Finally, in
addition to these informative priors, we adopt a Jeffreys prior for
the spectral parameters in order to suppress prior volume effects
\citep{jeffreys1946,eriksen2008,dunkley2009}.

\subsection{Gibbs sampling and posterior maximization}
\label{sec:gibbs}

As described above, the posterior distribution contains many millions
of free (both non-Gaussian and strongly correlated) parameters for
\Planck, and mapping out this distribution poses a significant
computational problem. Indeed, no direct sampling algorithm exists for
the full distribution, and the only computationally efficient solution
currently known is that of Gibbs sampling, a well-known textbook
algorithm in modern statistical analysis
\citep[e.g.,][]{Gelman03}. The underlying idea of this method is that
samples from a complicated multivariate distribution may be drawn by
iteratively sampling over the corresponding conditional distributions,
which usually have much simpler, and often analytic, sampling
algorithms. This framework was originally introduced to the CMB
analysis field by \citet{jewell2004} and \citet{wandelt2004}, and subsequently
developed into a fully functional computer code called
\texttt{Commander} by \citet{eriksen2004,eriksen2008}.

For the problem in question in this paper, this algorithm may be
schematically translated into an explicit set of sampling steps
through the following Gibbs chain:
\begin{align}
\a_i &\leftarrow P(\a_i|\beta_i, g_{\nu}, \m_{\nu}, \Delta_{\nu},
C_{\ell}) \\
\beta_i &\leftarrow P(\beta_i|\a_i, g_{\nu}, \m_{\nu}, \Delta_{\nu},
C_{\ell}) \\
g_\nu &\leftarrow P(g_{\nu}|\a_i, \beta_i, \m_{\nu}, \Delta_{\nu},
C_{\ell}) \label{eq:gibbs_g}\\
\m_\nu &\leftarrow P(m_\nu|\a_i, \beta_i, g_{\nu}, \Delta_{\nu},
C_{\ell}) \\
\Delta_{\nu} &\leftarrow P(\Delta_\nu|\a_i, \beta_i, g_{\nu}, \m_{\nu},
C_{\ell}) \\
C_{\ell} &\leftarrow P(C_\ell|\a_i, \beta_i, g_{\nu}, \m_{\nu}, \Delta_{\nu}),
\end{align}
Here ``$\leftarrow$'' denotes drawing a sample from the distribution on
the right-hand side. After some burn-in period, the theory of Gibbs
sampling guarantees that the joint set of parameters is indeed drawn from
the correct joint distribution. For a full description of the various
steps in the algorithm, see \citet{eriksen2008}.

\begin{table*}[tmb]                                                                                                                                                 
\begingroup                                                                                                                                   
\newdimen\tblskip \tblskip=2pt
\caption{Overview of data sets. \label{tab:data} The top section lists all detector and detector set (``ds'') maps included in the temperature analysis, and the bottom section lists all frequency maps included in the polarization analysis.}
\nointerlineskip                                                                                                                                                                                     
\vskip -5mm
\footnotesize                                                                                                                                       
\setbox\tablebox=\vbox{                                                                                                                                                                           
\newdimen\digitwidth                                                                                                                             
\setbox0=\hbox{\rm 0}
\digitwidth=\wd0
\catcode`*=\active
\def*{\kern\digitwidth}
\newdimen\signwidth
\setbox0=\hbox{+}
\signwidth=\wd0
\catcode`!=\active
\def!{\kern\signwidth}
\newdimen\decimalwidth
\setbox0=\hbox{.}
\decimalwidth=\wd0
\catcode`@=\active
\def@{\kern\signwidth}
\halign{ \hbox to 1.4in{#\leaderfil}\tabskip=1.0em&
    \hfil#\hfil\tabskip=1em&
    \hfil#\hfil\tabskip=1em&
    \hfil#\hfil\tabskip=1em&
    \hfil#\hfil\tabskip=0.5em&
    \hfil#\hfil\tabskip=0.5em&
    \hfil#\hfil\tabskip=1em&
    #\hfil\tabskip=0pt\cr
\noalign{\doubleline}
\omit&  Frequency& Detector& & Noise rms&  Min smooth&&\cr
\omit\hfil Instrument\hfil& [GHz]& label& Resolution&  $\sigma^{T}(1\deg)$ or $\sigma^P(40\arcm)$& scale& Units&
   \omit\hfil Reference\hfil\cr
\noalign{\vskip 3pt\hrule\vskip 12pt}
\omit{\bf TEMPERATURE}\cr
\noalign{\vskip 8pt}
\hglue 1em\Planck\ LFI&*30&(all)& 32\parcm4& *2.8*&  40\arcm*& \muK& \citet{planck2014-a07}\cr
\noalign{\vskip 4pt}
\omit\hglue 1em&       *44&(all)& 27\parcm1& *3.0*&  40\arcm*& \muK&\cr
\noalign{\vskip 4pt}
\omit\hglue 1em&       *70&  ds1& 13\parcm6& *3.8*&  40\arcm*& \muK&\cr
\omit\hglue 1em&          &  ds2& 13\parcm3& *4.0*&  40\arcm*& \muK&\cr
\omit\hglue 1em&          &  ds3& 13\parcm0& *4.1*&  40\arcm*& \muK&\cr
\noalign{\vskip 4pt}
\hglue 1em\Planck\ HFI&100&ds1&*9\parcm7&*0.9*&40\arcm*& \muK&\citet{planck2014-a09}\cr
\omit\hglue 1em&          &  ds2&*9\parcm7& *0.8*&  40\arcm*& \muK&\cr
\noalign{\vskip 4pt}
\omit\hglue 1em&       143&  ds1&*7\parcm2& *0.7*& *7\parcm5& \muK&\cr
\omit\hglue 1em&          &  ds2&*7\parcm2& *0.7*& *7\parcm5& \muK&\cr
\omit\hglue 1em&          &    5&*7\parcm2& *0.9*& *7\parcm5& \muK&\cr
\omit\hglue 1em&          &    6&*7\parcm2& *1.1*& *7\parcm5& \muK&\cr
\omit\hglue 1em&          &    7&*7\parcm2& *1.0*& *7\parcm5& \muK&\cr
\noalign{\vskip 4pt}
\omit\hglue 1em&       217&    1&*5\parcm0& *1.8*& *7\parcm5& \muK&\cr
\omit\hglue 1em&          &    2&*5\parcm0& *1.9*& *7\parcm5& \muK&\cr
\omit\hglue 1em&          &    3&*5\parcm0& *1.7*& *7\parcm5& \muK&\cr
\omit\hglue 1em&          &    4&*5\parcm0& *1.8*& *7\parcm5& \muK&\cr
\noalign{\vskip 4pt}
\omit\hglue 1em&       353&  ds2&*4\parcm9& *4.5*& *7\parcm5& \muK&\cr
\omit\hglue 1em&          &    1&*4\parcm9& *3.5*& *7\parcm5& \muK&\cr
\noalign{\vskip 4pt}
\omit\hglue 1em&       545&    2&*4\parcm7& *0.01& *7\parcm5& MJy\,sr\mo&\cr
\omit\hglue 1em&          &    4&*4\parcm7& *0.01& *7\parcm5& MJy\,sr\mo&\cr
\noalign{\vskip 4pt}
\omit\hglue 1em&       857&    2&*4\parcm4& *0.01& *7\parcm5& MJy\,sr\mo&\cr
\noalign{\vskip 6pt}
\hglue 1emWMAP&        *23&    K& 53\arcm*& *5.9*&  60\arcm*& \muK&\citet{bennett2012}\cr
\noalign{\vskip 4pt}
\omit\hglue 1em&       *33&   Ka& 40\arcm*& *4.3*&  60\arcm*& \muK&\cr
\noalign{\vskip 4pt}
\omit\hglue 1em&       *41&   Q1& 31\arcm*& *5.3*&  60\arcm*& \muK&\cr
\omit\hglue 1em&          &   Q2& 31\arcm*& *5.1*&  60\arcm*& \muK&\cr
\noalign{\vskip 4pt}
\omit\hglue 1em&       *61&   V1& 21\arcm*& *6.4*&  60\arcm*& \muK&\cr
\omit\hglue 1em&          &   V2& 21\arcm*& *5.5*&  60\arcm*& \muK&\cr
\noalign{\vskip 4pt}
\omit\hglue 1em&       *95&   W1& 13\arcm*& *8.84&  60\arcm*& \muK&\cr
\omit\hglue 1em&          &   W2& 13\arcm*& 10.1*&  60\arcm*& \muK&\cr
\omit\hglue 1em&          &   W3& 13\arcm*& 10.6*&  60\arcm*& \muK&\cr
\omit\hglue 1em&          &   W4& 13\arcm*& 10.1*&  60\arcm*& \muK&\cr
\noalign{\vskip 6pt}
\hglue 1em Haslam et al.&  0.408&     & 56\arcm*& *1.1*&  60\arcm*&    K&\citet{haslam1982},\cr
\omit&  \omit&     & \omit& \omit&  \omit&    \omit&\citet{remazeilles2014}\cr
\noalign{\vskip 14pt}
\omit{\bf POLARIZATION}\cr
\noalign{\vskip 8pt}
\hglue 1em\Planck\ LFI&*30&   & 32\parcm4& *7.5*&  40\arcm*& \muK& \citet{planck2014-a07}\cr
\noalign{\vskip 4pt}
\omit\hglue 1em&       *44&   & 27\parcm1& *7.5*&  40\arcm*& \muK& \omit\cr
\noalign{\vskip 4pt}
\omit\hglue 1em&       *70&   & 13\parcm3& *4.8*&  40\arcm*& \muK& \omit\cr
\noalign{\vskip 0pt}
\hglue 1em\Planck\ HFI&100&&*9\parcm5&*1.3*&10\arcm*&\muK&\citet{planck2014-a09}\cr
\noalign{\vskip 4pt}
\omit\hglue 1em&       143&   &*7\parcm2& *1.1*&  10\arcm*& \muK&\omit\cr
\noalign{\vskip 4pt}
\omit\hglue 1em&       217&   &*5\parcm0& *1.6*&  10\arcm*& \muK&\omit\cr
\noalign{\vskip 4pt}
\omit\hglue 1em&       353&   &*4\parcm9& *6.9*&  10\arcm*& \muK&\omit\cr
\noalign{\vskip 2pt\hrule\vskip 2pt}
}}
\endPlancktable                                                                                                                                              
\endgroup
\end{table*}

While no fully functional alternatives to Gibbs sampling have been
established for this full joint distribution to date, Gibbs sampling
alone by no means solves all computational problems. In particular,
this algorithm is notorious for its slow convergence for nearly
degenerate parameters, since it by construction only moves through
parameter space parallel to coordinate axes. For this reason, we
implement an additional posterior maximization phase, in which we
search directly for the posterior maximum point rather than attempt to
sample from the full distribution. The resulting solution may then
serve either as a final product in its own right, by virtue of being a
maximum-posterior estimate, or as the starting position for a regular
Gibbs sampling analysis. The crucial point, though, is that
special-purpose nonlinear search algorithms can be used in this phase,
moving in arbitrary directions through parameter space, and individual
optimization combinations may be introduced to jointly probe
directions with particularly strong degeneracies. Perhaps the single
most important example in this respect is the parameter combination
between component amplitudes, detector calibrations, and bandpass
uncertainties, $\{\a_i, g_\nu, \Delta_\nu\}$, all three of which
essentially correspond to scaling parameters. However, since both
$g_\nu$ and $\a_i$ are conditionally linear parameters, and only
$\Delta_\nu$ is truly nonlinear, it is possible to solve analytically
for $g_\nu$ or $\a_i$, \emph{conditioning} on any given fixed value of
$\Delta_\nu$. Consequently, one can set up a non-linear Powell-type
search \citep{press2002} for $\Delta_\nu$, in which the optimal values
of either $g_\nu$ or $\a_i$ are quickly computed at each iteration in
the search. A second example is the electron temperature discussed
above, for which non-local optimization is feasible, whereas a
full-blown sampling algorithm is too expensive to converge
robustly. In this situation, fixing the parameter at its
maximum-posterior value is vastly preferable compared to adding an
unconverged degree of freedom in the full sampler.

Even with this optimization phase, however, there is always an
inherent danger of the algorithm being trapped in a local posterior
maximum. Indeed, with a distribution involving millions of highly
correlated parameters, it is exceedingly difficult to prove that the
derived solution is the true global posterior maximum. As a partial
solution to this problem, we initialize the search using different
starting positions, and carefully monitor the convergence properties
of the chains. 

\section{Data selection and processing}
\label{sec:data}

\begin{figure}
\begin{center}
\mbox{
\epsfig{figure=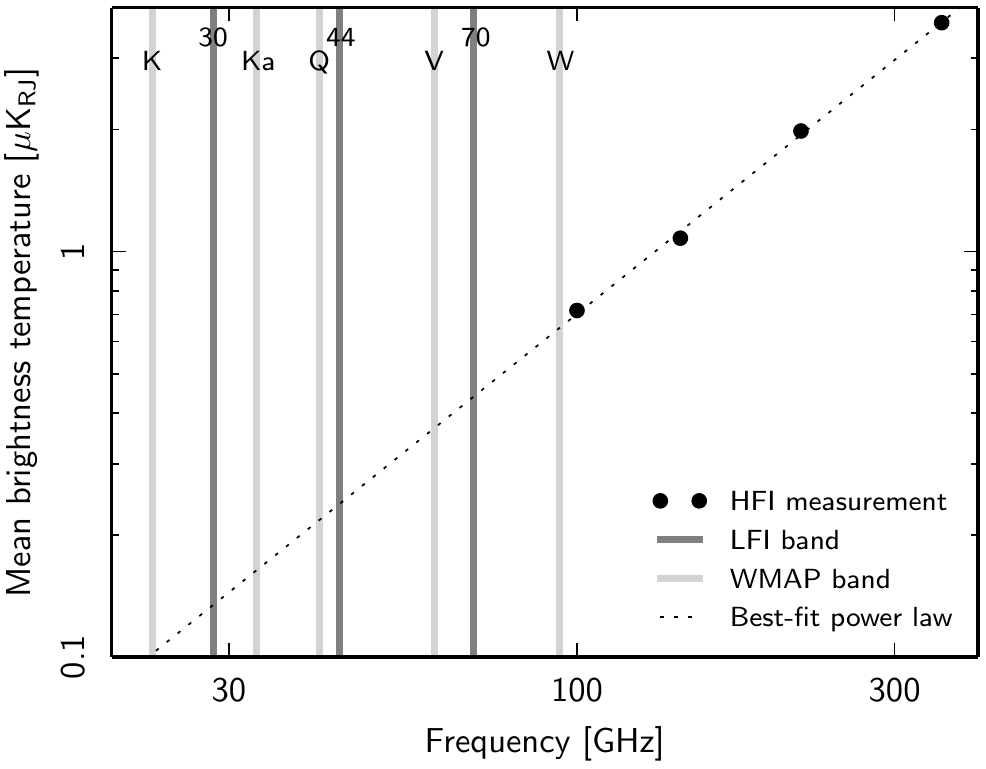,width=88mm,clip=}
}

\end{center}
\caption{Zodiacal light extrapolation from HFI to LFI and WMAP
  frequency channels in terms of full-sky mean brightness
  temperature. The dotted line shows the power-law fit to the HFI
  observations between 100 and 353\,GHz, $s(\nu) =
  0.70\,\mu\textrm{K}_{\textrm{RJ}}\, (\nu/100\,\textrm{GHz})^{1.31}$,
  and the vertical grey lines indicate the central frequencies of the
  LFI and WMAP frequency bands. The intersection between the dotted
  and grey lines defines the extrapolation to low frequencies.}
\label{fig:zodi}
\end{figure}

\begin{table}[tmb] 
\begingroup 
\newdimen\tblskip \tblskip=5pt
\caption{Zodiacal light template coefficients for WMAP and LFI frequencies relative to the 100\,GHz HFI zodical light template in thermodynamic temperature units.\label{tab:zodi}}
\nointerlineskip                                                                                                                                                                                     
\vskip -4mm
\footnotesize 
\setbox\tablebox=\vbox{ %
\newdimen\digitwidth 
\setbox0=\hbox{\rm 0}
\digitwidth=\wd0
\catcode`*=\active
\def*{\kern\digitwidth}
\newdimen\signwidth
\setbox0=\hbox{+}
\signwidth=\wd0
\catcode`!=\active
\def!{\kern\signwidth}
\newdimen\decimalwidth
\setbox0=\hbox{.}
\decimalwidth=\wd0
\catcode`@=\active
\def@{\kern\signwidth}
\def\s#1{\ifmmode $\rlap{$^{\rm #1}$}$ \else \rlap{$^{\rm #1}$}\fi}
\halign{ \hbox to 1.2in{#\leaderfil}\tabskip=2em&
    \hfil#\hfil\tabskip=0pt\cr
\noalign{\doubleline}
\omit\hfil Band\hfil& Amplitude\cr
\noalign{\vskip 2pt\hrule\vskip 4pt}
\Planck\ 30&                  0.15\cr
\phantom{\Planck\ }44&        0.28\cr
\phantom{\Planck\ }70&        0.56\cr
\noalign{\vskip 2pt}
WMAP K&                       0.11\cr
\phantom{WMAP }Ka&            0.19\cr
\phantom{WMAP }Q&             0.25\cr
\phantom{WMAP }V&             0.45\cr
\phantom{WMAP }W&             0.89\cr
\noalign{\vskip 4pt\hrule}
}}
\endPlancktable
\endgroup
\end{table}

\begin{figure*}
\begin{center}
\mbox{
\epsfig{figure=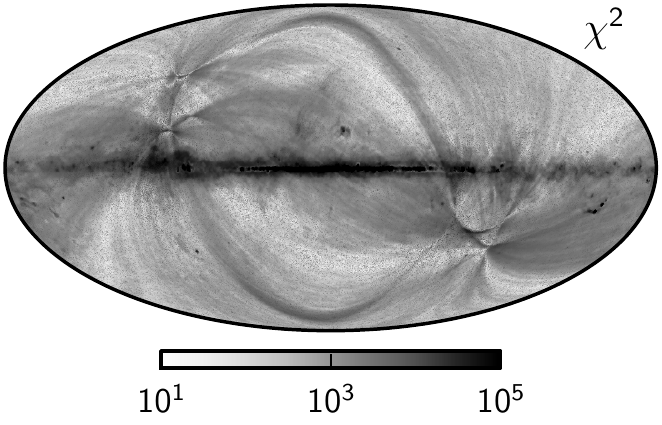,width=0.55\linewidth,clip=}
}
\mbox{
\epsfig{figure=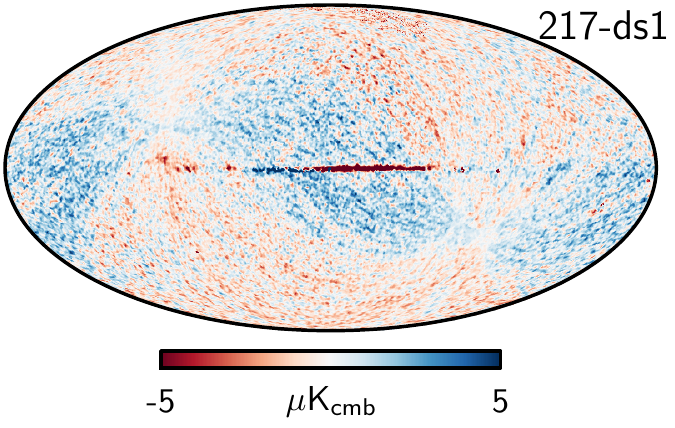,width=0.33\linewidth,clip=}
\epsfig{figure=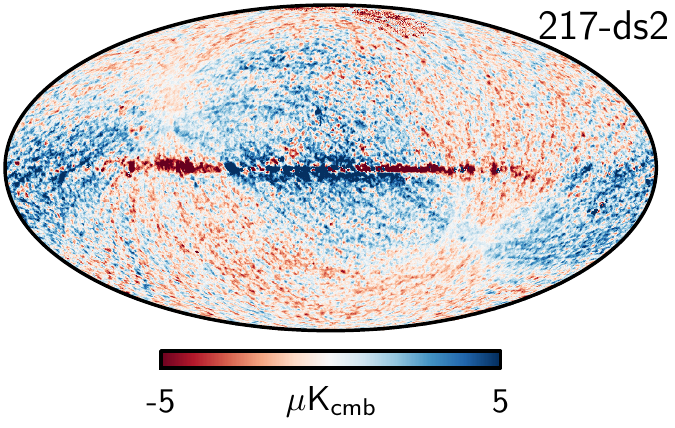,width=0.33\linewidth,clip=}
\epsfig{figure=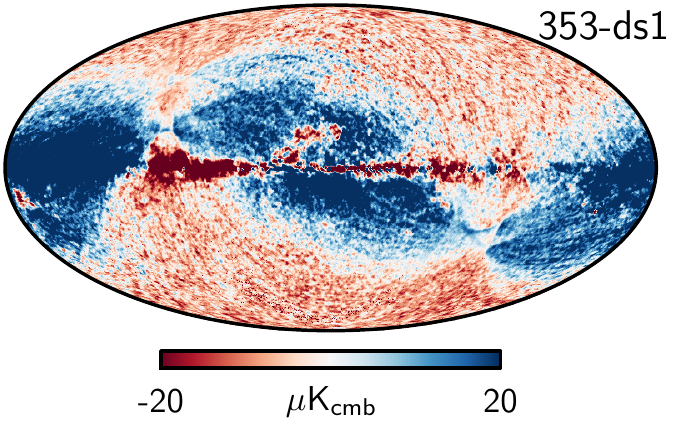,width=0.33\linewidth,clip=}
}
\mbox{
\epsfig{figure=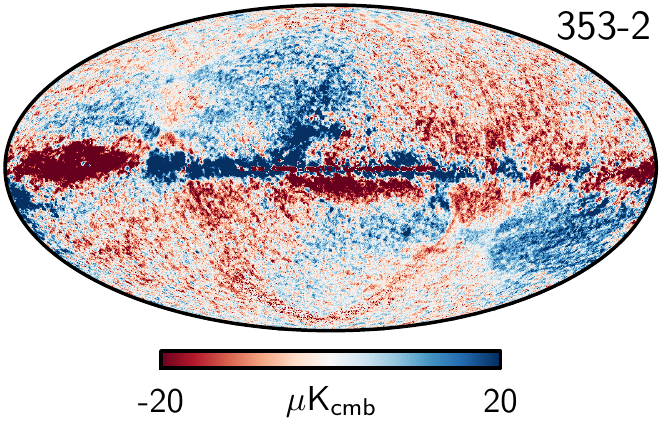,width=0.33\linewidth,clip=}
\epsfig{figure=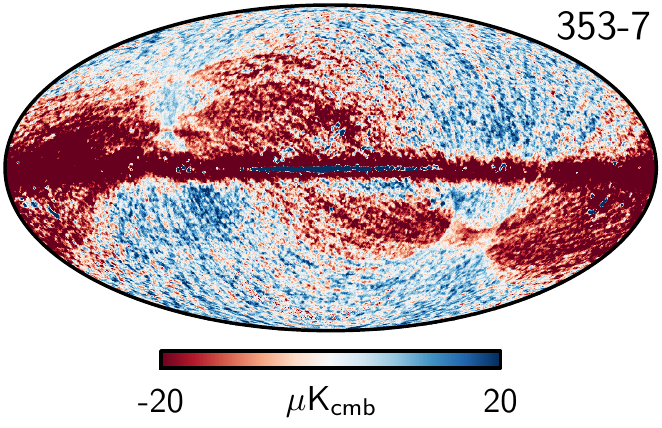,width=0.33\linewidth,clip=}
\epsfig{figure=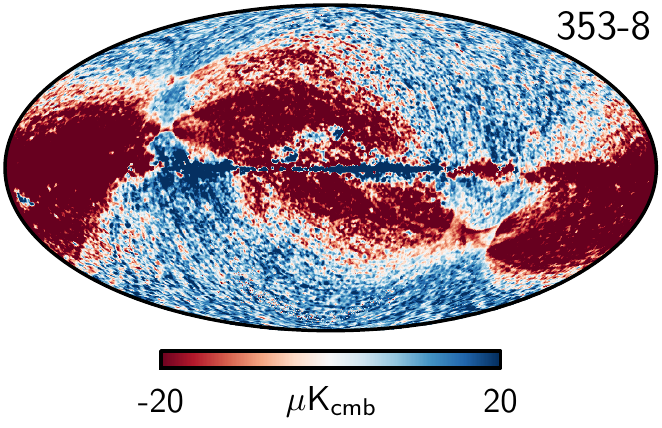,width=0.33\linewidth,clip=}
}
\mbox{
\epsfig{figure=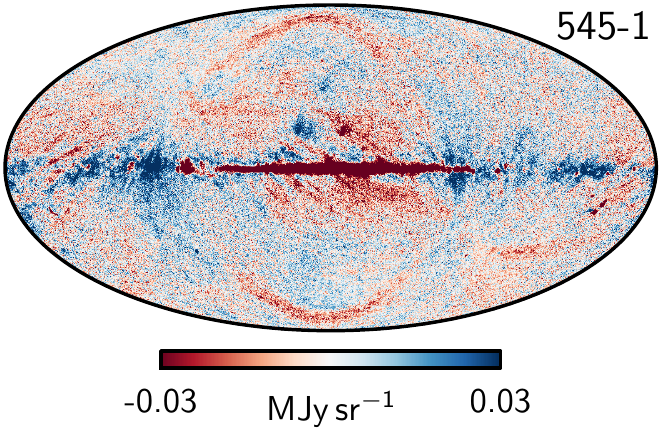,width=0.33\linewidth,clip=}
\epsfig{figure=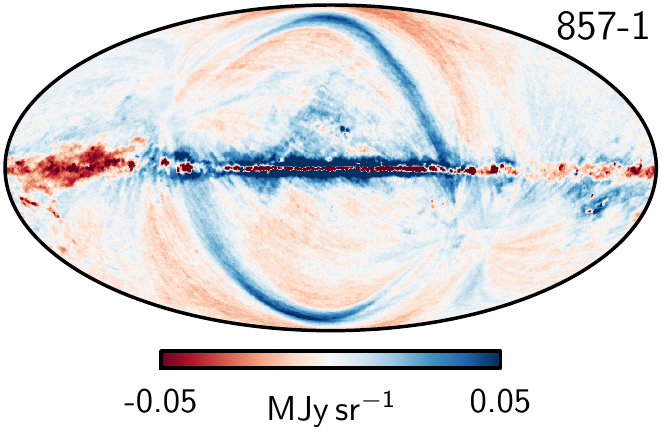,width=0.33\linewidth,clip=}
\epsfig{figure=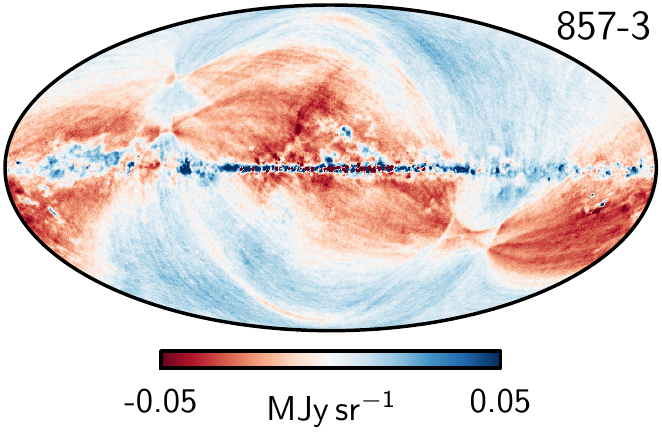,width=0.33\linewidth,clip=}
}
\mbox{
\epsfig{figure=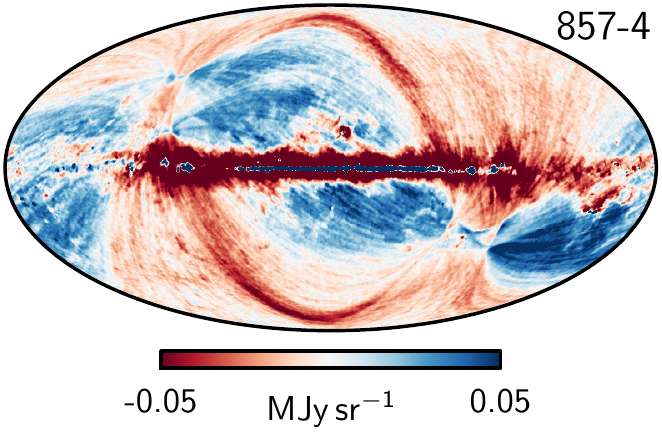,width=0.33\linewidth,clip=}
}
\end{center}
\caption{$\chi^2$ (\emph{top}) and residual maps, $\d_{\nu}-\s_{\nu}$
  (\emph{bottom}), for a \texttt{Commander} analysis that includes all
  \Planck\ channel maps. These residual maps correspond to channels
  that are rejected from the baseline analysis due to instrumental
  systematics. No regularization noise has been added to the
  high-frequency channels in this case. The sharp ring-like features
  at high Galactic latitudes correspond to far sidelobe residuals; the
  broad features extending between the north and south ecliptic poles
  correspond to destriping errors; and the Galactic plane features
  correspond to calibration, bandpass, and modelling residuals.}
\label{fig:bad_channels}
\end{figure*}

\begin{figure}
\begin{center}
\mbox{
\epsfig{figure=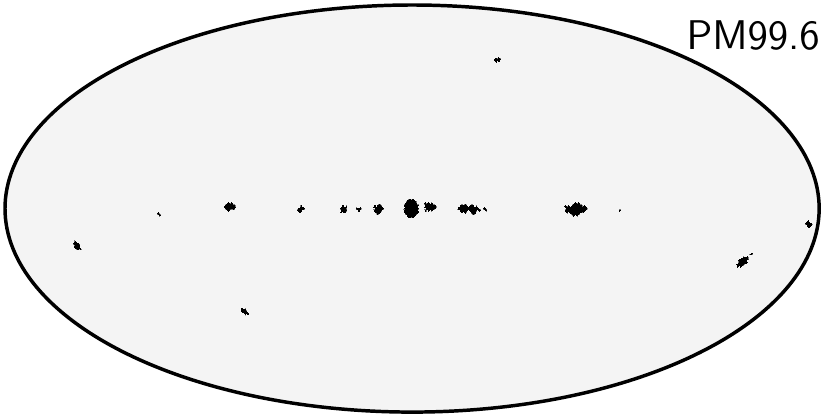,width=88mm,clip=}
}
\mbox{
\epsfig{figure=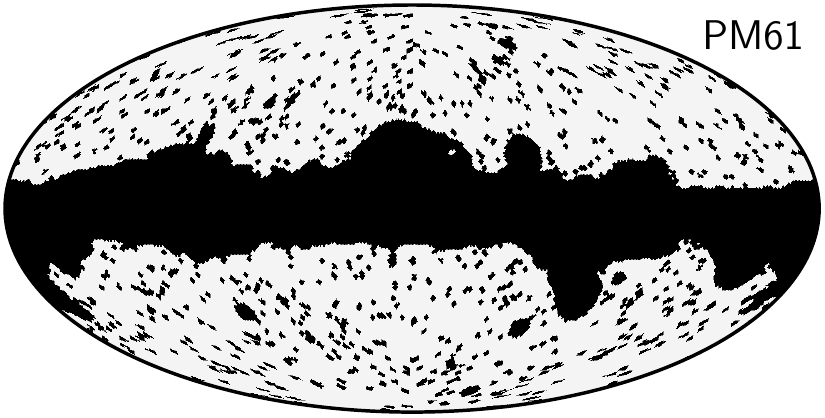,width=88mm,clip=}
}

\end{center}
\caption{Processing masks (PM) used in the joint temperature analysis,
  including 99.6\,\% and 61\,\% of the sky, respectively. The former is
  used for calibration and bandpass estimation, and the latter for
  monopole and dipole estimation.}
\label{fig:processing_masks}
\end{figure}

The primary data used in this paper are the 2015 \Planck\ temperature
and polarization sky maps \citep{planck2014-a07,planck2014-a09}. For
the temperature analysis we additionally include the 9-year
\WMAP\ observations\footnote{\url{http://lambda.gsfc.nasa.gov}}
\citep{bennett2012} and a full-sky 408\,MHz survey map
\citep{haslam1982}, with the goal of individually resolving
synchrotron, free-free, and spinning dust emission. For \WMAP, we
adopt the beam-symmetrized frequency maps for the foreground-dominated
K- and Ka-bands, to mitigate beam artifacts around compact sources,
but we use the standard maps for the CMB-dominated Q-, V-, and
W-bands, because of their more accurate noise description. At the
lowest frequency, we adopt the destriped version of the 408\,MHz
survey map recently published by \citet{remazeilles2014}.

In order to maximize our leverage with respect to bandpass measurement
uncertainties and line emission mechanisms, we employ individual
detector and detector set (``ds'') maps for all \Planck\ frequencies
between 70 and 857\,GHz, and differencing assembly (``DA'') maps for
\WMAP. However, the polarization analysis employs frequency maps in
order to maximize signal-to-noise ratio and to minimize correlated
noise from destriper mapmaking
uncertainties. Intensity-to-polarization leakage from bandpass
mismatch between detectors is suppressed through the use of
precomputed leakage templates \citep{planck2014-a03,planck2014-a09}.
For LFI, these templates are based on a preliminary version of the
foreground products presented in this paper. The full set of clean
channels used in this analysis is summarized in Table~\ref{tab:data}
in terms of centre frequencies, resolution, and noise levels. In
total, 32 individual detector and detector set maps\footnote{For
  uniformity, we refer to the 70\,GHz horn pair maps as ``detector set''
  maps in this paper, with the \{ds1, ds2, ds3\} maps corresponding to
  horns \{18+23, 19+22, 20+21\}, respectively.} and frequency maps are
included in the temperature analysis, and seven frequency maps in the
polarization analysis.

For the \Planck\ HFI channels, a model of zodiacal light emission is
subtracted from the time-ordered data prior to mapmaking
\citep{planck2014-a09}. In addition, in this paper we apply a small
correction to the low-frequency LFI and \WMAP\ channels by scaling the
effective HFI 100\,GHz zodiacal light correction map (i.e.,
uncorrected minus corrected map) to each frequency according to a
power law fitted to frequencies between 100 and 353\,GHz
\citep{planck2013-pip88}, as illustrated in Fig.~\ref{fig:zodi}; the
actual template amplitudes relative to the 100\,GHz correction map (in
thermodynamic units) are listed in Table~\ref{tab:zodi}. Although the
magnitude of this correction is small, with a maximum amplitude of
2$\,\mu\textrm{K}$ in the 70\,GHz map, applying no correction at all
below 100\,GHz results in a visually noticeable bias in the derived CO
$J$=1$\rightarrow$0 map at high Galactic latitudes, in the
characteristic form of the zodiacal light. Extending the zodiacal
light model to low frequencies efficiently eliminates this structure.

\begin{table*}[tmb] 
\begingroup 
\newdimen\tblskip \tblskip=5pt
\caption{Unit conversion coefficients between thermodynamic and
  brightness temperature and between thermodynamic temperature and
  flux density per unit area for each channel, with and without the
  bandpass corrections described in Sect.~\ref{sec:temperature}\label{tab:units}.}
\nointerlineskip
\vskip -2mm
\footnotesize 
\setbox\tablebox=\vbox{
\newdimen\digitwidth 
\setbox0=\hbox{\rm 0}
\digitwidth=\wd0
\catcode`*=\active
\def*{\kern\digitwidth}
\newdimen\signwidth
\setbox0=\hbox{+}
\signwidth=\wd0
\catcode`!=\active
\def!{\kern\signwidth}
\newdimen\decimalwidth
\setbox0=\hbox{.}
\decimalwidth=\wd0
\catcode`@=\active
\def@{\kern\signwidth}
\def\s#1{\ifmmode $\rlap{$^{\rm #1}$}$ \else \rlap{$^{\rm #1}$}\fi}
\halign{ \hbox to 1in{#\leaderfil}\tabskip=1.0em&
    \hfil#\hfil\tabskip=1em&
    \hfil#\hfil\tabskip=1em&
    \hfil$#$\hfil\tabskip=0em&
    \hfil$#$\hfil\tabskip=0em&
    \hfil$#$\hfil\tabskip=1em&
    \hfil$#$\hfil\tabskip=2em&
    \hfil$#$\hfil\tabskip=1em&
    \hfil$#$\hfil\tabskip=0pt\cr
\noalign{\doubleline}
\omit& \omit& \omit&\multispan3\hfil $U_c\,[\textrm{K}_{\textrm{CMB}}/\textrm{K}_{\textrm{RJ}}]$\hfil&\multispan3\hfil $U_c\,[\textrm{MJy}\,\textrm{sr}^{-1}/\textrm{K}_{\textrm{CMB}}]$\hfil\cr
\noalign{\vskip -3pt}
\omit&Frequency&Detector&\multispan3\hrulefill&\multispan3\hrulefill\cr
\noalign{\vskip 2pt}
\omit \hfil Instrument\hfil& [GHz]& label& \textrm{Nominal}& \textrm{Fitted}& \textrm{Change [\%]} & \textrm{Nominal}& \textrm{Fitted} & \textrm{Change [\%]}\hfil\cr                                                                                                                                                                                               
\noalign{\vskip 5pt\hrule\vskip 5pt}
\Planck\ LFI&          *30&\dots&      1.0212&     1.0217&   0.1&    23.510&    24.255&  3.2\cr
\noalign{\vskip 4pt}
\omit&                 *44&\dots&      1.0515&     1.0517&   0.0&    55.735&    56.094&  0.6\cr
\noalign{\vskip 4pt}
\omit&                 *70& ds1&       1.1378&     1.1360&  -0.2!&    132.07**&    129.77**& -1.7!\cr
\omit&                    & ds2&       1.1361&     1.1405&   0.4&    129.75**&    135.37**& 4.3\cr
\omit&                    & ds3&       1.1329&     1.1348&   0.2&    126.05**&    128.57**& 2.0\cr
\noalign{\vskip 4pt}
\Planck\ HFI&100&      ds1&1.3090&     1.3058&  -0.2!&    244.59**&    241.58**& -1.2!\cr
\omit&                    & ds2&       1.3084&     1.3057&  -0.2!&    243.77**&    241.22**& -1.1!\cr
\noalign{\vskip 4pt}
\omit&                 143& ds1&       1.6663&     1.6735&   0.4&    365.18**&    368.68**&  1.0\cr
\omit&                    & ds2&       1.6754&     1.6727&  -0.2!&    369.29**&    368.00**& -0.4!\cr
\omit&                    &   5&       1.6961&     1.6910&  -0.3!&    380.12**&    377.68**& -0.6!\cr
\omit&                    &   6&       1.6829&     1.6858&   0.2&    373.34**&    374.74**& 0.4\cr
\omit&                    &   7&       1.6987&     1.6945&  -0.2!&    381.25**&    379.26**& -0.5!\cr
\noalign{\vskip 4pt}
\omit&                 217&   1&       3.2225&     3.2203&  -0.1!&    486.02**&    485.87**& -0.0!\cr
\omit&                    &   2&       3.2378&     3.2336&  -0.1!&    486.39**&    486.10**& -0.1!\cr
\omit&                    &   3&       3.2296&     3.2329&   0.1&    486.89**&    485.85**& -0.2!\cr
\omit&                    &   4&       3.2183&     3.2161&  -0.1!&    486.01**&    487.11**& 0.2\cr
\noalign{\vskip 4pt}
\omit&                 353& ds2&      14.217**&   14.261**&  0.3&    287.89**&   287.62**&  -0.1!\cr
\omit&                    &   1&      14.113**&   14.106**& -0.1!&    288.41**&   288.45**& 0.0\cr
\noalign{\vskip 4pt}
\omit&                 545&   2&      164.54****&  168.94****&  2.7&  58.880&  57.953& -1.6!\cr
\omit&                    &   4&      167.60****&  174.13****&  3.9&  58.056&  56.732& -2.3!\cr
\noalign{\vskip 4pt}
\omit&                 857&   2&     9,738.8*******&  10,605.*********& 8.9&  **2.3449&  **2.1974& -6.3!\cr
\noalign{\vskip 6pt}
WMAP&                  *23&   K&      1.0134&  \cdots& \cdots&   14.985&    \cdots&  \cdots\cr
\noalign{\vskip 4pt}
\omit&                 *33&  Ka&      1.0284&  \cdots& \cdots&   31.556&    \cdots&  \cdots\cr
\noalign{\vskip 4pt}
\omit&                 *41&  Q1&      1.0437&  \cdots& \cdots&   47.880&    \cdots&  \cdots\cr
\omit&                    &  Q2&      1.0433&  1.0439&  0.1&   47.179&    48.226& 2.2\cr
\noalign{\vskip 4pt}
\omit&                 *61&  V1&      1.0974&  \cdots& \cdots&   97.343&    \cdots&  \cdots\cr
\omit&                    &  V2&      1.1006&  1.1001&  -0.1!& 101.87**&    101.20**& -0.7!\cr
\noalign{\vskip 4pt}
\omit&                 *94&  W1&      1.2473&  \cdots& \cdots&  209.61**&    \cdots& \cdots\cr
\omit&                    &  W2&      1.2500&  1.2458& -0.3!&  213.65**&    209.38**& -2.0!\cr
\omit&                    &  W3&      1.2440&  1.2460& 0.1&  207.88**&    209.92**& 1.0\cr
\omit&                    &  W4&      1.2488&  1.2453& -0.3!&  211.78**&    208.18**& -1.7!\cr
\noalign{\vskip 6pt}
Haslam&              0.408&\dots&     1.0000&  \cdots& \cdots&  **0.0051&    \cdots& \cdots\cr
\noalign{\vskip 5pt\hrule\vskip 5pt}
}}
\endPlancktablewide
\endgroup
\end{table*}

In our 2013 release, colour corrections and unit conversions for all
\Planck\ channels were based on individual bandpass profiles as
measured on the ground before launch
\citep{planck2013-p02b,planck2013-p03d}.  However, as discussed in
detail in Sects.~\ref{sec:method}, \ref{sec:instrumental_effects}, and
\ref{sec:temperature}, during the component separation process we find
that systematic uncertainties in the nominal bandpasses induce
significant residuals between data and model, and it is necessary to
fit for these bandpass uncertainties in order to obtain statistically
acceptable fits. For \WMAP, we adopt the nominal
bandpasses\footnote{As described by \citet{bennett2012}, the
  \WMAP\ bandpasses evolved during the 9 years of \WMAP\ observations,
  resulting in slightly lower effective full-mission frequencies as
  compared to the nominal bandpasses. We correct for these small
  shifts in the present analysis by shifting the bandpasses
  accordingly. } for the first DA within each frequency band, and fit
for the remaining DA bandpasses within each frequency. For the
408\,MHz survey, we adopt a delta function response at the nominal
frequency. The unit conversion factors between thermodynamic and
brightness temperatures and flux density per area are tabulated for
both nominal and fitted bandpass profiles in Table~\ref{tab:units}.

Some \Planck\ detector maps are affected more strongly by systematic
errors than others \citep{planck2014-a09}, and in the following
temperature analysis we exclude the worst channels in order not to
compromise the overall solution. Out of a total of 31 potential
\Planck\ detector and detector set maps, 10 are removed from further
analysis, while the remaining 21 are listed in Tables~\ref{tab:data}
and \ref{tab:units}.  The 10 removed maps are shown in
Fig.~\ref{fig:bad_channels} in the form of a
($\mathbf{d}_\nu-\mathbf{s}_\nu$) residual map, where
$\mathbf{s}_{\nu}$ is a signal model based on the global and spectral
parameters derived in Sect.~\ref{sec:temperature}, but with amplitudes
re-fitted to all 42 channels. The top panel shows the corresponding
$\chi^2$ map, defined as
\begin{equation}
\chi^2(p) = \sum_{\nu}
\left(\frac{\mathbf{d}_{\nu}(p)-\mathbf{s}_{\nu}(p)}{\sigma_{\nu}(p)}\right)^2.
\label{eq:chisq}
\end{equation}
Including any one of these ten maps in the full joint analysis
increases the $\chi^2$ of the total fit far beyond what is allowed by
random statistical fluctuations.

Several different classes of systematic effects may be seen in these
plots. Starting with the 217-ds1 map, we see a large-scale red-blue
pattern, aligned with the \Planck\ scanning strategy and crossing
through the Ecliptic poles \citep{planck2013-p01}. The same spatial
pattern is seen in the 217-ds2, 353-ds1, 353-2, 353-7, 353-8, 857-3,
and 857-4 maps as well, with varying amplitudes and signs. These are
due to low-amplitude destriping errors induced in the mapmaking
process \citep{planck2014-a09}, and only become visually apparent
after component separation removes the dominant Galactic signal. These
residuals may therefore, at least in principle, be suppressed by
iterating between mapmaking and component separation, essentially
performing a joint mapmaking/component separation $\chi^2$ fit for
the destriping offsets.

A second example of residual systematics is seen in the 857-1 and
857-4 maps, and to a lesser extent in the 545-1 map, in the form of
sharp features at high Galactic latitudes. These correspond to far
sidelobe (FSL) contamination at the 0.05\,MJy\,sr$^{-1}$ level. Third,
and somewhat more subtly, one can see the effect of the very-long time
constants (VLTC) discussed in \mbox{\citet{planck2014-a09}}, particularly
south of the Galactic plane in 353-2, 353-7, and 545-1. In each of
these maps, the Galactic plane appears to have been smeared along the
scanning direction. As in the case of destriping errors, combining
mapmaking and component separation should prove very powerful in
suppressing both FSL and VLTC errors in future analyses.

While several channels are omitted from the analysis, it is important
to note that the \Planck\ HFI temperature observations are strongly 
signal-dominated at all angular scales above $7\arcm$ FWHM.  Our data
cuts therefore have very little effect on the full error budget, 
which is dominated by modelling and systematic errors for most
parameters. The main cost of these cuts comes in the form of reduced
internal redundancy.  In particular, having access to only one
clean 857\,GHz channel limits our ability to determine its bandpass
and calibration coefficients. This situation will improve in the next
\Planck\ data release, when the remaining three 857\,GHz channels
are better cleaned at the time domain level.

The current analysis is carried out in a number of stages according to
angular resolution, in which higher-resolution stages implement a
simpler foreground model than lower-resolution stages. For
temperature, each full-resolution map is downgraded from its native
resolution to $1\deg$ (all channels), $40\arcm$ (all Planck channels),
and $7\parcm5$ (HFI channels above 100\,GHz) by deconvolving the
intrinsic instrumental beam profile and convolving with a Gaussian
beam of the appropriate size before repixelizing at
\healpix\footnote{\url{http://healpix.sourceforge.net/}} resolutions
$N_{\textrm{side}}=256$, 256, and 2048, respectively
\citep{gorski2005}. For polarization, the corresponding smoothing
scales are $40\arcm$ and $10\arcm$ FWHM, pixelized at
$N_{\textrm{side}}=256$ and 1024, respectively. 

The instrumental noise is assumed to be spatially uncorrelated and
Gaussian for all channels, with a spatially-varying rms given by the
scanning strategy of the experiment and the instantaneous sensitivity of
each detector. The low-resolution noise rms map is found by convolving
the high-resolution rms with the appropriate smoothing kernel,
properly accounting for its matrix-like nature, and retaining only the
diagonal element of the resulting covariance matrix. For the 545 and
857\,GHz channels, we additionally add 0.01\,MJy\,sr$^{-1}$ of uniform
white noise, to prevent known residual far sidelobe and destriping
contamination from propagating to lower frequencies through the
thermal dust temperature and spectral index and 
contaminating both the CMB and CO solutions. Similarly, for the
408\,MHz channel we add regularization noise, equal to 1\,\% of the
amplitude of the map, to account for low-level residuals not captured
by the white noise model described above.

Finally, two different processing masks are employed in the
temperature analysis (PM99.6 and PM61), removing 0.4\,\% and 39\,\% of the
sky, respectively, as shown in Fig.~\ref{fig:processing_masks}. PM99.6
is generated by thresholding the (smoothed) $\chi^2$ map of a
preliminary analysis\footnote{The preliminary analysis was similar to
  the one presented in this paper, but without application of any
  processing mask for calibration purposes.} at $10^4$, removing only
the very brightest outliers in the data set; this mask is used for
bandpass estimation and gain calibration of the 545-4 and 857-2
channels. PM61 is generated as the product of the $\chi^2$ map
resulting from an analysis without bandpass corrections thresholded at
the $5\,\sigma$ level, and the 9-year \WMAP\ point source mask. This
mask is used for calibration estimation of the CMB frequencies, and
for monopole and dipole estimation.

\section{Model survey}
\label{sec:model}

\begin{figure*}
\begin{center}
\mbox{
\epsfig{figure=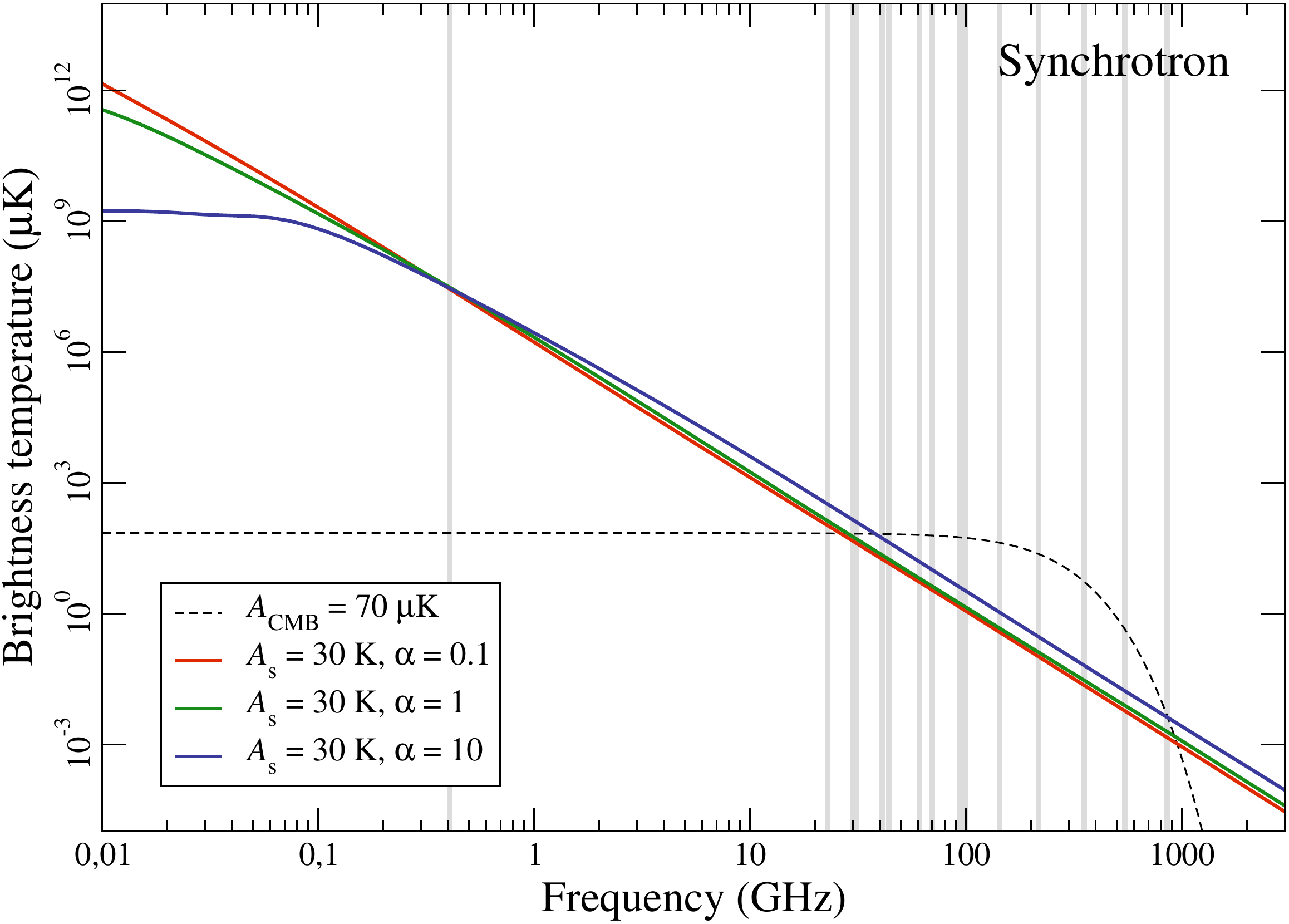,width=80mm,clip=}
\hspace*{3mm}
\epsfig{figure=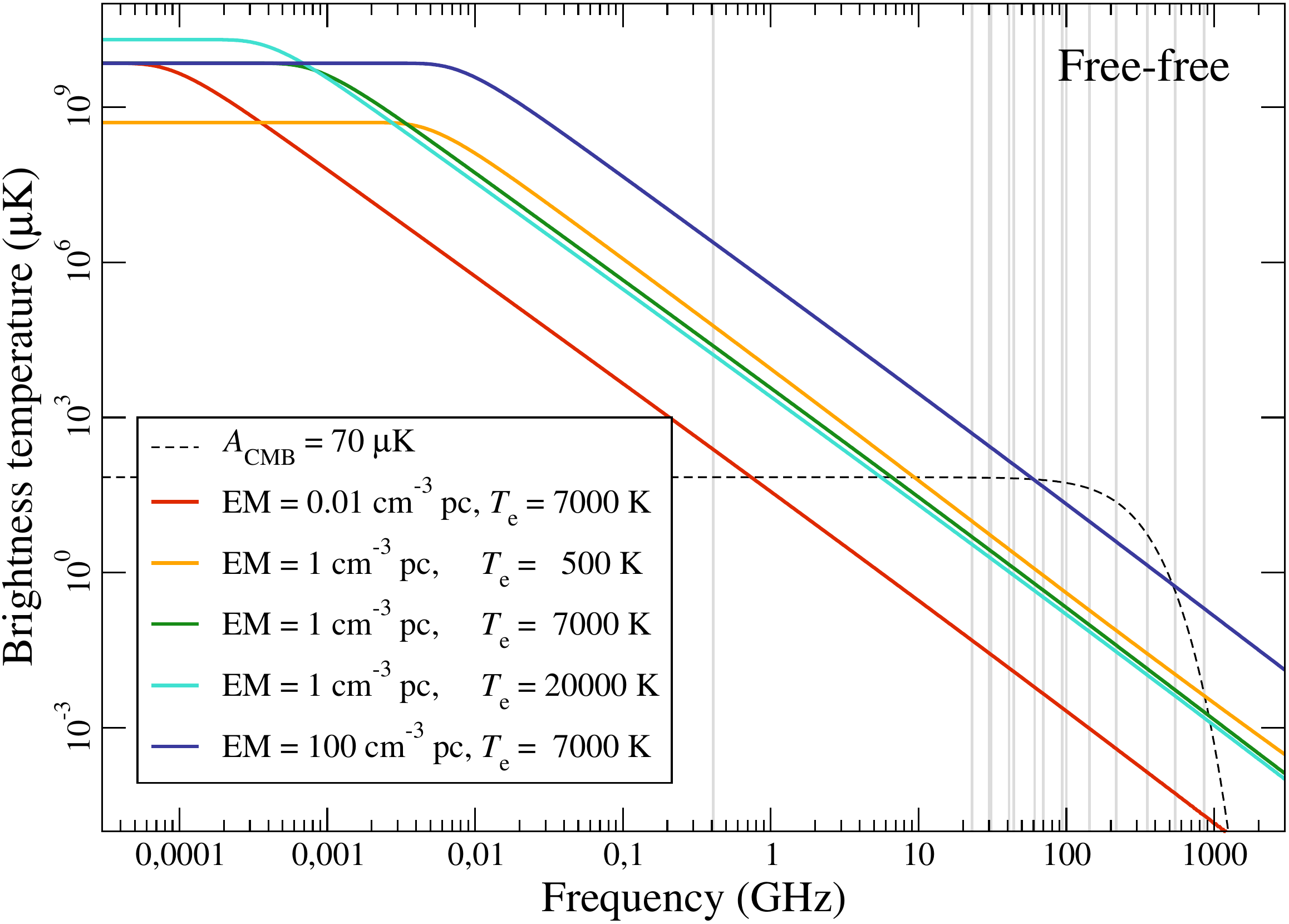,width=80mm,clip=}
}
\vspace*{3mm}
\mbox{
\epsfig{figure=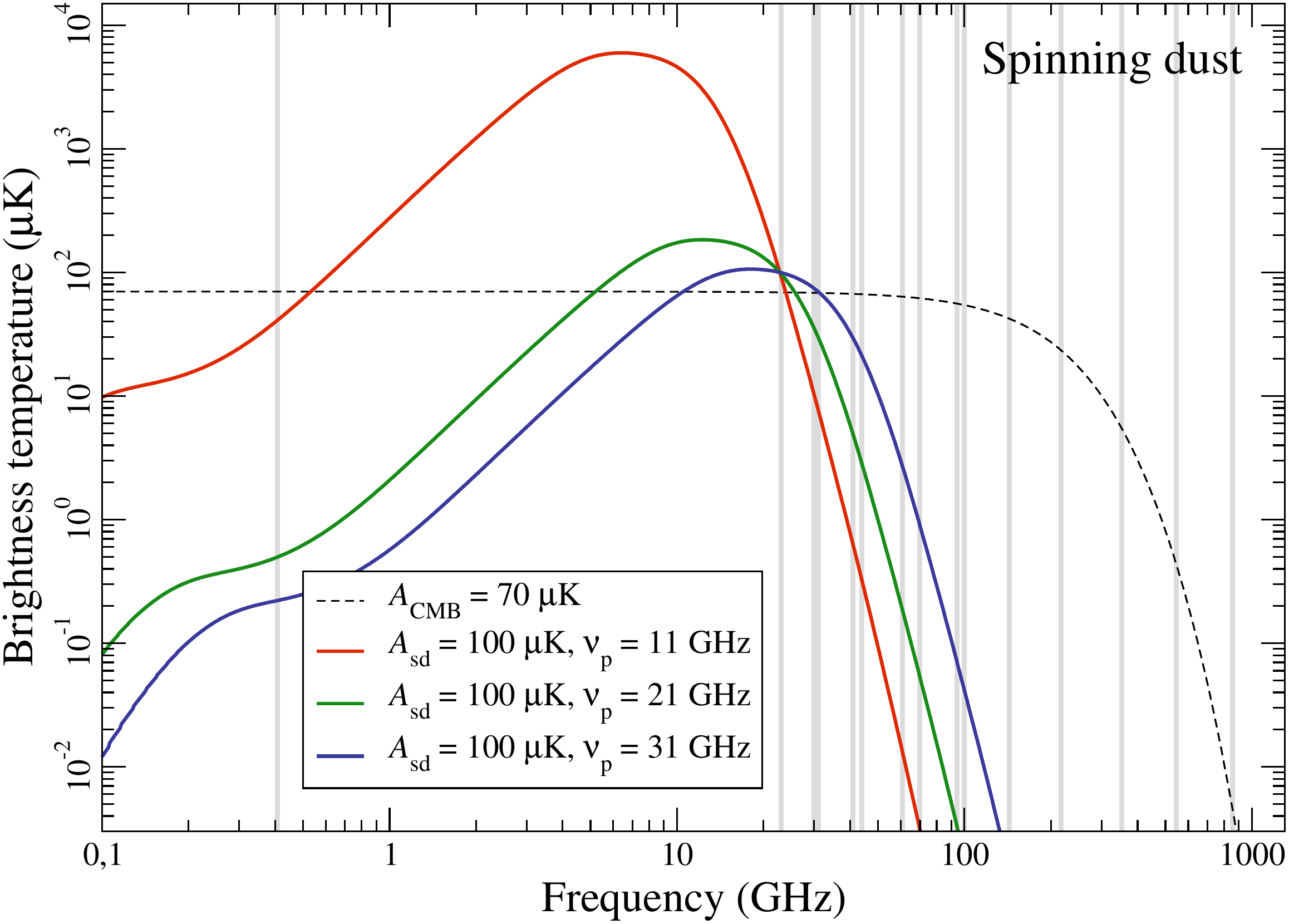,width=80mm,clip=}
\hspace*{3mm}
\epsfig{figure=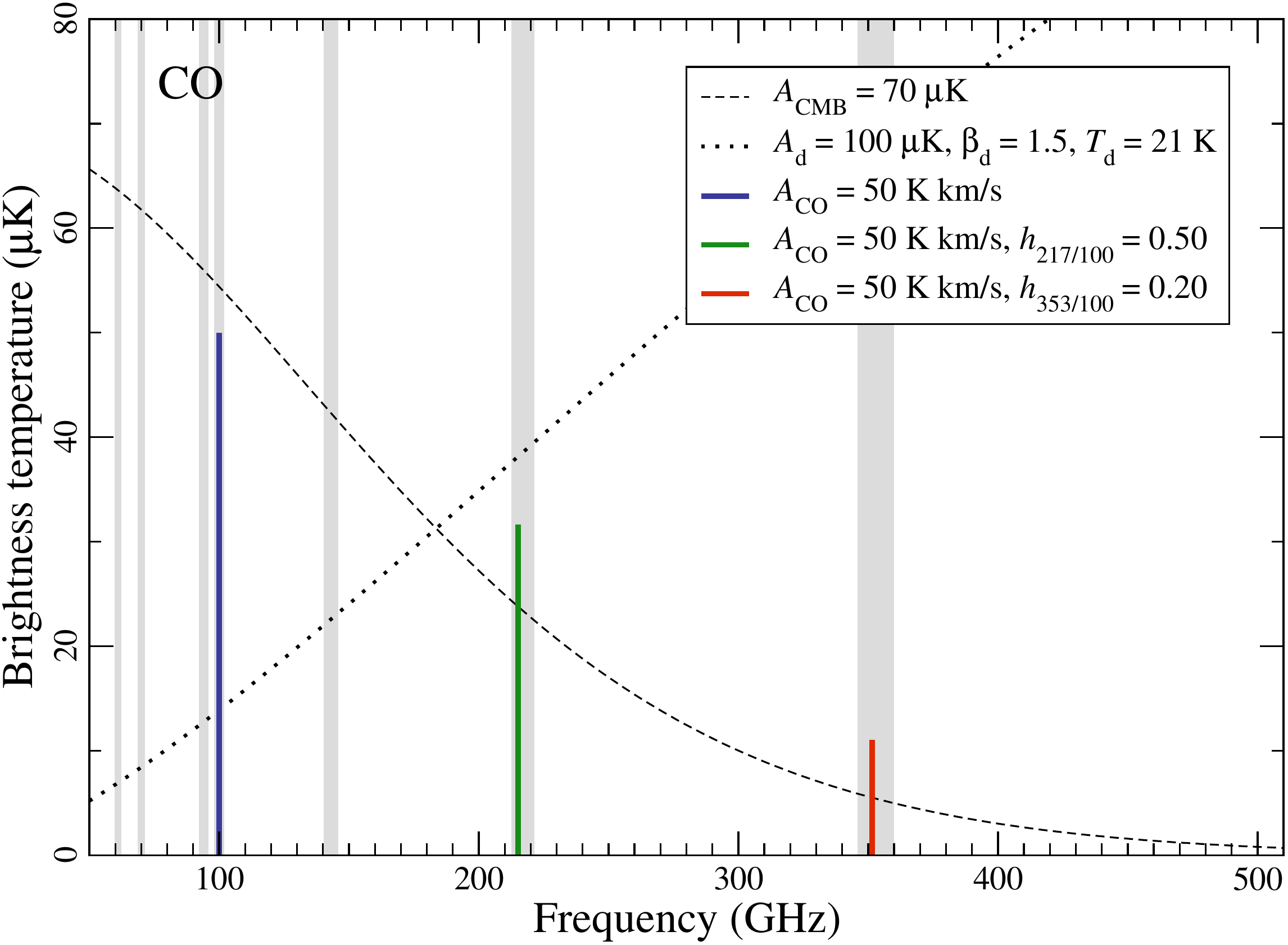,width=80mm,clip=}
}
\vspace*{3mm}
\mbox{
\epsfig{figure=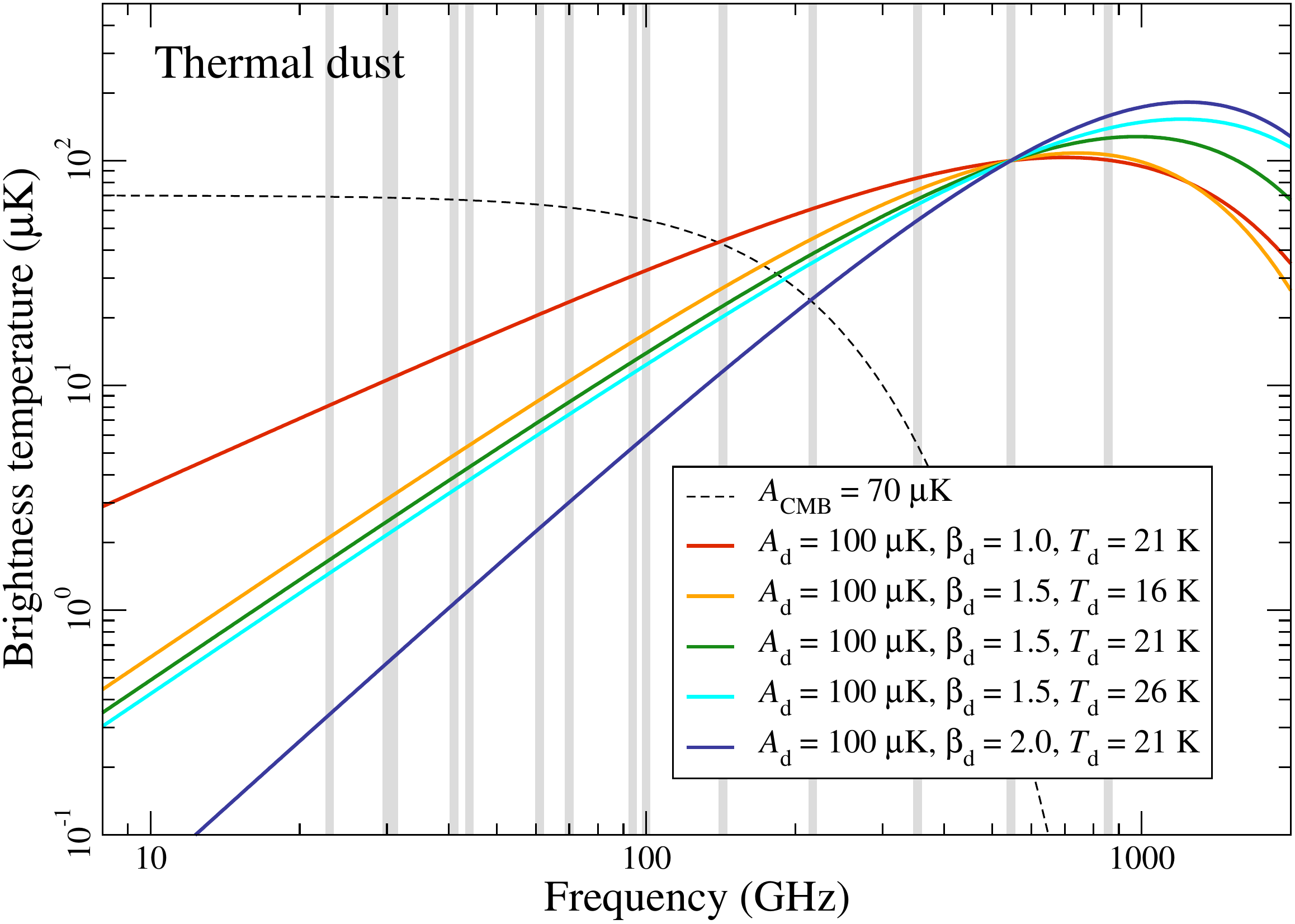,width=80mm,clip=}
\hspace*{3mm}
\epsfig{figure=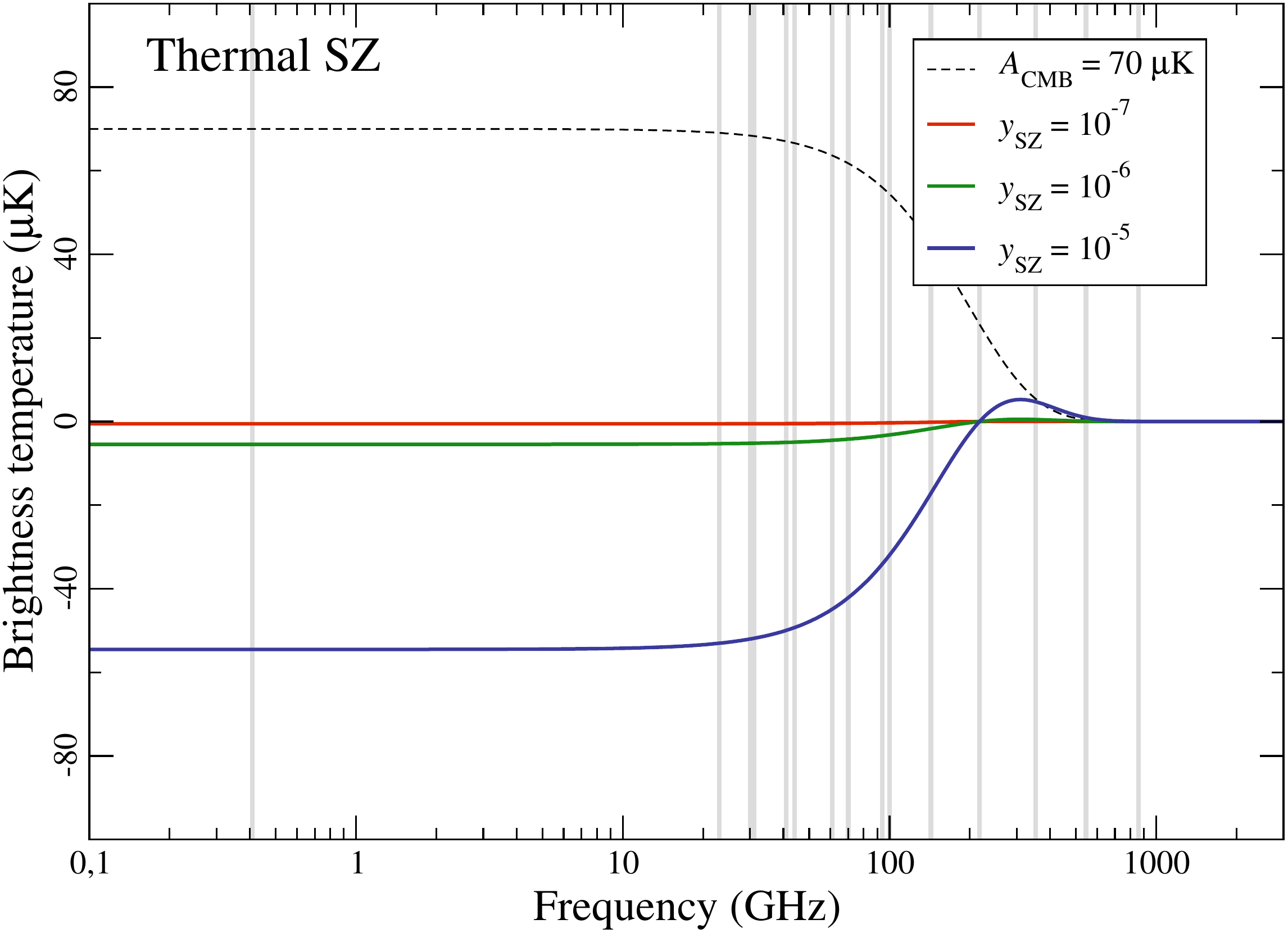,width=80mm,clip=}
}
\end{center}
\centering
\caption{Spectral energy densities (SEDs) for the main
  astrophysical components included in the present analysis, in
  brightness temperature. From left to right and top to bottom, panels
  show: (1) synchrotron emission; (2) free-free emission; (3) spinning
  dust emission; (4) CO line emission; (5) thermal dust emission; and (6)
  the thermal Sunyaev-Zeldovich effect. For each case, several
  parameter combinations are shown to illustrate their effect on the
  final observable spectrum. Vertical grey bands indicate the centre
  frequencies of the observations listed in Table~\ref{tab:data}, but
  for clarity true bandwidths are suppressed. In each panel, the black
  dashed line shows the CMB brightness temperature corresponding to a
  thermodynamic temperature of $70\,\mu\textrm{K}$, the CMB rms at
  $1\deg$ FWHM angular scale.}
\label{fig:components}

\end{figure*}

\begin{table*}[tmb]                                                                                                                                                   
\begingroup                                                                            
\newdimen\tblskip \tblskip=5pt
\caption{Summary of main parametric signal models for the temperature
  analysis. For polarization, the same parametric functions are
  employed, but only CMB, synchrotron, and thermal dust emission are
  included in the model, with spectral parameters fixed to the result
  of the temperature analysis. The symbol ``$\sim$'' implies that the
  respective parameter has a prior as given by the right-hand side
  distribution; Uni denotes a uniform distribution within the indicated
  limits, and $N$ denotes a (normal) Gaussian distribution with the
  indicated mean and standard devation.\label{tab:model}}
\nointerlineskip                                                                                                                                                                                     
\vskip -4mm
\footnotesize                                                                                                                                      
\setbox\tablebox=\vbox{ %
\newdimen\digitwidth                                                                                                                          
\setbox0=\hbox{\rm 0}
\digitwidth=\wd0
\catcode`*=\active
\def*{\kern\digitwidth}
\newdimen\signwidth
\setbox0=\hbox{+}
\signwidth=\wd0
\catcode`!=\active
\def!{\kern\signwidth}
\newdimen\decimalwidth
\setbox0=\hbox{.}
\decimalwidth=\wd0
\catcode`@=\active
\def@{\kern\signwidth}
\halign{ \hbox to 1.3in{#\leaderfil}\tabskip=1.0em&
    $#$\hfil\tabskip=1em&
    $#$\hfil\tabskip=1em&
    $#$\hfil\tabskip=0em\cr
\noalign{\doubleline}
\omit\hfil Component\hfil&\omit\hfil Free parameters and priors\hfil&\omit\hfil Brightness temperature, $s_{\nu}\,[\muK_{\rm RJ}]$\hfil&\omit\hfil Additional information\hfil\cr
\noalign{\vskip 5pt\hrule\vskip 3pt}
CMB$^{\rm a}$         & A_{\rm cmb} \sim {\rm Uni}(-\infty,\infty)  & \begin{array}{rcl} x &=& \frac{h\nu}{k_{\rm B} T_{\rm CMB}} \\
                                                                        g(\nu) &=& (\exp(x)-1)^2 / (x^2 \exp(x)) \\
                                                                   s_{\rm CMB} &=& A_{\rm CMB} / g(\nu) \\\end{array}& \begin{array}{rcl} T_{\rm CMB} &=& 2.7255\,\textrm{K}\end{array}\cr
\noalign{\vskip 10pt}
Synchrotron$^{\rm a}$ & \begin{array}{rcl} A_{\rm s} &>& 0 \\ \alpha &>& 0, \textrm{spatially constant} \end{array}& 
              s_{\rm s} = A_{\rm s} \left(\frac{\nu_0}{\nu}\right)^2 \frac{f_{\rm s}(\frac{\nu}{\alpha})}{f_{\rm s}(\frac{\nu_{0}}{\alpha})}& 
              \begin{array}{rcl} \nu_{0} &=& 408\,{\rm MHz} \\ f_{\rm s}(\nu)&=&{\rm Ext\ template} \end{array}\cr
\noalign{\vskip 10pt}
Free-free& \begin{array}{rcl} \log \EM &\sim& {\rm Uni}(-\infty,\infty) \\ \Te &\sim& N(7000\pm500\,{\rm K})\end{array}& 
           \begin{array}{rcl} g_{\rm ff} &=& \log\left\{\exp\left[5.960 -\sqrt{3}/\pi\log(\nu_9 * T_{4}^{-3/2})\right] + e\right\}\\ \tau &=& 0.05468 * \Te^{-3/2} * \nu_{9}^{-2} * \EM * g_{\rm ff}\\s_{\textrm{ff}} &=& 10^{6}\,\Te \,(1-e^{-\tau})\\ \end{array}&
\begin{array}{rcl}
T_4 &=& \Te/10^4 \\
\nu_9 &=& \nu/{(10^9\,{\rm Hz})}\\
\end{array}\cr
\noalign{\vskip 10pt}
Spinning dust& \begin{array}{rcl} A_{\rm sd}^1, A_{\rm sd}^2 &>& 0\\ \nu_{\rm p}^{1} &\sim& N(19\pm3\,{\rm GHz}) \\ \nu_{\rm p}^2 &>& 0, \textrm{spatially constant}\end{array}& 
               \begin{array}{rcl} s_{\rm sd} &=& A_{\rm sd} \cdot \left(\frac{\nu_0}{\nu}\right)^2 \frac{f_{\rm sd}(\nu\cdot\nu_{\rm p0}/\nu_{\rm p})}{f_{\rm sd}(\nu_{0}\cdot\nu_{{\rm p}0}/\nu_{\rm p})}\\ \end{array}& 
               \begin{array}{rcl} \nu_{0}^1 &=& 22.8\,{\rm GHz} \\ \nu_{0}^2 &=& 41.0\,{\rm GHz} \\ \nu_{\rm p0} &=& 30.0\,{\rm GHz}\\ f_{\rm sd}(\nu)&=&{\rm Ext\ template} \end{array}\cr
\noalign{\vskip 10pt}
Thermal dust$^{\rm a}$& \begin{array}{rcl} A_{\rm d} &>& 0 \\ \beta_{\rm d} &\sim& N(1.55\pm0.1)  \\ T_{\rm d} &\sim& N(23\pm3\,{\rm K})\end{array}& 
              \begin{array}{rcl} \gamma &=& \frac{h}{k_{\rm B}T_{\rm d}} \\ s_{\rm d} &=& A_{\rm d} \cdot \left(\frac{\nu}{\nu_0}\right)^{\beta_{\rm d}+1} \frac{\exp(\gamma\nu_0)-1}{\exp(\gamma\nu)-1}\\ \end{array}& 
              \begin{array}{rcl} \nu_{0} &=& 545\,{\rm GHz}\end{array}\cr
\noalign{\vskip 10pt}
SZ&            y_{\rm sz} > 0& 
               s_{\rm sz} = 10^6\,y_{\rm sz}/g(\nu)\,T_{\rm CMB} \left(\frac{x(\exp(x)+1)}{\exp(x)-1}-4\right)\cr
\noalign{\vskip 10pt}
Line emission& \begin{array}{rcl} A_{i} &>& 0 \\ h_{ij} &>& 0, \textrm{spatially constant}  \end{array} & 
               s_{i} = A_{i} h_{ij} \frac{F_{i}(\nu_j)}{F_{i}(\nu_0)} \frac{g(\nu_0)}{g(\nu_j)}& \begin{array}{rcl}
                 i &\in& \left\{\begin{array}{l} {\rm CO}\,J\!=\!1\!\rightarrow\!0\\ {\rm CO}\, J\!=\!2\!\rightarrow\!1\\ {\rm CO}\, J\!=\!3\!\rightarrow\!2\\ 94/100  \end{array}\right.\\
                 j &=& {\rm detector\ index} \\
               F &=& {\rm unit\ conversion}\end{array}\cr
\noalign{\vskip 5pt\hrule\vskip 3pt}
}}
\endPlancktablewide                                                                                                                                            
\tablenote {{\rm a}} Polarized component.\par
\endgroup
\end{table*}

Before presenting the results of our analysis, it is useful to review
the various model components that are relevant for this work, with the
goal of building intuition concerning important degeneracies and
residuals that may be observed in various goodness-of-fit
statistics. We first review each of the astrophysical sky signal
components for both temperature and polarization, as summarized in
Table \ref{tab:model} and Fig.~\ref{fig:components}, and then each of
the most important instrumental effects.

\subsection{Sky components in intensity}
\label{sec:sky}

\vskip 2mm\noindent{\bf CMB---}The CMB is given by a perfect blackbody with only a single 
spectral parameter, namely the CMB temperature. We adopt a
mean value of $T_{\textrm{cmb}}=2.7255\pm0.0006\,\textrm{K}$
\citep{fixsen2009}, and note that the uncertainty in this value is
sufficiently small to justify its use as a delta function prior. In
the current paper, we neglect both Rayleigh scattering and
higher-order relativistic effects
\citep{planck2013-pipaberration,lewis2013}; these could 
be accounted for in future work. The resulting CMB brightness
temperature is plotted as a dashed line in each panel of
Fig.~\ref{fig:components}, with an amplitude of 70\muK, corresponding
to the CMB rms at $1^{\circ}$ FWHM resolution, providing a useful
consistent visual reference for other components.

\vskip 2mm\noindent{\bf Synchrotron---}Diffuse synchrotron emission is
generated by relativistic cosmic-ray electrons spiraling in the
Galactic magnetic field.  This radiation may be highly polarized, with
a maximum polarization fraction of 0.75. Both theoretical models and
observations suggest that the synchrotron spectrum is well
approximated by a power law with an index of $\beta_{\rm s}\approx-3$
at frequencies above 20\,GHz, but with significant flattening at low
frequencies. In this paper we adopt a fixed spectral template for the
synchrotron spectrum. We use a spectrum extracted from the
\texttt{GALPROP} z10LMPD\_SUNfE synchrotron model from
\mbox{\citet{orlando2013}}, as described in \citet{planck2014-a31}. For
brevity we will refer to this as the \texttt{GALPROP} model from now on. A
description of the
\texttt{GALPROP}\footnote{\url{http://galprop.stanford.edu/}}\footnote{\url{https://sourceforge.net/projects/galprop}} code can be found
in \citet{moskalenko1998,strong2007,orlando2013} and references
therein.  We allow this spectrum to be rigidly shifted in
$\log\nu$--$\log S$ space with a single global frequency shift
parameter, $\alpha$, for the full sky; see Table~\ref{tab:model} for
the explicit definition. The main effect of such translation, however,
is to modify the synchrotron amplitude at 408\,MHz, leaving
$\beta_{\rm s}$ at frequencies above 20\,GHz essentially constant and
equal to $-3.11$ for any realistic shift parameter; see
Fig.~\ref{fig:components}.  Thus, with the current data the
synchrotron amplitude is determined almost exclusively by the 408\,MHz
survey, while the frequency spectrum is determined by the
\texttt{GALPROP} model, with the only free spectral parameter being
the relative normalization between the 408\,MHz and higher frequency
channels. Several alternative models (straight, broken, or
logarithmically-curved power laws) were also considered in the
preparation of this paper, but we found no statistically robust
evidence in favour of any of them over the \texttt{GALPROP}
spectrum. With only one very low frequency channel, the current data
set contains very little information about the synchrotron spectral
index. To constrain the synchrotron component, and break remaining
degeneracies between the synchrotron, free-free, and spinning dust
components, data from up-coming low-frequency experiments such as
S-PASS (2.3\,GHz; \citealp{carretti2009}), C-BASS (5\,GHz;
\citealp{king2010,king2014}), and QUIJOTE (10--40\,GHz;
\citealp{rubino-martin2012}) are critically important.  For now, we
prefer the \texttt{GALPROP} model over various power-law variations,
simply because it provides an acceptable fit to the data with
essentially no free additional parameters, and that it is based on a
well-defined physical model. For further discussion of synchrotron
emission in the \Planck\ observations, see \citet{planck2014-a31}.

\vskip 2mm\noindent{\bf Free-free---}Free-free emission, or
bremsstrahlung, is radiation from electron-ion collisions, and
consequently has a frequency spectrum that is well-constrained by
physical considerations \citep{dickinson2003}. We adopt the recent
two-parameter model of \cite{draine2011}, in which the free parameters
are the (effective beam-averaged) emission measure, $\EM$ (i.e., the
integrated squared electron density along a line of sight), and the
electron temperature, $\Te$. As seen in Fig.~\ref{fig:components}, the
spectrum is close to a power law at frequencies above 1\,GHz, but
exhibits a sharp break at lower frequencies, where the medium becomes
optically thick and the brightness temperature becomes equal to the
electron temperature. Intuitively, over our frequency range, the $\EM$
determines the amplitude of the free-free emission, scaling the
spectrum up or down, while the electron temperature changes the
effective power law index very slightly, from $\beta=-2.13$ for $\Te =
500\,\textrm{K}$ to $\beta=-2.15$ for $\Te=20\,000\,\textrm{K}$. At
low frequencies, free-free emission is significantly degenerate with
synchrotron emission, but because its power-law index is flatter than
for synchrotron, it extends into the high signal-to-noise HFI
channels. As a result, it is in fact possible to measure the electron
temperature for particularly high signal-to-noise regions of the sky
when imposing the smoothness prior discussed in
Sect.~\ref{sec:posterior}. The main difficulty regarding free-free
emission lies in a four-way degeneracy between CMB, free-free,
spinning dust, and CO emission.

\vskip 2mm\noindent{\bf Thermal dust---}At frequencies above 100\,GHz,
the dominant radiation mechanism is thermal dust emission
\citep{planck2013-p06b,planck2013-pip88,planck2013-p12,planck2014-XIX}. The
characteristic frequency is determined by the temperature of the dust
grains, and therefore varies with dust population and
environment. Empirically, thermal dust emission may be accurately
described across the \Planck\ frequencies as a modified blackbody with
a free emissivity index, $\beta_{\rm d}$, and dust temperature,
$T_{\rm d}$, per pixel, often referred to as a greybody,
\citep{planck2013-p06,planck2013-XVII}. At frequencies above 857\,GHz,
extending into the \COBE-DIRBE frequencies \citep{hauser1998}, the
dust physics quickly becomes far more complicated, and the
instrumental systematics more difficult to contain, and we therefore
restrict our current analysis to frequencies up to
857\,GHz. Correspondingly, we emphasize that the model derived here is
only expected to be accurate at frequencies up to 857\,GHz, and its
main application is extrapolation to low, not high, frequencies.
However, alternative parametrizations of thermal dust have been
suggested in the literature, for instance in terms of a two-component
grey-body model \citep[e.g.,][]{finkbeiner1999,meisner2015}, in which
each of the two components has a spatially constant emissivity and
temperature. Similar to the alternative synchrotron models mentioned
above, we also considered such models in the preparation of this
paper. Without higher-frequency observations we found that the data
are not able to discriminate between the various scenarios at any
statistically significant level. For simplicity, we therefore adopt
the one-component greybody model with a varying spectral index as a
phenomenological parametric thermal dust model, noting that it
provides a highly efficient representation of the true sky for
frequencies up to 857\,GHz, with no additional free global
parameters. Finally, we point out that cosmic infrared background
fluctuations (CIB) will be interpreted as thermal dust emission in our
model, by virtue of having a frequency spectrum similar to a greybody
component \citep{planck2013-pip56}. However, because CIB is relatively
weak compared to the Galactic signal, this is only a significant issue
in low-foreground regions of the sky.

\vskip 2mm\noindent{\bf Spinning dust---}Dust grains not only vibrate,
but they can also rotate. If they have a non-zero electric dipole
moment, this rotation will necessarily lead to microwave emission, as
first demonstrated by \cite{Erickson1957} and, in detail, by
\cite{draine1998}. The first hints of such radiation in real data were
found by cross-correlating CMB data with far-IR observations
\citep{kogut1996,oliveira-costa1997,netterfield1997}, but at the time
this correlation was thought to be dominated by free-free emission,
coupled to dust due to star formation. The first identification of a
new anomalous component was made by \citet{leitch1997}, and the
observed excess emission was simply referred to as ``anomalous
microwave emission'', in order not to make premature conclusions
regarding its physical nature. However, recent observational progress
has made the physical association much more firm \citep[e.g.,][and
  references
  theirin]{Davies2006,planck2011-7.2,Gold2011,planck2011-7.3,bennett2012,planck2013-XV}
and we will refer to this component simply as spinning dust. We adopt
a spinning dust spectral template as our spectral model, as evaluated for a
cold neutral medium with \texttt{SpDust2}
\citep{ali-haimoud2009,ali-haimoud2010,silsbee2011}. To fit this
template to the data, we introduce a free peak frequency,
$\nu_{\textrm{p}}$, and allow rigid translations the
$\log\nu$--$\log S$ space. As explicitly demonstrated by
\citet{bennett2012}, this simple two-parameter model can accommodate a
large number of spinning dust model variations to high precision.

To facilitate comparison with previously reported results, we report
the peak frequency for a spinning dust spectrum defined in flux
density units, and note that the conversion between brightness
temperature and flux density is $\propto \lambda^{-2}$; see
Table~\ref{tab:model}. The exact value of $\nu_{\textrm{p}}$ is highly
uncertain, and depends on the physical properties of the dust grains,
but is typically found to lie between 20 and 30\,GHz from dedicated
observations of individual objects \citep{planck2011-7.2}, in some
objects reaching as high as $\approx 50$\,GHz \citep{planck2013-XV}. The exact
shape of the spectrum is also uncertain, but we note that we have been
unable to fit this component with a single \texttt{spdust} component,
always finding significant residuals with dust-like morphology at
20--30\,GHz when forcing a high prior on $\nu_{\textrm{p}}$, or at
40--70\,GHz when forcing a low prior. For now, therefore, we adopt a
two-component spinning dust model in which one component has a free
peak frequency, $\nu_{\textrm{p}}$, per pixel and the other component
has a free, but spatially constant, $\nu_{\textrm{p}2}$, for a total
of three free spinning dust parameters per pixel and one global
parameter. We emphasize that this is a purely phenomenological model,
and we attach no physical reality to it.  The second component is in
effect only a computationally convenient parameterization of the width
of the spinning dust peak (see Fig.~\ref{fig:components}). We
therefore show only the sum of the two spinning dust components in the
following, evaluated at 30\,GHz. However, the set of released foreground data
products includes the individual component parameters, allowing
external uses to reproduce all results.

As listed in Table~\ref{tab:model}, we adopt a Gaussian prior of
$N(19\,\textrm{GHz}\pm3\,\textrm{GHz})$ on the peak frequency of the primary spinning
dust component, while no informative prior is enforced on the
spatially uniform secondary component. At first sight, the former
prior might look discrepant with respect to other targeted studies of
spinning dust, which often report best-fit peak frequencies between
25 and 30\,GHz. However, when adding the two spinning dust
components, one with $\nu_{\textrm{p}}\approx20$GHz and another
with $\nu_{\textrm{p}}\approx33\,$GHz, the total is indeed a
component with a peak frequency around 25\,GHz. 

Finally, we emphasize that the spinning dust component exhibits
significant correlations with both synchrotron and free-free emission,
and its individual parameters are sensitive to both instrumental
bandpasses and calibration. Thus, the reported parametric best-fit
values are associated with large correlated systematic uncertainties,
and additional low-frequency observations are required to make the
spinning dust model robust. As already mentioned, low-frequency
observations from experiments like S-PASS, C-BASS, and QUIJOTE are
critically needed to break these degeneracies.

\vskip 2mm\noindent{\bf CO and 94/100\,GHz line emission---}As
reported in \citet{planck2013-p03a}, line emission from carbon
monoxide (CO) is strongly detected in the \Planck\ 100, 217, and
353\,GHz frequency channels, and two sets of individual CO
$J$=1$\rightarrow$0, 2$\rightarrow$1, and 3$\rightarrow$2 line maps
(so-called ``Type-1'' and ``Type-2'' maps, produced by the
\texttt{MILCA} and \texttt{Commander} algorithms, respectively) as
well as a frequency co-added line map (``Type-3'', also produced with
\texttt{Commander}) were provided in the 2013 \Planck\ data
release. In Sect.~\ref{sec:CO} we present updated CO maps, based on
the full \Planck\ mission data set, plus a new general line emission
map, which captures a combination of emission lines that are detected
with the HFI 100\,GHz and WMAP W bands. An important contributor to
this map is the HCN line at 88.6\,GHz, providing about 23\,\% (63\,\%)
of the 100\,GHz (W-band) amplitude towards the Galactic Circumnuclear
Disk (GCD) and Sgr A$^*$ \citep{takekawa2014}. Several other lines
(CN, HCO+, CS etc.)  contribute at a level of 5--10\,\% each. In
addition, since we account for neither velocity effects nor detailed
cloud physics (opacity, local thermal equilibrium state, etc.), there
is also a non-negligible amount of CO leakage in this new map. We
therefore refer to the new component simply as ``94/100\,GHz line
emission''.

Generally speaking, the fact allowing us to separate line emission
from other diffuse components is that the bandpass filters of
individual detectors have different transmission levels at the line
frequency. Therefore, while components with continuous spectra very
nearly vanish in detector difference maps within single frequencies,
line emission does not. For this reason, we employ individual detector
maps in the 2015 temperature analysis, as opposed to co-added
frequency maps in the corresponding 2013 analysis. However, for the
highest resolution maps, for which neither LFI nor \WMAP\ are able to
provide useful information, there is a strong degeneracy between CMB,
free-free, and CO $J$=1$\rightarrow$0 emission. The high-resolution CO
maps are therefore produced with dedicated and special-purpose
methods, as described in Sect.~\ref{sec:CO}. Furthermore, we note that
the CO $J$=3$\rightarrow$2 emission map is significantly degenerate
with the brighter thermal dust component, and consequently subject to
large systematic uncertainties.

We describe the line emission maps parametrically in terms of an
amplitude map at the line frequency, $a(p)$, normalized relative to
one specific detector map, and with a rigid frequency scaling given by
the product of a spatially constant line ratio, $h_{ij}$, where $i$
corresponds to a spectral line index, and $j$ denotes detector
index. $F(\nu)$ denotes some unit conversion factor, converting for
example between $\mu\textrm{K}_{\textrm{RJ}}$ and
$\textrm{K}_{\textrm{RJ}}\,\textrm{km\,s}^{-1}$ (see
Table~\ref{tab:model} for the exact mathematical definition). The same
formalism applies whether $h_{ij}$ refers to line ratios between
frequencies or only to bandpass ratios within frequencies. For
example, in 2013 this formalism was used to construct the so-called CO
Type-3 map \citep{planck2013-p03a}, which is defined by assuming
spatially constant CO $J$=2$\rightarrow$1/$J$=1$\rightarrow$0 and
$J$=3$\rightarrow$2/$J$=1$\rightarrow$0 line ratios. In the current
analysis, we only assume that the same approximation holds between
detectors within single frequencies, i.e., that the ratio between the
CO signals observed by the 100-ds1 and 100-ds2 detectors is spatially
constant. Note that these assumptions are not strictly correct in
either case, both because of real variations in local physical
properties such as opacity and temperature, and because of non-zero
velocities of molecular clouds that either red- or blueshift the
intrinsic line frequency. Since the derivative of the bandpass profile
evaluated at the line frequency also varies between detectors, the
effective observed line ratio along a given line of sight also varies
on the sky. As we shall see in Sect.~\ref{sec:temperature}, this
effect represents the dominant residual systematic in a few of our
frequency channels after component separation.

\vskip 2mm\noindent{\bf Thermal Sunyaev-Zeldovich---}The last of the
main astrophysical components included in the following temperature
analysis is the thermal Sunyaev-Zeldovich (SZ) effect, which is caused
by CMB photons scattering off hot electrons in clusters
\citep{sun72}. After such scattering, the effective spectrum no longer
follows a perfect blackbody, but is rather given by the
expression\footnote{For simplicity, we adopt the non-relativistic
  expression for the thermal SZ effect in this paper.} listed in
Table~\ref{tab:model}. The only free parameter for this effect is the
Compton parameter, $y_{\textrm{sz}}$, which for our purposes acts as a
simple amplitude parameter. Note that the effective SZ spectrum is
negative below 217\,GHz and positive above this frequency, and this
feature provides a unique observational signature. Still, the effect
is weak for all but the very brightest clusters on the sky, and the
$y_{\textrm{sz}}$ map is therefore particularly sensitive to both
modelling and systematic errors. In this paper, we only fit for the
thermal SZ effect in two separate regions around the Coma and Virgo
clusters, which are by far the two strongest SZ objects on the sky, in
order to prevent these from contaminating the other components. While
the SZ decrement for the Coma cluster is as large as
$-400\,\mu\textrm{K}$ below 100\,GHz on an angular scale of a few
arcminutes \citep{battistelli2003,planck2012-X}, and for the Virgo
cluster a few tens of microkelvin on a few degrees scale
\citep{diego2008}, it is only a few microkelvin for most other objects
after convolution with the large beams considered in this
paper. Further, full-sky SZ reconstruction within the present global
analysis framework requires significantly better control of systematic
effects than what is achieved in the current analysis, in particular
at high frequencies.  For dedicated full-sky SZ reconstruction, see
\citet{planck2013-p05b} and \citet{planck2014-a28}.

\vskip 2mm\noindent{\bf Monopoles and dipoles---}In addition to the
above astrophysical components, the microwave sky exhibits important
signal contributions in the form of monopoles and dipoles. The prime
monopole examples are the CMB monopole of $2.7255\,\textrm{K}$ and the
average CIB amplitude \citep{planck2013-pip56}. The main dipole
contribution also comes from the CMB, which has an amplitude of
3.3655 (3.3640) $\textrm{K}$ as measured by LFI (HFI). The
difference between the LFI and HFI measurements of 1.5\muK\ is within
quoted uncertainties \citep{planck2014-a01}.

Ideally, the CMB dipole contribution should be removed during map
making \citep{planck2014-a07,planck2014-a09}, but because the
estimated dipole has a non-zero uncertainty, and different dipoles are
subtracted from the \Planck\ and \WMAP\ sky maps, it is necessary to
account for residual dipoles in each individual map. In this paper, we
adopt as delta function priors the nominal \Planck\ dipole estimates 
for a small subset of reference frequencies (namely the 30, 100-ds1,
143-ds1, 545, and 857 channels), as well as the parameters derived by
\citet{wehus2014} for the 408\,MHz map. That is, the dipoles of these
six channels are fixed at their nominal values, and, together, they
anchor the dipole solutions for all astrophysical component maps.

Likewise, we adopt the nominal monopole parameters for a subset of
frequency channels, anchoring the effective offsets of each component
to external values. Specifically, we adopt the HFI CIB monopole
predictions, as listed in Table 5 of \citet{planck2014-a09}, for the
100-ds1, 143-ds1 and 545-2 monopoles, and the zero-points derived by
\citet{wehus2014} for the \WMAP\ Ka and 408\,MHz bands. 

Thus the following analysis does not
derive independent absolute estimates for either monopoles or dipoles,
but relies critically on external priors for these values. If, say,
the CIB monopole prediction is significantly updated through
additional observations, this error will translate directly into an
error in the zero level of our thermal dust map. On the other hand,
our analysis does provide an independent internal consistency check
among the adopted priors, and, as we shall see, no major anomalies are
found.

\vskip 2mm\noindent{\bf Other components---}Finally, we mention
briefly some other sub-dominant effects that are either neglected or
only indirectly accounted for in our model. First, similar to the
thermal SZ effect, the kinetic Sunyaev-Zeldovich effect is caused by
CMB photons scattering on hot electrons \citep[e.g.,][and references
  therein]{rephaeli1995,planck2014-a28}. However, in this case a
non-zero bulk velocity in the electron population is the defining
feature. When interacting with moving electrons, the photons
effectively receive a Doppler shift proportional to the bulk
velocity. In this process, the CMB blackbody spectrum is conserved,
but its temperature is slightly changed. The kSZ effect is therefore
fully degenerate with the primordial CMB anisotropies when considered
pixel-by-pixel, and can only be disentangled using spatial information
and cross-correlations. However, this effect is small, and we neglect
it in the rest of the paper.

Second, the cosmic infrared background
\citep{planck2011-6.6,planck2013-pip56} consists of redshifted thermal
dust radiation from distant galaxies. By virtue of being a
cosmological signal, it is statistically isotropic in the sky, and can
therefore be described analogously to the CMB in terms of a sum of a
constant offset at each frequency plus small fluctuations imprinted on
top, tracing the column density of the emitting matter integrated
over redshift. In this paper, we account for the mean temperature
through the monopole parameter of each frequency map, but neglect the
spatially varying component. Since the CIB frequency spectrum is very
similar to the Galactic thermal dust spectrum, with typical mean
greybody parameters of $\beta_{\textrm{CIB}}=1.4\pm0.2$ and
$T_{\textrm{CIB}}=(13.6\pm1.5)\,\textrm{K}$ \citep{gispert2000}, these
fluctuations are effectively absorbed by our thermal dust
model. Future work will attempt to break this degeneracy by exploiting
the different spatial power spectra of the two components, and
through the use of external priors.

Third, there has been some discussion in the literature concerning the
existence of magnetic dipole emission from dust grains
\citep{draine2013}, with tentative evidence reported in
\citet{planck2014-XXII}. Suggestions have even been made that such
emission might contaminate previously published CMB maps at
significant levels \citep{liu2014}. \citet{planck2014-a31} comments
briefly on this question, while in this paper we simply note that the
current model is able to reproduce the data with statistically
acceptable accuracy without such a component, and it is neglected in
the following baseline analysis.

Fourth and finally, extra-Galactic point sources contribute 
significant power at both low and high frequencies
\citep{planck2014-a31}. In the following analysis, in which we fit for
foreground parameters pixel-by-pixel, these sources end up in the
component map with the most similar frequency spectrum, for instance
the free-free or thermal dust emission maps. In an attempt to minimize
such contamination, we preprocessed each input map by subtracting
catalogues of known resolved sources \citep{bennett2012,planck2013-p05}
before performing the analysis. However, due to source
variability, beam asymmetries, and uncertainties in the actual
catalogues, we invariably found worse results with such
pre-processing. A more promising approach is to fit for each source
jointly with the diffuse components, as for instance described by
\citet{bull2014}. This, however, is left for future work.

\subsection{Sky components in polarization}
\label{sec:sky_pol}

Whereas the microwave sky in total intensity exhibits very rich
astrophysics, in principle requiring between 10 and 20 different
physical components for a proper model (depending on the level of
detail required), the same is not true for polarization. At the
sensitivity level of current experiments, only three diffuse
components have been clearly detected, namely CMB, synchrotron, and
thermal dust.

\vskip 2mm\noindent{\bf CMB---}Thomson scattering between electrons
and photons is an intrinsically polarized process. If the incident
radiation surrounding a given electron is fully isotropic, no net
polarization will be observed from that electron. However, if the
radiation forms a local quadrupole, a net non-zero CMB polarization
signal results \citep{zaldarriaga1997,kamionkowski1997,hu1997}. Thus, any
process that generates quadrupolar structures at the time of
recombination or during the epoch of reionization will result in a CMB
polarization signal. The scalar fluctuations that are responsible for
creating the CMB temperature anisotropies will form so-called
``$E$-mode'' patterns, in which the polarization direction is either
orthogonal or parallel to the wave direction.  Inflationary
gravitational waves and weak gravitational lensing of CMB $E$-modes
produce so-called ``$B$-mode'' patterns, in which the polarization
direction is rotated $-45\deg$ or $+45\deg$ with respect to the wave
direction.  These two are the only known sources of such $B$-modes of
primordial or high redshift origin.

\vskip 2mm\noindent{\bf Synchrotron---}Radiation emitted from
relativistic electrons spiraling in a magnetic field is intrinsically
highly polarized, with about seven times more energy being emitted in
the plane of the electron's motion than in the orthogonal
direction. For a perfectly regular magnetic field, the synchrotron
polarization fraction may exceed 70\,\% \citep{pacholczyk1970},
although for realistic fields it is usually significantly
less. \WMAP\ found a polarization fraction of about 3\,\% in the
Galactic plane and about 20\,\% at high Galactic latitudes
\citep{page2007}.  Other analyses report polarization fractions as
high as 50\,\% on large angular scales
\citep{kogut2007,vidal2014}. The new \Planck\ measurements at 30 and
44\,GHz complement these observations, and allow us to put independent
constraints on this quantity \citep{planck2014-a31}.

\vskip 2mm\noindent{\bf Thermal dust---}Magnetic fields also have an
important effect on aspherical dust grains, in the sense that they
tend to align the grains' major axes with the local magnetic field
direction. This alignment naturally results in net microwave
polarization with a thermal dust spectrum. Furthermore, with the
frequency coverage and sensitivity of the HFI channels, \Planck\ is
ideally suited to measure this signal with very high accuracy, as has
already been demonstrated through a series of recent papers, including
\citet{planck2014-XIX}, \citet{planck2014-XX}, \citet{planck2014-XXI},
\citet{planck2014-XXII}, and \citet{planck2014-XXX}. One important
goal of the current paper is to provide direct access to these
observations in a computationally convenient form.

\vskip 2mm\noindent{\bf Neglected components---}Considering the other
components that are relevant for intensity analysis, we first note
that the free-free emission process is intrinsically independent of 
direction, and therefore naturally unpolarized, although some
polarization may arise due to Thomson scattering on electrons near the
edges of strong \ion{H}{ii} regions. This effect, however, is smaller than 10\,\%
for edges of optically thick clouds \citep{keating1998}, and
negligible away from any sharp edges. Averaged over the full sky,
free-free emission is observationally constrained to be less than
1\,\% polarized \citep{macellari2011}.

Next, spinning dust emission is generated by small dust grains, and
these generally align weakly with the local magnetic field. However,
rotational energy level splitting dissipates energy, and this leads to
a low level of grain alignment \citep{lazarian2000}. As a result, the
polarization fraction may be up to 1--3\,\% between 10 and 30\,GHz,
but falling to less than 1\,\% at higher frequencies. For a recent
review of current constraints, see \citet{rubino-martin2012}.

Third, polarized CO emission was first detected by \citet{greaves1999}
at the roughly 1\,\% level near the Galactic centre and in molecular
clouds. This signal may therefore in principle be detected in the very
sensitive \Planck\ 100 and 217\,GHz frequency channels. However, as
described in \citet{planck2014-a09}, one of the most important
systematic effects in the \Planck\ polarization data set is
temperature-to-polarization leakage, one component of which is
precisely leakage of the CO signal. For now, therefore, we
conservatively mask any regions with strong CO intensity detections
from all polarization maps (see Sect.~\ref{sec:polarization} for
further discussion).  This issue will be revisited in the future.

Contributions from the CIB, extragalactic point sources
\citep{tucci2012}, the Sunyaev-Zeldovich effect, etc., are all
small compared to the noise level of \Planck, and neglected in the
following.

\subsection{Instrumental effects}
\label{sec:instrumental_effects}

In addition to astrophysical components, our model includes two 
important instrumental effects, relative calibration between
detectors and bandpass uncertainties.

\vskip 2mm\noindent{\bf Relative calibration---}A calibration
error scales a map up or down in amplitude, affecting all
components equally. For astrophysical foregrounds, such errors are
therefore strongly degenerate with spectral indices. The CMB, however, 
is a nearly perfect blackbody across the relevant frequency range, and 
provides a powerful relative calibration source. In the following, we
(arbitrarily) adopt the 143-ds1 channel as our absolute calibration
reference. For any channel between 30 and 353\,GHz, we then use
CMB fluctuations between multipoles $\ell_{\textrm{min}}=25$ and 100
to determine the relative calibration by means of a simple
cross-correlation,
\begin{equation}
\g_i =
\frac{1}{76}\sum_{\ell=25}^{100}
\frac{\sum_{m=-\ell}^{\ell} |a_{\ell m}^i (a_{\ell
    m}^{\textrm{cmb}})^*|}{\sum_{m=-\ell}^{\ell} |a_{\ell m}^{\textrm{cmb}}|^2},
\end{equation}
where $\s_{\textrm{cmb}} = \sum_{\ell,m} a_{\ell m} Y_{\ell m}$ is
the usual spherical harmonic transform.  This equation effectively
replaces Eq.~(\ref{eq:gibbs_g}) in the Gibbs sampling chain for the
relevant channels.  The cross-correlation coefficient is evaluated
over pixels admitted by the PM61 processing mask
(Fig.~\ref{fig:processing_masks}), and the multipole range
$25\le\ell\le100$ is choosen to minimize contamination from diffuse
foregrounds on the low end and confusion with beam uncertainties on
the high end.

Strictly speaking, this estimation step is a violation of the Gibbs
sampling algorithm as defined in Eq.~(\ref{eq:gibbs_g}), since that
formally should include all signal components over the full sky, not
just the CMB component evaluated over parts of the sky. However, in
this special case we trade mathematical rigour for the sake of
increased robustness with respect to systematic effects.

The 545 and 857\,GHz channels are treated as special cases. As 
discussed in Sect.~\ref{sec:data}, these channels
are significantly affected by far sidelobe and destriping error
contamination.  We therefore exclude the 545-1, 857-1, 857-3, and
857-4 channels entirely, and add white regularization noise of
$0.01\,\textrm{MJy}\,\textrm{sr}^{-1}$ to the remaining channels.  Unfortunately,
this noise addition has the side-effect that any CMB fluctuations
at 545\,GHz are obscured by random white noise, to the extent that it
is no longer possible to calibrate the 545\,GHz channel with CMB
fluctuations within the usual Gibbs chain. As a partial solution to
this problem, we perform a dedicated pre-analysis that is identical 
to the main analysis, except that no regularization
noise is added to the high-frequency channels. The 545-2 calibration
is then fixed at the resulting value in the subsequent main analysis,
while the other 545-4 and 857-2 calibrations are fitted using the full
Gibbs expression, including full-sky (up to the PM99.6 processing
mask) and all-component observations. We estimate that the uncertainty
on the 545-1 recalibration factor derived in this paper is 1--2\,\%,
which translates into a 3--6\,\% uncertainty on the 857\,GHz calibration
through the thermal dust scaling.  Note, however, that the 857\,GHz
calibration is almost perfectly correlated with the 545\,GHz
calibration. These two values represent the most significant sources
of systematic error in the entire temperature analysis.

\begin{figure}
\begin{center}
\mbox{
\epsfig{figure=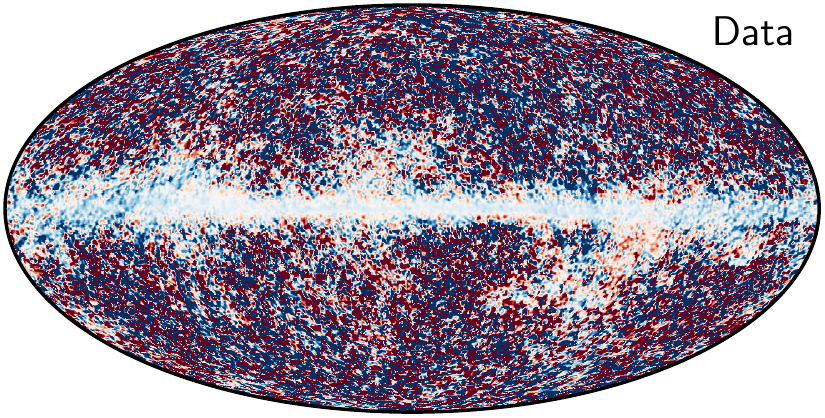,width=88mm,clip=}
}
\mbox{
\epsfig{figure=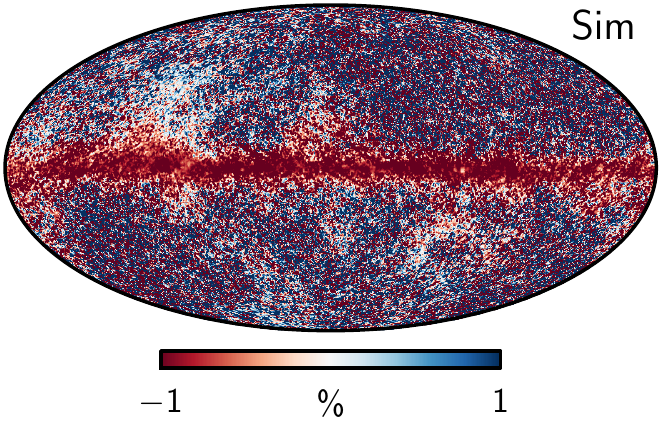,width=88mm,clip=}
}
\end{center}
\caption{Fractional difference maps on the form
  $(\d_{\textrm{143-ds1}}-\d_{\textrm{143-ds2}})/\d_{\textrm{143-ds1}}$,
  as evaluated from the real data (\emph{top}) and from the FFP8
  simulation (\emph{bottom}), both smoothed to $40\arcm$ FWHM. The 1\,\%
  difference observed along the Galactic plane is caused by a mismatch
  between the measured bandpass profiles (which are used to construct
  the FFP8 simulations) and those corresponding to in-flight
  observations. See Sect.~\ref{sec:instrumental_effects} for further discussion,
  and Sect.~\ref{sec:baseline} for explicit corrections.}
\label{fig:bandpass}
\end{figure}

Finally, the calibration of the two lowest frequencies (408\,MHz and
\WMAP\ K-band) are fixed to their nominal values. The estimated
calibration uncertainty of the 408\,MHz channel is 10\,\%
\citep{haslam1982}, and as a result our synchrotron model also has a
10\,\% calibration uncertainty.  Note that the 408\,MHz calibration is
almost perfectly degenerate with the \texttt{GALPROP} scale frequency,
$\nu_{\textrm{p}}^{\textrm{s}}$, a degeneracy that is only partially
broken by the presence of free-free emission. For the \WMAP\ K-band,
the main problem lies in the considerable model uncertainty of the
spinning dust component, which prevents an effective CMB
cross-correlation at the required sub-percent accuracy level. For now,
we consider the default CMB dipole-based K-band calibration to be more
robust than a CMB cross-correlation-based estimate.

\vskip 2mm\noindent{\bf Bandpass errors---}\label{sec:bandpass}The bandpass profile of a
detector provides the information required to convert between a given
foreground spectrum and the actually measured signal, both in terms of
unit conversion and effective colour corrections
\citep{planck2013-p02b,planck2013-p03d}. Therefore, a measurement
error in the bandpass profile essentially translates into a 
spectral-index-dependent multiplicative scale error. As such, bandpass
measurement errors are almost perfectly degenerate with calibration
errors, except in one crucial aspect. For maps that are calibrated on
the CMB dipole, the amplitude of the CMB fluctuations is fully
independent of any bandpass assumptions. This fact holds the key to
separating bandpass from calibration effects. While calibration errors
affect both foreground and CMB components equally, bandpass
measurement errors only affect foreground components.

As shown in the following, bandpass measurement uncertainties are
important for both LFI and HFI in high-precision component
separation. Figure~\ref{fig:bandpass} provides a simple illustration
of this, in the form of the fractional difference between the 143-ds1
and 143-ds2 detector maps,
\begin{equation}
r = \frac{d^{143\textrm{ds-1}}-d^{143\textrm{ds-2}}}{d^{143\textrm{ds-1}}}.
\end{equation}
The top panel shows the fractional difference map computed from the
observed \Planck\ 2015 data and smoothed to $40\arcm$ FWHM, and the
bottom panel shows the same from the 2015 FFP8 simulation
\citep{planck2014-a14}, which is based on the nominal bandpass
profiles. The effective frequencies for these two channels, based on
laboratory measurements, are listed in Table 4 of
\citet{planck2013-p03d}, and are $(144.22\pm0.02)$ and
$(145.05\pm0.02)\,\textrm{GHz}$ for a spectral index of $\alpha=2$. The
nominal 143-ds2 effective frequency is therefore 0.8\,GHz higher than
the corresponding 143-ds1 effective frequency, and this holds almost
independently of the assumed spectral index. Given these numbers, and
recognizing that the 143\,GHz channel is dominated by thermal dust,
which has an effective spectral index of $\beta\approx1.4$ at this
frequency, one would expect the above fractional ratio to be
$(1-(145.05/144.22)^{1.4})\approx -0.8\,\%$, as is indeed observed in
the simulation. However, in the real data, we see that the fractional
ratio is approximately $+0.2\,\%$, strongly suggesting that 143-ds1 in fact has a
higher effective frequency than 143-ds2, not lower.

An alternative explanation for this discrepancy between the model and
the data might be relative calibration errors. However, a 1\,\%
relative calibration error in the 143\,GHz channel is strongly ruled
out by CMB cross-correlations, since the internal relative calibration
of this channel is accurate to about 0.1\,\% (see
Sect.~\ref{sec:temperature}, \citealt{planck2014-a01}, and
\citealt{planck2014-a09}). A second possibility is that the 143\,GHz
channel might be dominated by some other component than thermal dust,
such as free-free emission. However, such a hypothesis can only
explain the observed difference for one particular detector
combination, while similar differences are observed between many
detector maps. Furthermore, it does not match the morphology of the
observed fractional difference map. The only viable explanation we
have found is indeed bandpass errors with magnitudes at the
$\lesssim1\,$GHz level. It is worth noting that significant
discrepancies among the 143\,GHz detector bandpasses were noted
already in \citet{planck2013-p03d}, and these qualitative findings are
therefore not new.

Fortunately, there is sufficient information in the current data to
self-consistently constrain and mitigate these errors. First, by
employing detector-set maps instead of full-frequency maps, bandpass
errors within frequency bands can be constrained with high
signal-to-noise levels, as illustrated in
Fig.~\ref{fig:bandpass}. Second, because the true sky contains
multiple foreground components with varying spectral indices, while
there is only one true bandpass per detector that applies to all
components and all pixels on the sky, even the mean bandpass
correction per frequency is constrained to some level, although this
dependency is conditional on the overall foreground model. Therefore,
the bandpass corrections presented in the following are strongly
constrained with respect to relative bandpass shifts \emph{within}
frequency bands, but uncertain with respect to frequency shifts
\emph{across} frequency bands. Intuitively, it is easy to conclude
with high significance that the relative difference between the
143-ds1 and 143-ds2 effective frequencies is +0.2\,GHz, not
$-0.8\,$GHz, but it is difficult to say with confidence whether both
should be shifted jointly by an additional 0.2 or $-0.2$\,GHz. That
depends on the spectral index of thermal dust emission, and
consequently on the 545 and 857\,GHz calibration factors, which are
still associated with significant uncertainties.

As for an explicit parameterization of the bandpass errors, we adopt
in this paper a simple and purely phenomenological model given by a
rigid bandpass translation in frequency space:
\begin{equation}
b(\nu) = b_0(\nu+\Delta_\nu).
\end{equation}
Here $b_0(\nu)$ denotes the nominal bandpass, $b(\nu)$ is the fitted
bandpass, and $\Delta_\nu$ is the effective frequency shift between
the two. We emphasize that this is not to be taken as a physical model
of the true bandpass errors, and more realistic models should include
information specifically regarding the tails of the bandpasses, the
separation between main bandpass modes, and overall large-scale tilts,
not only centre frequency shifts.

\section{Temperature analysis}
\label{sec:temperature}

\subsection{Model motivation}
\label{sec:motivation}

\begin{figure*}
\begin{center}
\mbox{
\hspace*{-2mm}
\epsfig{figure=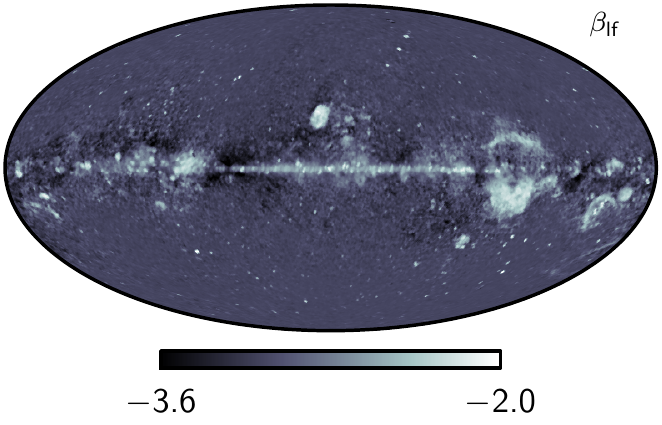,width=88mm,clip=}
\epsfig{figure=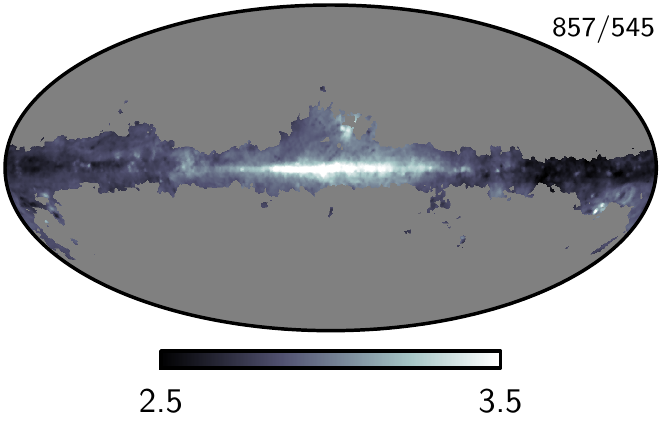,width=88mm,clip=}
}
\mbox{
\epsfig{figure=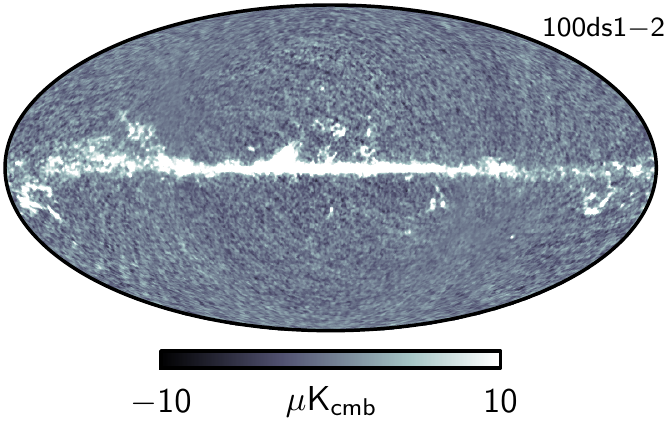,width=88mm,clip=}
\epsfig{figure=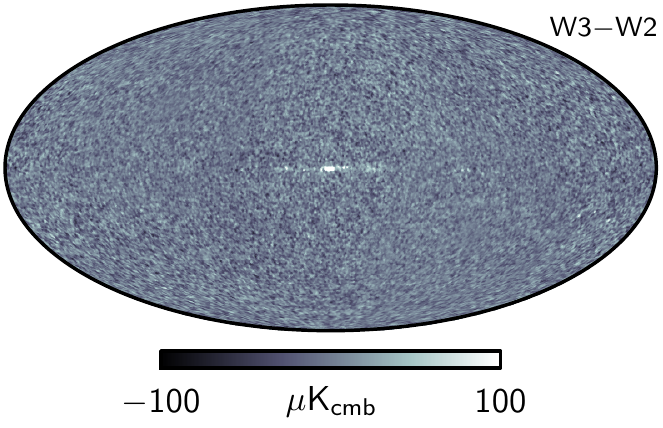,width=88mm,clip=}
}

\end{center}
\caption{\emph{Top left}: effective low-frequency foreground spectral
  index as measured from the combination of \Planck, \WMAP, and
  408\,MHz, with no attempt to disentangle synchrotron, free-free, and
  spinning dust emission into separate components. However,
  higher-frequency components (CMB, CO, thermal dust, etc.) are fitted
  component-by-component, as in the baseline model. Note the very
  steep spectral indices, $\beta_{\textrm{lf}}\lesssim-3.6$, near the
  Galactic plane, with dust-like morphology. These can only be
  reasonably explained in terms of spinning dust.\quad \emph{Top right}:
  ratio between the 857 and 545\,GHz frequency maps, smoothed to
  $1\deg$ FWHM, highlighting the spatially varying temperature of
  thermal dust. The mask is defined by any region for which the
  545\,GHz amplitude is smaller than ten times the 545\,GHz
  monopole.\quad \emph{Bottom left}: difference between the 100-ds1 and
  100-ds2 detector maps, smoothed to $1\deg$, demonstrating the
  presence of CO $J$=1$\rightarrow$0 emission in these
  channels.\quad \emph{Bottom right}: difference between the \WMAP\ W3 and
  W2 differencing assembly maps, smoothed to $1\deg$ FWHM. The excess
  signal near the Galactic centre is due to line emission in the
  94\,GHz channels. The peak amplitude of the difference map is
  740\muK.}
\label{fig:intuition}
\end{figure*}

Our baseline \Planck\ 2015 astrophysical model, derived from a joint
analysis of the \Planck, \WMAP, and 408\,MHz observations, includes
the following components: (1)~CMB; (2)~synchrotron emission;
(3)~free-free emission; (4)~two spinning dust emission components;
(5)~CO line emission at 100, 217, and 353\,GHz; (6)~94/100\,GHz line
emission; (7)~thermal dust emission; and (8)~thermal SZ emission
around the Coma and Virgo clusters. In addition, the model includes
monopole, dipole, recalibration, and bandpass corrections for each
channel.

Many model variations were considered in the preparatory stages of
this analysis, including: (1)~broken or smoothly-curved synchrotron
frequency spectra; (2)~free-free models with fixed power-law indices;
(3)~one-component spinning dust models, based both on physical
\texttt{SpDust} templates and phenomenological second-order polynomial
frequency spectra; (4)~two-component thermal dust models; (5)~thermal
dust models with flattening at low frequencies; (6)~Type-3 CO line
emission; (7)~no 94/100\,GHz line emission component; (8)~full-sky SZ
reconstruction; (9)~no bandpass corrections; and (10)~various priors
for relevant components. It is impossible to fully describe all of
these topics within a single paper, and we will not attempt to do
so. Some low-frequency model variations are, however, discussed in
\citet{planck2014-a31}.

In this paper our focus is the baseline model, which is the simplest
sufficient model considered to date, as defined in terms of three
criteria. First and foremost, the baseline model has to provide a
statistically acceptable fit to the observations over the bulk of the
sky, including most of the Galactic plane. Second, it should have the
smallest possible number of free degrees of freedom that still result
in a statistically acceptable fit to the data. Third and finally, in
cases where there are free choices, physically well-motivated models
are preferred over phenomenological models.

\begin{table*}[tmb]                                                                                                                                              
\begingroup                                                                                                                                   
\newdimen\tblskip \tblskip=5pt
\caption{Summary of full-sky foreground products available from the PLA. Each entry in the first column corresponds to one multi-column and (optionally) multi-extension FITS file, named \texttt{COM\_CompMap\_\{label\}-commander\_\{nside\}\_R2.00.fits}. The various columns in each extension list the posterior maximum, mean, and rms maps, in that order, when available. The values reported in columns 5 to 7 in this table are the mean and standard deviations of these posterior statistic maps.\label{tab:products}} 
\nointerlineskip                                                                                                                                                                                     
\vskip -5mm
\footnotesize                                                                                                                                        
\setbox\tablebox=\vbox{ %
\newdimen\digitwidth                                                                                                                             
\setbox0=\hbox{\rm 0}
\digitwidth=\wd0
\catcode`*=\active
\def*{\kern\digitwidth}
\newdimen\signwidth
\setbox0=\hbox{+}
\signwidth=\wd0
\catcode`!=\active
\def!{\kern\signwidth}
\newdimen\decimalwidth
\setbox0=\hbox{.}
\decimalwidth=\wd0
\catcode`@=\active
\def@{\kern\signwidth}
\halign{ \hbox to 1.5in{#\leaderfil}\tabskip=1.5em&
    \tabskip=0.5em\hfil#\hfil\tabskip=2em&
    \hfil#\hfil\tabskip=1.5em&
    \hfil#\hfil\tabskip=1.5em&
    \hfil#\hfil\tabskip=1.5em&
    \hfil#\hfil\tabskip=1.5em&
    \hfil#\hfil\tabskip=1.5em&
    \hfil#\hfil\tabskip=0pt\cr                                                                                                                                                               
\noalign{\doubleline}
\omit& \omit& \omit&                 \omit& \multispan3\hfil Posterior outside LM93  \hfil& \omit\cr
\noalign{\vskip -3pt}
\omit&\lower3pt\hbox{FITS}&&\lower3pt\hbox{$\nu_{\textrm{ref}}$}&\multispan3\hrulefill&\omit\cr
\noalign{\vskip 2pt}
\omit\hfil File\hfil&extension& Parameter& [GHz/band]& $P_{\textrm{max}}$& Mean& RMS& Unit\cr
\noalign{\doubleline}
\noalign{\vskip 3pt}
\multispan{8}{\bf TEMPERATURE AT $\mathbf{1^{\circ}}$ FWHM, $\mathbf{N_{\textrm{side}}=256}$}\hfil\cr
\noalign{\vskip 5pt}
\hglue 1em\texttt{AME}& 0& $ A_{\textrm{sd1}}$& 22.8& $      *93\pm118 $& $   *92\pm118 $& $    11\pm 3* $& $ \mu\textrm{K}_{\textrm{RJ}}$\cr
\omit\hglue 1em& \omit& $ \nu_{\textrm{sd1}}$& $\cdots$& $    19\pm1*  $& $    19\pm1*  $& $   2.0\pm 0.8$& GHz\cr
\omit\hglue 1em& 1& $ A_{\textrm{sd2}}$& 41& $                14\pm21  $& $    18\pm22  $& $   4.1\pm 2.8$& $ \mu\textrm{K}_{\textrm{RJ}}$\cr
\noalign{\vskip 5pt}
\hglue 1em\texttt{CMB}& 0& $ A_{\textrm{cmb}}$& $\cdots$& $   *3\pm67  $& $    *3\pm67  $& $   1.5\pm 0.8$& $ \mu\textrm{K}_{\textrm{cmb}}$\cr
\noalign{\vskip 5pt}
\hglue 1em\texttt{CO}& 0& $ A_{\textrm{CO10}}$& 100-ds1& $    0.3\pm1.3 $& $   0.4\pm1.3 $& $  0.06\pm0.05$& $ \textrm{K}_{\textrm{RJ}}\,\textrm{km}\,\textrm{s}^{-1}$\cr
\omit\hglue 1em& 1& $ A_{\textrm{CO21}}$& 217-1**& $         0.22\pm0.57$& $  0.29\pm0.57$& $  0.04\pm0.01$& $ \textrm{K}_{\textrm{RJ}}\,\textrm{km}\,\textrm{s}^{-1}$\cr
\omit\hglue 1em& 2& $ A_{\textrm{CO32}}$& 353-ds2& $         0.16\pm0.21$& $  0.26\pm0.26$& $  0.05\pm0.01$& $ \textrm{K}_{\textrm{RJ}}\,\textrm{km}\,\textrm{s}^{-1}$\cr
\noalign{\vskip 5pt}
\hglue 1em\texttt{dust}& 0& $ A_{\textrm{d}}$& 545& $         163\pm228 $& $   163\pm228 $& $  0.66\pm0.11$& $ \mu\textrm{K}_{\textrm{RJ}}$\cr
\omit\hglue 1em& \omit& $ T_{\textrm{d}}$& $\cdots$& $        21\pm2*  $& $    21\pm2*  $& $   1.1\pm0.7 $& K\cr
\omit\hglue 1em& \omit& $ \beta_{\textrm{d}}$& $\cdots$& $  1.53\pm0.05$& $  1.51\pm0.06$& $  0.05\pm0.03$& $ \cdots$\cr
\noalign{\vskip 5pt}
\hglue 1em\texttt{freefree}& 0& $ \EM$& $\cdots$& $            15\pm35  $& $    13\pm35  $& $   2.3\pm2.4 $& $ \textrm{cm}^{-6}\,\textrm{pc}$\cr
\omit\hglue 1em& \omit& $ T_{e}$& $\cdots$& $               7000\pm11**$& $  7000\pm11**$& $   \cdots    $& K\cr
\noalign{\vskip 5pt}
\hglue 1em\texttt{Synchrotron}& 0& $ A_{\textrm{s}}$& 0.408& $ 20\pm15  $& $    20\pm15  $& $   1.1\pm0.2 $& $ \textrm{K}_{\textrm{RJ}}^{\rm a}$\cr
\noalign{\vskip 5pt}
\hglue 1em\texttt{SZ}& 0& $ A_{\textrm{sz}}$& $\cdots$& $     \phantom{^{\rm b}}1.4\pm1.4$$^{\rm b} $& $    \phantom{^{\rm b}}2.0\pm1.3$$^{\rm b}$& $  \phantom{^{\rm b}}0.8\pm0.2$$^{\rm b}$& $ 10^{-6}\,y_{\textrm{sz}}$\cr
\noalign{\vskip 5pt}
\hglue 1em\texttt{xline}$^{\rm c}$& 0& $ A_{\textrm{94/100}}$& 100-ds1& $    0.09\pm0.06$& $    0.9\pm0.8$& $    0.7\pm0.6$& $ \mu\textrm{K}_{\textrm{cmb}}$\cr
\noalign{\vskip 5pt}
\noalign{\vskip 3pt}
\multispan{8}{\bf TEMPERATURE AT $\mathbf{7\parcm5}$ FWHM, $\mathbf{N_{\textrm{side}}=2048}$}\hfil\cr
\noalign{\vskip 5pt}
\hglue 1em\texttt{CO21}$^{\rm d}$& 0& $ A_{\textrm{CO21}}$& 217-1& $    0.2\pm0.8$& $\cdots$& $\cdots$& $ \textrm{K}_{\textrm{RJ}}\,\textrm{km}\,\textrm{s}^{-1}$\cr
\noalign{\vskip 5pt}
\hglue 1em\texttt{ThermalDust}$^{\rm d}$& 0& $ A_{\textrm{d}}$& 545& $    0.2\pm0.8$& $\cdots$& $\cdots$& $ \mu\textrm{K}_{\textrm{RJ}}$\cr
\omit\hglue 1em& \omit& $ \beta_{\textrm{d}}$& $\cdots$& $    1.54\pm    0.07$& $\cdots$& $\cdots$& $ \cdots$\cr
\noalign{\vskip 5pt}
\noalign{\vskip 3pt}
\multispan{8}{\bf POLARIZATION AT $\mathbf{40'}$ FWHM, $\mathbf{N_{\textrm{side}}=256}$}\hfil\cr
\noalign{\vskip 5pt}
\hglue 1em\texttt{SynchrotronPol}$^{\rm d}$& 0& $ P_{\textrm{s}}^{\rm e}$& 30& $   12\pm9*$& $\cdots$& $\cdots$& $ \mu\textrm{K}_{\textrm{RJ}}$\cr
\noalign{\vskip 5pt}
\noalign{\vskip 3pt}
\multispan{8}{\bf POLARIZATION AT $\mathbf{10'}$ FWHM, $\mathbf{N_{\textrm{side}}=1024}$}\hfil\cr
\noalign{\vskip 5pt}
\hglue 1em\texttt{DustPol}$^{\rm d}$& 0& $ P_{\textrm{d}}^{\rm e}$& 353& $    *8\pm10$& $\cdots$& $\cdots$& $ \mu\textrm{K}_{\textrm{RJ}}$\cr
\noalign{\vskip 5pt}
\noalign{\doubleline}
}}
\endPlancktablewide 
\tablenote {{\rm a}} The data file unit is $ \mu\textrm{K}_{\textrm{RJ}}$ but for convenience we list numbers in  $\textrm{K}_{\textrm{RJ}}$ in this table.\par
\tablenote {{\rm b}} Evaluated only over the Coma and Virgo regions.\par
\tablenote {{\rm c}} This is the 94/100\,GHz line emission component. \par
\tablenote {{\rm d}} Only the full-mission maps are summarized in this table, but the data files also include corresponding maps for half-mission, half-year, and half-ring data splits.\par
\tablenote {{\rm e}} Data files contains Stokes $Q$ and $U$ parameters, not the polarization amplitude, $P=\sqrt{Q^2+U^2}$, listed here.\par
\endgroup
\end{table*}

\begin{table*}[h] 
\begingroup 
\newdimen\tblskip \tblskip=5pt
\caption{Monopoles, dipoles, calibration factors and bandpass corrections derived within the baseline temperature model\label{tab:monopole}.}
\nointerlineskip
\vskip -6mm
\footnotesize 
\setbox\tablebox=\vbox{
\newdimen\digitwidth 
\setbox0=\hbox{\rm 0}
\digitwidth=\wd0
\catcode`*=\active
\def*{\kern\digitwidth}
\newdimen\signwidth
\setbox0=\hbox{+}
\signwidth=\wd0
\catcode`!=\active
\def!{\kern\signwidth}
\newdimen\decimalwidth
\setbox0=\hbox{.}
\decimalwidth=\wd0
\catcode`@=\active
\def@{\kern\signwidth}
\def\s#1{\ifmmode $\rlap{$^{\rm #1}$}$ \else \rlap{$^{\rm #1}$}\fi}
\halign{ \hbox to 1in{#\leaderfil}\tabskip=1.0em&
    \hfil#\hfil\tabskip=1em&
    \hfil#\hfil\tabskip=1em&
    \hfil$#$\hfil\tabskip=1em&
    \hfil$#$\hfil\tabskip=1em&
    \hfil$#$\hfil\tabskip=1em&
    \hfil$#$\hfil\tabskip=1em&
    \hfil$#$\hfil\tabskip=1em&
    \hfil$#$\hfil\tabskip=0pt\cr
\noalign{\doubleline}
\omit& Frequency& Detector&\omit\hfil Monopole\hfil&\omit\hfil X dipole\hfil&\omit\hfil Y dipole\hfil&\omit\hfil Z dipole\hfil&
     \omit\hfil Calibration\hfil&\omit\hfil Bandpass shift\hfil\cr
\omit \hfil Survey\hfil& [GHz]& label& [\muK]& [\muK]& [\muK]& [\muK]& [\%]&\omit\hfil[GHz]\hfil\cr                                                                                                                                                                                               
\noalign{\vskip 5pt\hrule\vskip 5pt}
\Planck\ LFI&          *30&\dots&       -17\pm1!*&       0\s{a}&         0\s{a}&       0\s{a}&  -0.3\pm0.1\s{f}& ! 0.3\pm0.1\cr
\omit&                 *44&\dots&         11\pm1*&    0.5\pm0.2&    -0.3\pm0.1!&    0.5\pm0.1& ! 0.3\pm0.1\s{f}& ! 0.1\pm0.1\cr
\omit&                 *70& ds1&          16\pm1*&    0.5\pm0.1&    -1.1\pm0.1!&    1.1\pm0.1& ! 0.0\pm0.1\s{f}&  -0.4\pm1.0\cr
\omit&                    & ds2&          16\pm1*&    0.5\pm0.1&    -1.0\pm0.1!&    1.1\pm0.1& ! 0.1\pm0.1\s{f}& ! 1.1\pm1.0\cr
\omit&                    & ds3&          16\pm1*&  -0.1\pm0.1!&    -0.9\pm0.1!&    0.8\pm0.1&  -0.1\pm0.1\s{f}& ! 0.5\pm1.0\cr
\Planck\ HFI&          100&ds1&            9\s{a}&       0\s{a}&         0\s{a}&       0\s{a}& ! 0.11\pm0.02&    -0.5\pm0.7\cr
\omit&                    & ds2&            8\pm1&    0.0\pm0.1&    -0.1\pm0.1!&    0.3\pm0.2& ! 0.08\pm0.02&    -0.4\pm0.6\cr
\omit&                 143& ds1&          21\s{a}&       0\s{a}&         0\s{a}&       0\s{a}&        0\s{a}& !   0.7\pm0.2\cr
\omit&                    & ds2&          21\pm1*&    0.0\pm0.1&      0.0\pm0.1&  -0.1\pm0.1!&  -0.04\pm0.02&    -0.2\pm0.2\cr
\omit&                    &   5&          21\pm1*&  -0.5\pm0.1!&      0.0\pm0.1&  -0.1\pm0.1!& ! 0.09\pm0.02&    -0.5\pm0.2\cr
\omit&                    &   6&          21\pm1*&  -0.4\pm0.1!&      0.0\pm0.1&  -0.1\pm0.1!& ! 0.12\pm0.02& !   0.3\pm0.2\cr
\omit&                    &   7&          20\pm1*&  -0.2\pm0.1!&      0.0\pm0.1&  -0.0\pm0.1!& ! 0.01\pm0.02&    -0.4\pm0.2\cr
\omit&                 217&   1&          68\pm1*&  -0.8\pm0.1!&    -2.6\pm0.1!&    2.9\pm0.1&        0\s{a}&    -0.1\pm0.1\cr
\omit&                    &   2&          68\pm1*&  -0.7\pm0.1!&    -2.8\pm0.1!&    3.1\pm0.1& ! 0.01\pm0.03&    -0.1\pm0.1\cr
\omit&                    &   3&          67\pm1*&  -1.0\pm0.1!&    -2.6\pm0.1!&    3.0\pm0.1&  !0.00\pm0.03& !   0.1\pm0.1\cr
\omit&                    &   4&          68\pm1*&  -0.4\pm0.1!&    -2.7\pm0.1!&    3.0\pm0.1& ! 0.04\pm0.03&    -0.1\pm0.1\cr
\omit&                 353& ds2&        447\pm5**&      -3\pm1!&        -6\pm1!&        6\pm1& !   0.3\pm0.1& !   0.3\pm0.1\cr
\omit&                    &   1&        449\pm6**&      -4\pm1!&      -16\pm1!*&      15\pm1*& !   0.8\pm0.1&    -0.0\pm0.1\cr
\omit&                 545&   2&      0.37\s{ac}&       0\s{a}&         0\s{a}&       0\s{a}&     -2.8\s{e}&     2.0 \s{e}\cr
\omit&                    &   4&*0.36\pm0.01\s{c}&       0\s{a}&         0\s{a}&       0\s{a}&     -3.2\s{e}&     2.8 \s{e}\cr
\omit&                 857&   2&*0.62\pm0.01\s{c}&       0\s{a}&         0\s{a}&       0\s{a}& !    1.7\s{e}&     5.8 \s{e}\cr
\noalign{\vskip 2pt}
WMAP&                  *23&   K&          -8\pm1!&  -4.5\pm2.0!&      1.6\pm0.5&  -3.7\pm0.4!&        0\s{a}&        0\s{a}\cr
\omit&                 *33&  Ka&           3\s{b}&  -0.7\pm0.6!&    -4.7\pm0.2!&    3.8\pm0.1& !   0.1\pm0.1&        0\s{a}\cr
\omit&                 *41&  Q1&            2\pm1&    0.5\pm0.3&    -4.6\pm0.1!&    3.5\pm0.1&    -0.1\pm0.1&        0\s{a}\cr
\omit&                    &  Q2&            2\pm1&    0.4\pm0.3&    -4.8\pm0.1!&    3.7\pm0.1& !   0.2\pm0.1& !   0.3\pm0.1\cr
\omit&                 *61&  V1&            1\pm1&    0.2\pm0.1&    -5.5\pm0.1!&    4.2\pm0.1& !   0.1\pm0.1&        0\s{a}\cr
\omit&                    &  V2&            1\pm1&    0.0\pm0.1&    -5.5\pm0.1!&    4.2\pm0.1& !   0.3\pm0.1&    -0.1\pm0.1\cr
\omit&                 *94&  W1&          -5\pm2!&    0.3\pm0.1&    -5.3\pm0.2!&    4.1\pm0.2& !   0.3\pm0.1&        0\s{a}\cr
\omit&                    &  W2&          -5\pm2!&    0.1\pm0.1&    -5.0\pm0.1!&    3.9\pm0.2& !   0.4\pm0.1&    -0.7\pm0.3\cr
\omit&                    &  W3&          -6\pm2!&    0.2\pm0.1&    -6.0\pm0.2!&    4.3\pm0.3&    -0.1\pm0.1& !   0.3\pm0.3\cr
\omit&                    &  W4&          -5\pm2!&  -0.0\pm0.1!&    -6.1\pm0.1!&    5.2\pm0.2& !   0.1\pm0.1&    -0.6\pm0.3\cr
\noalign{\vskip 5pt}
Haslam&              0.408&\dots&       8.9\s{bd}&    3.2\s{bd}&      0.7\s{bd}&   -0.8\s{bd}&        0\s{a}&        0\s{a}\cr
\noalign{\vskip 5pt\hrule\vskip 5pt}
}}
\endPlancktablewide
\tablenote {{\rm a}} Fixed at reference value. \par
\tablenote {{\rm b}} Fixed at values derived by Wehus et al.\ (2014). \par
\tablenote {{\rm c}} Unit is MJy/sr. \par
\tablenote {{\rm d}} Unit is K. \par
\tablenote {{\rm e}} For a detailed discussion of bandpass and calibration uncertainties at 545 and 857\,GHz, see Sect.~\ref{sec:instrumental_effects}.\par
\tablenote {{\rm f}} Adjusted for the well-understood LFI ``beam normalization factor'' (see \citealt{planck2014-a03}).\par
\endgroup
\end{table*}

Three examples will illustrate our approach. First, for synchrotron emission we have
found that a broken power-law model, i.e., a power-law model with
fixed but different spectral indices above and below some break
frequency (say, $3\,\textrm{GHz}$) provides an equally good
$\chi^2$ fit as the \texttt{GALPROP} model do.  A straight power-law 
spectrum, on the other hand, does not fit the data, because the
408\,MHz map appears dimmer than expected from a straight
extrapolation from 23\,GHz to 408\,MHz, assuming the spectral index
of $\beta\approx-3.0$ to $-3.2$ required to fit higher
frequencies. The reasons for preferring the \texttt{GALPROP} model are
therefore not data-driven, but rather that it requires less tuning
(e.g., no choice of break frequency, no free spectral indices, very
weak dependency on $\nu_{\textrm{p}}^{\textrm{s}}$), and that it is
based on a physically well-motivated calculation
\citep{strong2011,orlando2013}. The cost, however, is less flexibility for
tracing real spatial variations in the synchrotron spectral index, and
thereby possibly greater cross-talk between synchrotron, free-free, and
spinning dust. Nevertheless, in the absence of sufficiently strong data to
disentangle these variations, we consider this a lesser evil than
introducing a very strong statistical degeneracy between synchrotron,
free-free, and spinning dust.

A second important example is the spinning dust component, which is
currently implemented in terms of two independent $\texttt{SpDust}$
components, one with a free peak frequency, $\nu_{\textrm{p}}$, per
pixel, and one with a spatially constant peak frequency, for a total
of three free effective spinning dust parameters per pixel. While the
introduction of the first component is unavoidable when combining the
\Planck\ and \WMAP\ observations, the necessity of the second is more
subtle.  Without it we invariably find highly significant residuals
(many tens to a few hundreds of microkelvin in the Galactic plane) in
the 60--70\,GHz frequency range, with dust-like morphology. This
strongly suggests a model problem with either spinning or thermal dust
(or both), but so far the only acceptable solution we have found is
effectively to ``widen'' the \texttt{SpDust} spectrum slightly, which
is most easily implemented by adding a second independent component.
Introducing, say, a flatter thermal dust index at around 100\,GHz
tends to exacerbate rather than ameliorate the problem. We emphasize,
however, that the current two-component spinning dust model is a
purely phenomenological model introduced in the absence of physically
well-motivated alternatives.  We fully anticipate that additional
low-frequency data or further theoretical developments will improve
this model significantly in the future.

Third, we currently adopt a one-component greybody model with free
emissivity index and temperature for thermal dust
emission. Experimenting with various two-component alternatives, we
find only one absolute requirement on the thermal dust model for
frequencies up to 857\,GHz, namely that at least three free thermal
dust parameters per pixel are required to achieve an acceptable fit to
the high-frequency observations. Whether those are
$\{A_{\textrm{d}},\beta_{\textrm{d}},T_{\texttt{d}}\}$, 
$\{A^1_{\textrm{d}},T^1_{\texttt{d}}, A^{2}_{\textrm{d}}\}$, or even
$\{A^1_{\textrm{d}},\beta^1_{\texttt{d}}, A^{2}_{\textrm{d}}\}$(!) is
not clear from the current data set\footnote{Superscripts 1 and 2
refer to two independent greybody components.}. On the other hand,
it \emph{is} clear that additional parameters are not required to
model the current data set. For now, we prefer the one-component model
simply because it has fewer global parameters than a corresponding
two-parameter model, i.e., it requires no spatially fixed
$\beta_{\textrm{d}}$ and $T_{\textrm{d}}$ for a second component, and
therefore requires less tuning. This is not to be taken as a statement
on the relative physical merits of the two models, however.  Additional
high-frequency observations are required to distinguish between them.

\subsection{Data preview}
\label{sec:preview}

Before presenting the actual baseline results, it is useful to
visually consider a few specific data combinations in order to gain
some intuition regarding the power of these data for constraining
specific parameters. In Fig.~\ref{fig:intuition}, we show four
different sky maps, each of which highlights an important and distinct
feature in the data. A similar discussion based on an Internal Linear
Combination (ILC) method is presented in Sect.~3 of
\citet{planck2014-a31}.

Starting with the top panel, we plot the effective power-law index of
the combined \Planck, \WMAP, and 408\,MHz data, when fitting only a
single power-law model at low-frequencies, as opposed to fitting
individual synchrotron, free-free, and spinning dust components. All
other components are defined by the usual baseline model. On the
low-frequency side, this approach is thus identical to the 2013 model
\citep{planck2013-p06}, although the data volume and frequency range
are significantly increased. Despite these differences, the 2013 and
2015 low-frequency spectral index maps agree reasonably well, with a
mean and standard deviation difference of $0.16\pm0.14$.

The main features in this spectral index map can be broadly
characterized into two types of spatial signatures. First, we see many
distinct regions (e.g., the Gum nebula, Orion, Zeta Oph, and the Cygnus
complex) with a shallow spectral index of $\beta\approx-2$ and 
morphology as expected from various free-free tracers
\citep[e.g.,][]{alves2014}. Second, there are extended dark regions
surrounding the Galactic plane, with very steep spectral indices of
$\beta\lesssim-3.6$, and morphology similar to that of thermal dust
emission. In addition, it is possible to see some weak hints of the
North Galactic Spur, a strong synchrotron emission region,
but since we adopted a synchrotron-like prior of $\beta\sim N(-3,0.2)$
for the low-frequency power-law index in this analysis, it is
difficult to distinguish this region from the prior-dominated
background. The main point, however, is that even with such minimal
modelling, there is strong evidence for at least three distinct
physical components at low frequencies, namely: synchrotron, with
$\beta\approx-3$; free-free, with $\beta \approx -2$; and a dust-like
component, with $\beta\lesssim-3.6$. Among all the spectra shown in
Fig.~\ref{fig:components}, only a spinning dust component with
$\nu_{\textrm{p}}\lesssim30\,\textrm{GHz}$ can reasonably account for
the latter.

In the second panel we show the ratio between the 857 and 545\,GHz
frequency maps, masking all pixels for which the 545\,GHz amplitude is
smaller than ten times its monopole. The spatial variations seen in
this map cannot be explained either in terms of calibration or
bandpass errors (because it is a ratio map, and either of those errors
primarily affects the scale, not the morphology) or in terms of
absolute offsets (because of the high mask threshold).  They are
robust features of the Galactic thermal dust emission, and must be
explained by any Galactic model that include these observations. In
fact, within our baseline model, these features can only be explained
in terms of a spatially varying dust temperature. To be specific, the
thermal dust appears hotter (i.e., has larger 857-to-545 ratio; see
Fig.~\ref{fig:components}) near the Galactic centre than in the outer
Galaxy (quadrants 2 and 3).  Adopting a spatially constant dust
temperature is no longer possible, and the only reason that the
corresponding 2013 analysis could produce meaningful results with such
an assumption was that it considered frequencies only up to 353\,GHz,
and also focused primarily on high Galactic latitudes
\citep{planck2013-p06}.

The third panel shows the straight difference between the 100-ds1 and
100-ds2 maps. Because of the very short lever arm between these two
frequencies, essentially all continuous emission mechanisms cancel in
this difference, leaving only the sharp line emission from CO
$J$=1$\rightarrow$0, as well as instrumental noise. This demonstrates the
power of employing detector maps in the following analysis as opposed
to frequency maps; using a fine-grained data set vastly increases our
ability to extract line emission.

\begin{table}[t]                                                                                                                                             
\begingroup                                                                                                                                     
\newdimen\tblskip \tblskip=5pt
\caption{Posterior mean and rms values for spatially constant parameters in the temperature model. The uncertainties on these values are dominated by modelling errors that are difficult to quantify properly, and are omitted here; see main text for further discussion. The quoted line ratios refer to amplitudes relative to the respective reference band.}
\label{tab:Tglobal}
\nointerlineskip                                                                                                                                                                                     
\vskip -4mm
\footnotesize                                                                                                                                        
\setbox\tablebox=\vbox{                                                                                                                                                                             
\newdimen\digitwidth                                                                                                                            
\setbox0=\hbox{\rm 0}
\digitwidth=\wd0
\catcode`*=\active
\def*{\kern\digitwidth}
\newdimen\signwidth
\setbox0=\hbox{+}
\signwidth=\wd0
\catcode`!=\active
\def!{\kern\signwidth}
\newdimen\decimalwidth
\setbox0=\hbox{.}
\decimalwidth=\wd0
\catcode`@=\active
\def@{\kern\signwidth}
\halign{ \hbox to 2.23in{#\leaderfil}\tabskip=1em&
    \hfil#\hfil\tabskip=1em&
    \hfil#\hfil\tabskip=0pt\cr                                                                                                                                                                 
\noalign{\doubleline}
\omit&Detectors\cr
\omit\hfil Quantity\hfil&or Band&Value\cr
\noalign{\vskip 4pt\hrule\vskip 5pt}
Synchrotron freq scale factor, $\alpha$&\dots& 0.26$^{\textrm{b}}$\cr
\noalign{\vskip 2pt}
Spinning dust secondary peak freq, $\nu_{\textrm{p}}^{\textrm{sd2}}$&\dots& 33.35\,GHz$^{\textrm{b}}$\cr
\noalign{\vskip 2pt}
CO $J$=1$\rightarrow$0 line ratio& 100-ds1& $1^{\textrm{a}}$\cr
\omit&  100-ds2&    $1.02\pm0.01$\cr
CO $J$=2$\rightarrow$1 line ratio& 217-1& $1^{\textrm{a}}$\cr
\omit&  217-2& $1.07\pm0.01$\cr
\omit&  217-3& $1.15\pm0.01$\cr
\omit& 217-4& $1.18\pm0.01$\cr
CO $J$=3$\rightarrow$2 line ratio& 353-ds2& $1^{\textrm{a}}$\cr
\omit&  353-1& $1.3\pm0.1$\cr
\noalign{\vskip 2pt}
94/100 GHz line ratio& 100-ds1& $1^{\textrm{a}}$\cr
\omit&  100-ds2& $1.4\pm0.3$\cr
\omit&  W1& $4.6\pm3.2$\cr
\omit&  W2& $4.2\pm2.9$\cr
\omit&  W3& $5.3\pm3.7$\cr
\omit&  W4& $4.3\pm3.0$\cr
\noalign{\doubleline}
}}
\endPlancktable 
\tablenote {{\rm a}} Reference channel. \par
\tablenote {{\rm b}} Only the maximum-likelihood value is provided; see main text. \par 
\endgroup
\end{table}

Whereas CO line emission was studied extensively in the \Planck\ 2013
release, an additional new 94--100\,GHz line feature is included in
the current release (see Sect.~\ref{sec:sky}). The primary
contribution to this component is visualized in the bottom panel of
Fig.~\ref{fig:intuition}, in terms of the \WMAP\ W3$-$W2 difference
map. The small but bright features near the Galactic centre (the
maximum amplitude is 740$\,\mu\textrm{K}$) are line emission within
the W-band, as discussed in Sect.~\ref{sec:sky}. This line is 
also covered by the two \Planck\ 100\,GHz channels. To prevent this
additional emission from contaminating the main CO estimates, and to
achieve an acceptable fit for the \WMAP\ W-channels, we fit for this
additional line emission component at 94/100$\,\textrm{GHz}$.

\subsection{Baseline model} 
\label{sec:baseline}

\begin{figure*}[p]
  \begin{center}
    \mbox{
      \epsfig{figure=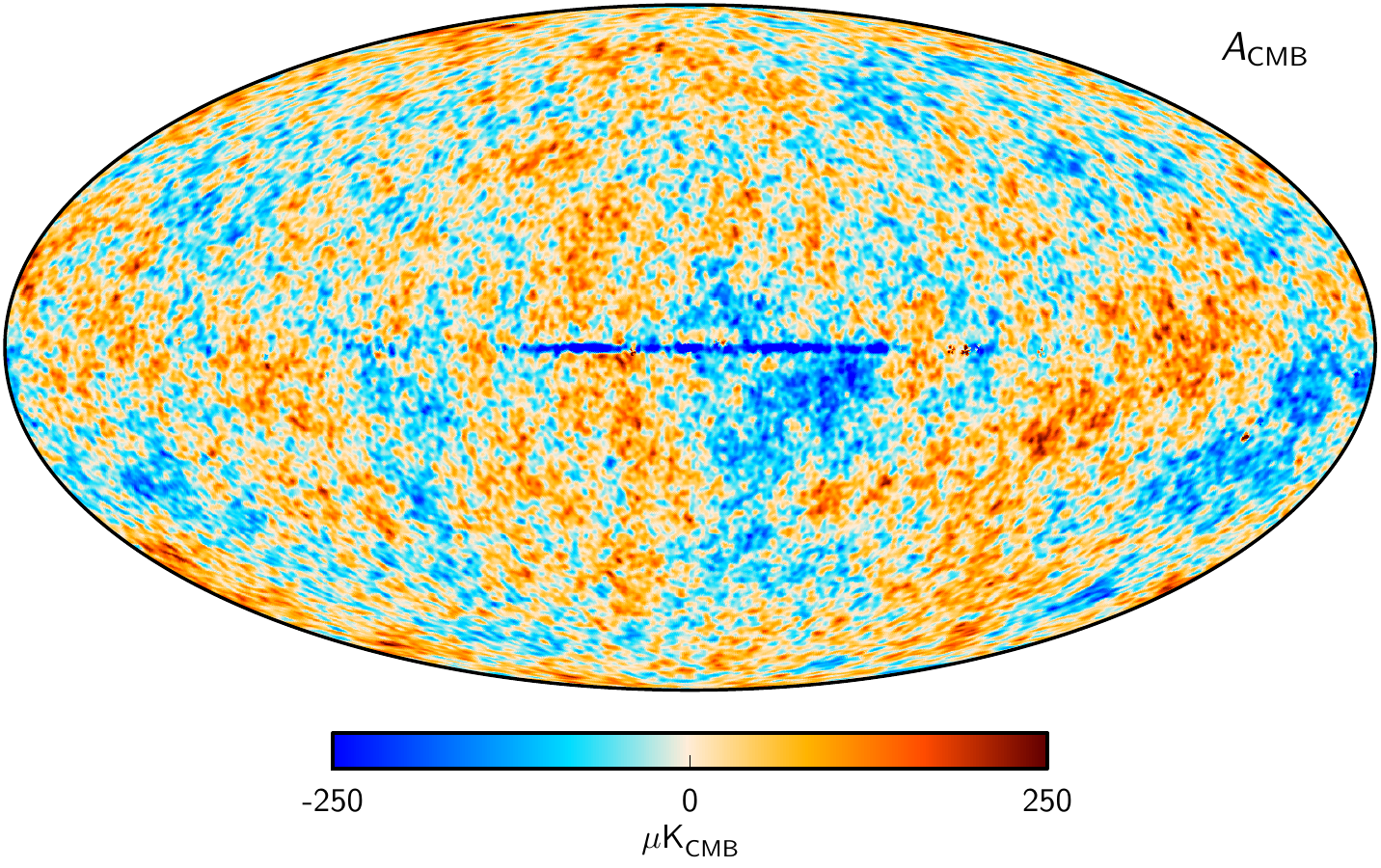,width=1.00\linewidth,clip=}
    }
    \vskip 2mm
    \mbox{
      \epsfig{figure=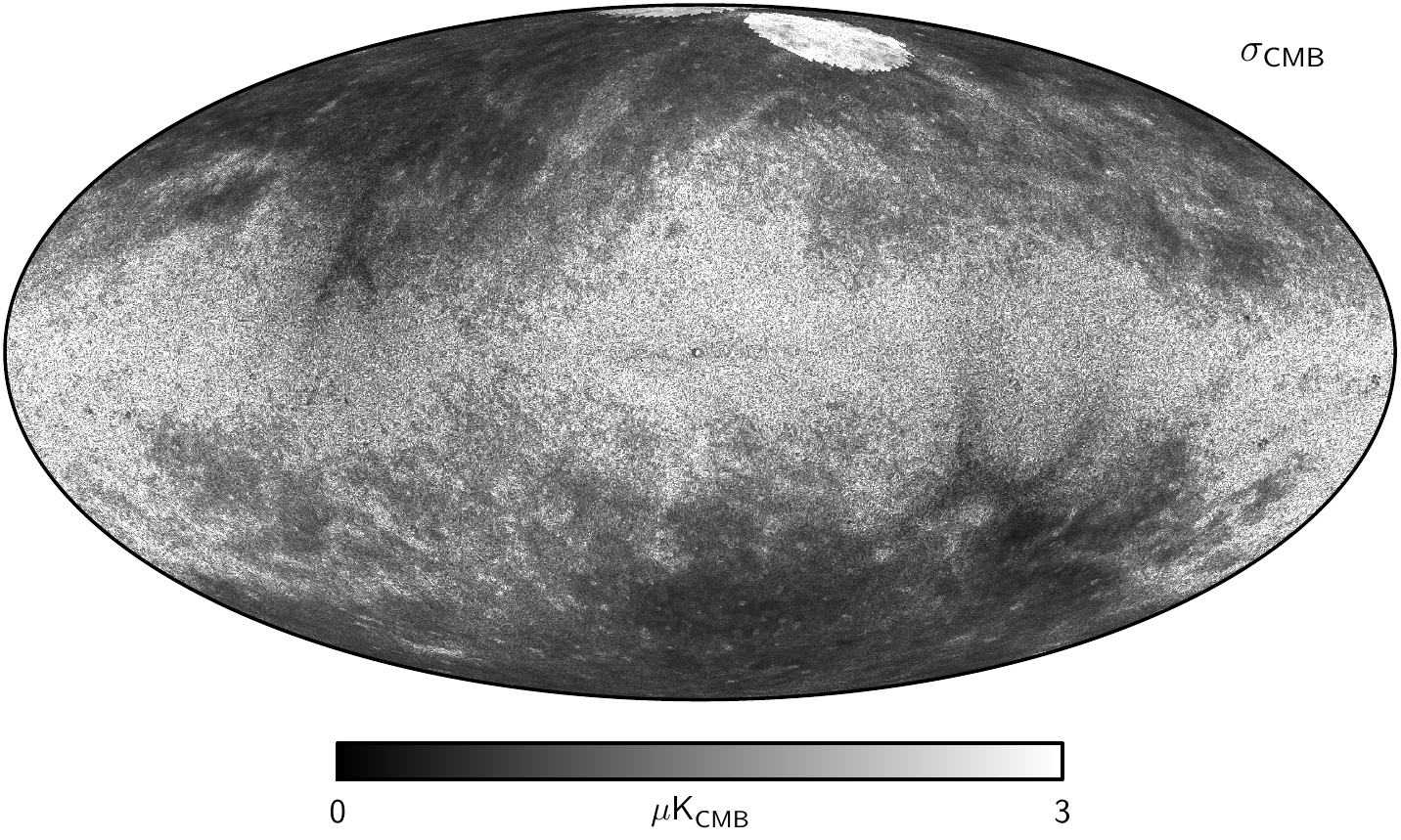,width=1.00\linewidth,clip=}
    }
  \end{center}
  \caption{Maximum posterior (\emph{top}) and posterior rms
    (\emph{bottom}) CMB intensity maps derived from the joint baseline
    analysis of \Planck, \WMAP, and 408\,MHz observations. The two
    circular regions close to the North Galactic Pole in the rms map
    correspond to the Coma and Virgo clusters, for which the thermal
    SZ efect is fitted together with the primary diffuse
    components. Note also that the rms map includes statistical errors
    alone, not modelling errors, and they are therefore only
    meaningful in regions where the corresponding $\chi^2$ is
    acceptable; see Fig.~\ref{fig:chisq_map}.  }
  \label{fig:cmb_amp_map}
  
\end{figure*}

\begin{figure*}[p]
  \begin{center}
    \mbox{
      \epsfig{figure=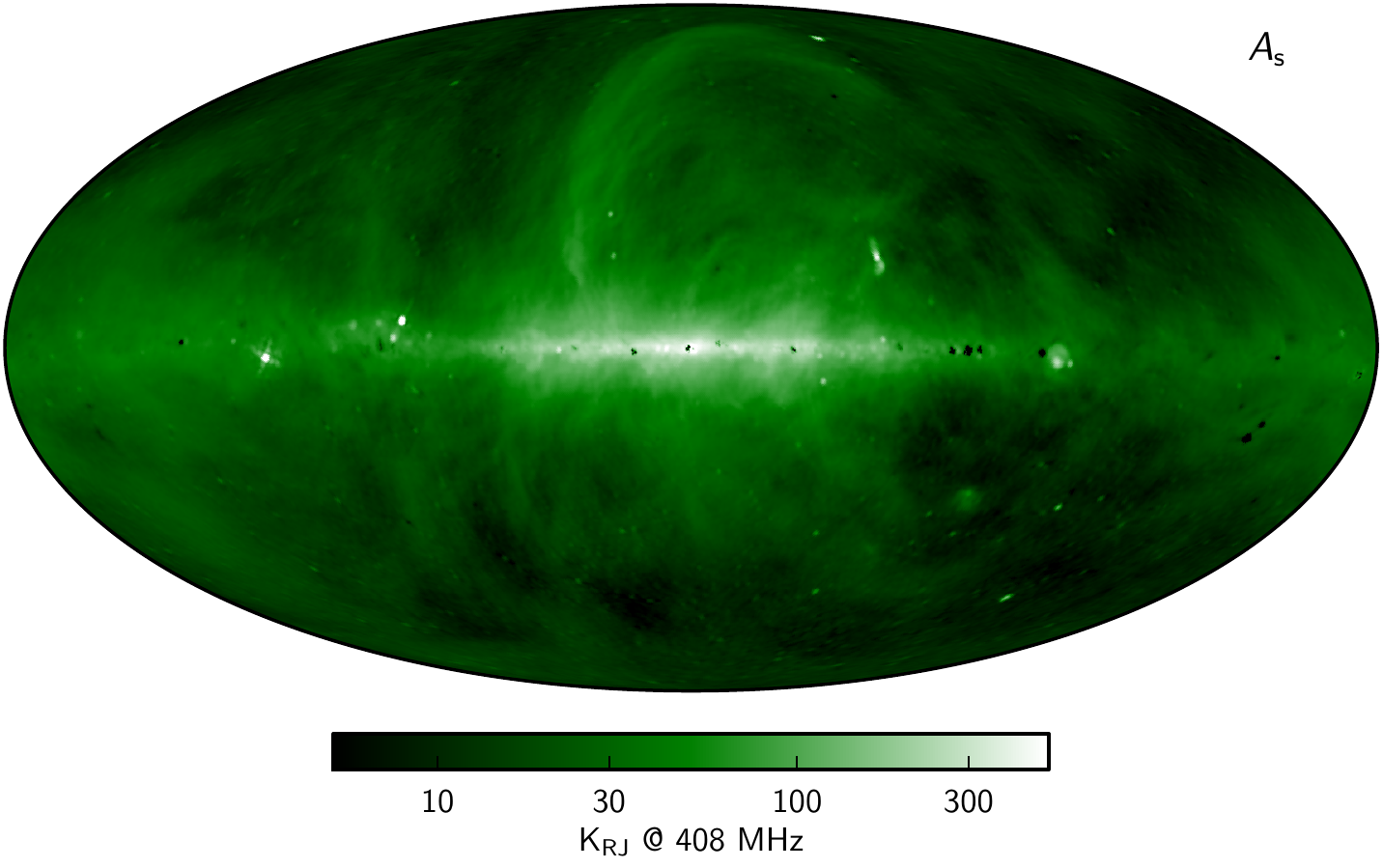,width=1.00\linewidth,clip=}
    }
    \vskip 4mm
    \mbox{
      \epsfig{figure=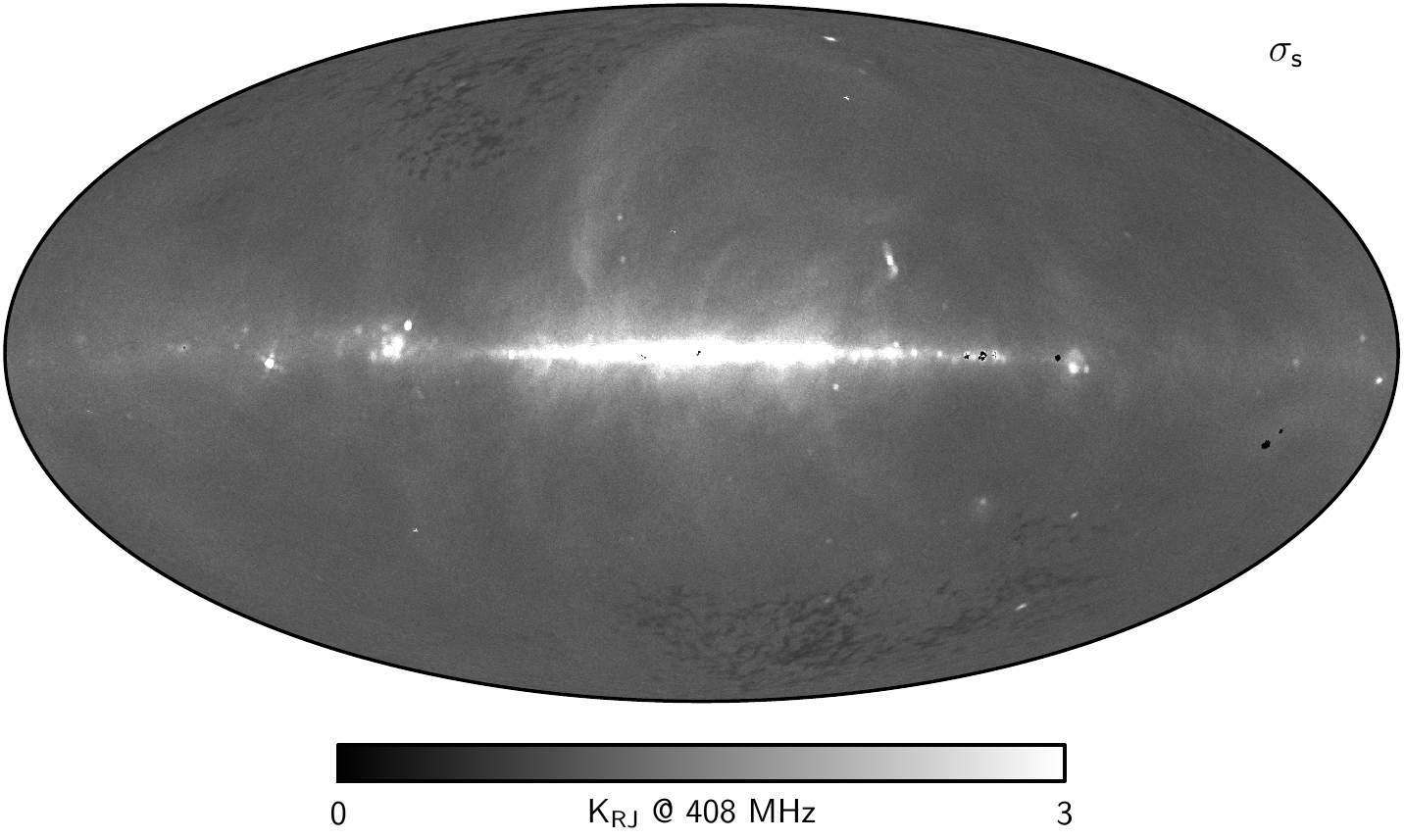,width=1.00\linewidth,clip=}
    }
  \end{center}
  \caption{Maximum posterior (\emph{top}) and posterior rms
    (\emph{bottom}) synchrotron intensity maps derived from the joint baseline
    analysis of \Planck, \WMAP, and 408\,MHz observations.}
  \label{fig:synch_amp_map}
  
\end{figure*}

\begin{figure*}[p]
  \begin{center}
    \mbox{
      \epsfig{figure=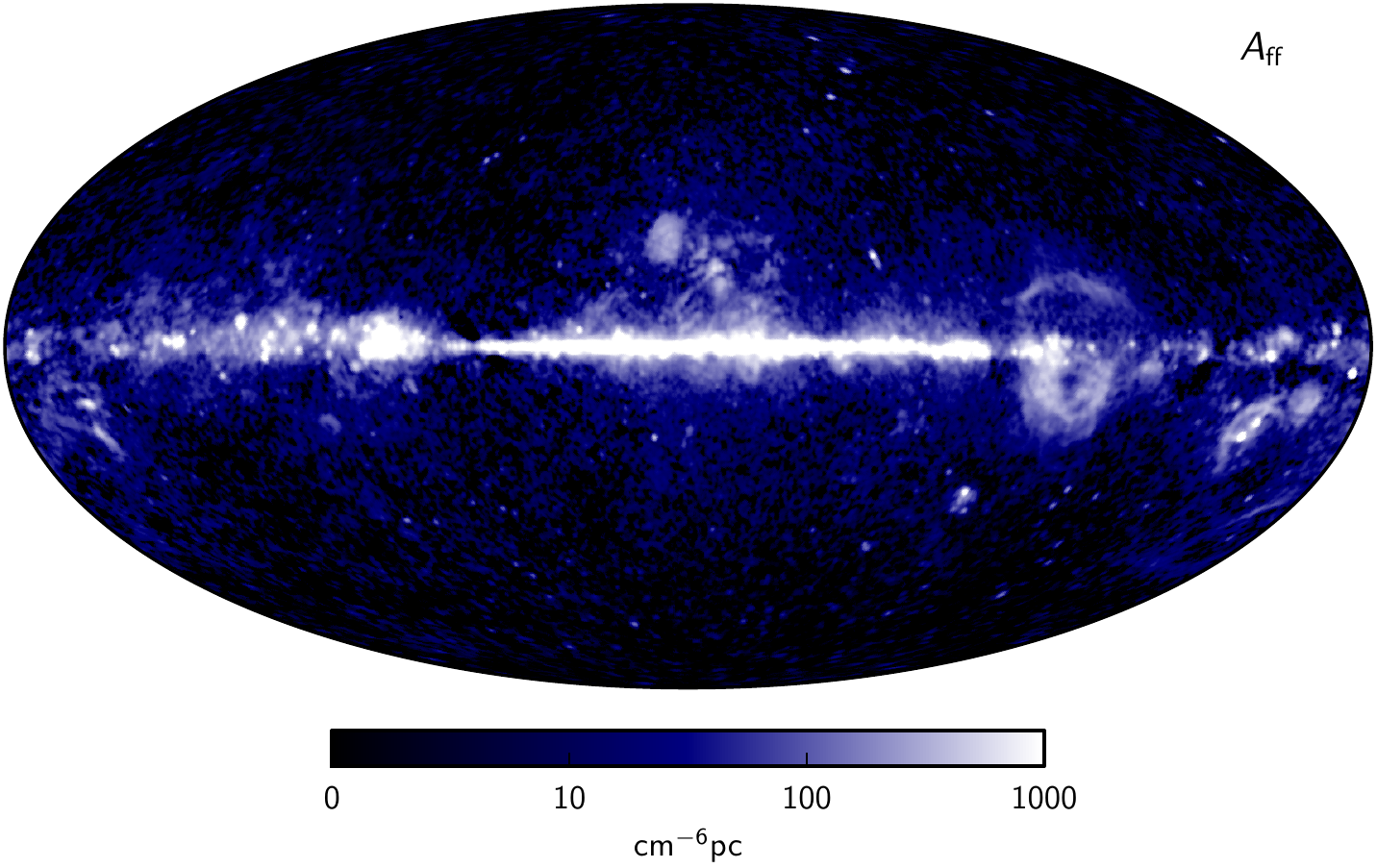,width=1.00\linewidth,clip=}
    }
    \vskip 4mm
    \mbox{
      \epsfig{figure=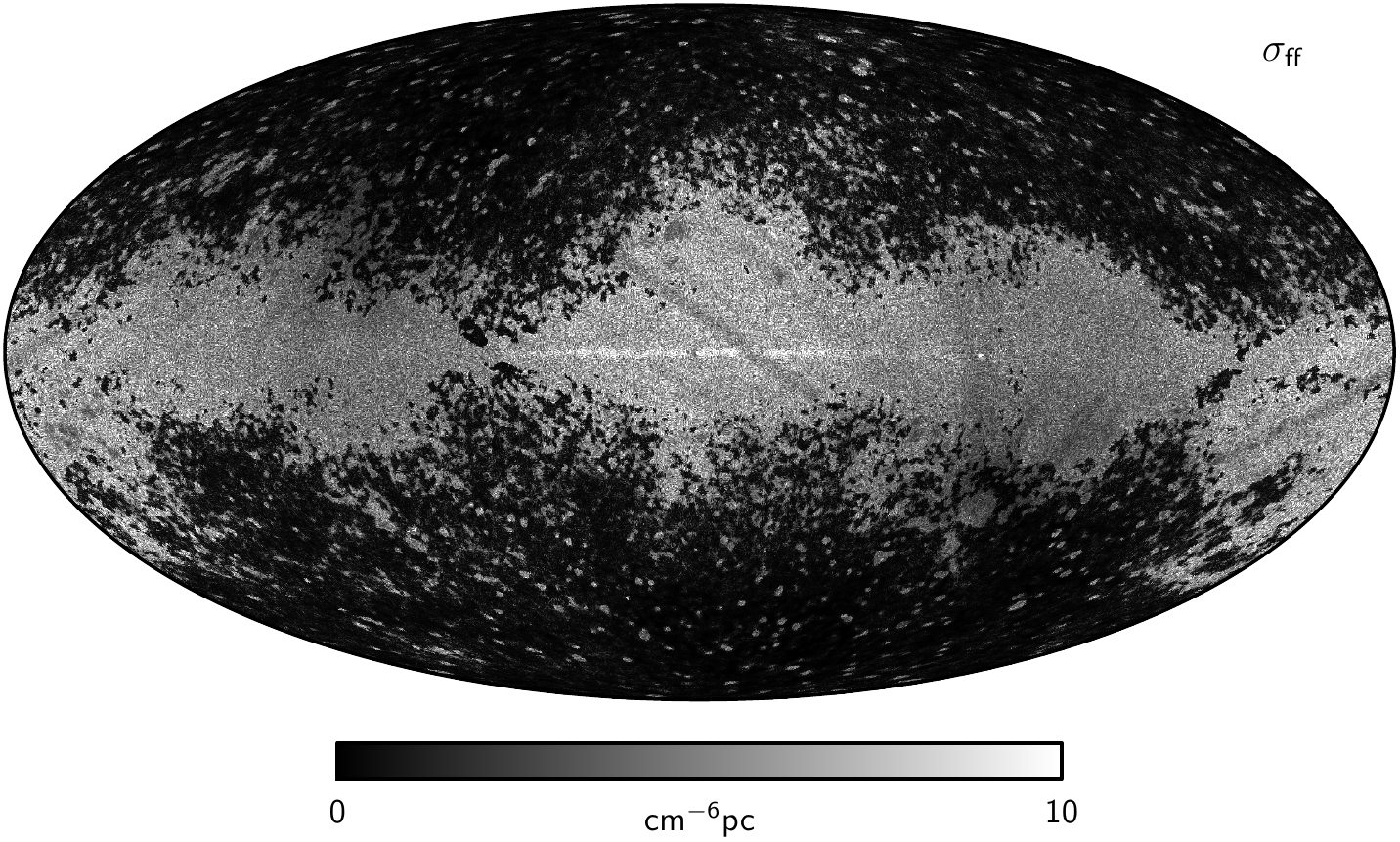,width=1.00\linewidth,clip=}
    }
  \end{center}
  \caption{Maximum posterior (\emph{top}) and posterior rms
    (\emph{bottom}) free-free emission measure maps
    derived from the joint baseline analysis of \Planck, \WMAP, and
    408\,MHz observations.}
  \label{fig:ff_EM_map}
  
\end{figure*}

\begin{figure*}[p]
  \begin{center}
    \mbox{
      \epsfig{figure=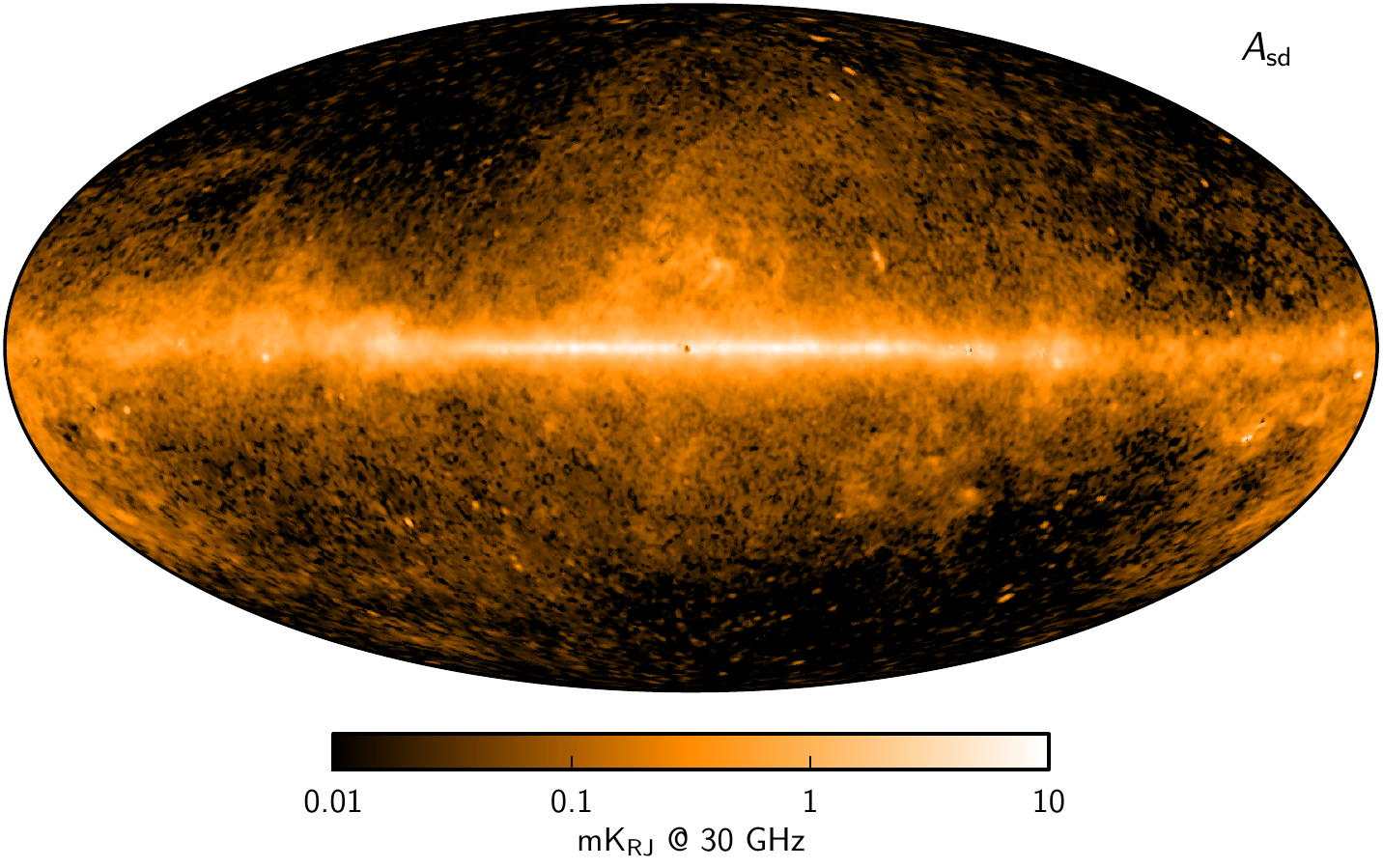,width=1.00\linewidth,clip=}
    }
    \vskip 4mm
    \mbox{
      \epsfig{figure=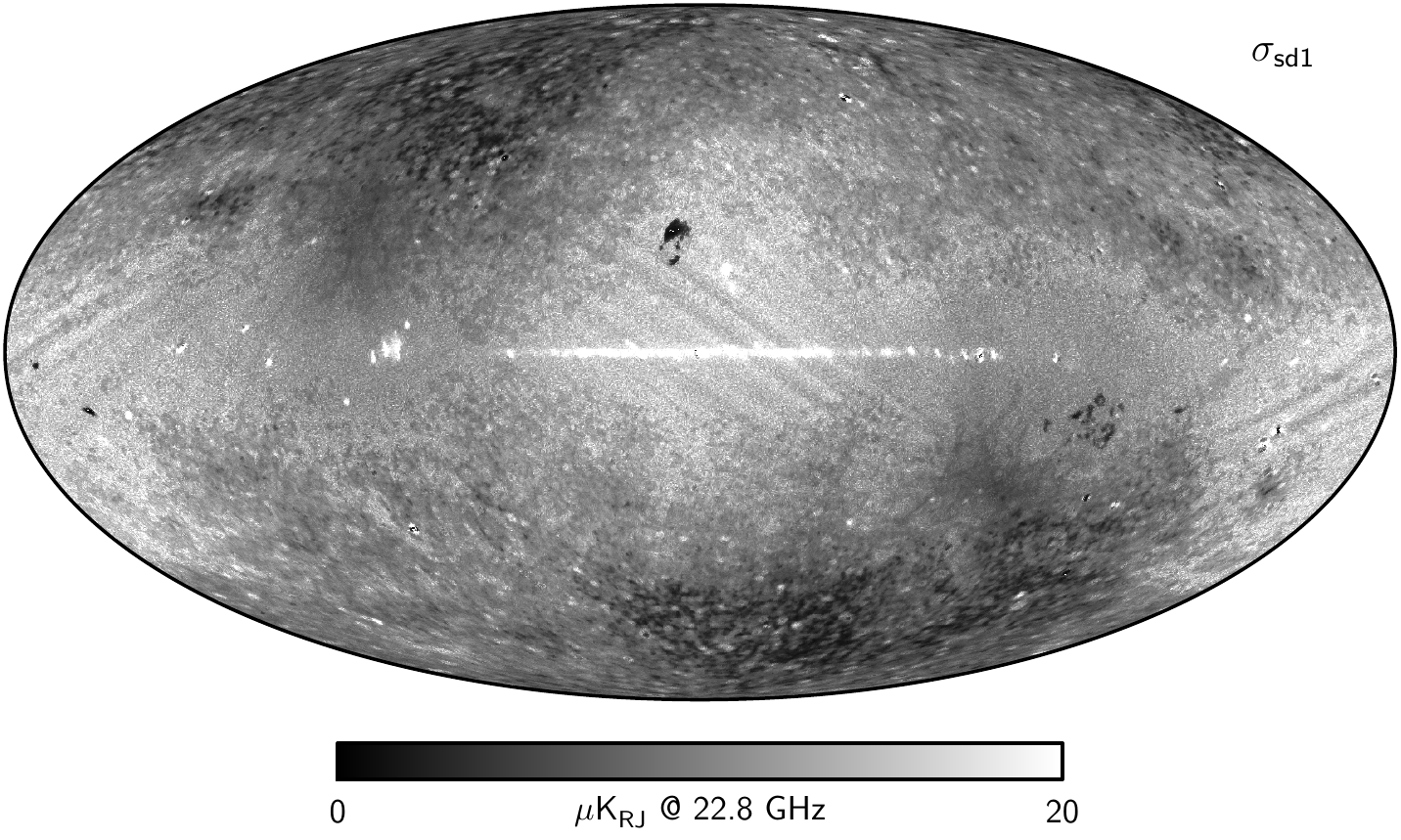,width=1.00\linewidth,clip=}
    }
  \end{center}
  \caption{Maximum posterior (\emph{top}) and posterior rms
    (\emph{bottom}) spinning dust intensity maps derived from the joint baseline
    analysis of \Planck, \WMAP, and 408\,MHz observations. The top
    panel shows the sum of the two spinning dust components in the
    baseline model, evaluated at 30\,GHz, whereas the bottom shows
    the standard deviation of only the primary spinning dust
    component, evaluated at 22.8\,GHz.}
  \label{fig:ame_amp_map}
  
\end{figure*}

\begin{figure*}[p]
  \begin{center}
    \mbox{
      \epsfig{figure=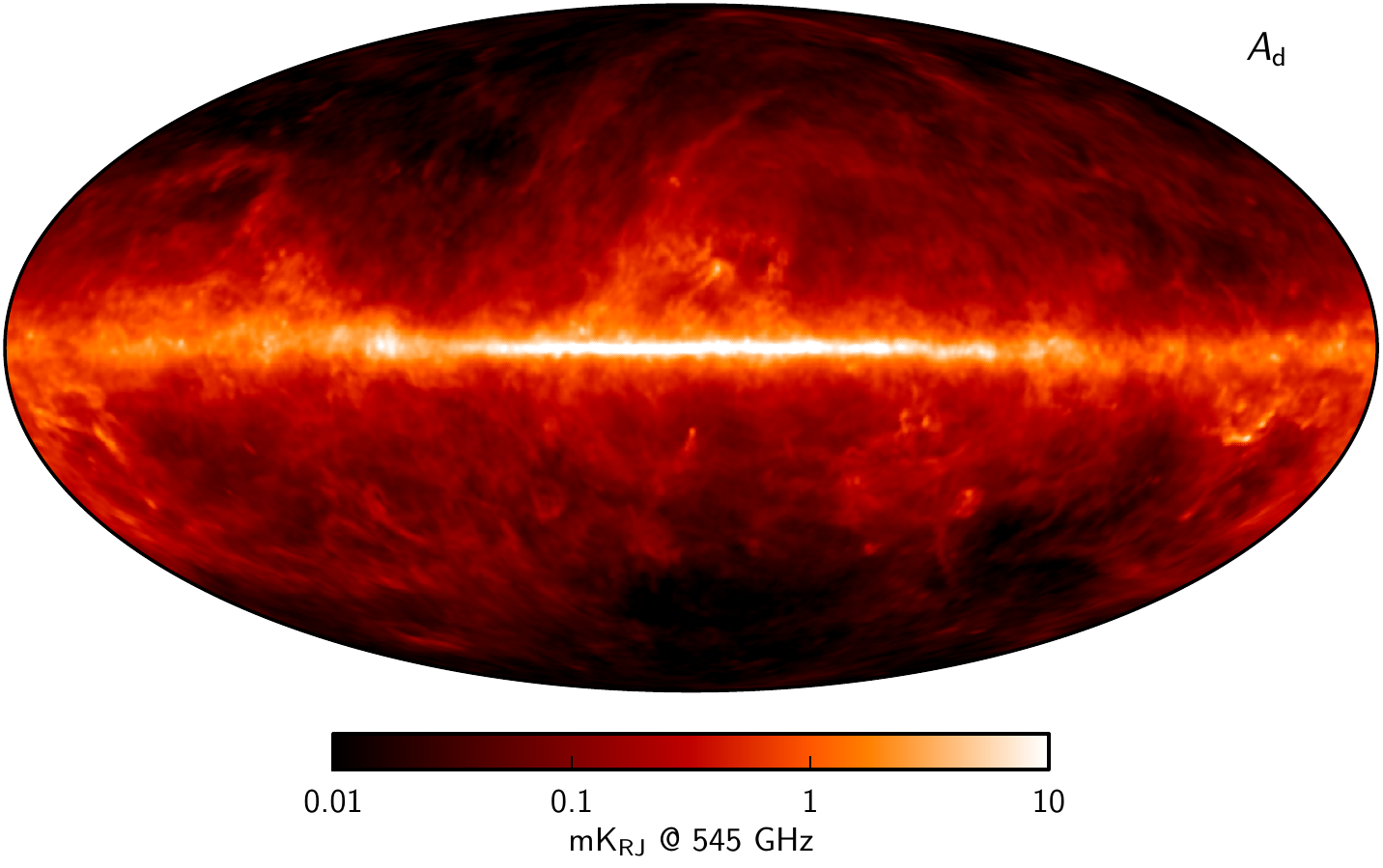,width=1.00\linewidth,clip=}
    }
    \vskip 4mm
    \mbox{
      \epsfig{figure=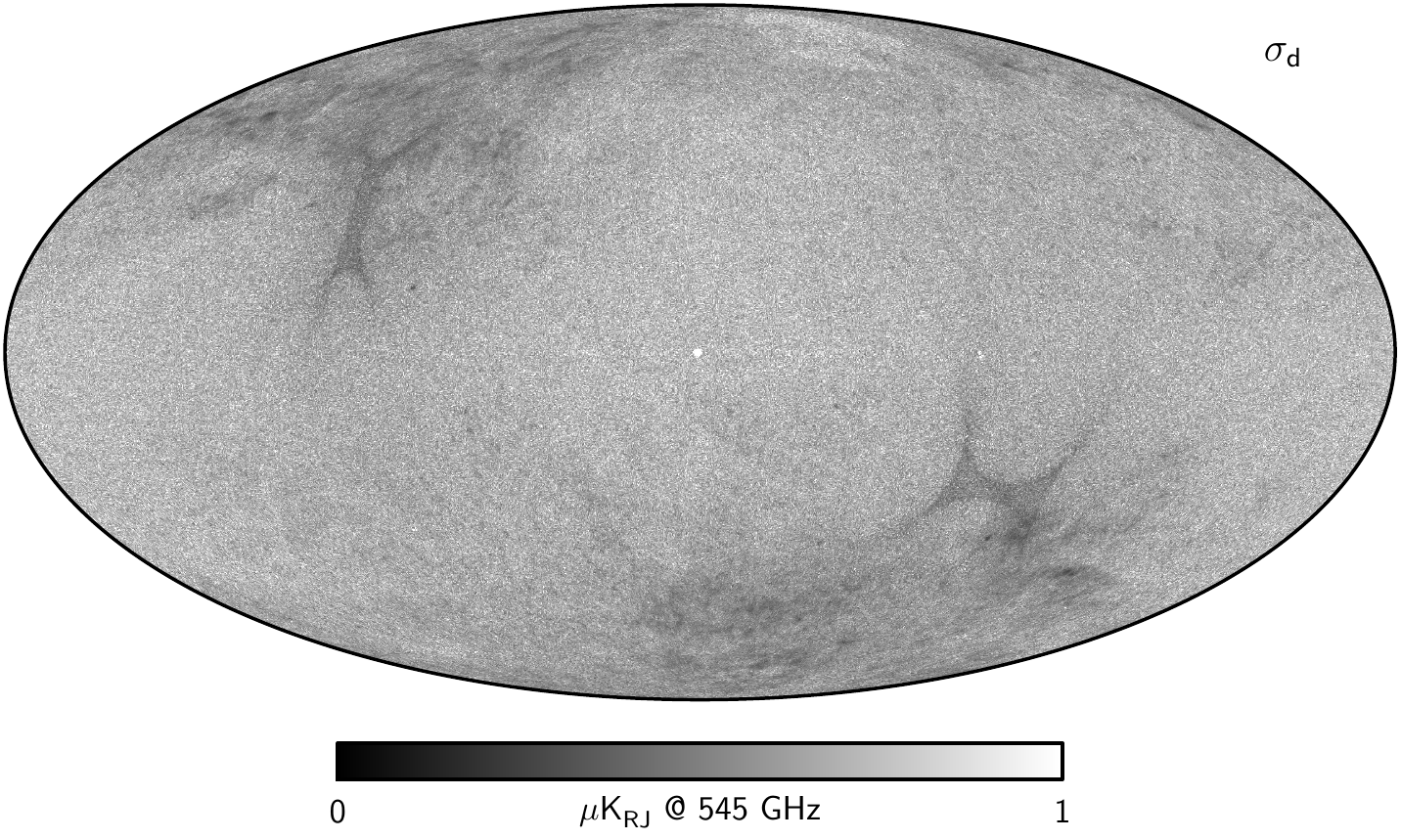,width=1.00\linewidth,clip=}
    }
  \end{center}
  \caption{Maximum posterior (\emph{top}) and posterior rms
    (\emph{bottom}) thermal dust intensity maps derived from the joint baseline
    analysis of \Planck, \WMAP, and 408\,MHz observations.}
  \label{fig:dust_amp_map}
\end{figure*}

\begin{figure*}[p]
  \begin{center}
    \mbox{
      \epsfig{figure=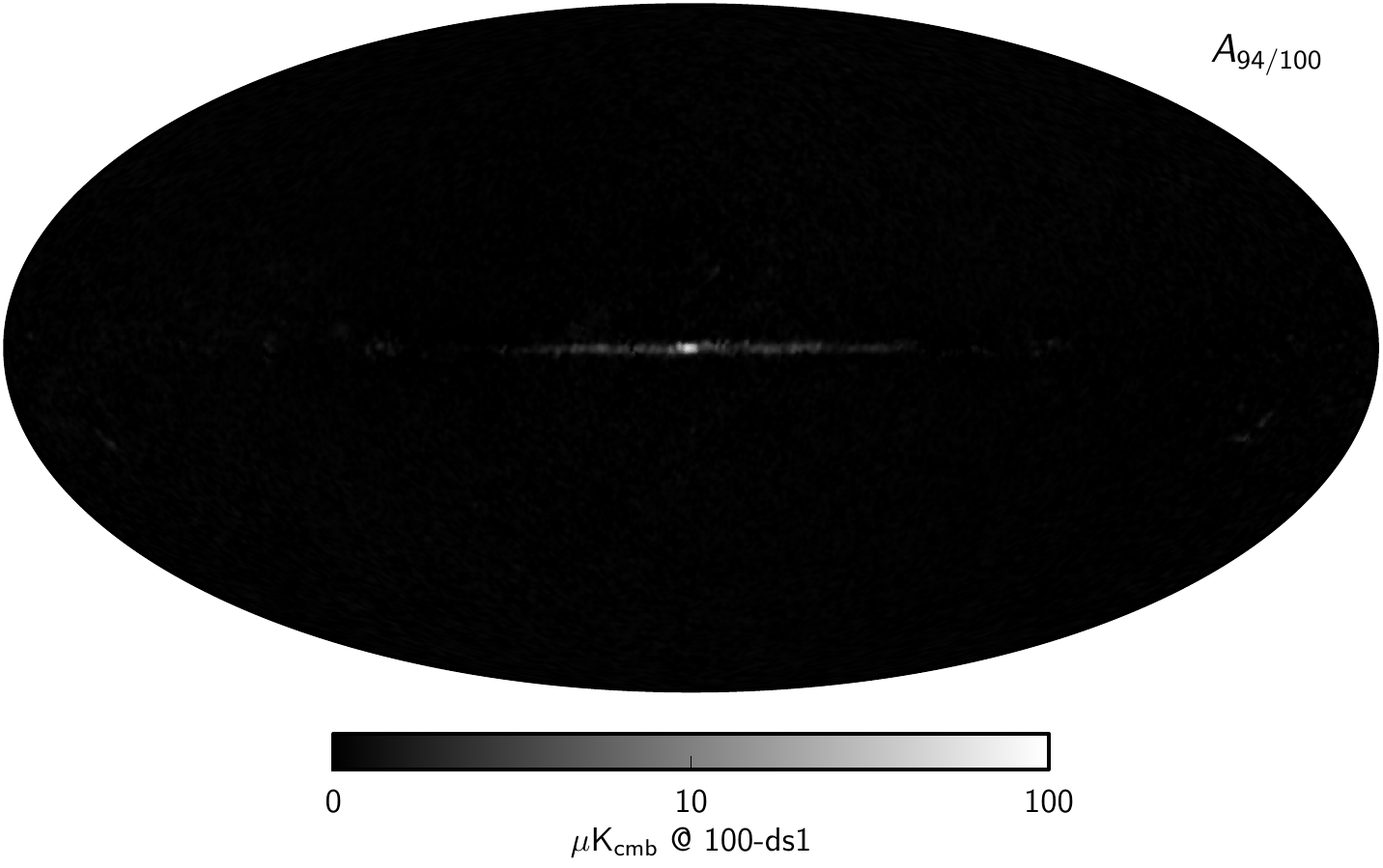,width=1.00\linewidth,clip=}
    }
    \vskip 4mm
    \mbox{
      \epsfig{figure=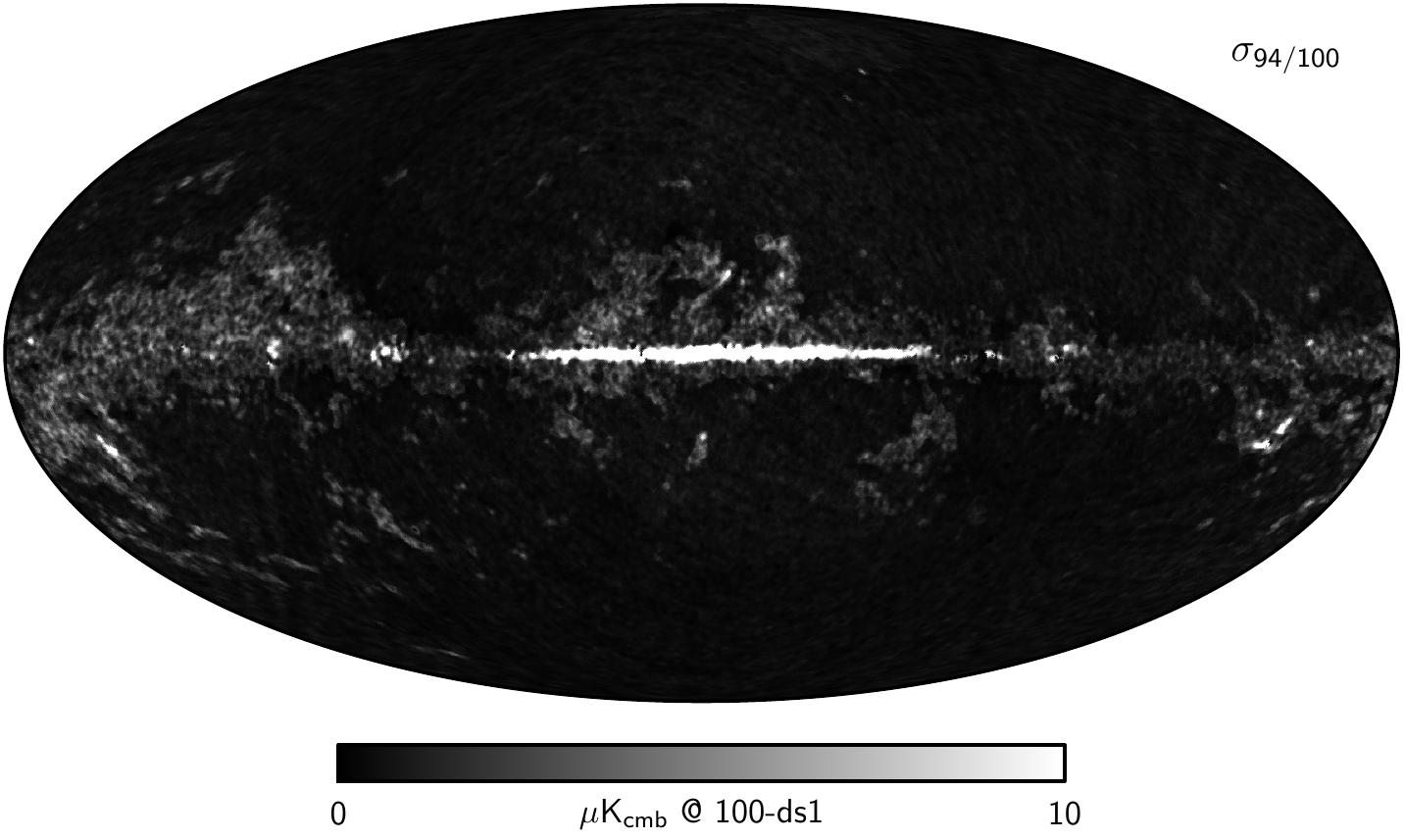,width=1.00\linewidth,clip=}
    }
  \end{center}
  \caption{Maximum posterior (\emph{top}) and posterior rms
    (\emph{bottom}) 94/100\,GHz line emission maps derived from the joint baseline
    analysis of \Planck, \WMAP, and 408\,MHz observations.}
  \label{fig:hcn_amp_map}
\end{figure*}

\begin{figure*}[p]
  \begin{center}
    \mbox{
      \epsfig{figure=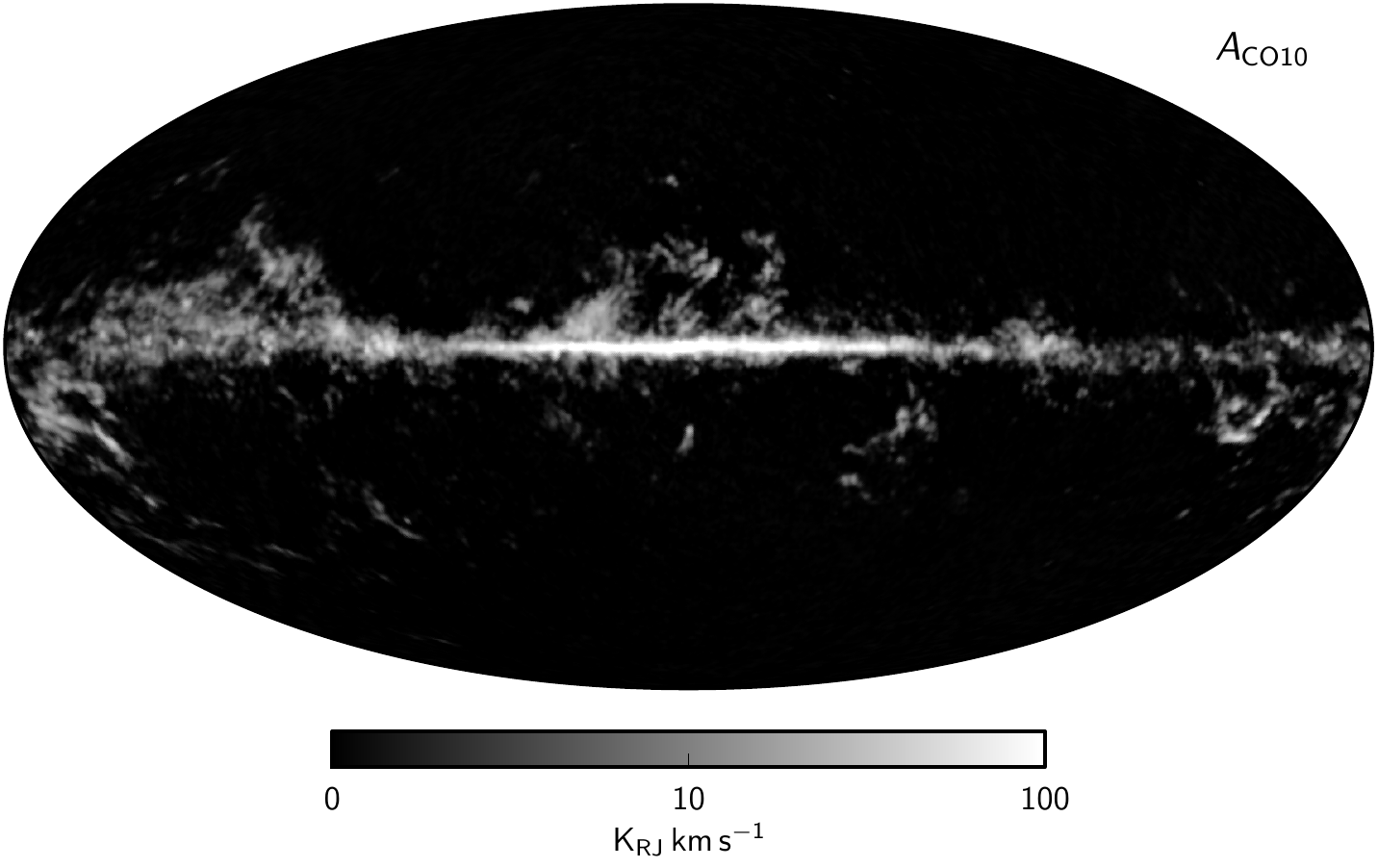,width=1.00\linewidth,clip=}
    }
    \vskip 4mm
    \mbox{
      \epsfig{figure=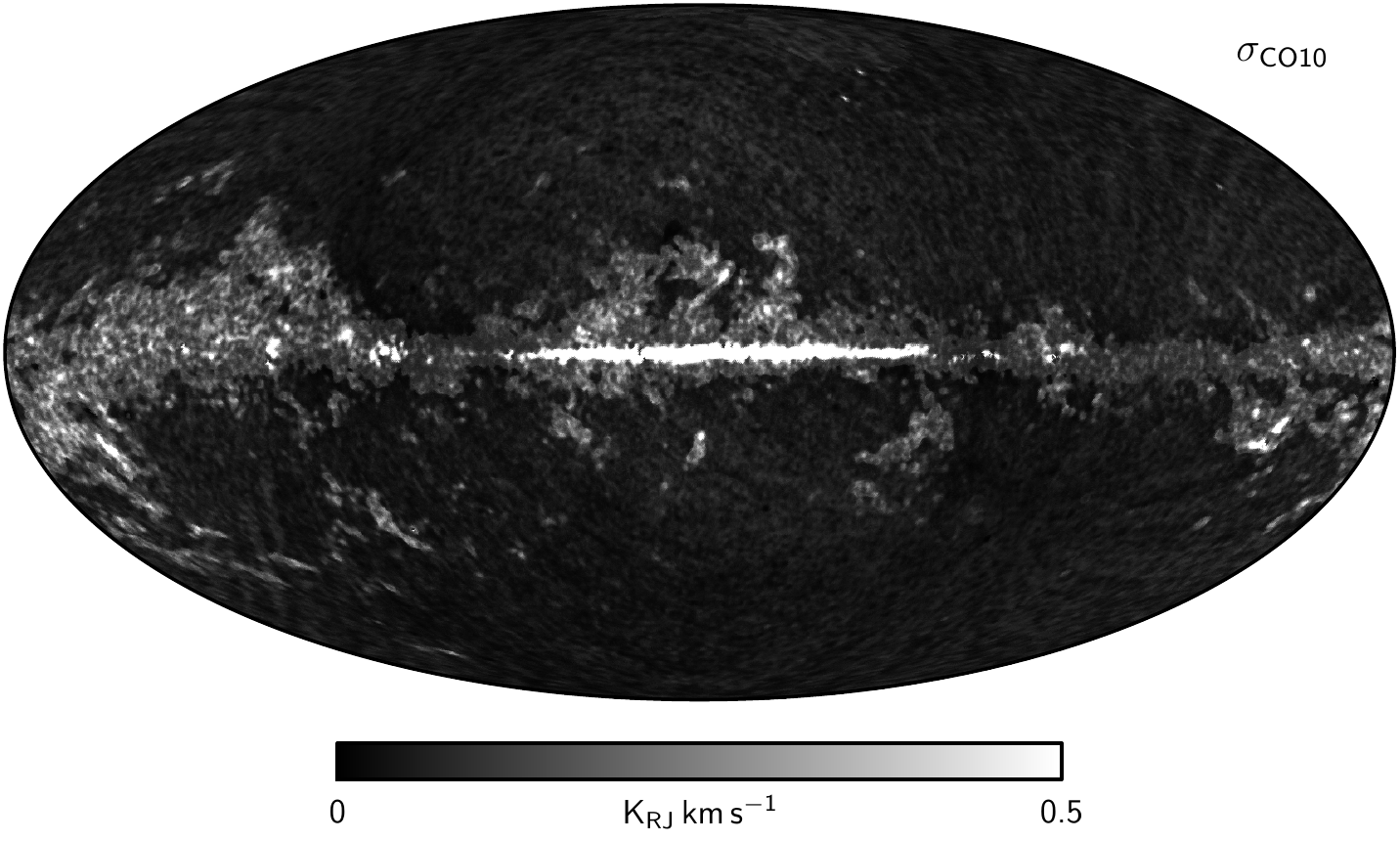,width=1.00\linewidth,clip=}
    }
  \end{center}
  \caption{Maximum posterior (\emph{top}) and posterior rms
    (\emph{bottom}) CO $J$=1$\rightarrow$0 line emission maps derived from the joint baseline
    analysis of \Planck, \WMAP, and 408\,MHz observations.}
  \label{fig:co10_amp_map}
\end{figure*}

\begin{figure*}[p]
  \begin{center}
    \mbox{
      \epsfig{figure=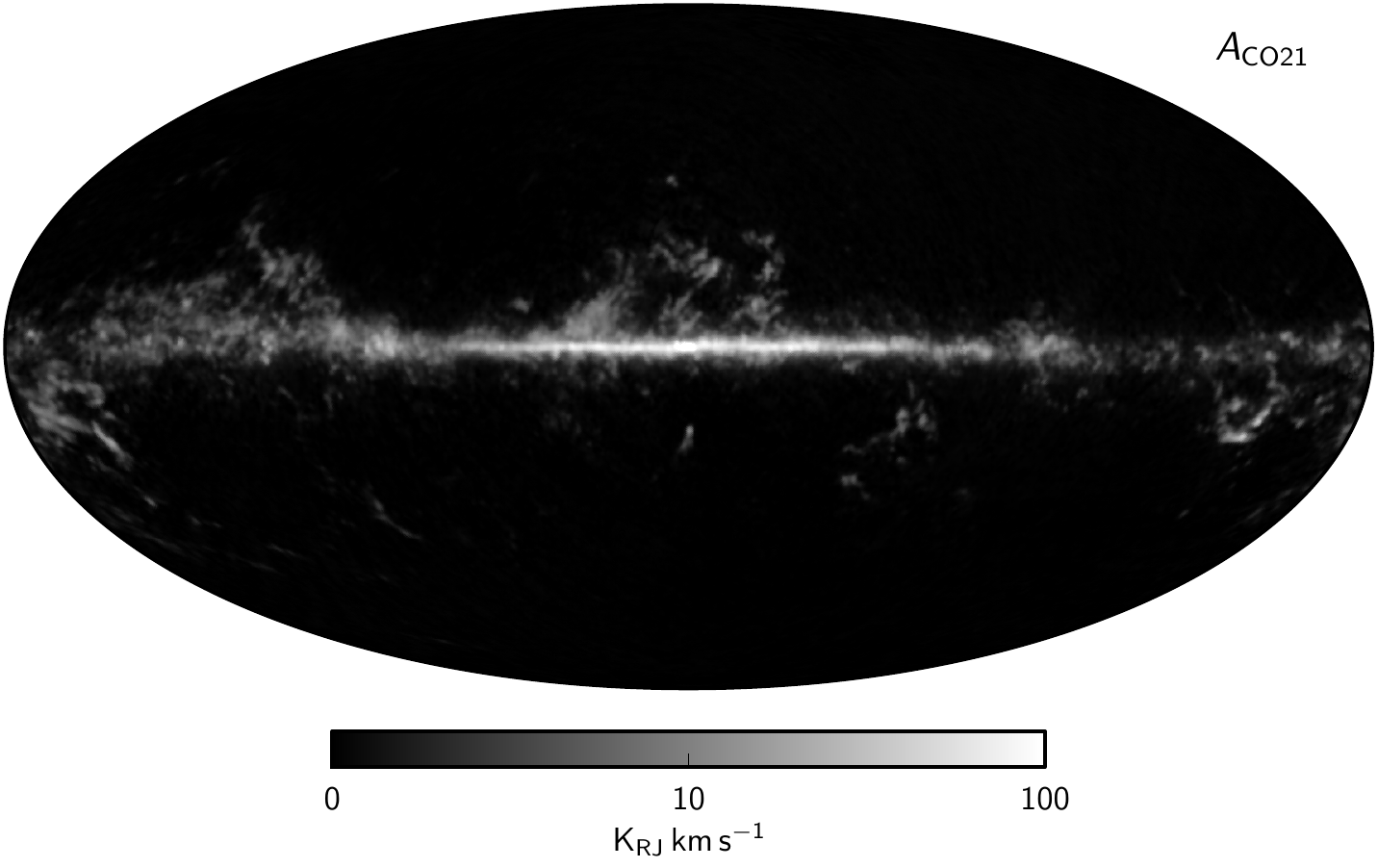,width=1.00\linewidth,clip=}
    }
    \vskip 4mm
    \mbox{
      \epsfig{figure=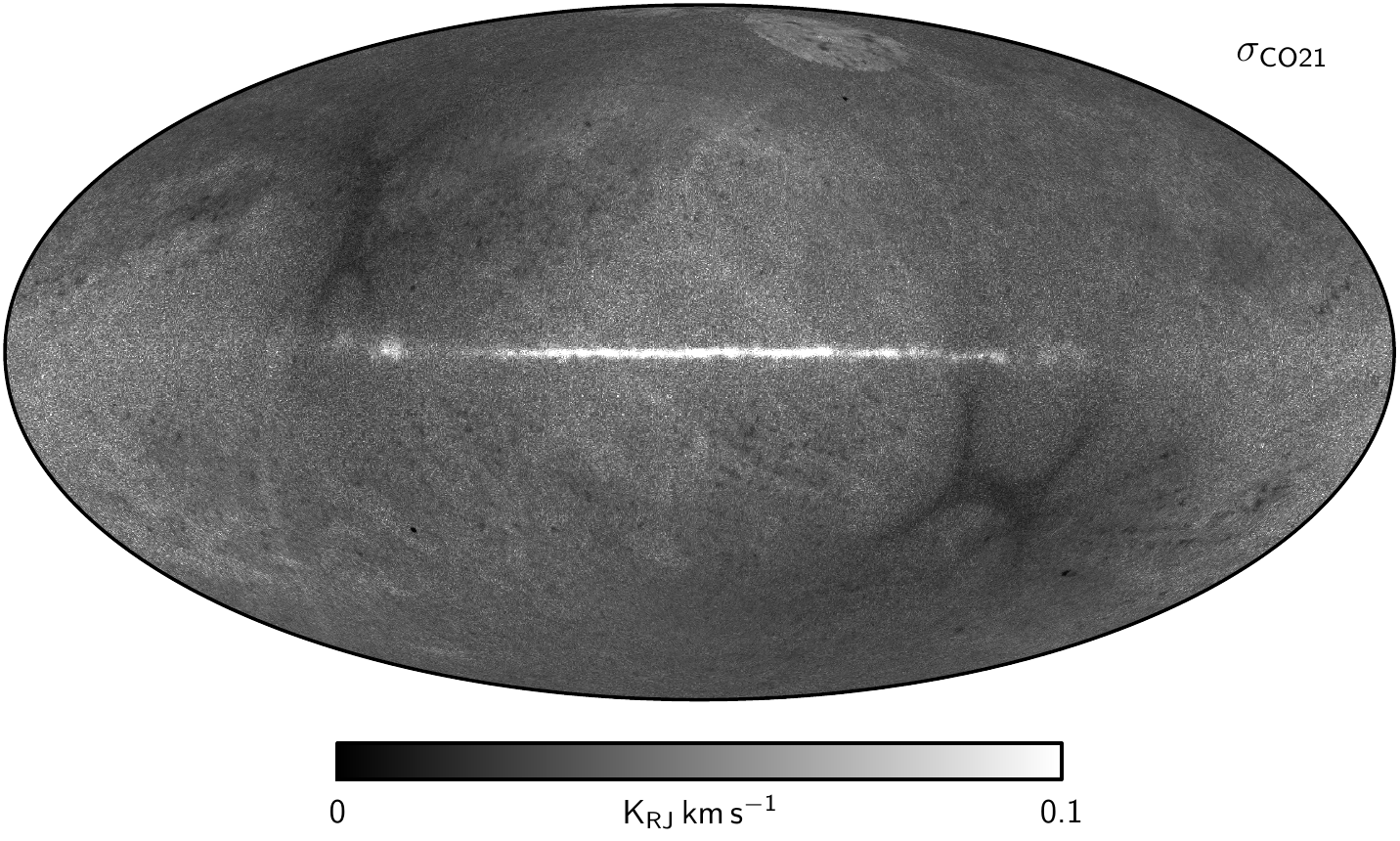,width=1.00\linewidth,clip=}
    }
  \end{center}
  \caption{Maximum posterior (\emph{top}) and posterior rms
    (\emph{bottom}) CO $J$=2$\rightarrow$1 line emission maps derived
    from the joint baseline analysis of \Planck, \WMAP, and 408\,MHz
    observations. The two circular regions close to the North Galactic
    Pole in the rms map correspond to the Coma and Virgo
    clusters, for which the thermal SZ effect is fitted together with the primary
    diffuse components.}
  \label{fig:co21_amp_map}
\end{figure*}

\begin{figure*}[p]
  \begin{center}
    \mbox{
      \epsfig{figure=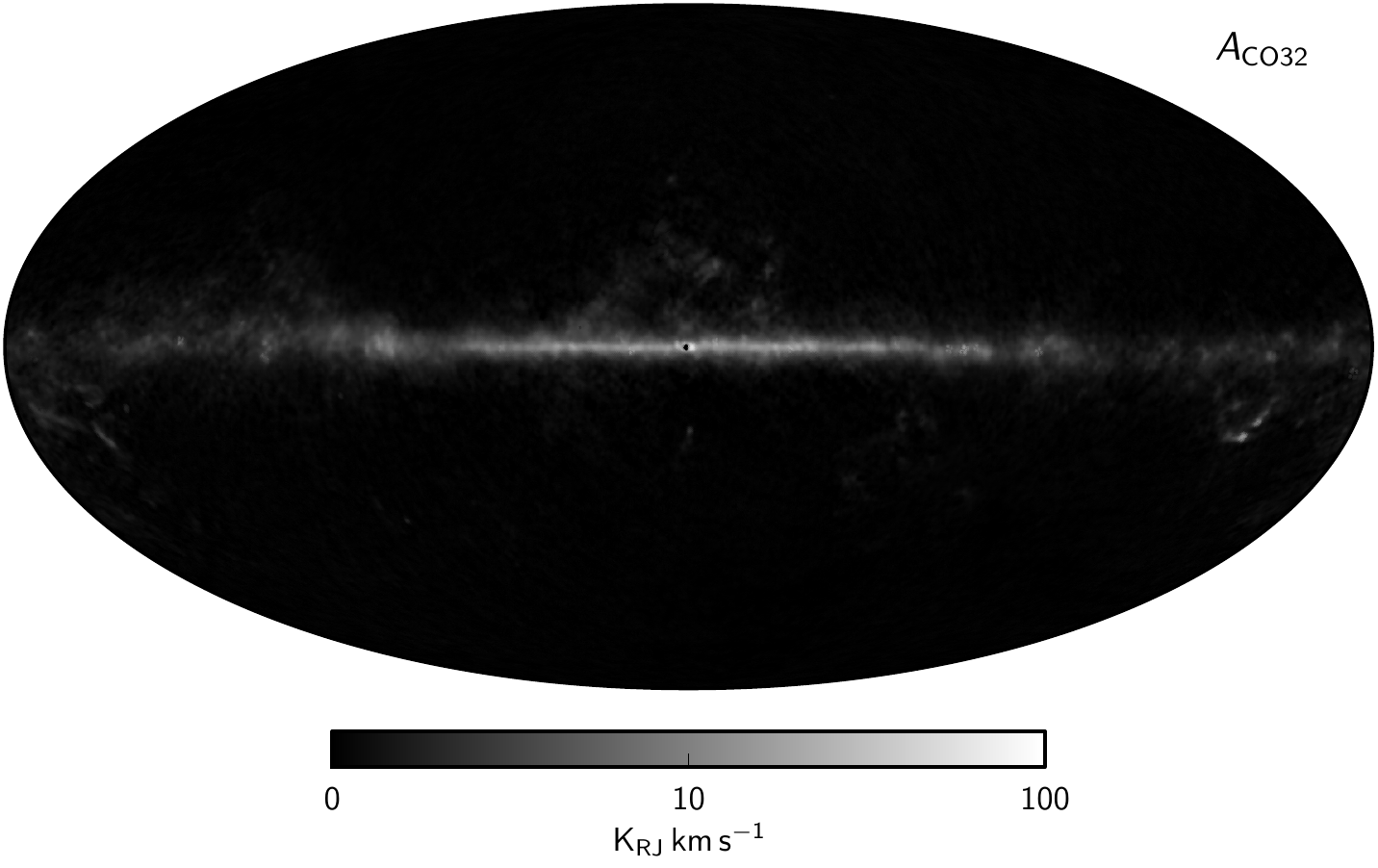,width=1.00\linewidth,clip=}
    }
    \vskip 4mm
    \mbox{
      \epsfig{figure=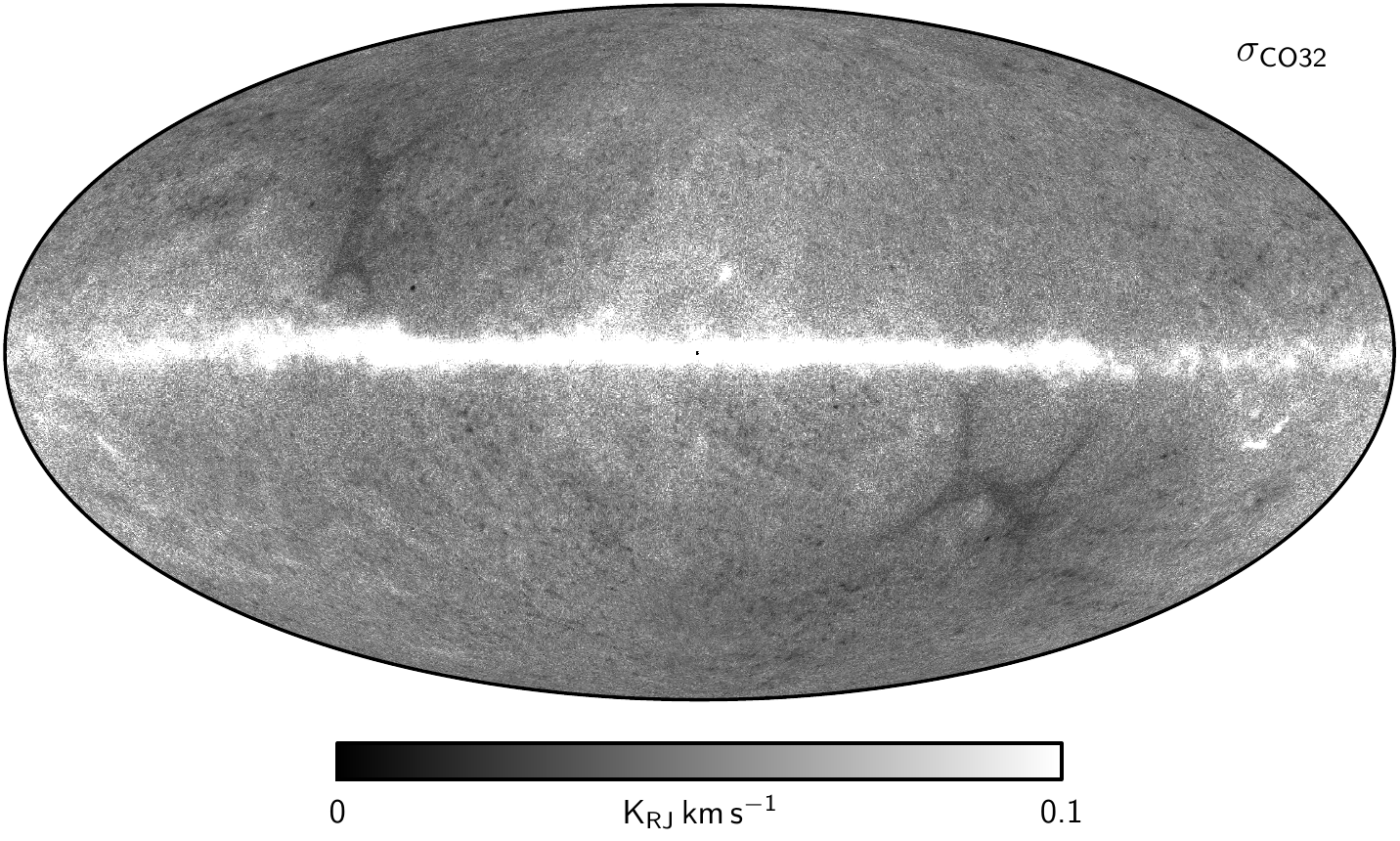,width=1.00\linewidth,clip=}
    }
  \end{center}
  \caption{Maximum posterior (\emph{top}) and posterior rms
    (\emph{bottom}) CO $J$=3$\rightarrow$2 line emission maps derived from the joint baseline
    analysis of \Planck, \WMAP, and 408\,MHz observations.}
  \label{fig:co32_amp_map}
 \end{figure*}

\begin{figure*}[p]
  \vskip 4mm
  \begin{center}
    \mbox{
      \epsfig{figure=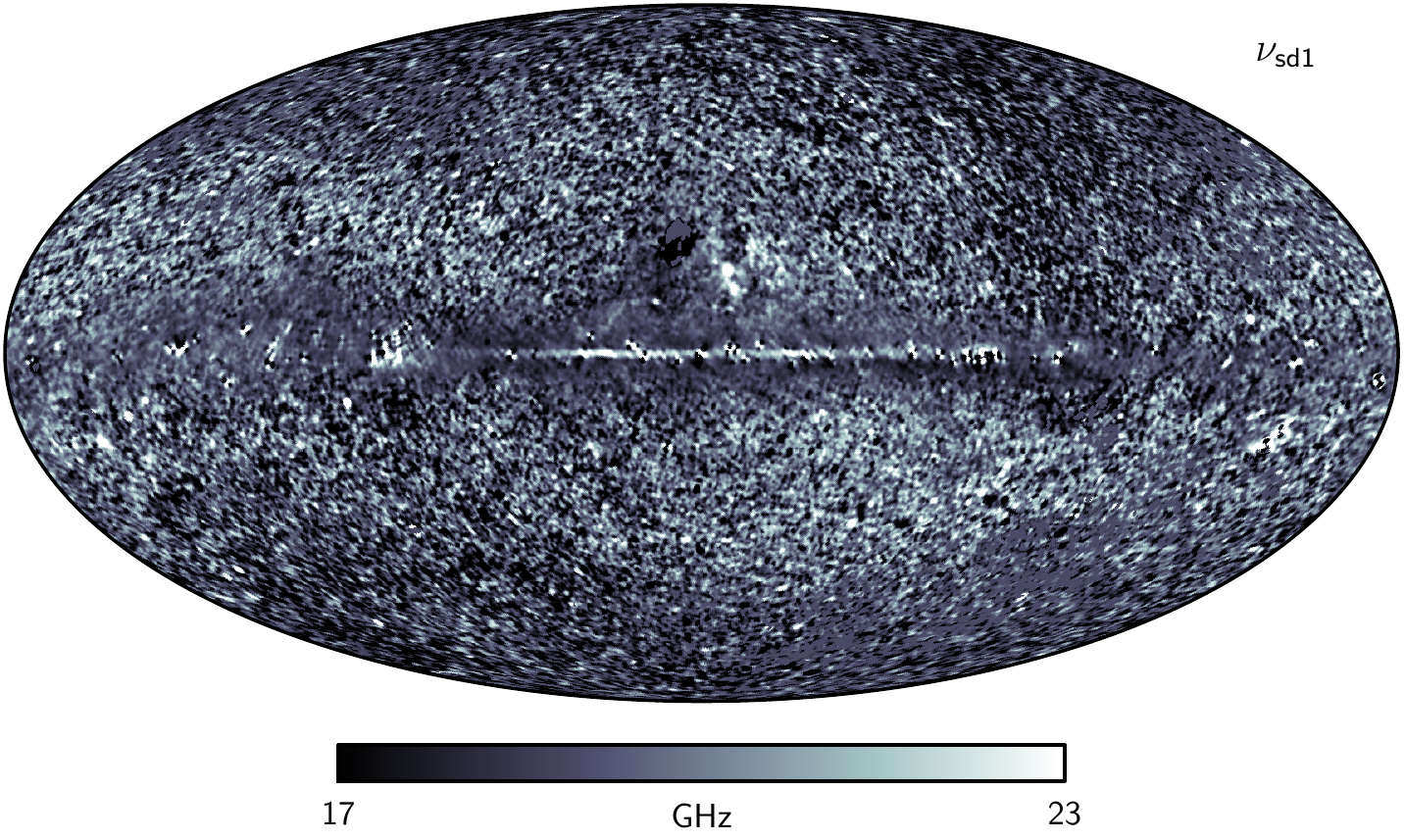,width=1.00\linewidth,clip=}
    }
    \vskip 4mm
    \mbox{
      \epsfig{figure=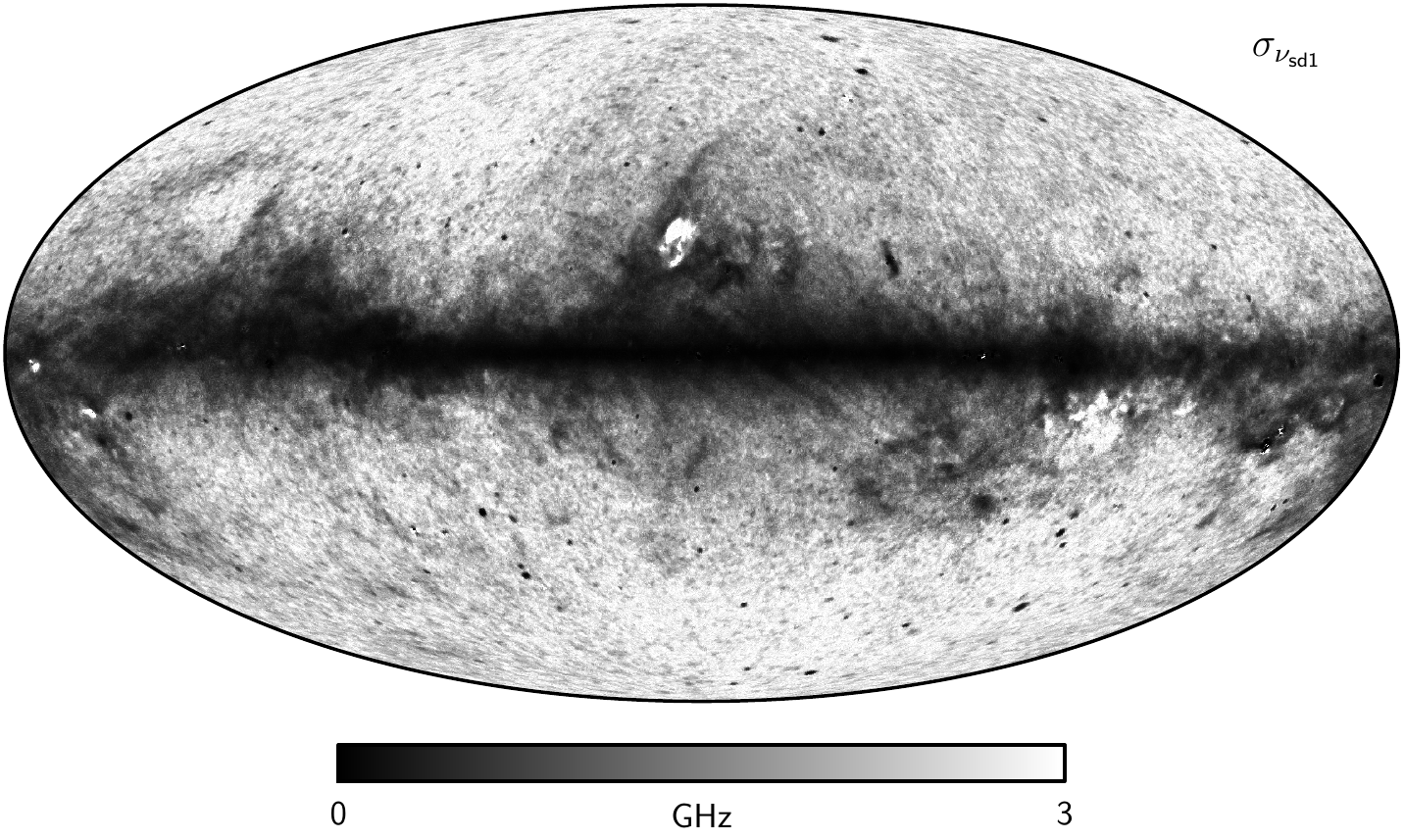,width=1.00\linewidth,clip=}
    }
  \end{center}
  \caption{Maximum posterior (\emph{top}) and posterior rms
    (\emph{bottom}) spinning dust peak frequency maps derived from the joint baseline
    analysis of \Planck, \WMAP, and 408\,MHz observations.}
  \label{fig:ame1_nup_map}
\end{figure*}

\begin{figure*}[p]
  \vskip 4mm
  \begin{center}
    \mbox{
      \epsfig{figure=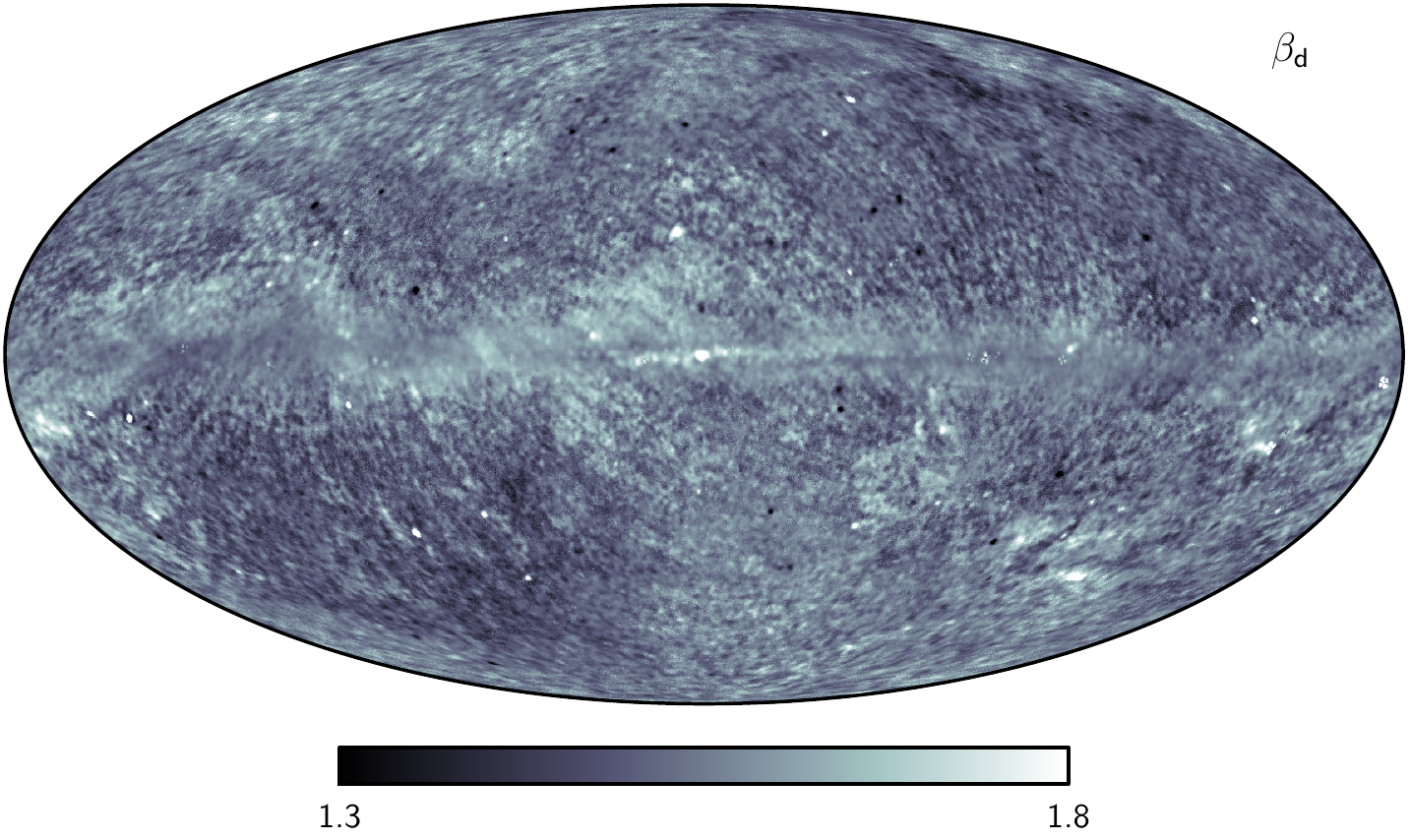,width=1.00\linewidth,clip=}
    }
    \vskip 4mm
    \mbox{
      \epsfig{figure=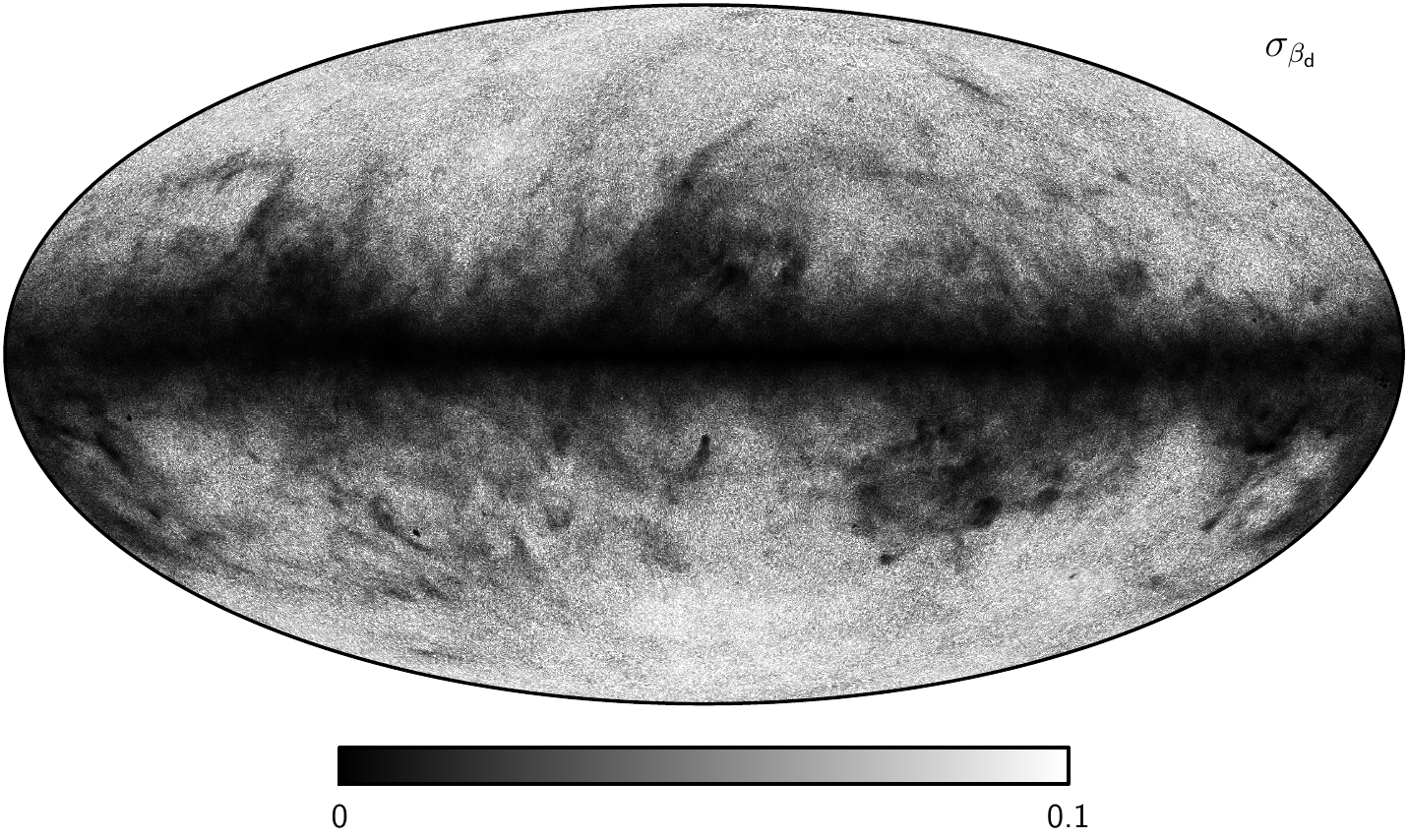,width=1.00\linewidth,clip=}
    }
  \end{center}
  \caption{Maximum posterior (\emph{top}) and posterior rms
    (\emph{bottom}) thermal dust spectral index maps derived from the joint baseline
    analysis of \Planck, \WMAP, and 408\,MHz observations.}
  \label{fig:dust_beta_map}
\end{figure*}

\begin{figure*}[p]
  \vskip 4mm
  \begin{center}
    \mbox{
      \epsfig{figure=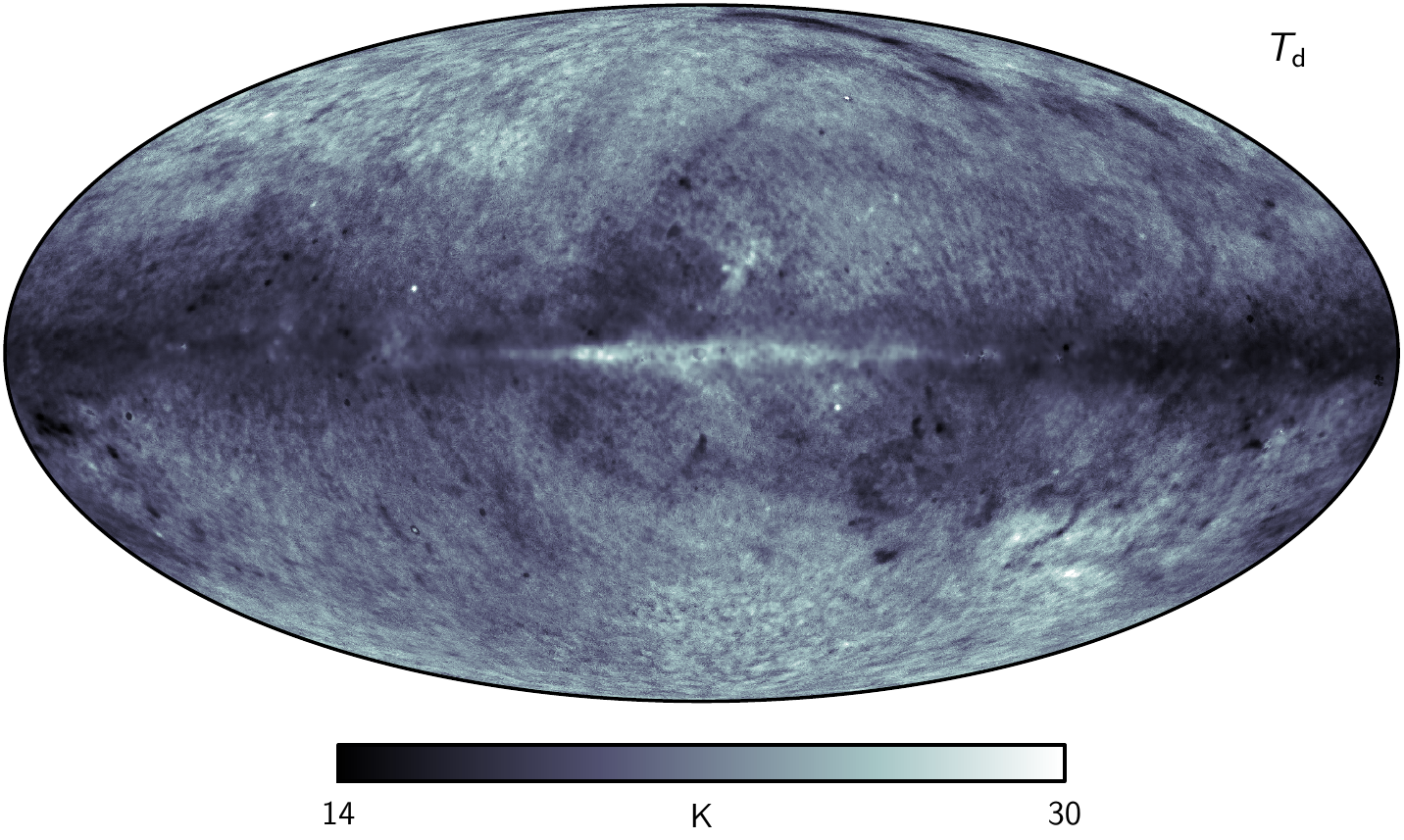,width=1.00\linewidth,clip=}
    }
    \vskip 4mm
    \mbox{
      \epsfig{figure=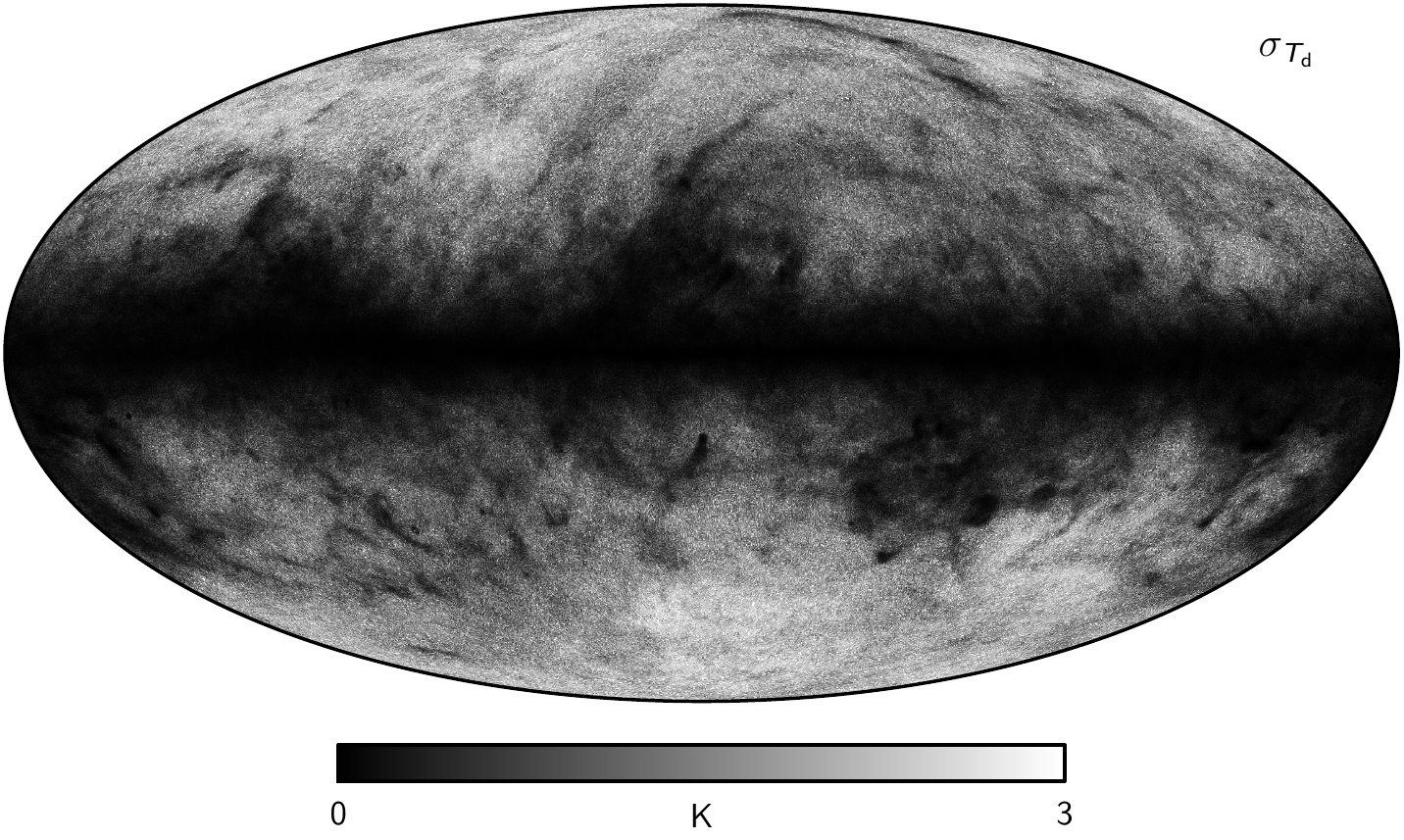,width=1.00\linewidth,clip=}
    }
  \end{center}
  \caption{Maximum posterior (\emph{top}) and posterior rms
    (\emph{bottom}) thermal dust temperature maps derived from the joint baseline
    analysis of \Planck, \WMAP, and 408\,MHz observations.}
  \label{fig:dust_Td_map}
\end{figure*}

\begin{figure*}[t]
  \begin{center}
    \mbox{
      \epsfig{figure=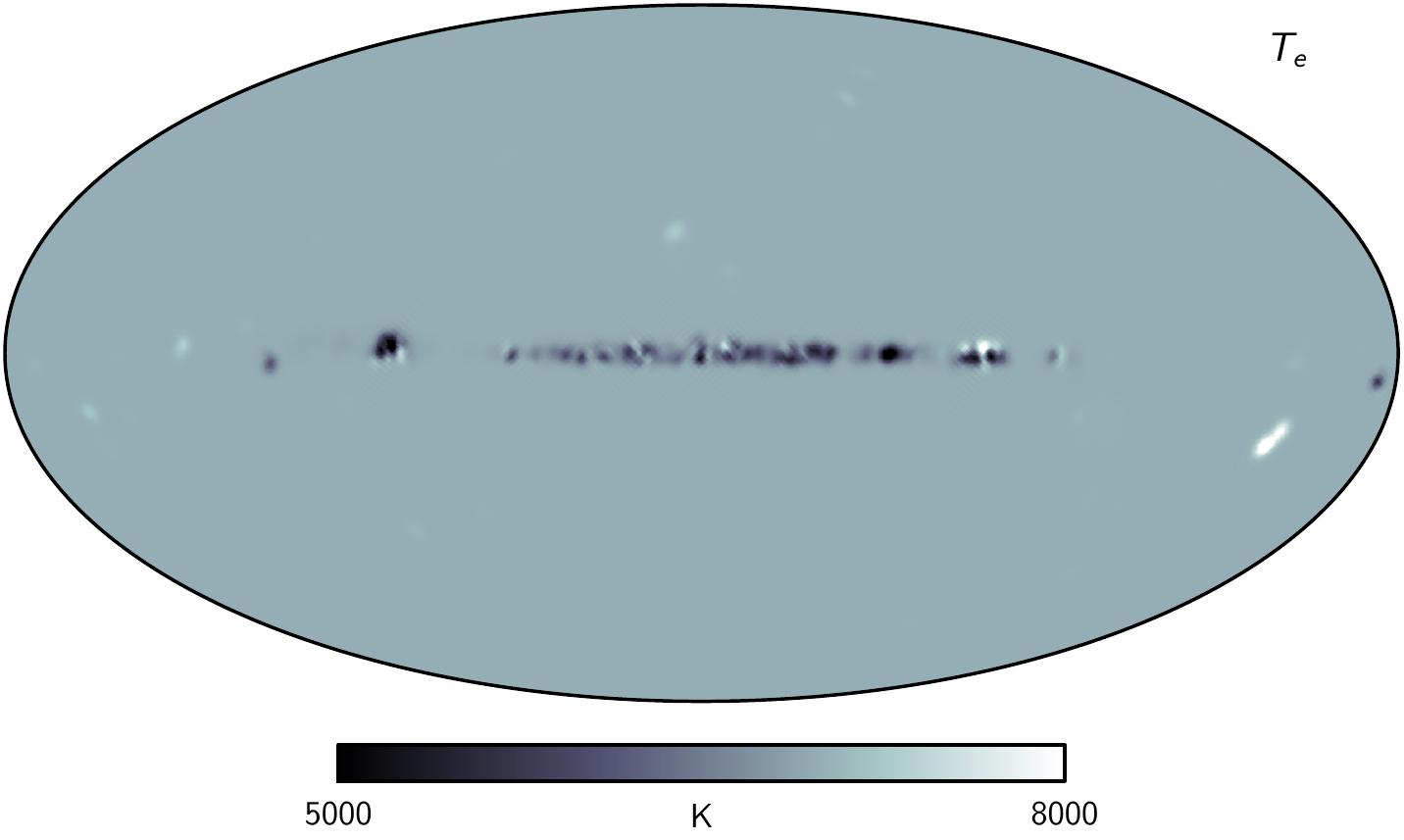,width=1.00\linewidth,clip=}
    }
  \end{center}
  \caption{Maximum posterior (free-free) electron temperature map
    derived from the joint baseline analysis of \Planck, \WMAP, and
    408\,MHz observations.}
  \label{fig:ff_Te_map}
\end{figure*}

We are now ready to present the \Planck\ 2015 baseline diffuse
astrophysical component model in temperature, as derived from the
\Planck, \WMAP, and 408\,MHz observations. An overview of the data
products delivered through the \Planck\ Legacy
Archive (PLA)\footnote{\url{http://pla.esac.esa.int/pla}} is given in
Table~\ref{tab:products}, including file names and FITS extension
numbering, as well as summary statistics in the form of posterior
maximum, mean, and rms values.

We start our review by considering data that are smoothed to a
common resolution of $1^{\circ}$ FWHM, representing the most complete
data set and model possible, given the current data set. Processing
these observations through the analysis pipeline outlined in
Sect.~\ref{sec:method}, and adopting the baseline model described in
Sect.~\ref{sec:sky}, we obtain the set of maximum-posterior
astrophysical parameter maps shown in the top panels of
Figs.~\ref{fig:cmb_amp_map}--\ref{fig:ff_Te_map}.  The bottom panels
show the corresponding rms maps found by computing the standard
deviation over the ensemble of Gibbs samples. Note that these rms maps
account only for statistical errors from instrumental noise, not for
systematic uncertainties due to modelling errors. They are therefore
only meaningful for pixels for which the goodness-of-fit is
acceptable. Most importantly, they do not represent the true errors in
the Galactic plane, where modelling errors dominate statistical errors.

\begin{figure}[t]
\begin{center}
\mbox{
\epsfig{figure=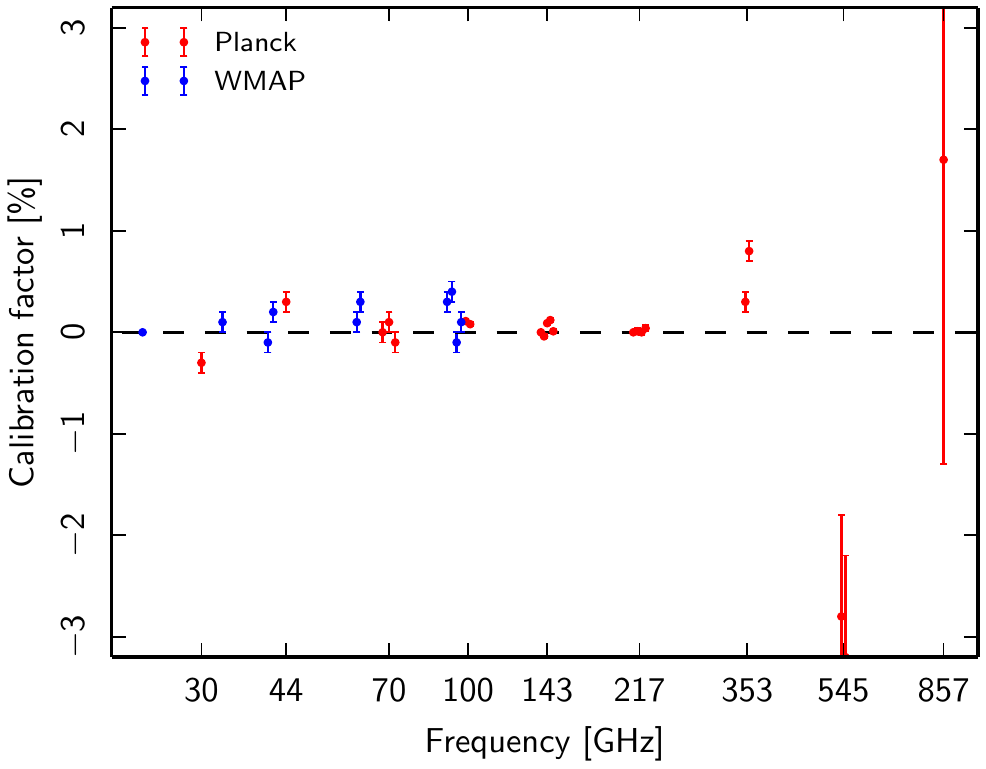,width=80mm,clip=}
}
\end{center}
\caption{Detector map and differencing assembly re-calibration factors
  for \Planck\ and \WMAP, as reported in Table~\ref{tab:monopole}.  }
\label{fig:calibration}
\end{figure}

Instrumental parameters, as well as monopole and dipole coefficients,
are listed in Table~\ref{tab:monopole}, and Fig.~\ref{fig:calibration}
provides a visual comparison of the calibration factors for each
\Planck\ and \WMAP\ detector map.\footnote{The re-calibration factors
  listed in Table~\ref{tab:monopole} for 545 and 857\,GHz refer to
  multiplicative factors in native map units,
  $\textrm{MJy}\,\textrm{sr}^{-1}$. When making comparison with
  similar calibration factors derived from corresponding maps defined
  in units of $\mu\textrm{K}$, one additionally has to account for
  differences in unit conversion due to bandpass shifts, as listed in
  Table~\ref{tab:units}.} Global (i.e., spatially uniform)
astrophysical parameters are listed in Table~\ref{tab:Tglobal}. The
full marginal uncertainties of these parameters are dominated by
modelling, not statistical errors. For this reason, the tabulated
uncertainties are computed through realistic end-to-end simulations,
as implemented in the \Planck\ 2015 FFP8 simulations
\citep{planck2014-a14}. These simulations were analysed blindly as
part of the CMB validation efforts \citep{planck2014-a11}, using the
exact same machinery as used in this paper. Further, they are based on
a significantly different foreground model than the baseline model
adopted here, and they therefore at least partially account for
modelling errors, as well as known systematic and mapmaking
uncertainties. The only differences in the fitted model compared to
the present baseline are that it includes only one spinning dust
component and no 94/100\,GHz line emission or SZ components.

Based on these simulations, we estimate the uncertainties on the
calibration and bandpass measurements directly from the
simulations, comparing output parameters against known inputs. The
monopole and dipole uncertainties are, however, first based on the
statistical errors derived from the Gibbs chains, and then rounded up
(where necessary) to be no smaller than the corresponding FFP8
simulation uncertainties.\footnote{Based on the FFP8 simulations we
  never report monopole (dipole) uncertainties smaller than
  $1\,\mu\textrm{K}$ ($0.1\,\mu\textrm{K}$).}  Thus they correspond to
the maximum of the instrumental and the modelling errors. Furthermore,
we reemphasize that the reported monopole and dipole values are
conditional with respect to the nominal \Planck, \ion{H}{i} and
synchrotron priors, as defined by \mbox{\citet{planck2014-a07}}, 
\citet{planck2014-a09}, and \citet{wehus2014}. 

The uncertainties in the CO line ratios listed in
Table~\ref{tab:Tglobal}, which are model-dominated, are also taken
from the FFP8 simulations. The 94/100\,GHz line ratios, however, are
noise dominated, and they are therefore taken directly as the standard
deviation of the posterior distribution, i.e., from the Gibbs
samples. Finally, we do not quote any uncertainties on the peak
frequency of the second spinning dust component or the frequency scale
factor of the synchrotron \texttt{GALPROP} model. Both of these are
completely dependent on other parameters in the model, and the full
joint distribution is highly non-Gaussian. As a result, we only quote
the maximum-likelihood values for these, and note that their full
marginal uncertainties are very large, possibly up to 50\,\% or more.

\begin{figure*}
  \begin{center}
    \mbox{
\epsfig{figure=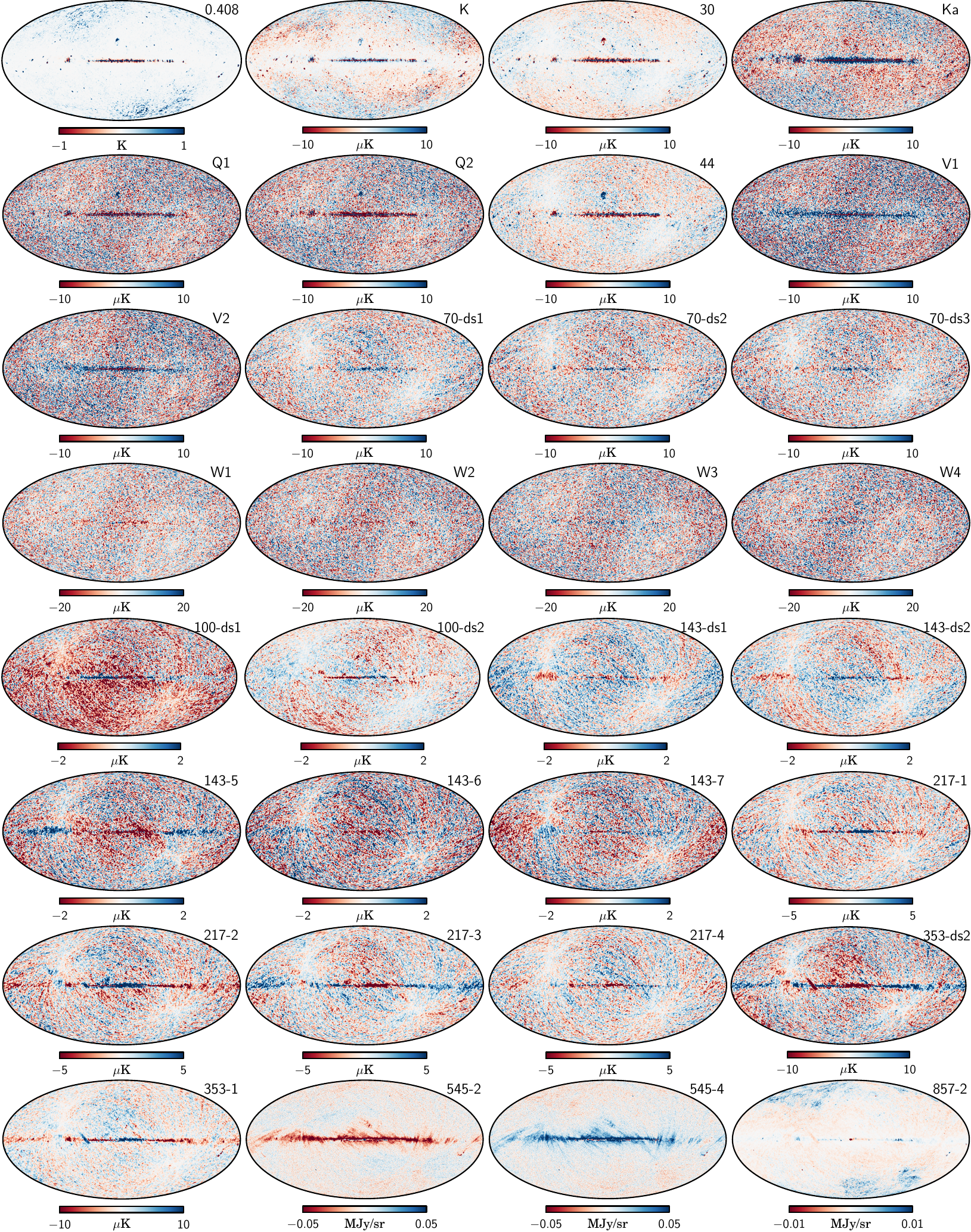,width=\linewidth,clip=}
}

\end{center}
\caption{Residual maps, $\d_{\nu}-\s_{\nu}$, for each detector data
  set included in the baseline joint \Planck, \WMAP, and 408\,MHz
  temperature analysis. All panels employ linear colour scales. }
\label{fig:map_residuals}
\end{figure*}

\begin{figure}
  \begin{center}
\mbox{
\epsfig{figure=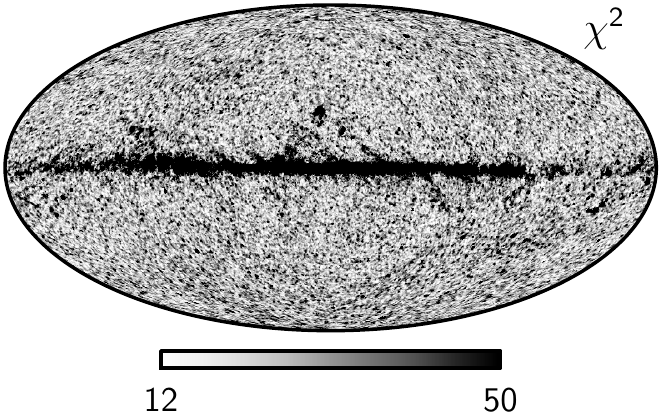,width=88mm,clip=}
}
\vspace*{2mm}
\mbox{
\epsfig{figure=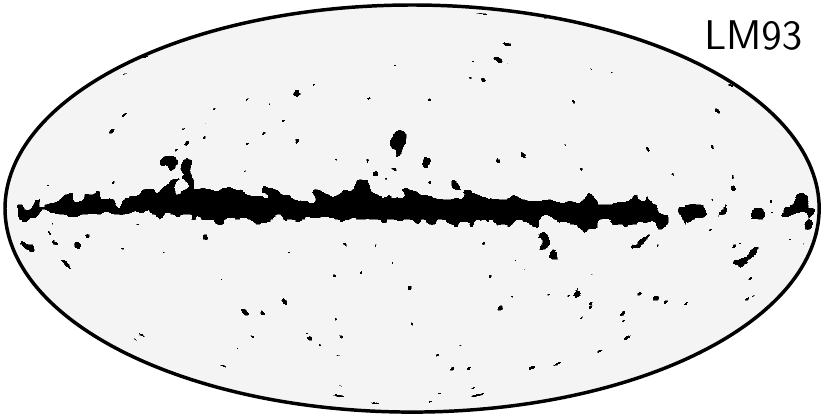,width=88mm,clip=}
}
\vspace*{2mm}
\mbox{
\epsfig{figure=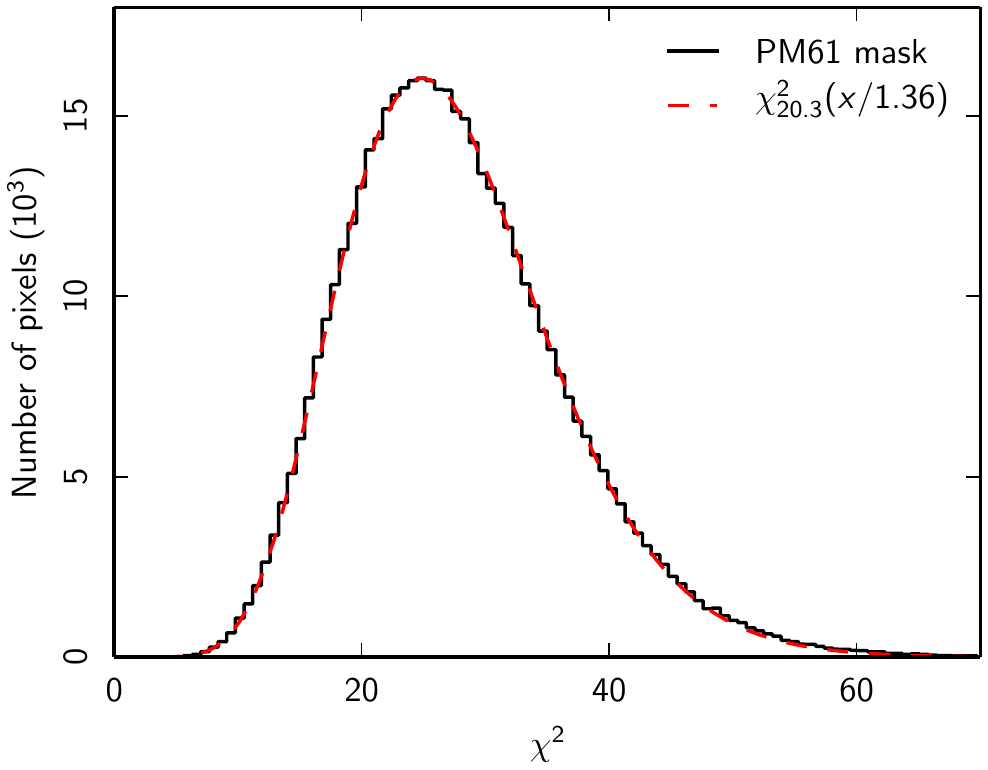,width=88mm,clip=}
}
\end{center}
\caption{\emph{Top}: $\chi^2$ per pixel for joint baseline \Planck,
  \WMAP, and 408\,MHz intensity analysis.\quad \emph{Middle}:
  confidence mask derived by smoothing the $\chi^2$ map to $1\deg$
  FWHM, and thresholding at a value of $\chi^2_{\textrm{max}}=50$. Its
  primary application is for the low-$\ell$ 2015 \Planck\ temperature
  likelihood, and it is accordingly denoted LM93 (93\,\% likelihood
  mask); see \citet{planck2014-a13} for further details.\quad
  \emph{Bottom}: histogram of $\chi^2$ values outside the conservative
  PM61 mask. The grey dashed line shows the best-fit $\chi^2$
  distribution with a variable degree of freedom and scaling, used to
  account for prior and noise modelling effects; see
  Sect.~\ref{sec:baseline} for further discussion.}
\label{fig:chisq_map}
\end{figure}


\subsubsection{Goodness-of-fit}

\begin{table}[tmb]                                                                                                                                                
\begingroup                                                                                                                                  
\newdimen\tblskip \tblskip=5pt
\caption{Goodness-of-fit statistics for the temperature analysis
  evaluated at $1\deg$ FWHM.\label{tab:Tgof} The second column shows
  the rms residual outside the 93\,\% \texttt{Commander} likelihood mask
  for each channel, while the third column shows the same, but
  normalized with respect to the instrumental noise rms listed in
  Table~\ref{tab:data}. The fourth column lists the median fractional
  residual in the complementary 7\,\% of the sky, covering the Galactic
  plane region. }
\nointerlineskip                                                                                                                                                                                     
\vskip -4mm
\footnotesize                                                                                                                                       
\setbox\tablebox=\vbox{                                                                                                                                                                           
\newdimen\digitwidth                                                                                                                      
\setbox0=\hbox{\rm 0}
\digitwidth=\wd0
\catcode`*=\active
\def*{\kern\digitwidth}
\newdimen\signwidth
\setbox0=\hbox{+}
\signwidth=\wd0
\catcode`!=\active
\def!{\kern\signwidth}
\newdimen\decimalwidth
\setbox0=\hbox{.}
\decimalwidth=\wd0
\catcode`@=\active
\def@{\kern\signwidth}
\def\s#1{\ifmmode $\rlap{$^{\rm #1}$}$ \else \rlap{$^{\rm #1}$}\fi}
\halign{ \hbox to 1.2in{#\leaderfil}\tabskip=2em&
    \hfil#\hfil\tabskip=2.4em&
    \hfil#\hfil\tabskip=1em&
    \hfil#\hfil\tabskip=0pt\cr
\noalign{\doubleline}
\omit&\multispan2\hfil Rms outside LM93\hfil& Frac. res.\cr
\noalign{\vskip -3pt}
\omit&\multispan2\hrulefill&\lower3pt\hbox{inside}\cr
\noalign{\vskip 2pt}
\omit\hfil Map\hfil& [$\mu\textrm{K}$]  & $\sigma^{\rm res}_{\nu}/\sigma^{\rm inst}_{\nu}$ &LM93 [\%]\cr
\noalign{\doubleline}
\Planck\ 30&                  *1.56& 0.55& 0.08\cr
\phantom{\Planck\ }44&        *2.51& 0.83& 0.42\cr
\phantom{\Planck\ }70-ds1&    *3.67& 0.96& 0.78\cr
\phantom{\Planck\ }70-ds2&    *3.88& 0.97& 0.82\cr
\phantom{\Planck\ }70-ds3&    *3.98& 0.97& 0.80\cr
\phantom{\Planck\ }100-ds1&   *1.00& 1.11& 0.18\cr
\phantom{\Planck\ }100-ds2&   *0.61& 0.76& 0.08\cr
\phantom{\Planck\ }143-ds1&   *0.72& 1.02& 0.08\cr
\phantom{\Planck\ }143-ds2&   *0.68& 0.97& 0.08\cr
\phantom{\Planck\ }143-5&     *1.03& 1.14& 0.15\cr
\phantom{\Planck\ }143-6&     *1.17& 1.06& 0.12\cr
\phantom{\Planck\ }143-7&     *1.04& 1.04& 0.10\cr
\phantom{\Planck\ }217-1&     *1.70& 0.94& 0.06\cr
\phantom{\Planck\ }217-2&     *1.90& 1.00& 0.09\cr
\phantom{\Planck\ }217-3&     *1.68& 0.98& 0.08\cr
\phantom{\Planck\ }217-4&     *1.64& 0.91& 0.05\cr
\phantom{\Planck\ }353-ds2&   *4.28& 0.95& 0.03\cr
\phantom{\Planck\ }353-1&     *2.11& 0.60& 0.02\cr
\phantom{\Planck\ }545-2&*8.89\s{a}& 0.88& 0.14\cr
\phantom{\Planck\ }545-4&*9.04\s{a}& 0.90& 0.13\cr
\phantom{\Planck\ }857-2&*1.39\s{a}& 0.13& 0.00\cr
\noalign{\vskip 2pt}
WMAP K&                       *2.26& 0.38& 0.03\cr
\phantom{WMAP }Ka&            *4.52& 1.05& 0.56\cr
\phantom{WMAP }Q1&            *5.22& 0.98& 0.64\cr
\phantom{WMAP }Q2&            *5.09& 0.99& 0.67\cr
\phantom{WMAP }V1&            *6.29& 0.98& 1.70\cr
\phantom{WMAP }V2&            *5.62& 1.02& 1.37\cr
\phantom{WMAP }W1&            *7.90& 0.89& 1.44\cr
\phantom{WMAP }W2&            *9.74& 0.96& 1.74\cr
\phantom{WMAP }W3&            *9.57& 0.90& 1.67\cr
\phantom{WMAP }W4&            10.15& 1.00& 1.77\cr
\noalign{\vskip 2pt}
Haslam 0.408&           *0.12\s{b}& 0.10&  $0.03$\cr
\noalign{\doubleline}
}}
\endPlancktable
\tablenote {{\rm a}} Unit is kJy\,sr\mo.\par
\tablenote {{\rm b}} Unit is K.\par
\endgroup
\end{table}

Next, the goodness-of-fit of the model is summarized in
Figs.~\ref{fig:map_residuals}--\ref{fig:chisq_map} and in
Table~\ref{tab:Tgof}. Starting with Fig.~\ref{fig:map_residuals},
residual maps for each channel are shown on the form
($\d_{\nu}-\s_{\nu}$), and these give the most complete view of the
performance of the model out of all the statistics considered in the
following. In addition, they provide a direct visual summary of
remaining low-level systematics in the data, which should prove useful
for future reprocessing of the same data. Overall, we see in these
maps that the model provides an excellent fit to all channels at high
Galactic latitudes. Note that the colour scales vary from
$\pm2\,\mu\textrm{K}$ at the CMB dominated \Planck\ HFI channels,
through $\pm10\,\mu\textrm{K}$ for most of the \WMAP\ channels, to
$\pm1\textrm{K}$ and $\pm0.05\,\textrm{MJy\,sr\mo}$ for the 408\,MHz and
545\,GHz channels. In fact, the solution is sufficiently free of
artifacts that the dominant systematic effect in the 100-ds2 residual
map is Galactic rotation projected into the CMB frequencies by the CO
$J$=1$\rightarrow$0 emission line. The same feature is also seen in
several other channels, although at slightly lower significance.

The top panel of Fig.~\ref{fig:chisq_map} shows the corresponding
$\chi^2$ map as defined by Eq.~\ref{eq:chisq}, but summed only over
the accepted frequency channels.  The bottom panel shows a histogram
of these $\chi^2$ values including all pixels admitted by the PM61
processing mask.

Together, these figures provide a useful qualitative summary of the
goodness-of-fit of the baseline model. However, providing a
corresponding rigorous statistical description is complicated, because
the effective number of degrees of freedom per pixel is not well
defined. The usual approach of subtracting the number of free
parameters from the number of data points is not applicable, because
informative priors (most notably the positivity prior) eliminate large
parts of the parameter space. A parameter that is prior-dominated
therefore does not reduce the number of degrees of freedom by unity,
but only a fraction of unity. Second, smoothing the data to a common
resolution of $1\deg$ FWHM introduces noise correlations between
pixels, and these are not accounted for in the noise description. The
overall $\chi^2$ distribution will therefore be broader than a
corresponding distribution with no correlated noise. To estimate the
effective number of degrees of freedom, we therefore fit a scaled
$\chi^2$ distribution to the empirical $\chi^2$ distribution,
including only the very cleanest parts of the sky, where actual
foreground contamination is minimal. We adopt the conservative PM61
mask for this. The resulting best-fit $\chi^2$ distribution reads
\begin{equation}
\chi^2_{20.3}(x) \propto \left(\frac{x}{1.37}\right)^{20.3/2-1} e^{-\frac{x}{2}},
\end{equation}
and we accordingly estimate that the empirical number of degrees of
freedom is 20, and the correlated noise scaling factor is 1.37. The
former of these suggests that our model effectively contains
$32-20=12$ free parameters, not 14 as obtained by naively counting
free parameters per pixel. In other words, the combined effect of all
priors is to remove 2 of 14 degrees of freedom, indicating that the
model is indeed highly data driven. The 99\,\% confidence $\chi^2$ range
from the analytic fit is $11 < \chi^2 < 59$.

The middle panel of Fig.~\ref{fig:chisq_map} shows the mask obtained
by thresholding the $\chi^2$ map smoothed with a $1^{\circ}$ Gaussian
kernel at a value of 50.\footnote{Because of the additional smoothing,
  $\chi_{\textrm{smooth}}^2>50$ corresponds roughly to a $5\,\sigma$ outlier, not
  $2.5\,\sigma$ as it does in the unsmoothed $\chi^2$ map.} This mask
is called the 93\,\% likelihood mask (LM93), and constitutes the primary
confidence mask for the \Planck\ 2015 model. Also, together with the
CMB solution in Fig.~\ref{fig:cmb_amp_map}, this mask defines the main
inputs to the \Planck\ 2015 low-$\ell$ CMB temperature likelihood
\citep{planck2014-a13}. Note that this mask removes many bright
extra-galactic point sources at high Galactic latitudes; the
algorithm adopted in this paper treats point sources and diffuse
emission identically through pixel-by-pixel fits, and any subsequent
analysis of the resulting component maps should take these sources
into account either through explicit masking, as done here, or by
post-processing fits.

\begin{figure*}[p]
  \vskip 10mm
  \begin{center}
    \mbox{
      \epsfig{figure=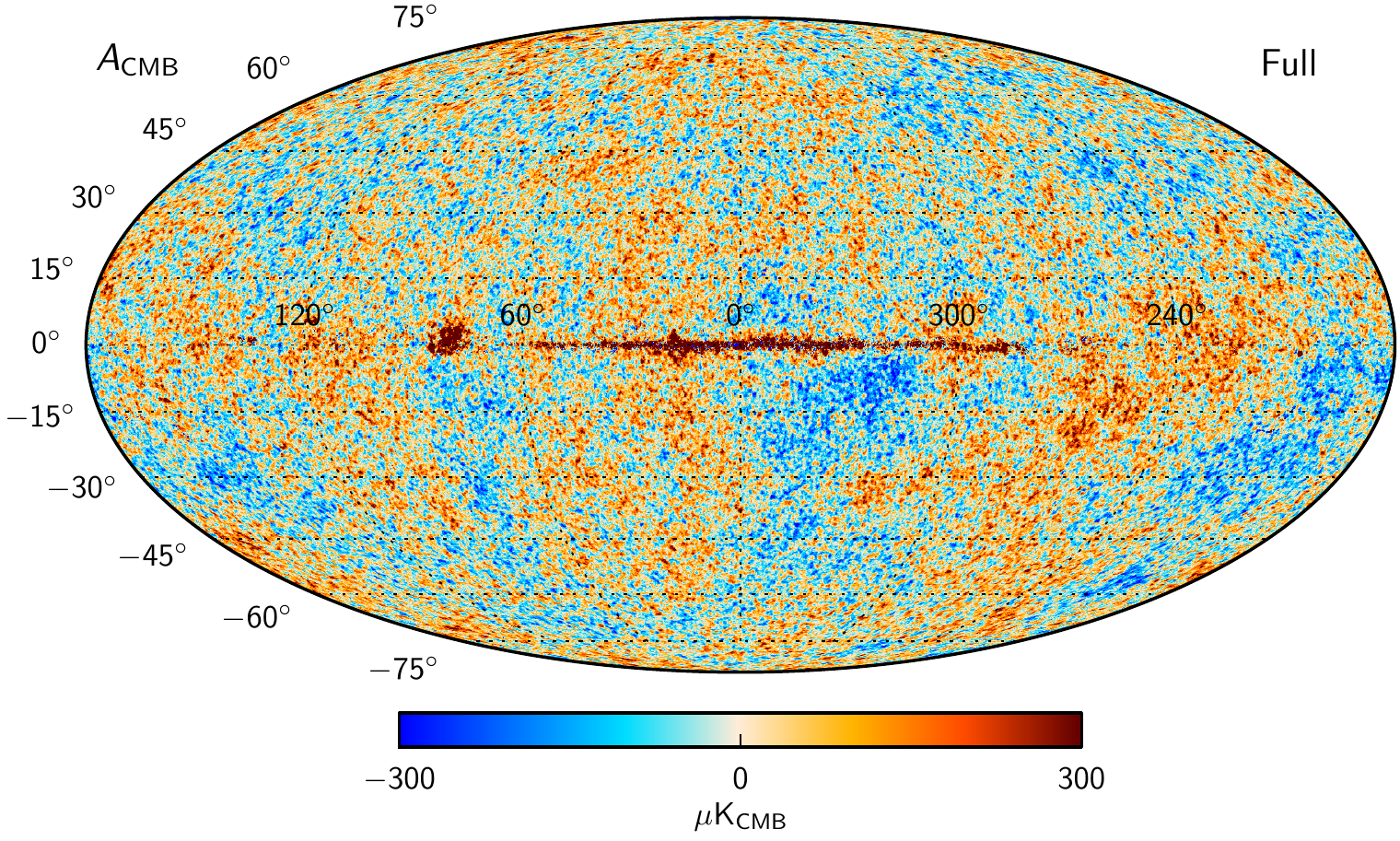,width=1.0\linewidth,clip=}
    }
    \vskip 4mm
    \mbox{
      \epsfig{figure=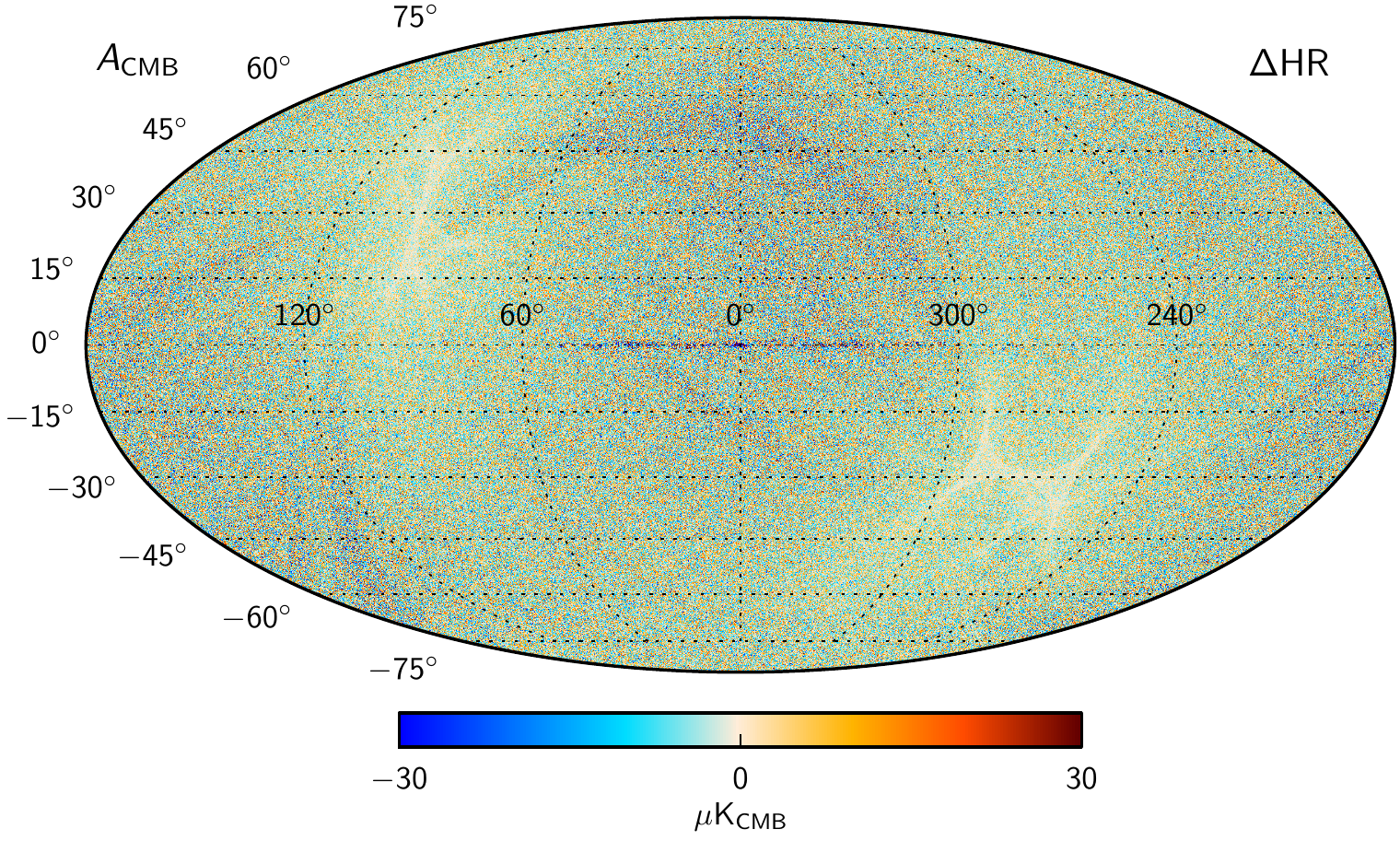,width=1.0\linewidth,clip=}
    }
  \end{center}
  \caption{High-resolution maximum-posterior (\emph{top}) and half-ring
    half-difference (\emph{bottom}) CMB amplitude maps. }
  \label{fig:highres_cmb_maps}
\end{figure*}

\begin{figure*}[p]
    \vskip 10mm
    \begin{center}
      \mbox{
        \epsfig{figure=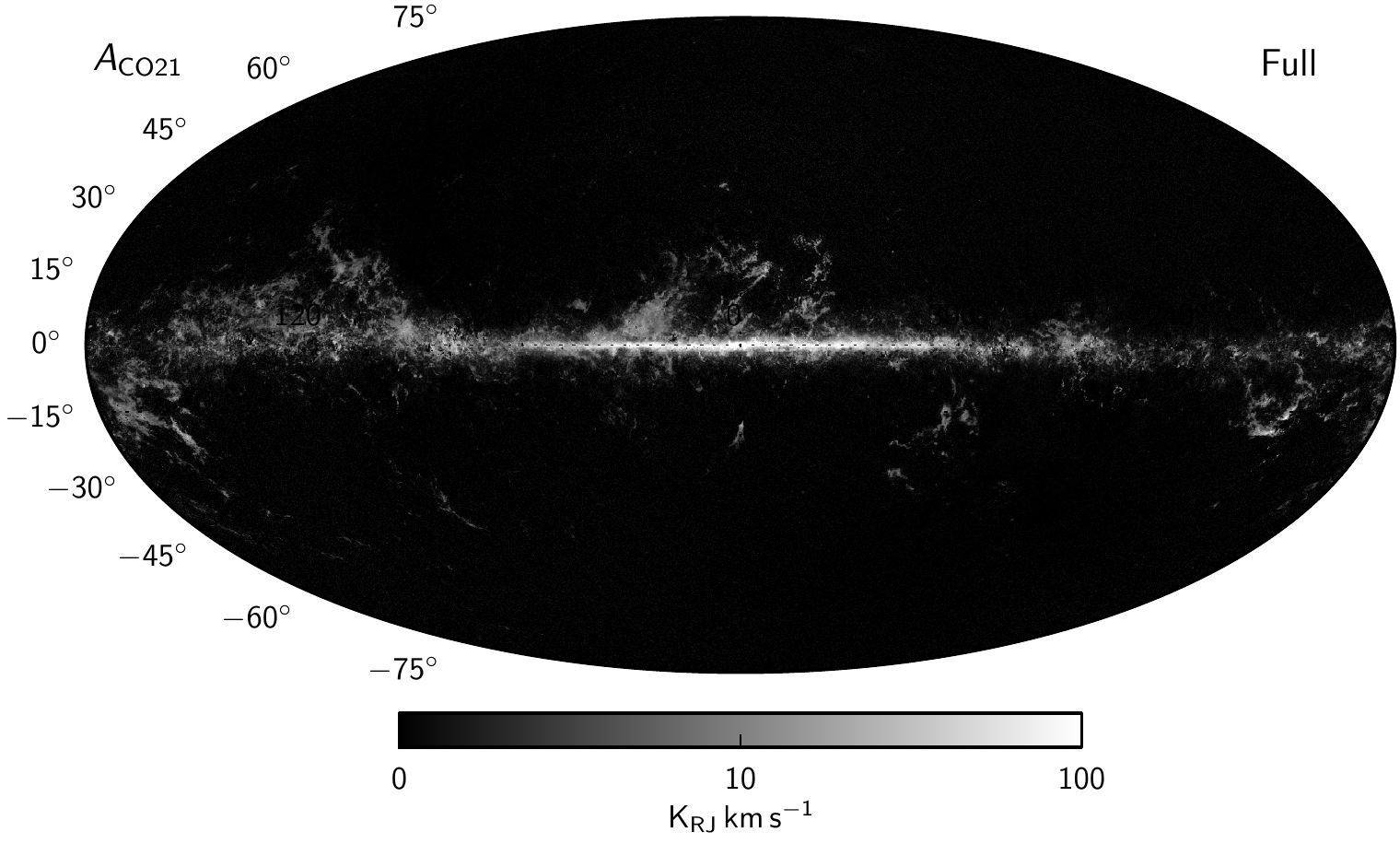,width=1.0\linewidth,clip=}
      }
      \vskip 4mm
      \mbox{
        \epsfig{figure=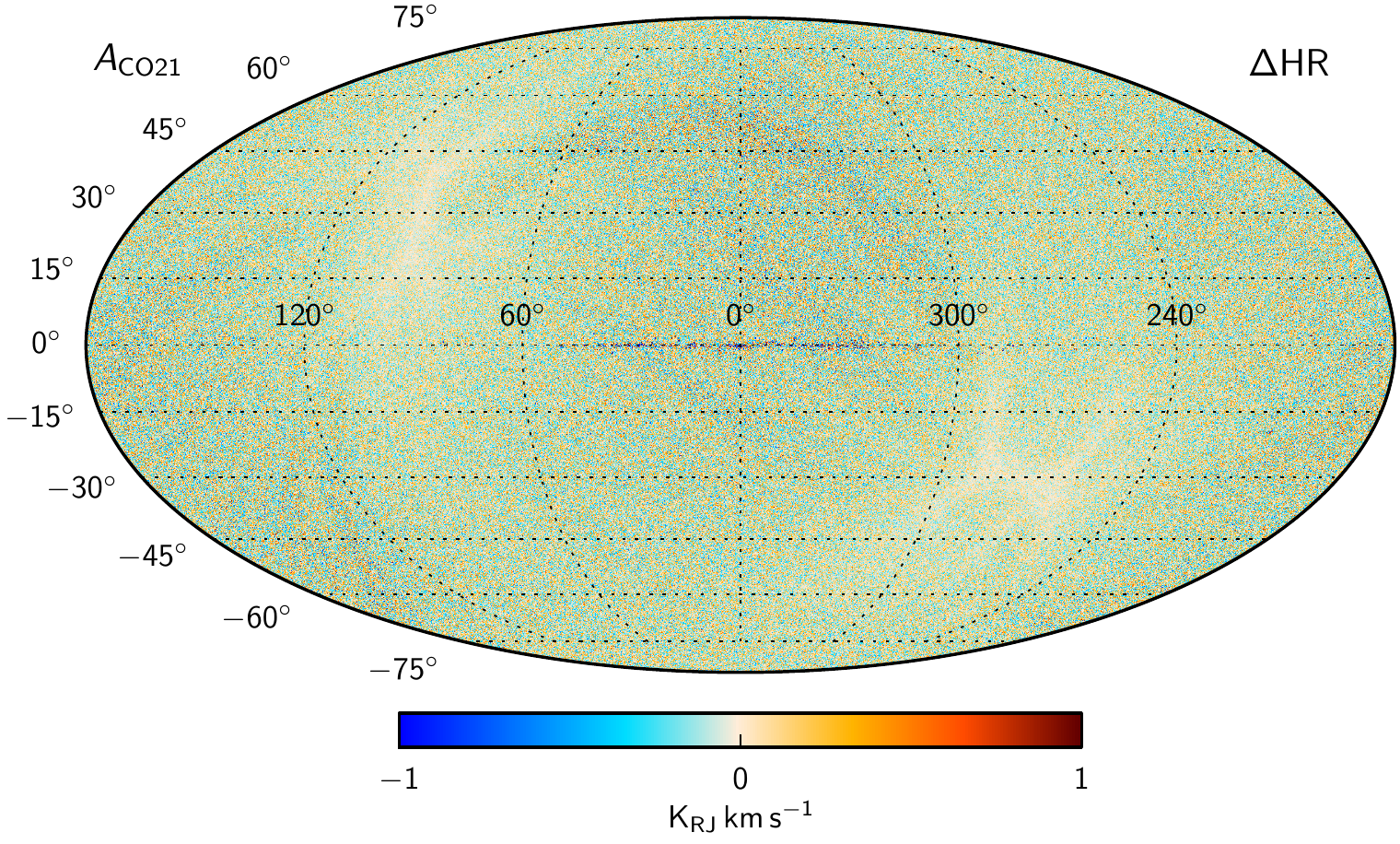,width=1.0\linewidth,clip=}
      }
    \end{center}
    \caption{High-resolution maximum-posterior (\emph{top}) and half-ring
      half-difference (\emph{bottom}) CO $J$=2$\rightarrow$1 amplitude maps. }
    \label{fig:highres_co21_maps}
  \end{figure*}
      
\begin{figure*}[p]
    \vskip 10mm
    \begin{center}
      \mbox{
        \epsfig{figure=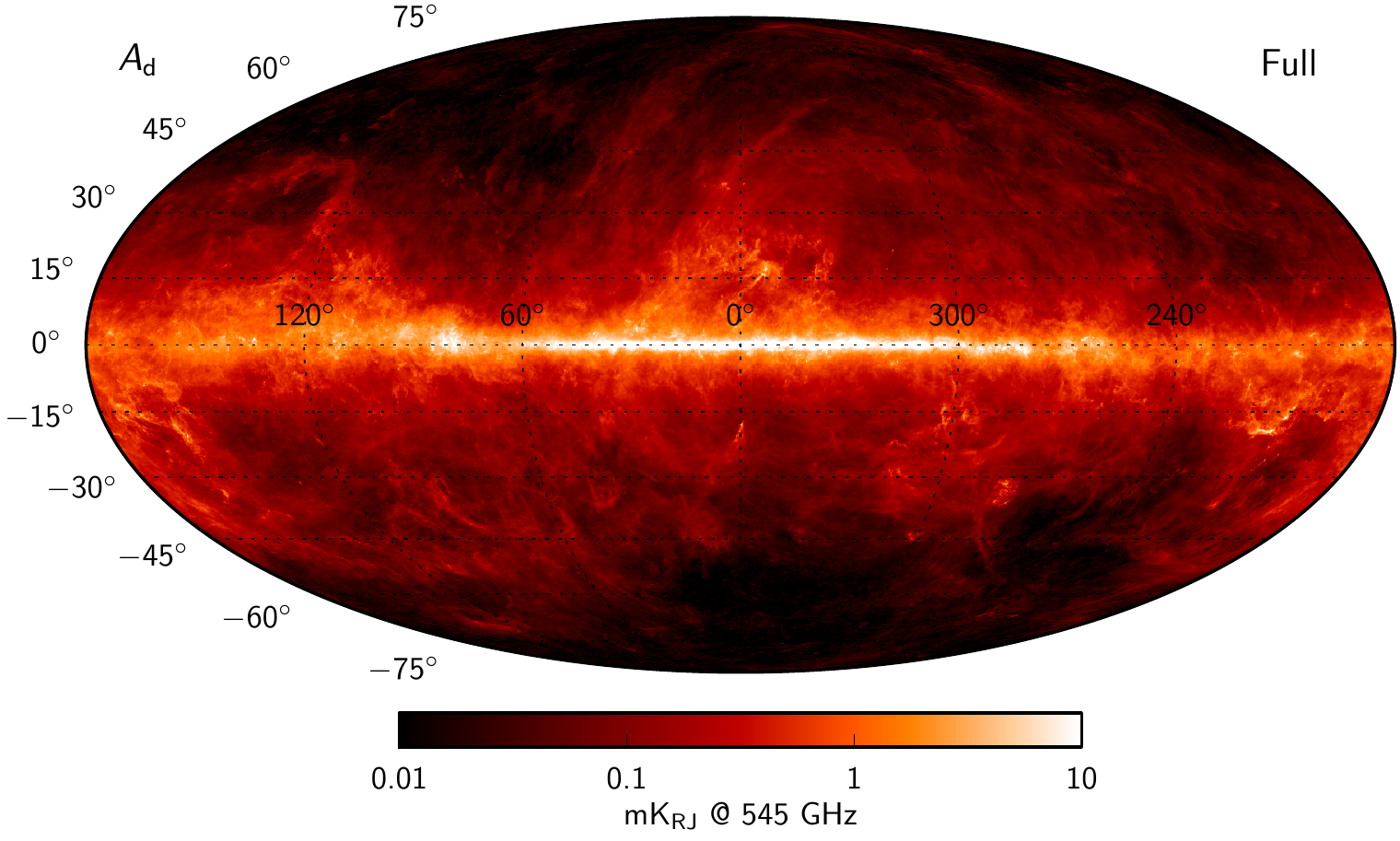,width=1.0\linewidth,clip=}
      }
      \vskip 4mm
      \mbox{
        \epsfig{figure=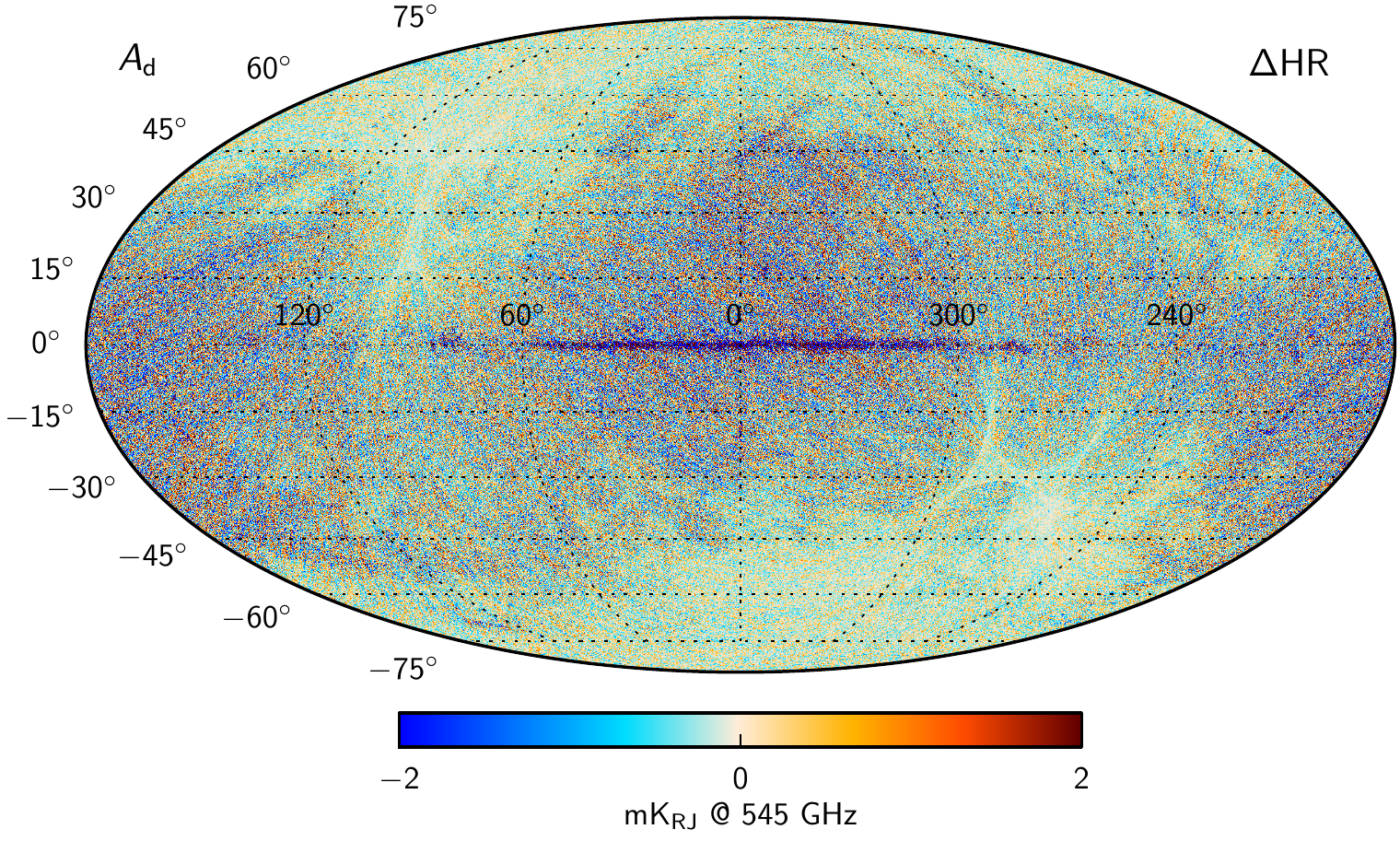,width=1.0\linewidth,clip=}
      }
    \end{center}
    \caption{High-resolution maximum-posterior (\emph{top}) and half-ring
      half-difference rms (\emph{bottom}) thermal dust amplitude maps. }
    \label{fig:highres_dust_maps}
  \end{figure*}

\begin{figure*}[p]
    \vskip 10mm
    \begin{center}
      \mbox{
        \epsfig{figure=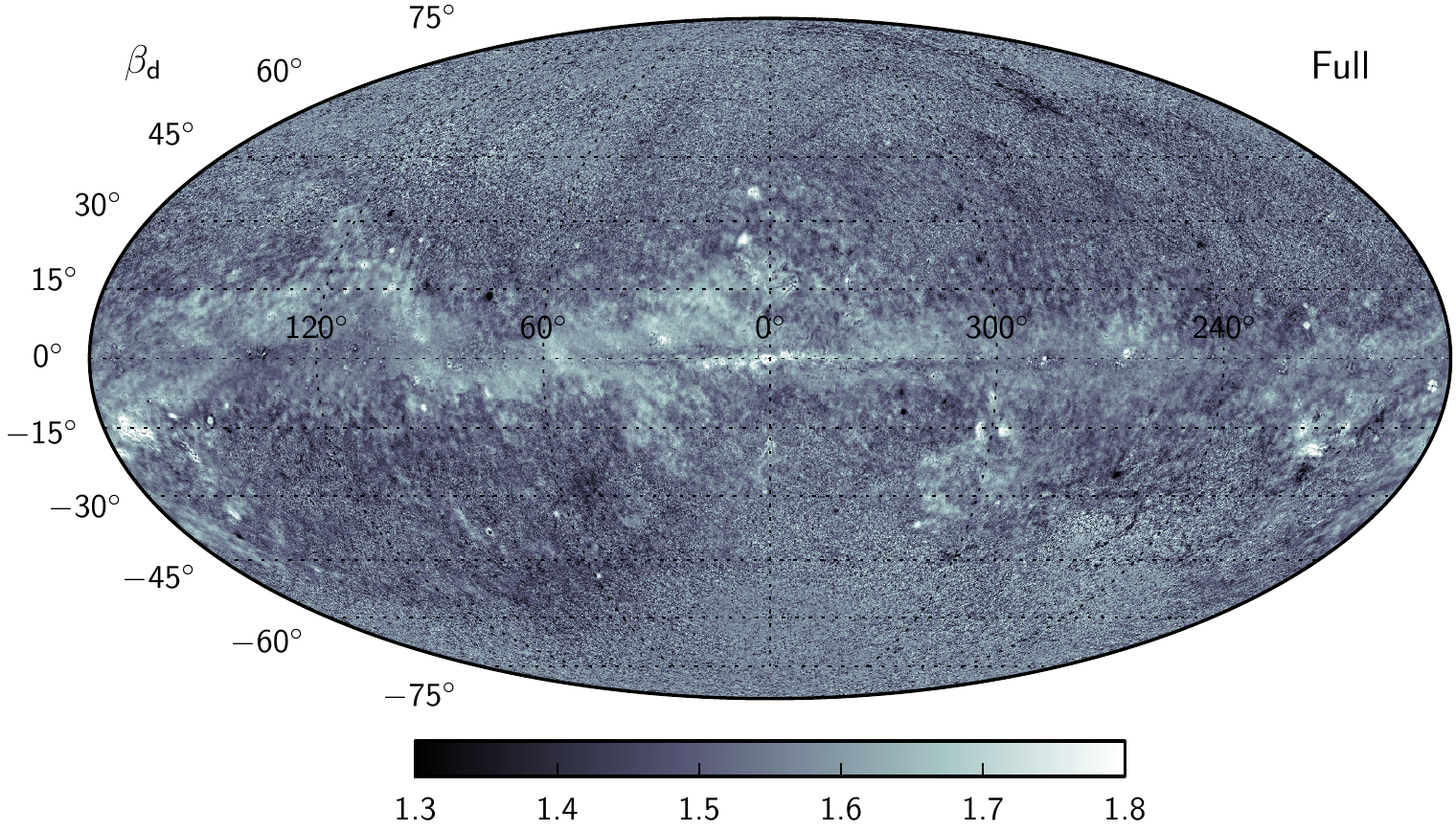,width=1.0\linewidth,clip=}
      }
      \vskip 6mm
      \mbox{
        \epsfig{figure=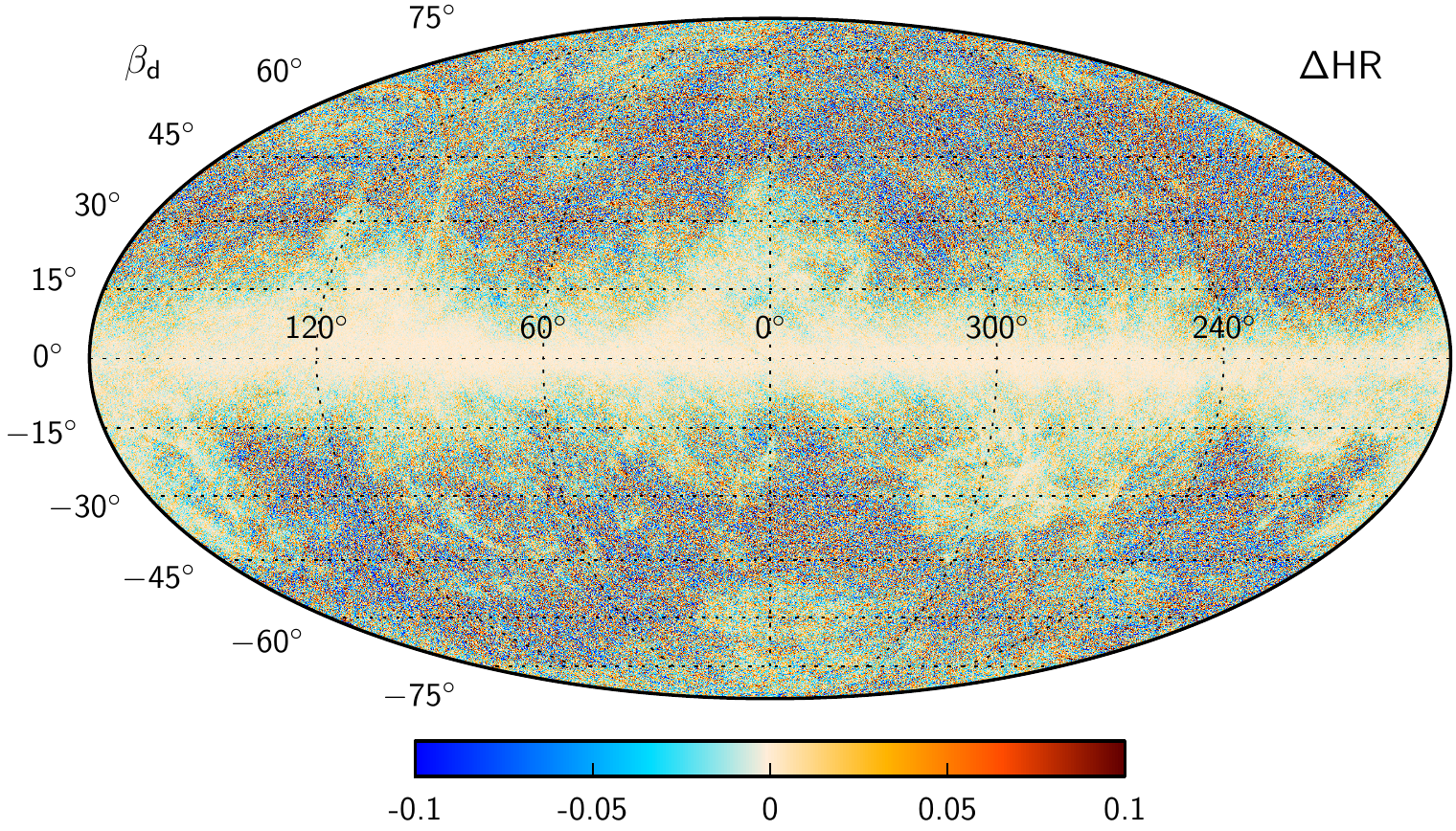,width=1.0\linewidth,clip=}
      }
    \end{center}
    \caption{High-resolution maximum-posterior (\emph{top}) and half-ring
      half-difference rms (\emph{bottom}) thermal dust spectral index maps. }
    \label{fig:highres_dust_beta_maps}
  \end{figure*}

The second column of Table~\ref{tab:Tgof} lists the rms of each
residual map outside the LM93 mask. The third column lists the ratio
between these rms values and the corresponding instrumental noise rms,
as listed in Table~\ref{tab:data}. The fourth column shows the
monopole and dipole corrected median fractional residual,
$(\d_{\nu}-\s_{\nu})/(\d_{\nu}-\T_{\nu}\m_{\nu})$, this time evaluated
inside the LM93 mask. As seen in the third column, the rms values of the
residual map are close to the instrumental noise for most channels,
indicating an excellent goodness-of-fit, not only in terms the
integrated $\chi^2$, but also channel-by-channel.

In accordance with the above discussion, we see that a number of
channels have residuals that are lower than the instrumental noise,
with the 408\,MHz and 857-2 channels being the two most striking
examples. This is expected, because the $\chi^2$ only has 20 degrees
of freedom, whereas there are 32 individual frequency channels; the
normalized mean residuals must therefore sum to less than one per
channel. However, a ratio much smaller than unity typically indicates
that the corresponding channel has a much higher effective
signal-to-noise ratio with respect to some signal parameter than all
other channels combined. In our case, the 408\,MHz and 857-2 channels
strongly dominate the synchrotron and thermal dust amplitudes,
respectively. At first sight, one might suspect these values to
indicate the presence of worrisome parameter degeneracies, which
typically also can result in residuals lower than the instrumental
noise. However, from the parameter maps shown in
Figs.~\ref{fig:cmb_amp_map}--\ref{fig:ff_Te_map}, it is visually
obvious that the synchrotron and thermal dust emission amplitude maps
are not degenerate with any components. Rather than degeneracies,
these low residual values indicate that the current data set is
\emph{non-redundant} with respect to these two amplitude maps; the
408\,MHz map determines almost exclusively the synchrotron amplitude,
and the 857-2 map determines almost exclusively the thermal dust
amplitude. The main problem with these low residuals is therefore not
degeneracies, but rather lack of robustness with respect to
systematics; any systematic error that may be present in the 408\,MHz
and 857-2 channels can and will propagate into the respective
foreground amplitude maps. In order to improve on this situation in
the future, recovering the currently systematics contaminated 857-1,
857-3, and 857-4 channels is imperative on the high-frequency side, and
incorporating additional low-frequency observations (between, say, 1
and 20\,GHz) is critical on the low-frequency side.

A similar effect is seen for a number of other channels, if not
equally strongly. For instance, we see that the WMAP K-band and
$\Planck$ 30\,GHz channels have rms ratios of 0.38 and 0.55,
respectively, and these dominate the two spinning dust
amplitudes, $A_{\textrm{sd1}}$ and $A_{\textrm{sd2}}$. The 100-ds2
channel has a ratio of 0.76, and defines the CO $J$=1$\rightarrow$0
amplitude together with 100-ds1. Finally, the 353-1 channel has a
ratio of 0.60, and this channel has the greatest pull on the dust
emissivity index, $\beta_{\textrm{d}}$.

The single most important conclusion from Table~\ref{tab:Tgof},
however, is that the baseline \Planck\ temperature sky model is an
excellent fit to the observed data, in agreement with the visual
impression of Fig.~\ref{fig:map_residuals}. The residuals are largely
consistent with instrumental noise over 93\,\% of the sky, and the
median fractional residual in the complementary 7\,\% of the sky is;
smaller than 0.2\,\% for all HFI channels; smaller than 1\,\% for all LFI
channels; and smaller than 2\,\% for all WMAP channels.

\subsection{High-resolution component maps}
\label{sec:temp_highres}

We now consider the high-resolution intensity maps derived using the
same pipeline as above, but with a reduced data set and astrophysical
model. Specifically, we only include channels from 143\,GHz and above,
all smoothed to a common resolution of $7\parcm5$ FWHM. The signal
model includes CMB, thermal dust, and CO $J$=2$\rightarrow$1 and
3$\rightarrow$2 emission lines coadded into one map, similar to the
Type-3 map in the 2013 data release.\footnote{Although our
  high-resolution CO map formally is a weighted avarage of
  $J$=2$\rightarrow$1 and $J$=3$\rightarrow$2 line emission, the
  former vastly dominates, and we therefore refer to the map as CO
  $J$=2$\rightarrow$1.} We fix all global instrumental parameters on
the values listed in Table~\ref{tab:monopole}, and the thermal dust
temperature, $T_{\textrm{d}}$, to the low-resolution solution,
upgraded in harmonic space (to avoid pixelization effects) to a
\healpix\ resolution of $N_{\textrm{side}}=2048$.  The only free
spectral parameter per pixel is now the thermal dust emissivity index,
$\beta_{\textrm{d}}$.

The resulting amplitude maps are shown in the top panels of
Figs.~\ref{fig:highres_cmb_maps}--\ref{fig:highres_dust_maps}, while
the bottom panels show the so-called half-ring half-difference maps,
i.e., half-difference between two maps derived from independent
half-ring maps \citep{planck2014-a07,planck2014-a09}; these provide a
direct estimate of the instrumental noise present in the full-mission
maps. Figure~\ref{fig:highres_dust_beta_maps} shows the same for the
high-resolution dust spectral index.

Note that while the Galactic centre appears negative in the
low-resolution CMB solution shown in Fig.~\ref{fig:cmb_amp_map}, it
appears positive in the corresponding high-resolution CMB shown in
Fig.~\ref{fig:highres_cmb_maps}. This qualitative difference demonstrates
the importance of modelling errors in the Galactic plane. At high
resolution, our model includes only CMB, CO and thermal dust, but no
dedicated low-frequency component. Any residual free-free
contributions to frequencies at or above 143\,GHz is therefore
necessarily interpreted as a combination of CMB and CO, both of which
have redder spectral shapes than thermal dust. In the low-resolution
solution, on the other hand, the main problem lies in the interplay
between CO and thermal dust modelling errors and high-frequency
calibration uncertainties.

\subsection{Comparison with independent data products}

To further validate the baseline model presented in
Sect.~\ref{sec:baseline}, we now compare the derived products with
similar maps produced either by independent observations or through
different analysis techniques, focusing in particular on spinning and
thermal dust and CO emission. Synchrotron and free-free (and spinning
dust) are addressed separately in a dedicated companion paper, and we
refer the interested reader to \citet{planck2014-a31} for full
details. A short summary of that analysis includes the following main
points.
\begin{enumerate}
  \item The synchrotron estimates derived by the \WMAP\ team
    \citep{bennett2012} using either Markov chain Monte Carlo (MCMC)
    or maximum entropy methods (MEM) typically have 50--70\,\% higher
    amplitudes at high Galactic latitudes compared to that derived in
    this paper, and this increases to factors of several when
    including the Galactic plane. This is compensated by about twice
    as much spinning dust in our model compared to the \WMAP\ models.
  \item The free-free model derived in the current analysis correlates
    well with H$\alpha$ observations. For instance, the scaling factor
    between the two maps in the Gum Nebula is
    $(8.2\pm1.3)\,\mu\textrm{K}\,\textrm{R}^{-1}$. For comparison,
    \citet{dickinson2003} found values ranging between 8.2 and
    $13.1\,\mu\textrm{K}\,\textrm{R}^{-1}$, depending on the exact
    position within the Gum Nebula.
  \item The free-free map also shows good morphological agreement
    with respect to radio recombination line (RRL) observations
    \citep{alves2014}, although the derived amplitude is notably
    higher in our map. The six brightest objects have a mean relative
    amplitude ratio of $1.14\pm0.04$, whereas the ten next have an
    amplitude ratio of $1.36\pm0.08$. Considering that RRLs are in
    principle a very clean probe of free-free emission, the most
    likely explanation for this discrepancy is leakage between
    synchrotron, spinning dust, CO, and free-free in the current
    solution. On the other hand, the derived RRL amplitudes also
    depends sensitively on the assumed electron temperature, and
    raising $/Te$ by $\approx$20\,\% would resolve much of the
    difference. In addition, it is worth noting that other component
    separation techniques, including FastMEM, CCA, and both the 9-year
    WMAP MCMC and MEM analyses, all derive free-free amplitudes
    consistent with the \texttt{Commander} result
    \citep{planck2014-XXIII}.
\end{enumerate}

\subsubsection{Dust template amplitude consistency by \ion{H}{i} cross-correlation}
\label{sec:ame}

We next perform an internal consistency test of the \texttt{Commander}
dust model in the range from 23 to 353 GHz, as defined by the sum of
the \texttt{SpDust2} components described in Sect.~\ref{sec:sky} and
thermal dust, by cross-correlating our dust model against GASS \ion{H}{i}
observations \citep{gass1,gass2} covering 18\,\% of the high Galactic latitude
sky near the South Galactic pole, following the procedure of
\citet{planck2013-XVII}. In particular, we compare the resulting
template amplitudes against the corresponding amplitudes derived
directly from cross-correlation with the raw \Planck\ and \WMAP\ sky
maps.

\begin{figure}
\begin{center}
\mbox{
\epsfig{figure=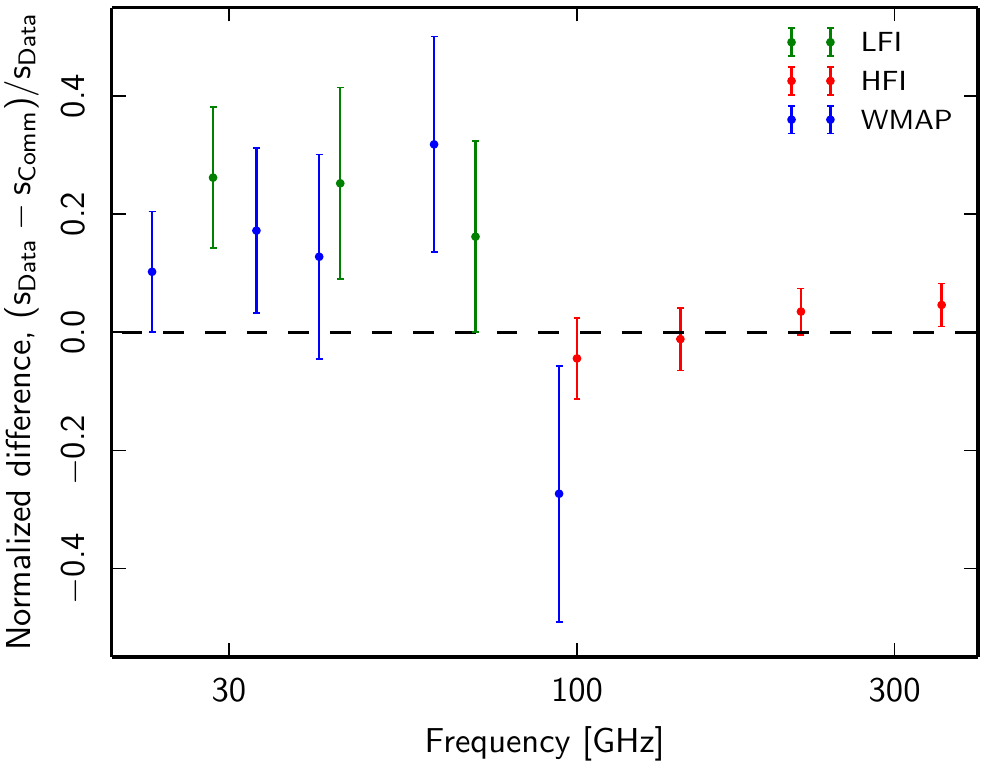,width=\linewidth,clip=}
}
\end{center}
\caption{Fractional difference between the template amplitudes derived
  when fitting the GASS \ion{H}{i} survey data at high Galactic
  latitudes to: (1) the raw \Planck\ and \WMAP\ temperature maps; and
  to (2) the sum of the spinning and thermal dust models derived by
  \commander\ in this paper. }
\label{fig:ame_SED}
\end{figure}

Figure~\ref{fig:ame_SED} shows the fractional difference between the
resulting template amplitudes for each frequency band. Overall, the
agreement is satisfactory with typically 20\,\% differences in
the 20--70\,GHz range, in which spinning dust provides a larger
contribution to the frequency spectrum than thermal dust. At higher
frequencies, where thermal dust emission starts to dominate, the
agreement improves further to around 5\,\%, and to within $1\,\sigma$ in
terms of statistical uncertainties.

\subsubsection{Dust SED consistency by \ion{H}{i} and internal \Planck\ cross-correlations}
\label{sec:dust}

Next, we consider the robustness and consistency of the thermal dust
SED model, as parametrized in terms of the two greybody parameters,
$\beta_{\textrm{d}}$ and $T_{\textrm{d}}$. Specifically, we compare
the new SED estimates derived in this paper with corresponding
estimates derived by \ion{H}{i} cross-correlation at high latitudes in
\citet{planck2013-XVII}, and by internal \Planck\ cross-correlations
at intermediate Galactic latitudes in \citet{planck2014-XXII}, both of
which have been updated with the latest \Planck\ 2015 sky maps.

\begin{figure}
\begin{center}
\mbox{
\epsfig{figure=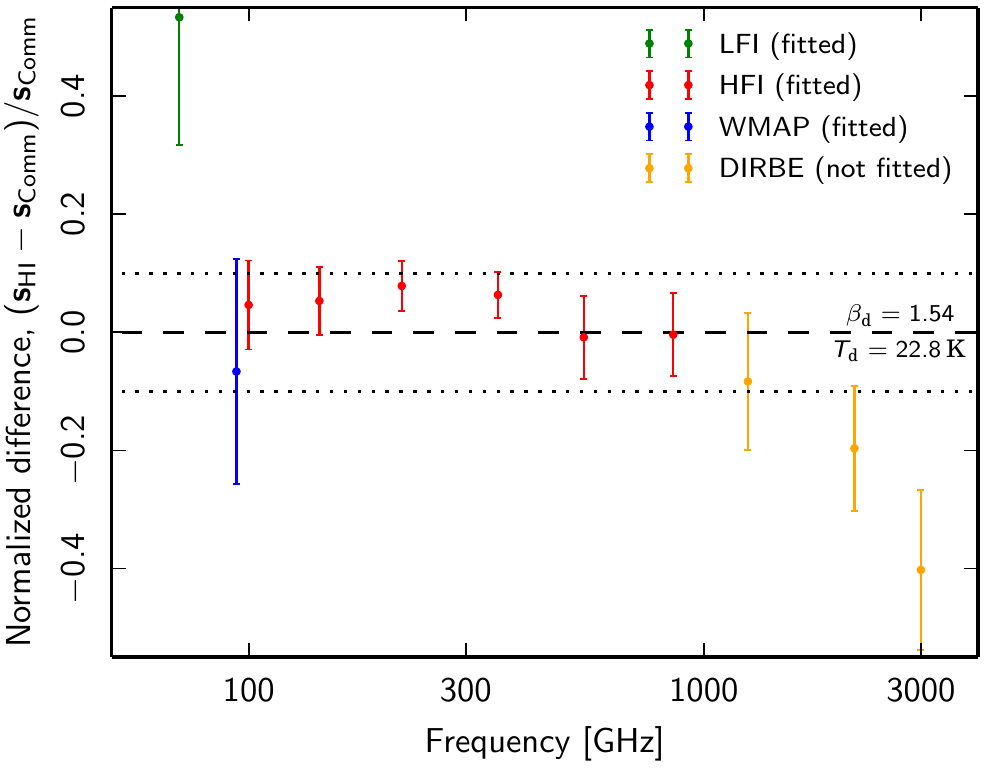,width=\linewidth,clip=}
}
\end{center}
\caption{Fractional difference of the mean thermal dust SEDs as
  derived by cross-correlation with the GASS \ion{H}{i} survey data at
  high Galactic latitudes, updated with the latest \Planck\ 2015
  temperature sky maps, \citep{planck2013-XVII} and by
  \texttt{Commander} in this paper. The dotted horizontal lines
  indicate fractional differences of $\pm10\,$\%. For comparison
  purposes, we also show the extrapolation to the 100, 140, and
  240$\,\mu\textrm{m}$ DIRBE frequencies. These observations are not
  included in the fits performed in this paper; see
  Sect.~\ref{sec:dust} for further discussion.}
\label{fig:dust_SED}
\end{figure}

Figure~\ref{fig:dust_SED} compares the mean thermal dust SED derived
from \ion{H}{i}--CMB cross-correlation and the \texttt{Commander} estimates at
high Galactic latitudes in terms of the fractional difference,
$(s_{\textrm{\ion{H}{i}}} -s_{\textrm{Comm}})/s_{\textrm{Comm}}$. The two sets
of best-fit thermal dust spectral parameters are
$(\beta_{\textrm{d}},T_{\textrm{K}})_{\textrm{Comm}}=(1.54,22.8\,\textrm{K})$
and
$(\beta_{\textrm{d}},T_{\textrm{K}})_{\textrm{\ion{H}{i}}}=(1.54,21.4\,\textrm{K})$,
respectively, and the two models agree point-by-point to 5--10\,\%
between 100 and 857\,GHz. At 70\,GHz the difference is 50\,\%, and
this is due to different spinning dust modelling; as already shown in
Fig.~\ref{fig:ame_SED}, the sum of spinning and thermal dust agree to
20\,\% in this range between the two methods. Note also that 70\,GHz
is very close to the foreground minimum, and these numerically large
relative differences therefore correspond to small absolute
differences.

\begin{figure}
\begin{center}
\mbox{
  \epsfig{figure=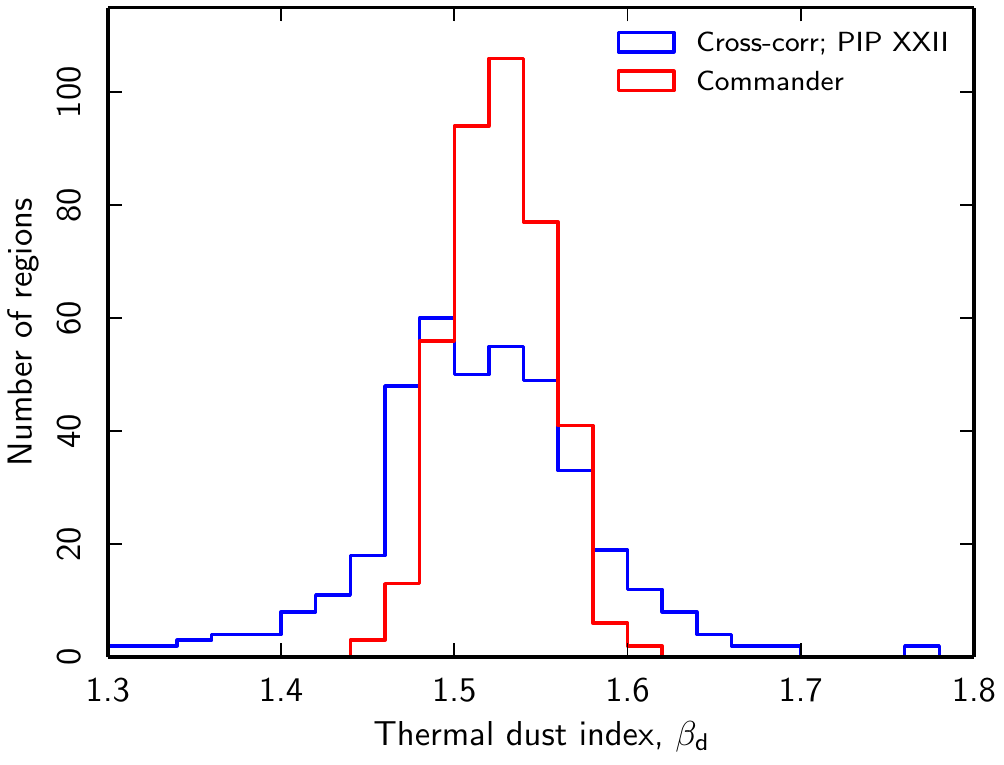,width=\linewidth,clip=}
}

\end{center}
\caption{Comparison of the thermal dust spectral index,
  $\beta_{\textrm{d}}$, estimated by internal \Planck\ map
  cross-correlations over \healpix\ $N_{\textrm{side}}=8$ pixels in
  \citet{planck2014-XXII} and those presented in this paper.  The
  best-fit Gaussian distributions to the two histograms have mean and
  standard deviations of $\beta_{\textrm{d}}^{\texttt{Comm}} =
  1.53\pm0.03$ and $\beta_{\textrm{d}}^{\textrm{cross-corr}} =
  1.51\pm0.06$, respectively. }
\label{fig:dust_par}
\end{figure}

This test provides a robust estimate of residual systematic errors in
the \texttt{Commander} thermal dust model from potential zero-level
and dipole uncertainties in the high-frequency HFI channels arising
from zodiacal light emission and CIB residuals, as discussed in
Sect.~\ref{sec:data}. Because the \ion{H}{i} analysis is insensitive
to such errors, we take the 1--2\,K difference between the two as an
estimate of the systematic uncertainty on the \commander\
thermal dust temperature at high Galactic latitudes.

At frequencies above 857\,GHz we also plot the extrapolation of the
new \texttt{Commander} model into the \COBE-DIRBE wavelengths of 240,
140, and 100$\,\mu\textrm{m}$ \citep{hauser1998}. Here we clearly see
that the current model breaks down beyond the \Planck\ frequencies,
with a fractional difference of 40\,\% between the \texttt{Commander}
model and the DIRBE 100$\,\mu$m observations. Including the DIRBE
channels in the fit would of course reduce these fractional residuals
dramatically, but only at a very significant cost of increasing the
residuals at lower frequencies between 70 and 353\,GHz. The simple
one-component greybody thermal dust model adopted in this paper is not
able to simultaneously fit this entire frequency range, because of
both intrinsic complexity of the dust particle population and because
of residual systematics and calibration uncertainties in the DIRBE and
high-frequency \Planck\ data. Integration of these channels requires
substantial additional work, and is beyond the scope of the current
paper. For a first analysis of similar type, see
\citet{planck2014-XXIX}.

Next, we turn to intermediate Galactic latitudes, for which the
signal-to-noise ratio is higher, but also the astrophysical
composition is richer. In this case we therefore compare our results
with the outputs from the internal \Planck\ template cross-correlation
analysis of \citet{planck2014-XXII}. In short, this analysis estimates
the SED parameters by cross-correlating the \Planck\ 353\,GHz channel
with lower frequencies over circular patches of $10\deg$
radius. Figure~\ref{fig:dust_par} compares the histogram of
$\beta_{\textrm{d}}$ derived using this method with the corresponding
\texttt{Commander} estimates over the same sky region. The agreement
is very good, and with averages and dispersions of
$\beta_{\textrm{d}}^{\textrm{Comm}}=1.53\pm0.03$ and
$\beta_{\textrm{d}}^{\textrm{cross-corr}}=1.51\pm0.06$,
respectively. When interpreting the widths of these distributions, it
is useful to return to the thermal dust spectral index maps shown in
Figs.~\ref{fig:dust_beta_map} and \ref{fig:highres_dust_beta_maps}.
These maps are quite uniform, and, indeed, at the current level of
leakage between thermal dust, CO, compact objects and residual
offsets, there is little convincing evidence for true spatial
variation in $\beta_{\textrm{d}}$ in the results presented here. If
this conjecture is true, the widths of the histograms shown in
Fig.~\ref{fig:dust_par} are primarily expressions of analysis
uncertainties in the form of instrumental noise, parameter
degeneracies and systematic errors, rather than true spatial
variation.

\subsubsection{CO line emission}
\label{sec:CO}

\begin{figure*}
\begin{center}
\mbox{
\epsfig{figure=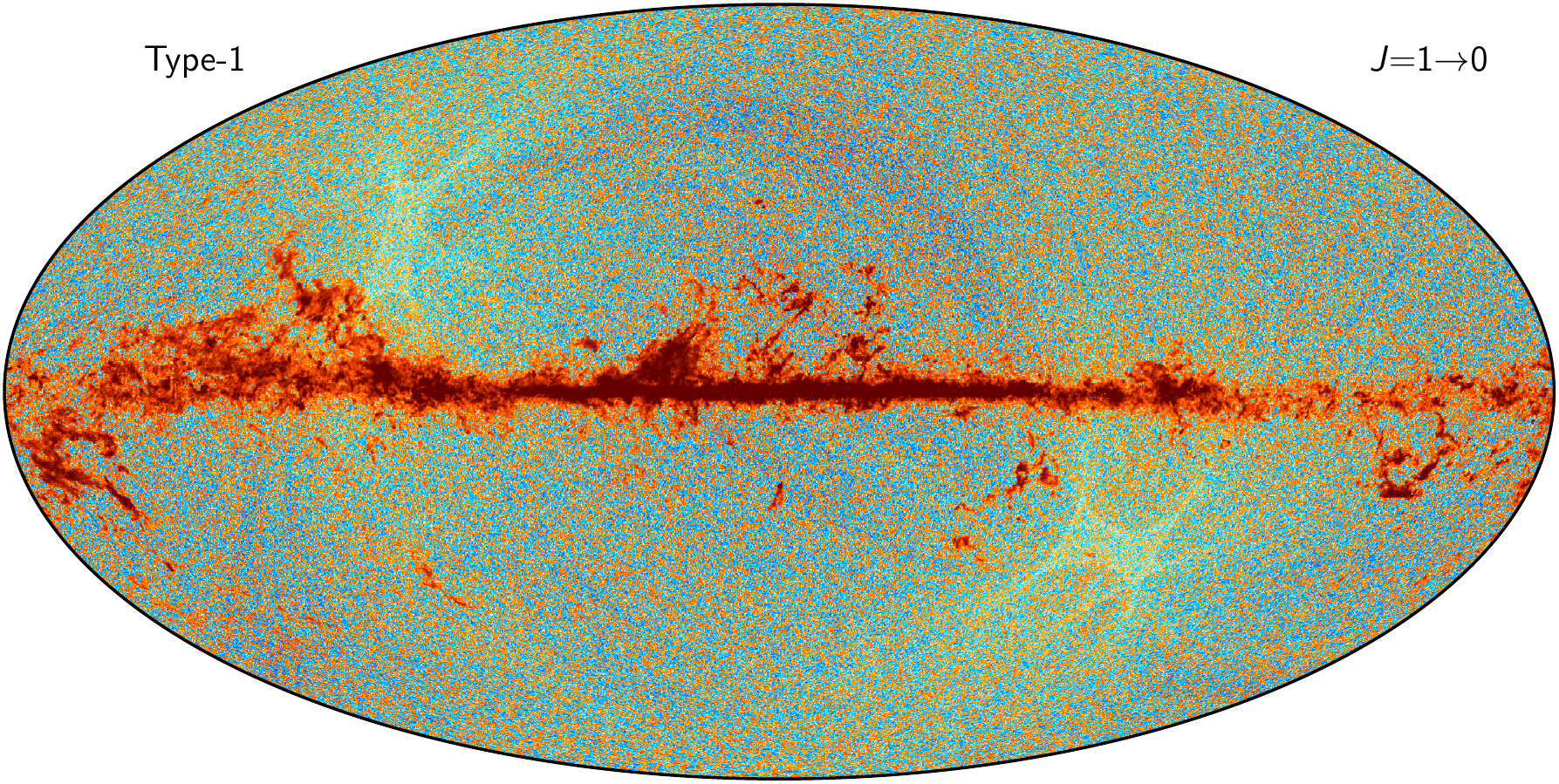,width=0.77\linewidth,clip=}
}
\mbox{
\epsfig{figure=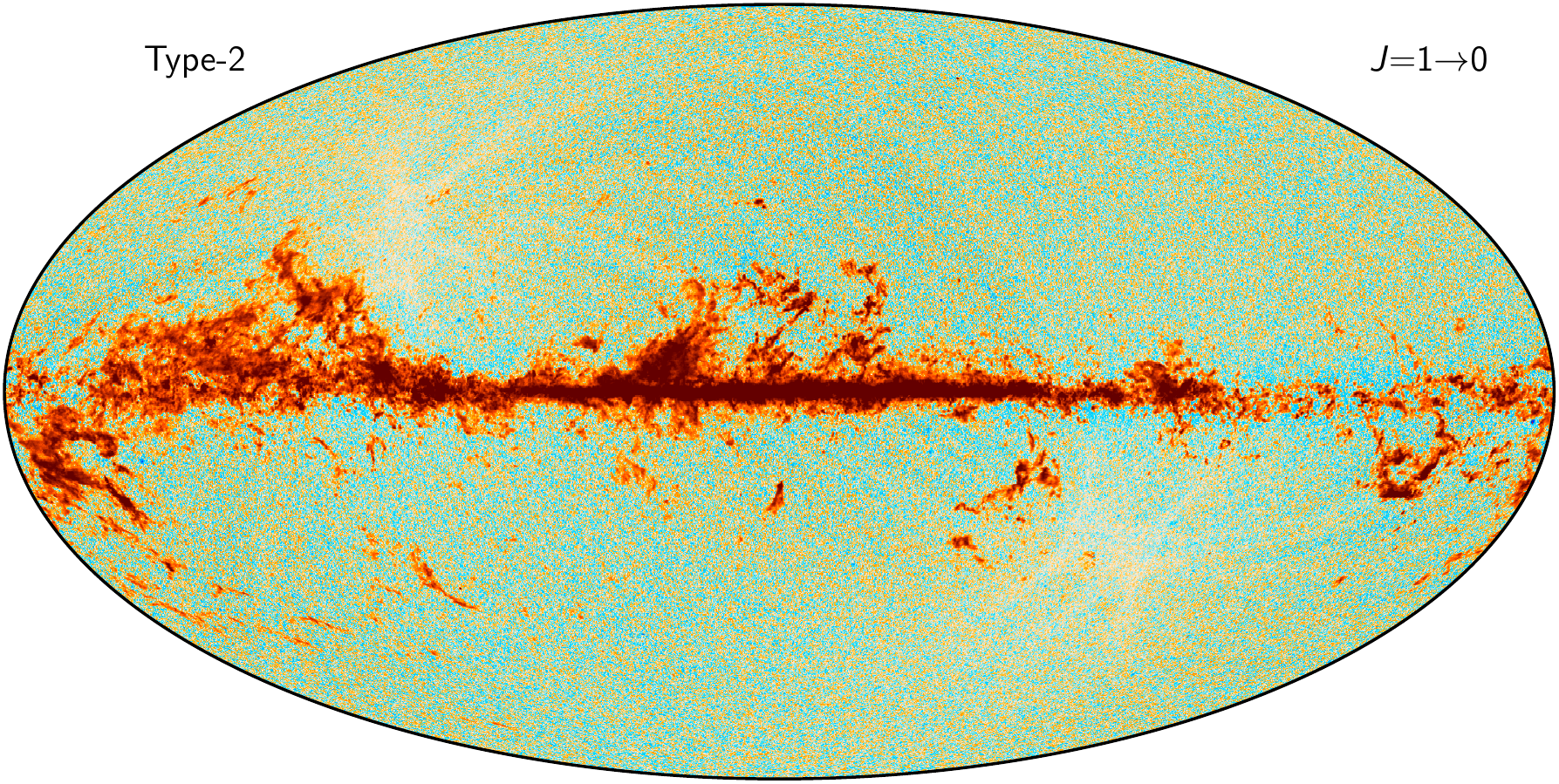,width=0.77\linewidth,clip=}
}
\mbox{
\epsfig{figure=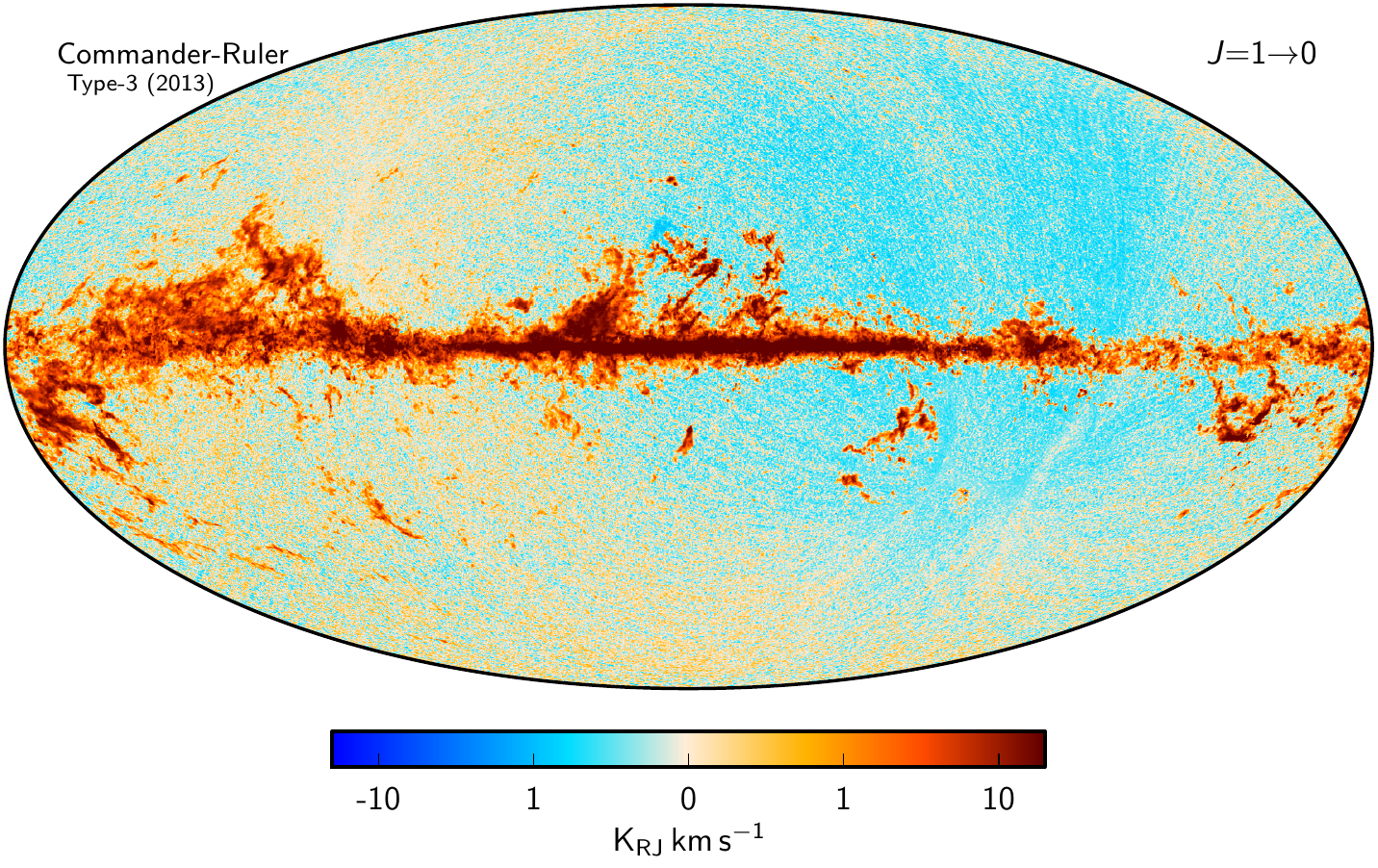,width=0.77\linewidth,clip=}
}
\end{center}
\caption{High-resolution CO $J$=1$\rightarrow$0 maps derived from the
  2015 \Planck\ sky maps with two different algorithms, denoted Type-1
  (\emph{top row}) and Type-2 (\emph{middle row}); and derived from the
  2013 \Planck\ sky maps with \texttt{Commander-Ruler} (\emph{bottom
    row}). Details can be found in \cite{planck2013-p03a}. All maps
  are smoothed to a common resolution of $15\arcm$ FWHM. }
\label{fig:co10_maps}
\end{figure*}

\begin{figure*}
\begin{center}
\mbox{
\epsfig{figure=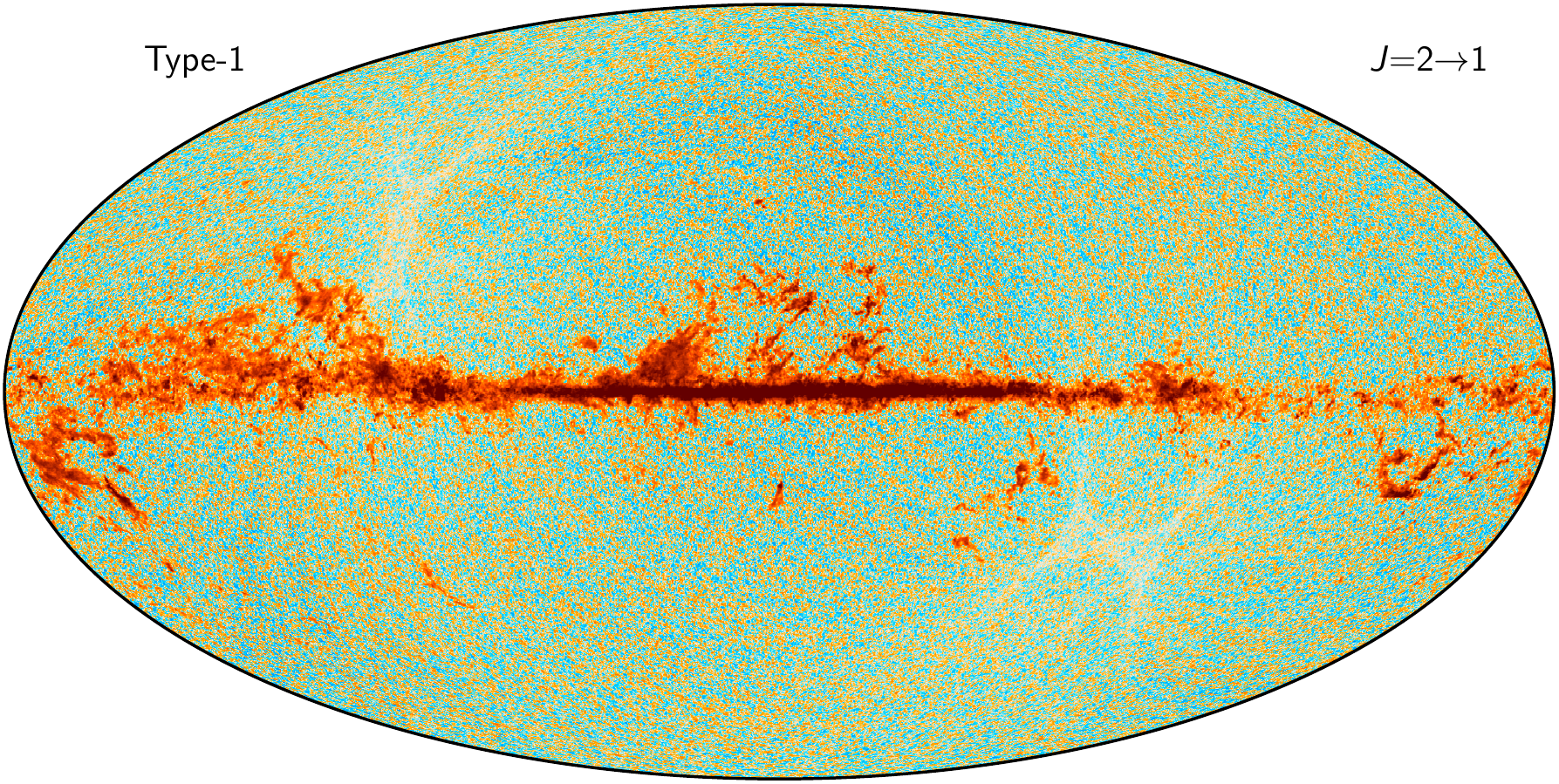,width=0.77\linewidth,clip=}
}
\mbox{
\epsfig{figure=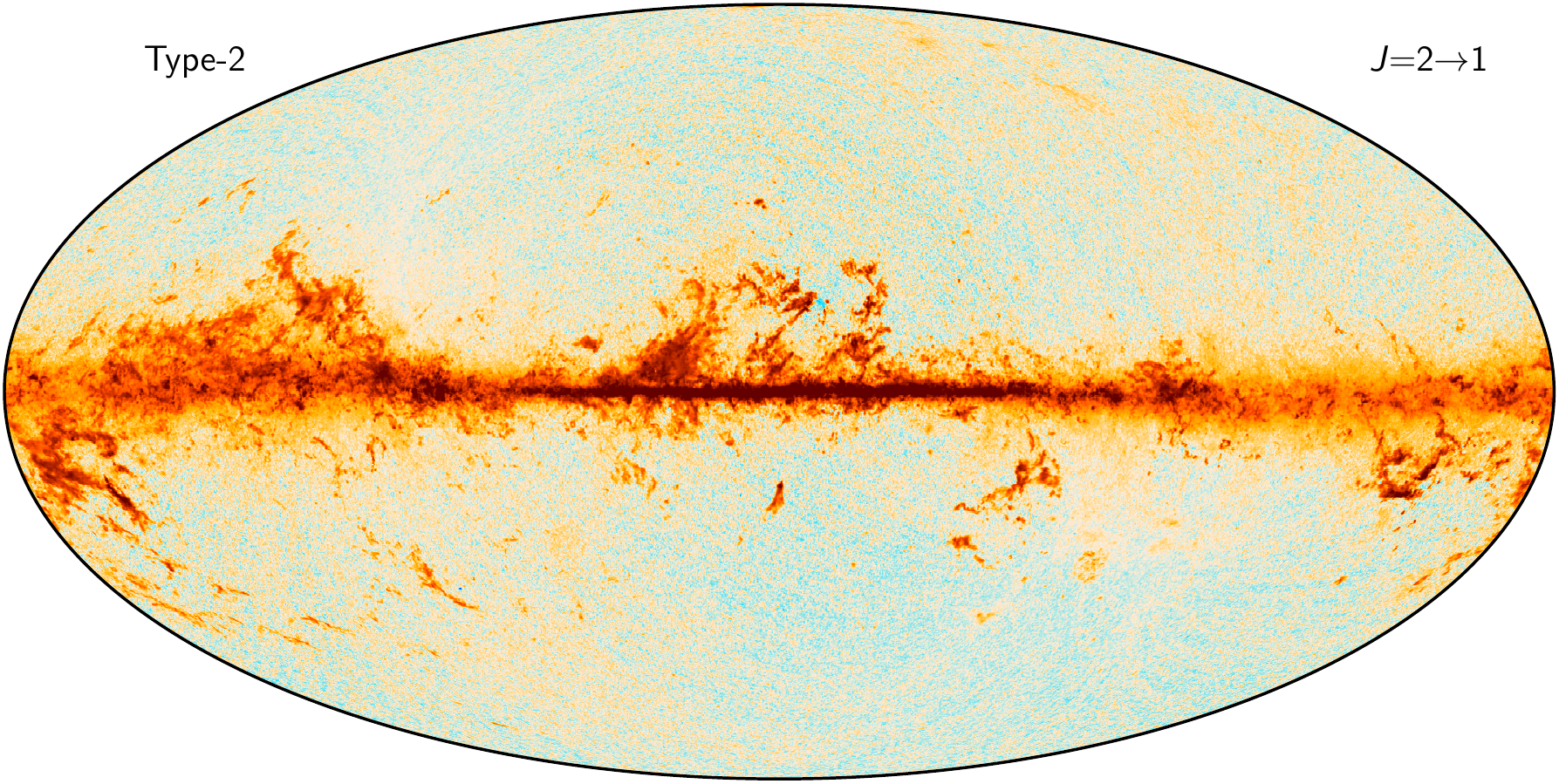,width=0.77\linewidth,clip=}
}
\mbox{
\epsfig{figure=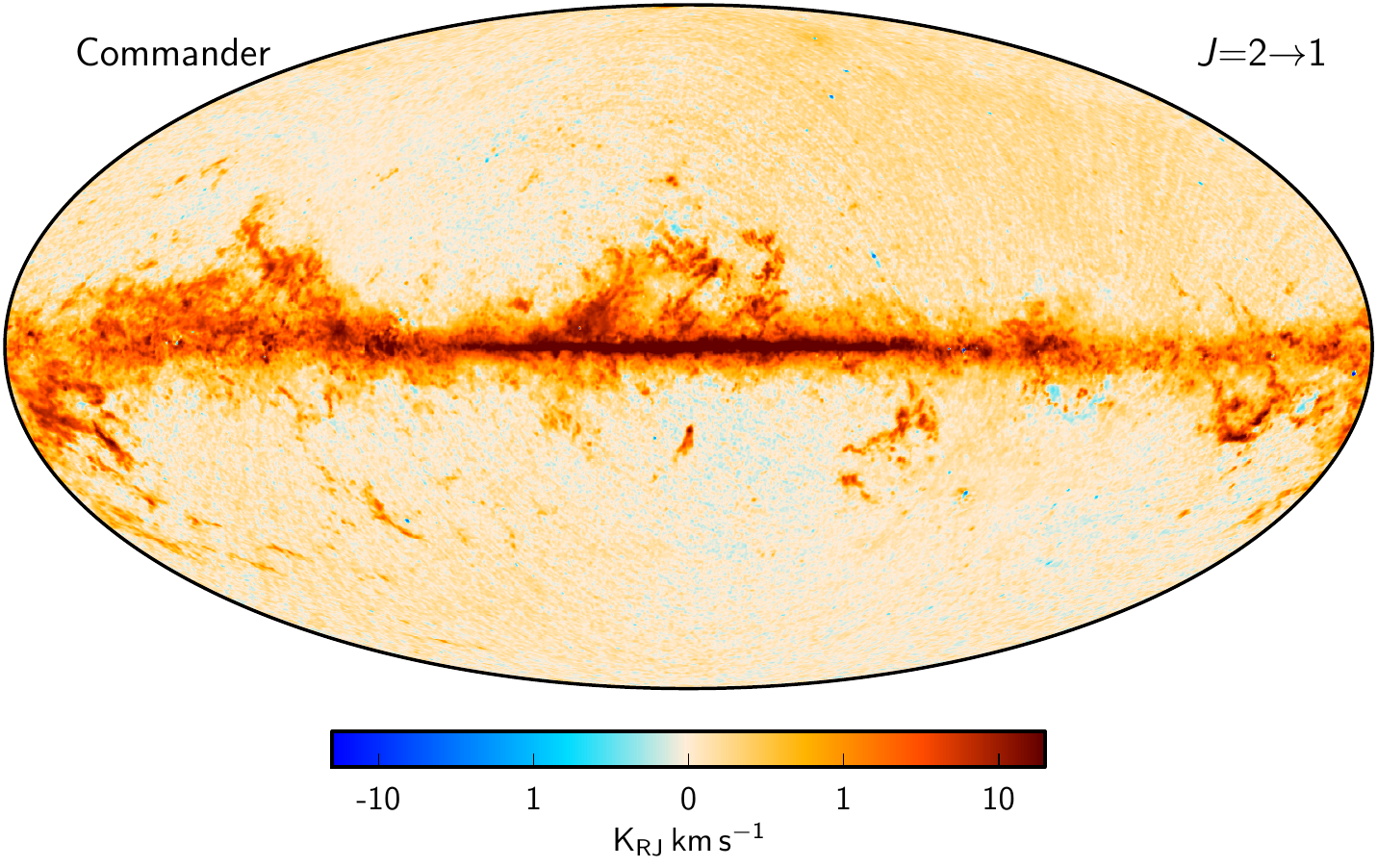,width=0.77\linewidth,clip=}
}
\end{center}
\caption{High-resolution CO $J$=2$\rightarrow$1 maps derived from the
  2015 \Planck\ sky maps with three different algorithms, denoted
  Type-1 (\emph{top row}), Type-2 (\emph{middle row}) and
  \texttt{Commander} (\emph{bottom row}), see \cite{planck2013-p03a}
  for details. All maps are smoothed to a common resolution of
  $15\arcm$ FWHM. }
\label{fig:co21_maps}
\end{figure*}

\begin{table*}[tb]                                                                                                                                                  
\begingroup                                                                                                                                      
\newdimen\tblskip \tblskip=5pt
\caption{Summary of main CO product characteristics.}
\label{tab:co_summary}
\vskip -8mm
\footnotesize                                                                                                                                        
\setbox\tablebox=\vbox{ %
\newdimen\digitwidth                                                                                                                              
\setbox0=\hbox{\rm 0}
\digitwidth=\wd0
\catcode`*=\active
\def*{\kern\digitwidth}
\newdimen\signwidth
\setbox0=\hbox{+}
\signwidth=\wd0
\catcode`!=\active
\def!{\kern\signwidth}
\newdimen\decimalwidth
\setbox0=\hbox{.}
\decimalwidth=\wd0
\catcode`@=\active
\def@{\kern\signwidth}
\halign{ \hbox to 0.7in{#\leaderfil}\tabskip=1em&
  \hfil#\hfil\tabskip=1em&
  \hfil#\hfil\tabskip=1em&
  \hfil#\hfil\tabskip=1em&
  \hfil#\hfil\tabskip=0.5em&
  \hfil#\hfil\tabskip=1em&
  \hfil#\hfil\tabskip=1.0em&
  \hfil#\hfil\tabskip=0pt\cr                                                                                                                                                                  
\noalign{\doubleline}
\omit& \omit&\omit&\omit& \multispan2\hfil Noise rms [$\textrm{K}_{\textrm{RJ}}\,\textrm{km}\,\textrm{s}^{-1}$] \hfil& \multispan2\hfil Analysis details\hfil\cr
\noalign{\vskip -3pt}
\omit&\omit&\omit&\lower3pt\hbox{Resolution}&\multispan2\hrulefill&\multispan2\hrulefill\cr
\noalign{\vskip 2pt}
\omit\hfil Map\hfil& Algorithm& CO line& [arcmin]& $15\arcm$ FWHM& $60\arcm$ FWHM& Frequencies [GHz]& Model\cr
\noalign{\doubleline}
\sc Type 1 & \texttt{MILCA}&$J$=1$\rightarrow$0& 9.6& 1.4*& 0.34*& 100 (bol maps)& CO, CMB\cr
\omit& \texttt{MILCA}&$J$=2$\rightarrow$1& 5.0& 0.53& 0.16*& 217 (bol maps)& CO, CMB, dust\cr
\omit& \texttt{MILCA}&$J$=3$\rightarrow$2& 4.8& 0.55& 0.18*& 353 (bol maps)& CO, dust\cr
\noalign{\vskip 5pt}
\sc Type 2 & \texttt{MILCA}& $J$=1$\rightarrow$0& 15& 0.39& 0.085& 70, 100, 143, 353& CO, CMB, dust, free-free\cr
\omit& \texttt{MILCA}&$J$=2$\rightarrow$1& 15& 0.11& 0.042& 70, 143, 217, 353& CO, CMB, dust, free-free\cr
\noalign{\vskip 5pt}
\omit & \texttt{Commander}&$J$=1$\rightarrow$0& 60& $\cdots$& 0.084& 0.408--857& Full; see Sect.~\ref{sec:temperature}\cr
\omit& \texttt{Commander}& $J$=2$\rightarrow$1& 60& $\cdots$& 0.037& 0.408--857& Full; see Sect.~\ref{sec:temperature}\cr
\omit& \texttt{Commander}& $J$=3$\rightarrow$2& 60& $\cdots$& 0.060& 0.408--857& Full; see Sect.~\ref{sec:temperature}\cr
\noalign{\vskip 5pt}
\sc Type 3& \texttt{Commander}&$J$=2$\rightarrow$1\rlap{$^{\rm a}$}& 7.5& *0.090& 0.031& 143--857& CO, CMB, dust\cr
\noalign{\vskip 5pt}
\omit& \texttt{Commander-Ruler}&$J$=1$\rightarrow$0\rlap{$^{\rm b,c}$}&5.5&0.19&0.082&30--353&CO, CMB, dust, low-freq\cr
\noalign{\vskip 3pt\hrule\vskip 3pt}}}
\endPlancktablewide
\tablenote {{\rm a}} Formally a weighted average of CO $J$=2$\rightarrow$1 and $J$=3$\rightarrow$2, but strongly dominated by CO $J$=2$\rightarrow$1.\par
\tablenote {{\rm b}} Formally a weighted average of CO $J$=1$\rightarrow$0, $J$=2$\rightarrow$1 and $J$=3$\rightarrow$2, but strongly dominated by CO $J$=1$\rightarrow$0.\par
\tablenote {{\rm c}} Only published in 2013.\par
\endgroup
\end{table*}

\begin{figure}
\begin{center}
\mbox{
\epsfig{figure=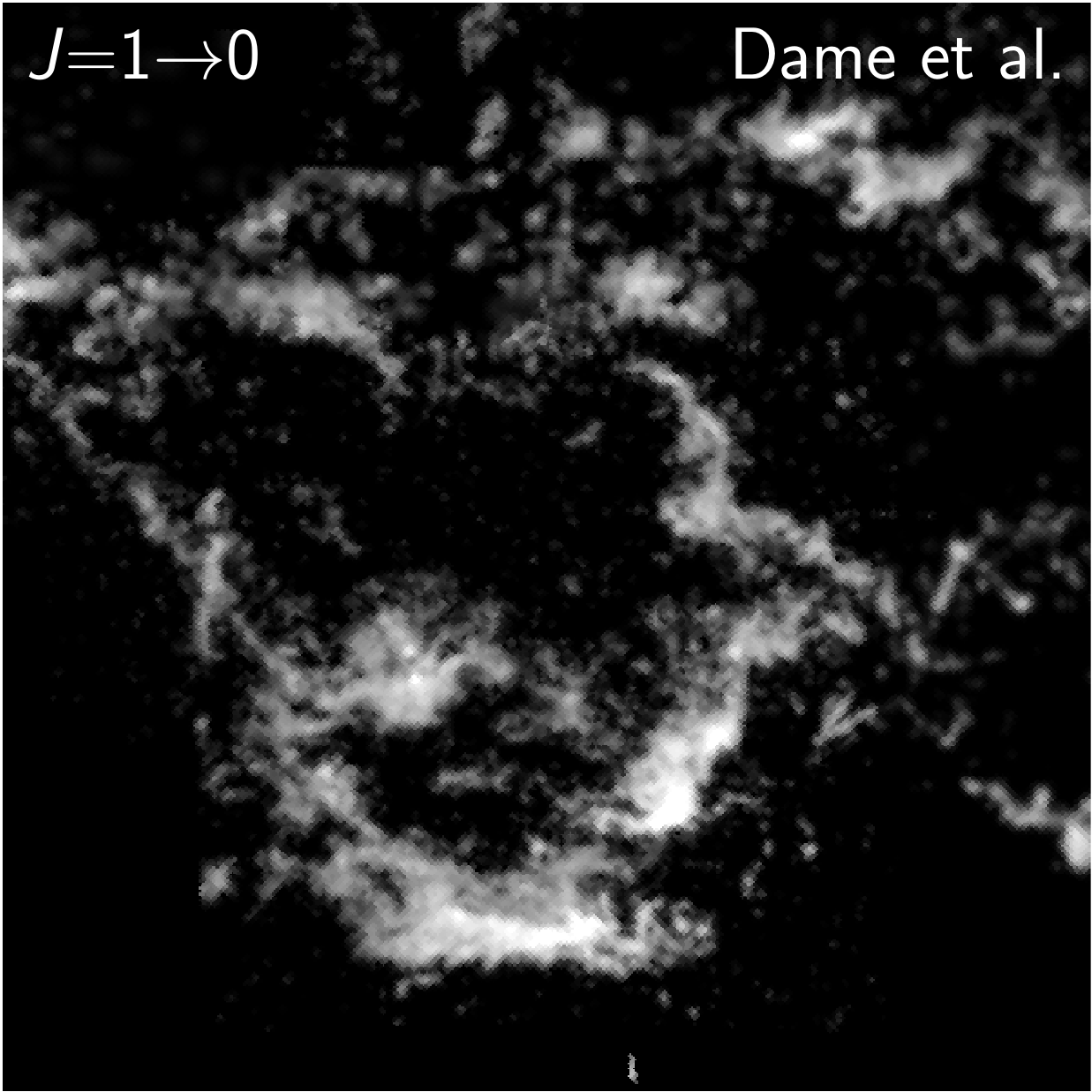,width=0.49\linewidth,clip=}
\epsfig{figure=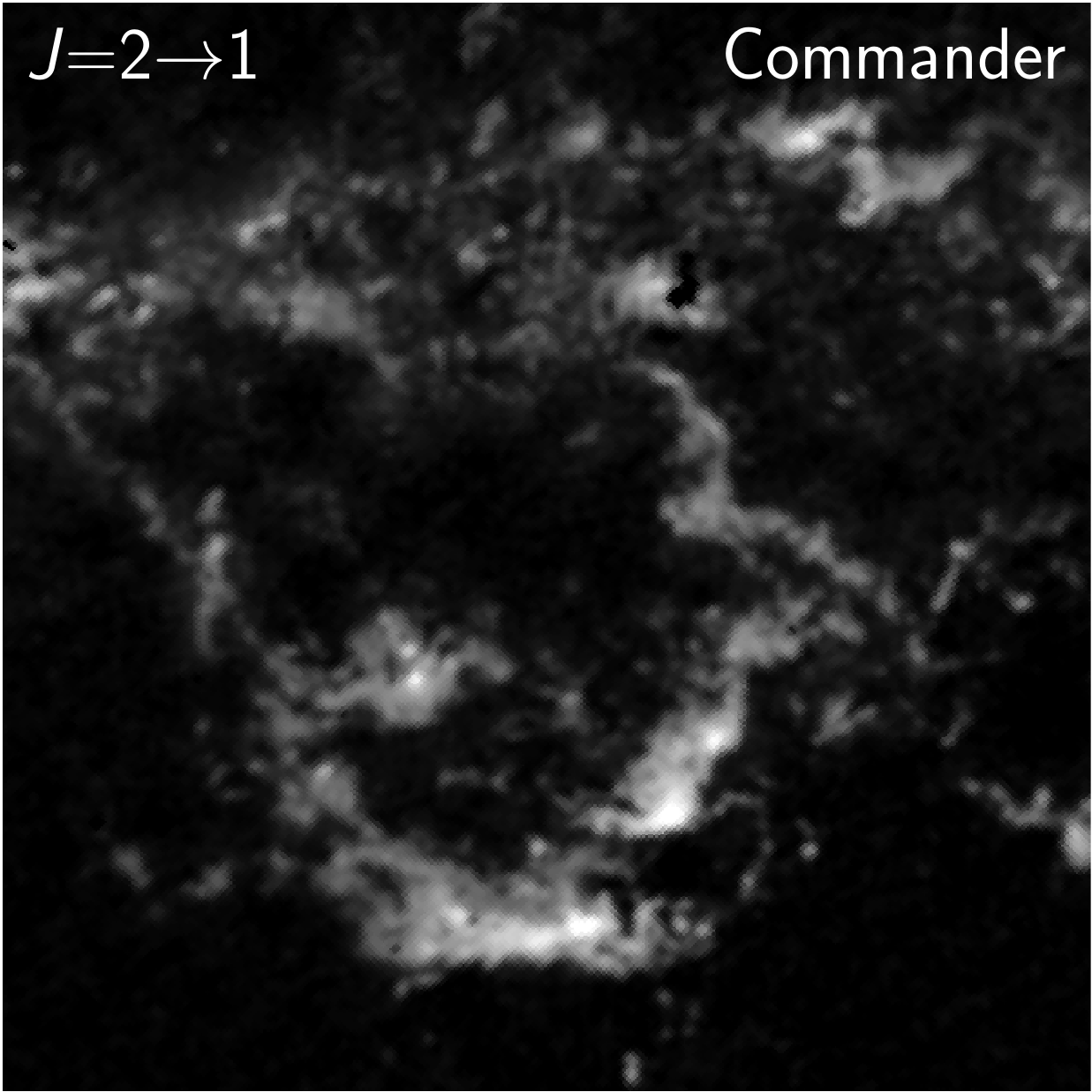,width=0.49\linewidth,clip=}
}
\mbox{
\epsfig{figure=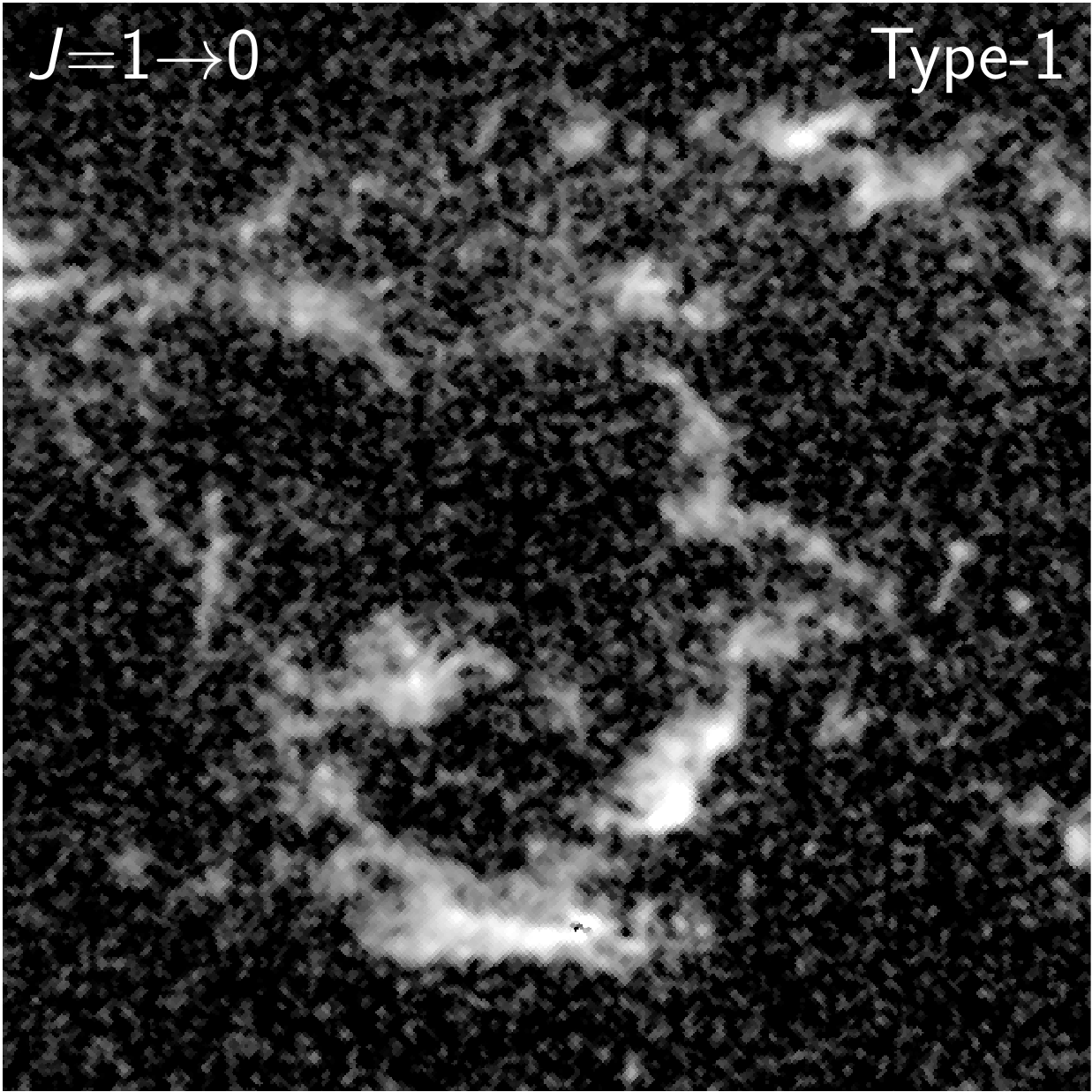,width=0.49\linewidth,clip=}
\epsfig{figure=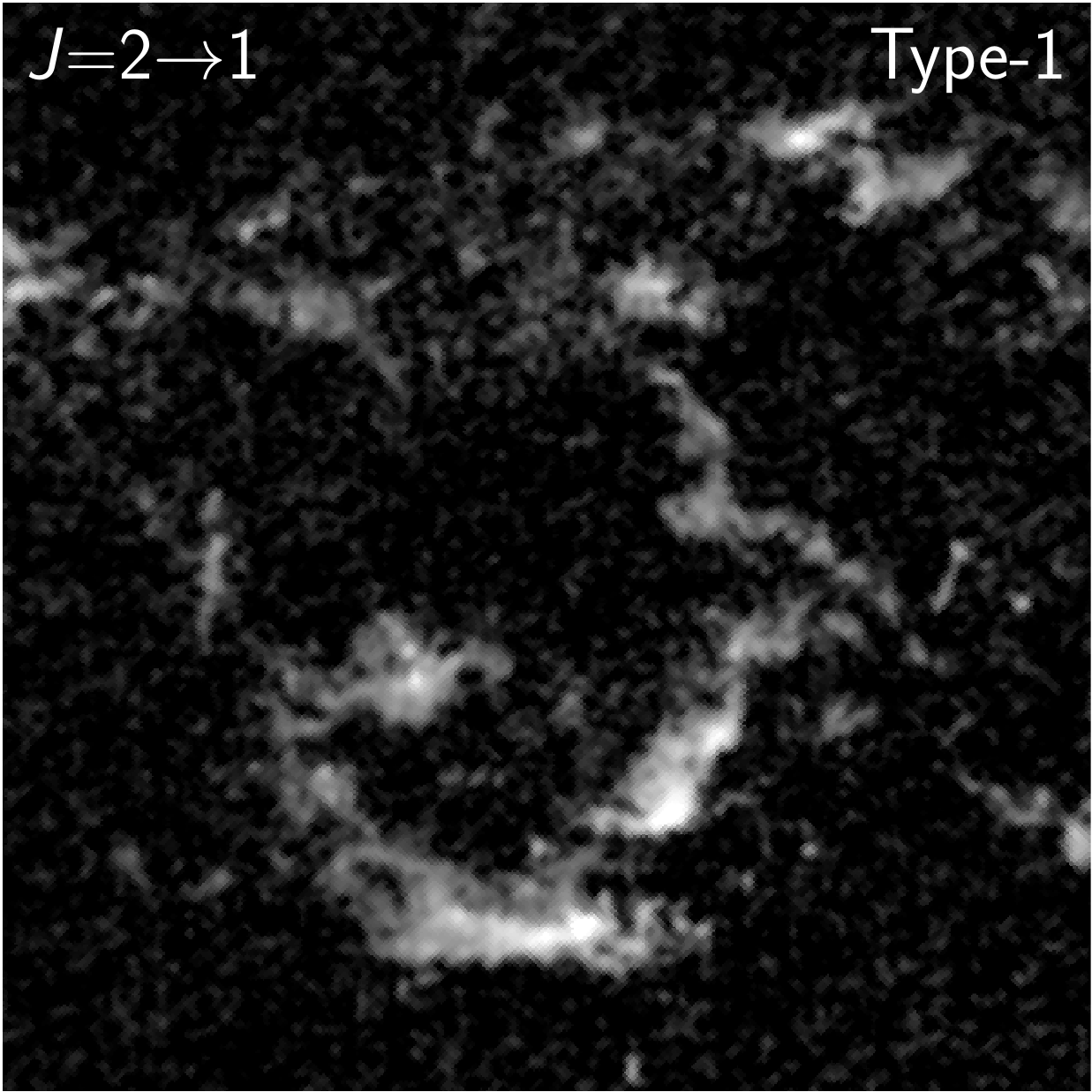,width=0.49\linewidth,clip=}
}
\mbox{
\epsfig{figure=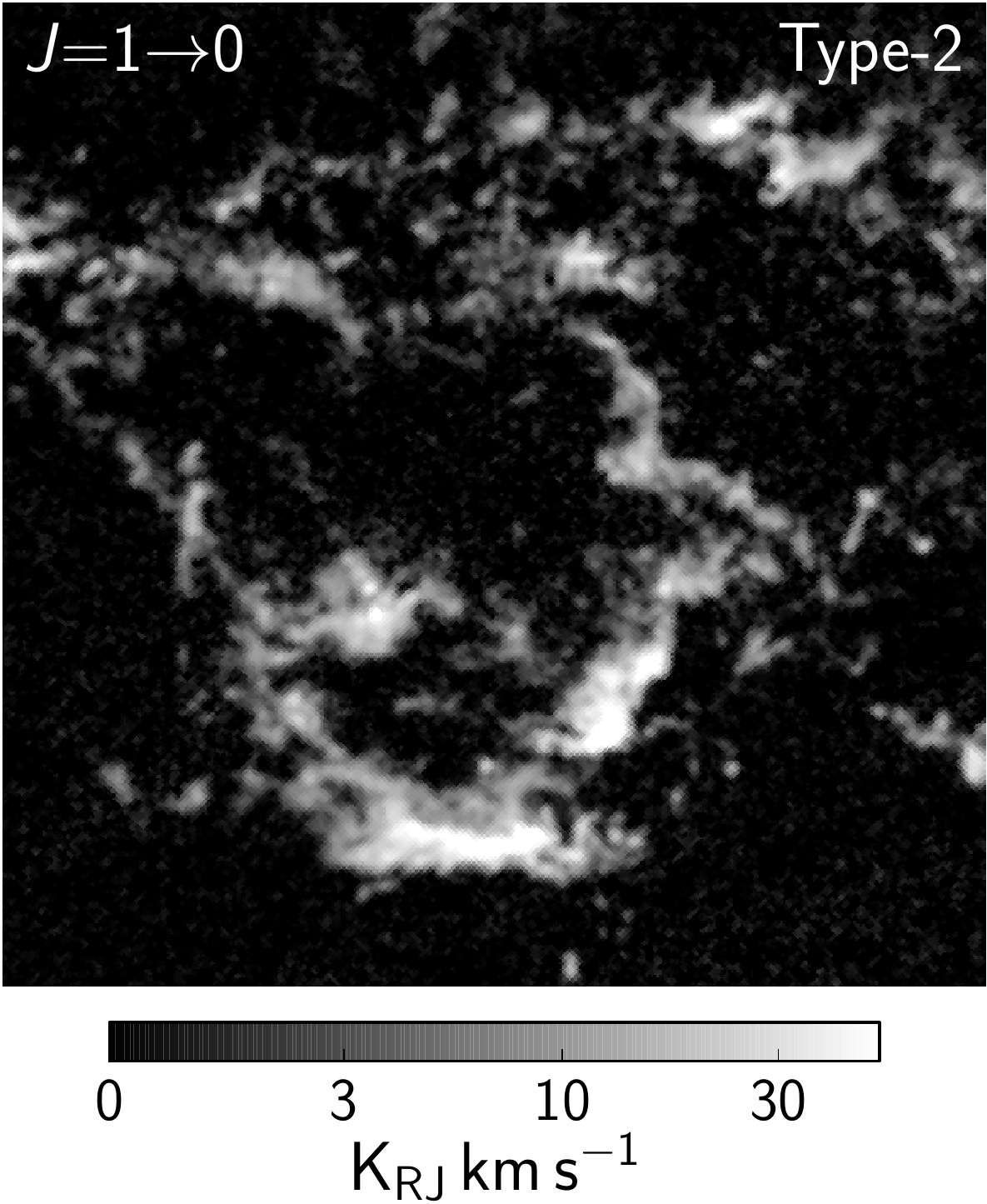,width=0.49\linewidth,clip=}
\epsfig{figure=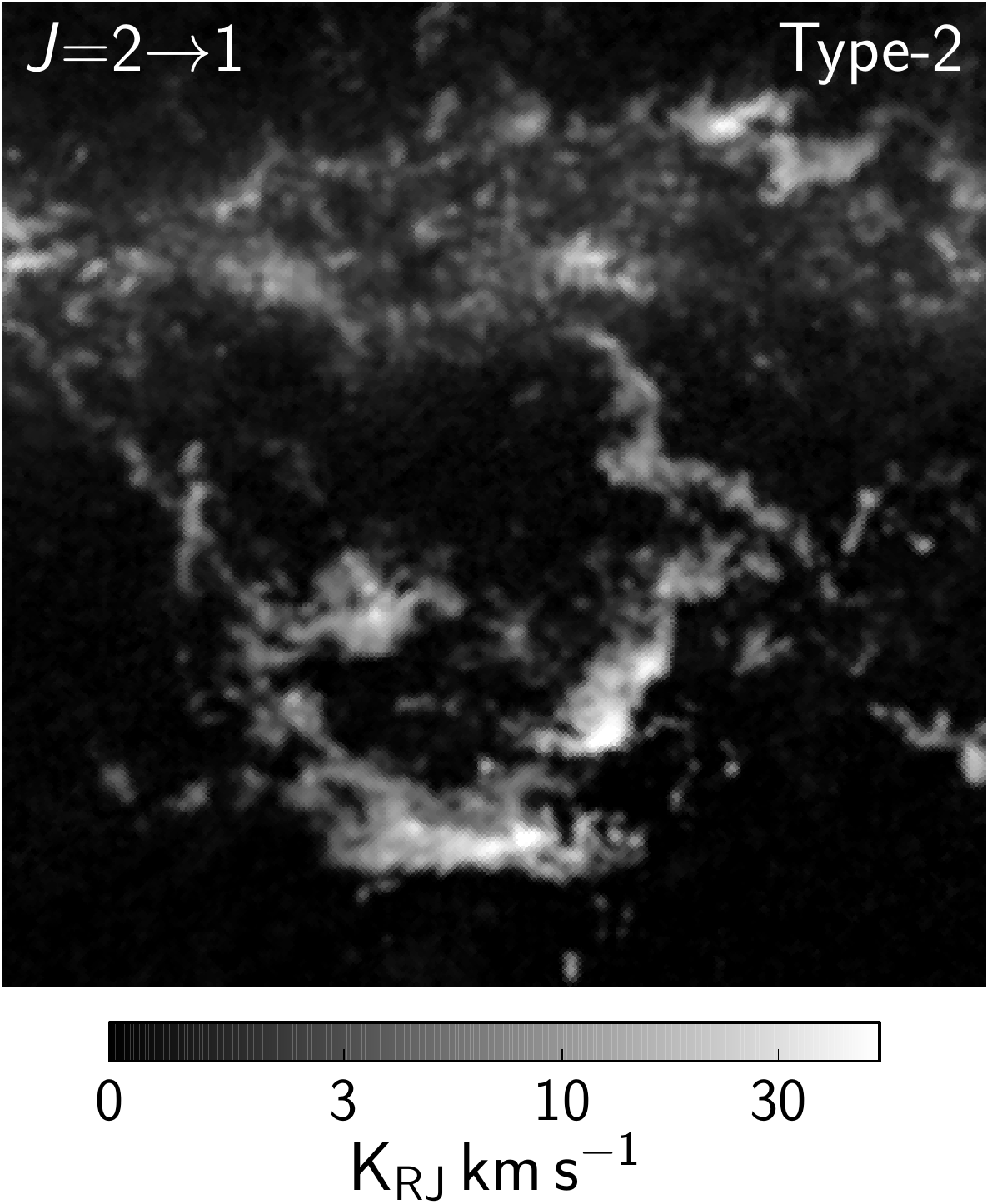,width=0.49\linewidth,clip=}
}
\end{center}
\vskip -4mm
\caption{$30\deg \times 30\deg$ zoom-in of various CO emission line
  maps. All maps smoothed to $15\arcm$ FWHM and centred on the Orion
  region, with Galactic coordinates $(l,b) = (201\deg,-9\deg)$.}
\label{fig:co_highres}
\end{figure}

\begin{figure*}
\begin{center}
\mbox{
\epsfig{figure=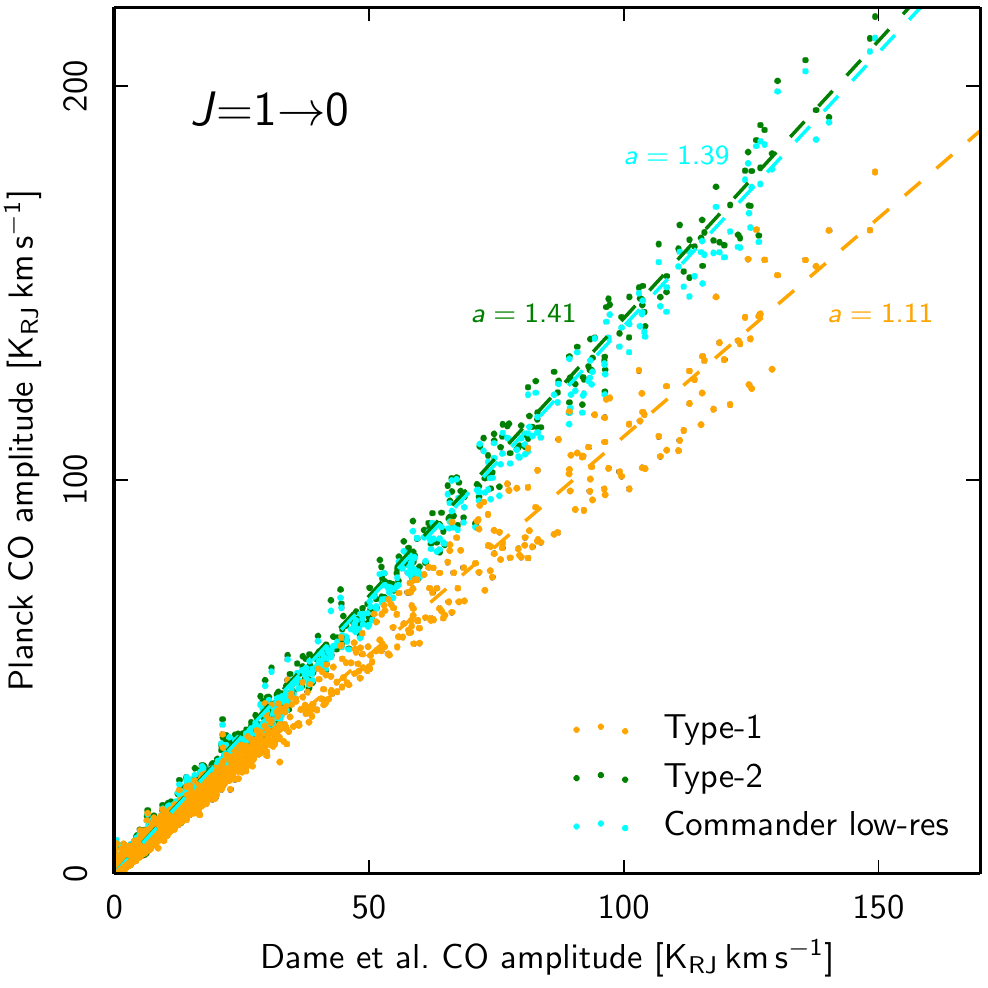,width=0.5\linewidth,clip=}
\epsfig{figure=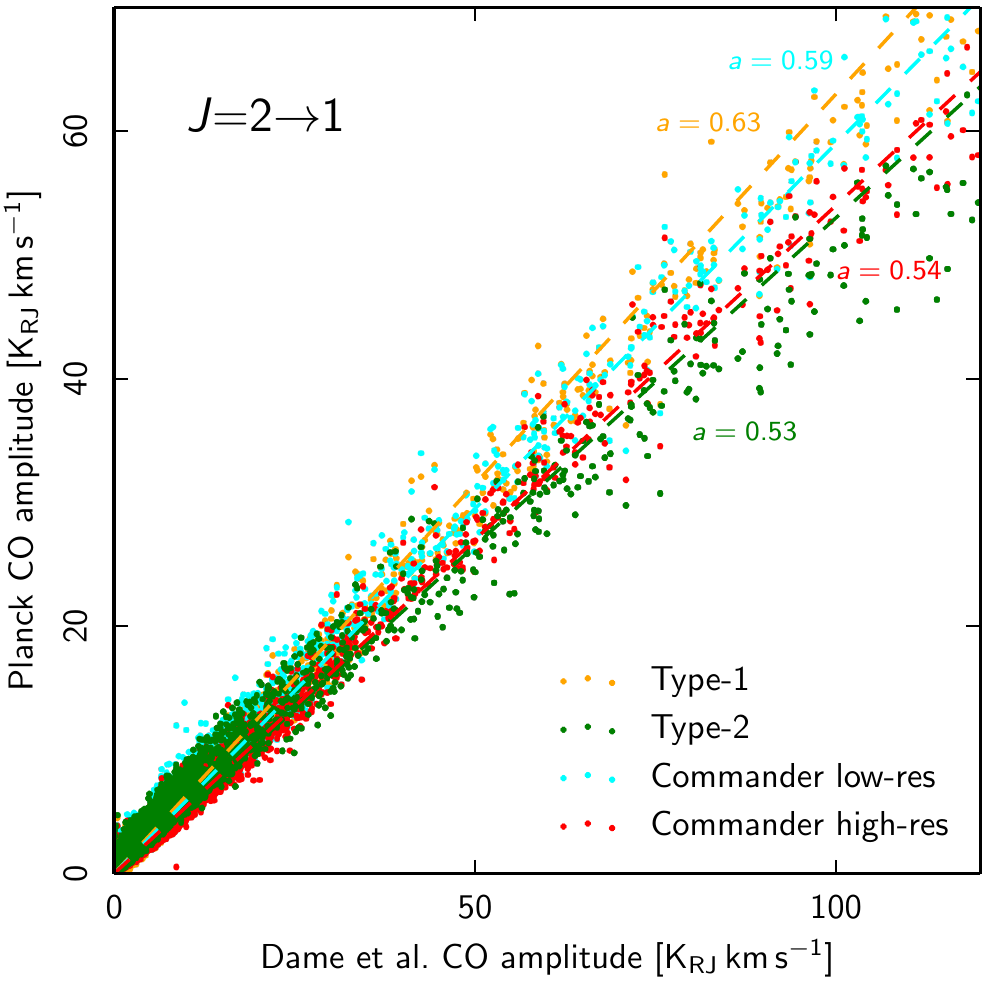,width=0.5\linewidth,clip=}
}
\mbox{
\epsfig{figure=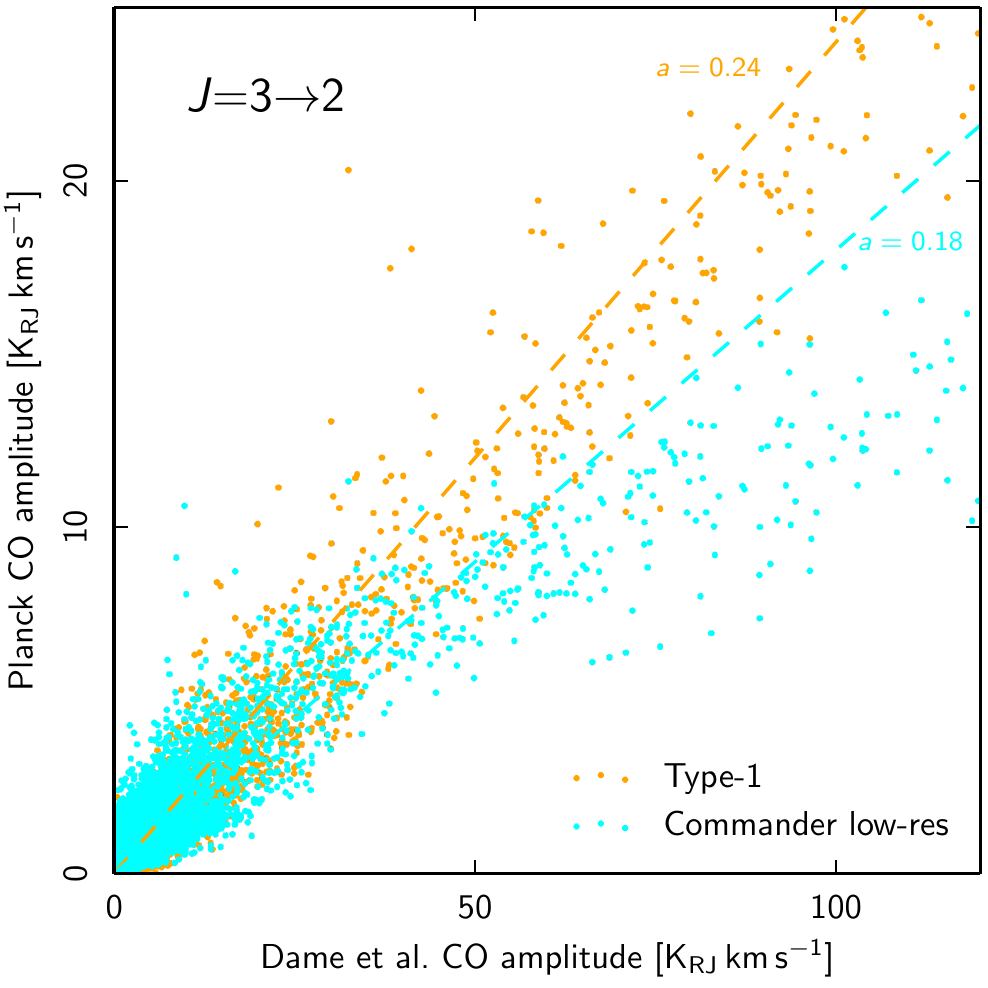,width=0.5\linewidth,clip=}
\epsfig{figure=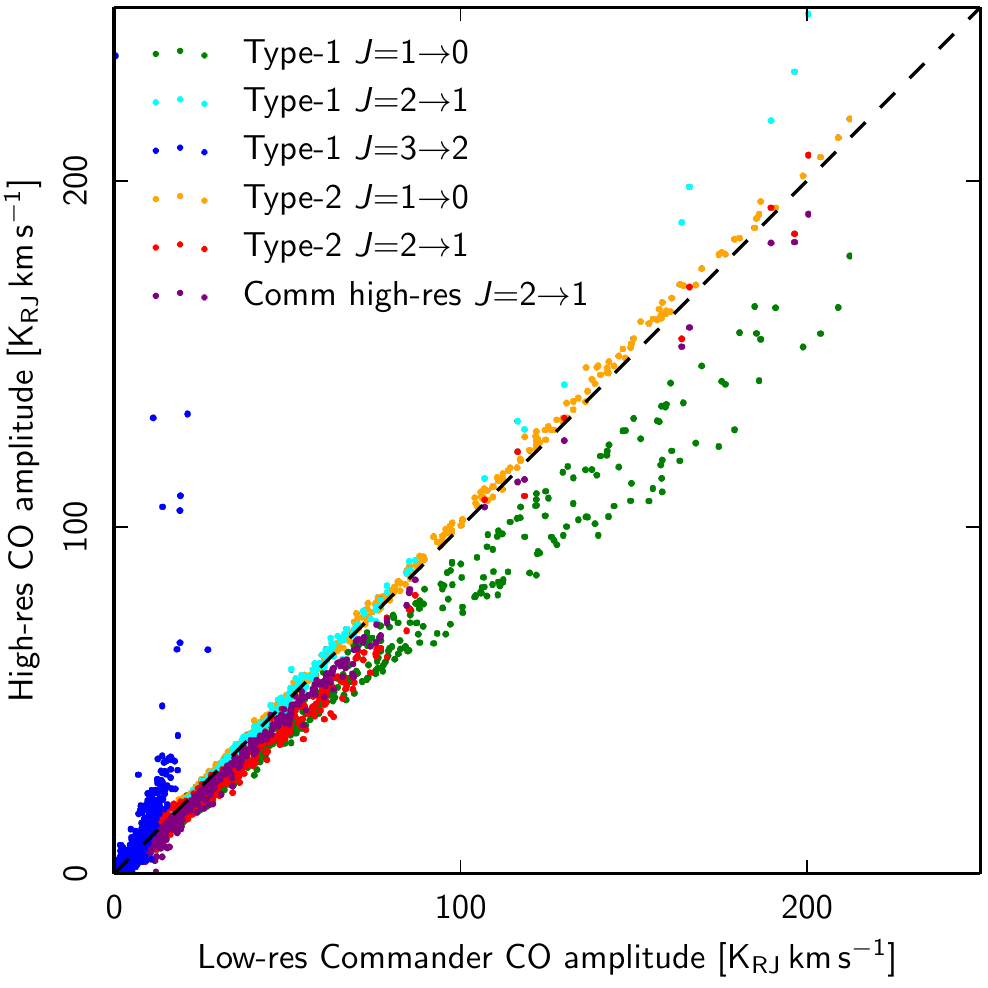,width=0.5\linewidth,clip=}
}
\end{center}
\caption{$T$--$T$ scatter plots between Type-1, Type-2, \texttt{Commander} CO
  maps and the \citet{dame2001} CO\,$J$=1$\rightarrow$0 map,
  smoothed to $1\deg$ FWHM and pixelized with a \healpix\ resolution
  parameter $N_{\textrm{side}}=64$. The panels show correlations for
  $J$=1$\rightarrow$0 (\emph{top left}), $J$=2$\rightarrow$1
  (\emph{top right}) and $J$=3$\rightarrow$2 (\emph{bottom left})
  line maps; the bottom right panel show correlations between the
  baseline $1\deg$ FWHM \texttt{Commander} CO maps and the (smoothed)
  Type-1, Type-2 and the high-resolution \texttt{Commander}
  $J$=2$\rightarrow$1 map.}
\label{fig:co_scatter}
\end{figure*}

\begin{figure}
\begin{center}
\mbox{
\epsfig{figure=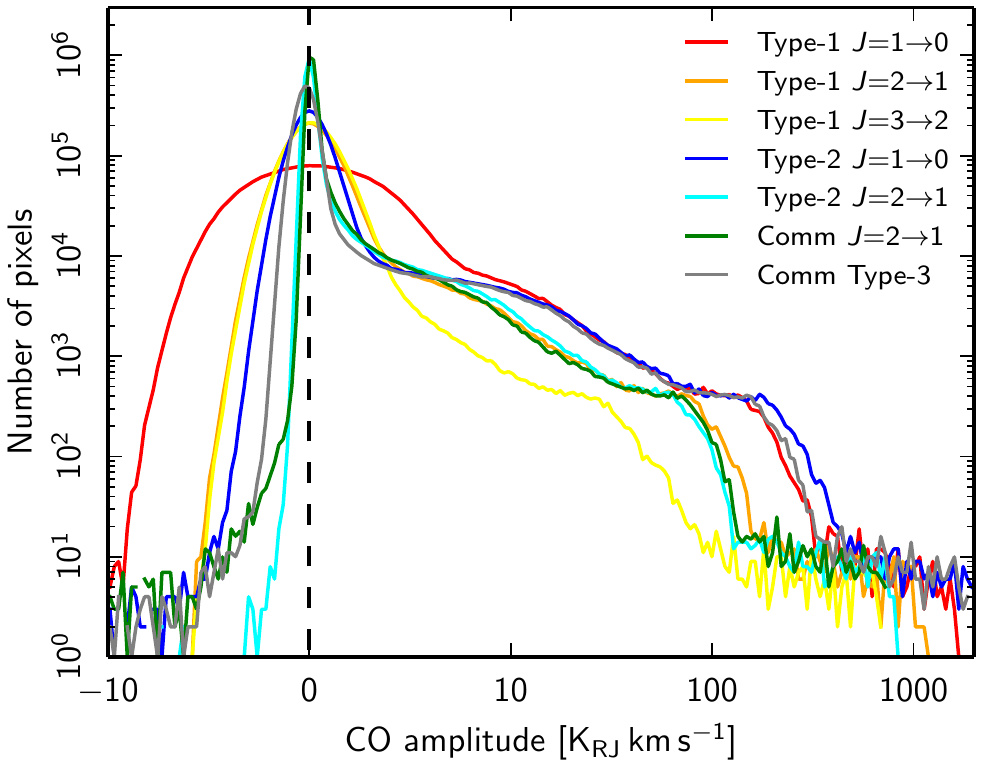,width=\linewidth,clip=}
}
\end{center}
\caption{Pixel histograms of high-resolution CO line emission maps,
  all smoothed to a common resolution of $15\arcm$ FWHM and
  re-pixelized at \healpix\ resolution $N_{\textrm{side}}=512$.}
\label{fig:co_noise}
\end{figure}

\begin{figure}
\begin{center}
\mbox{
\epsfig{figure=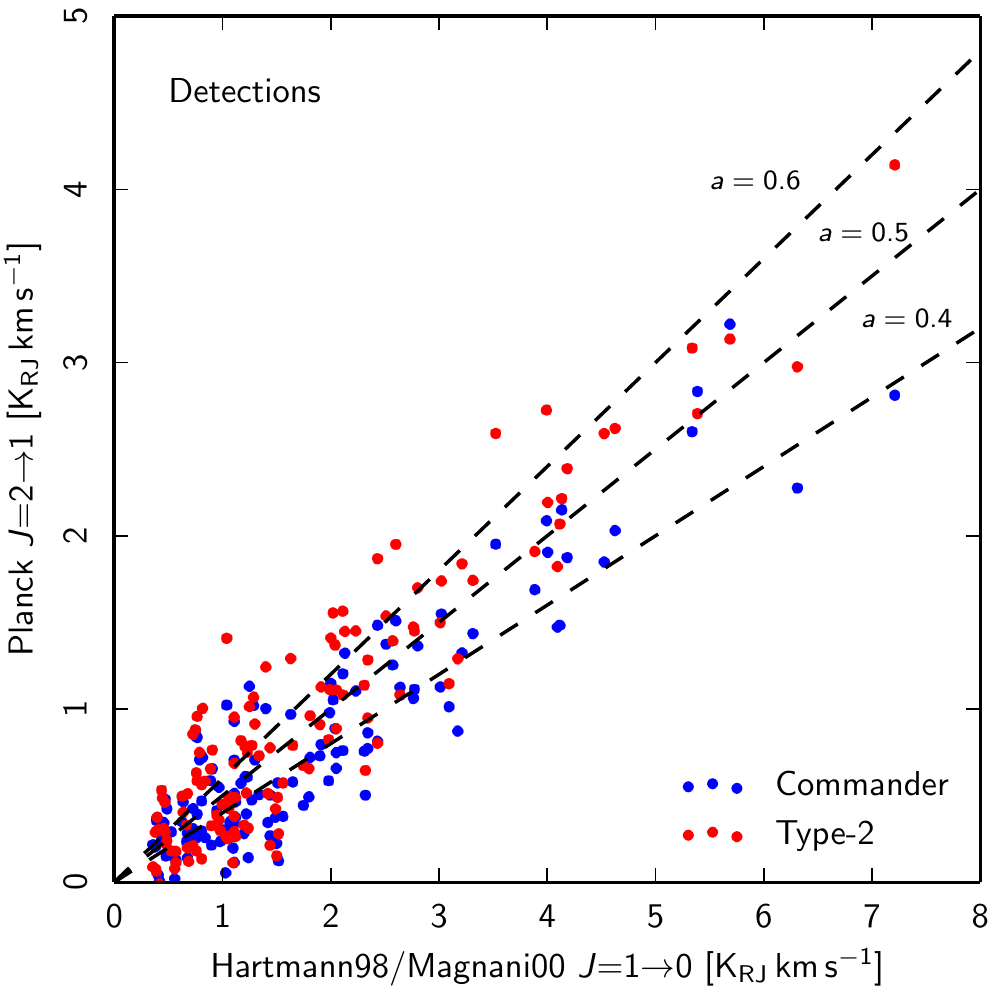,width=\linewidth,clip=}
}
\mbox{
\epsfig{figure=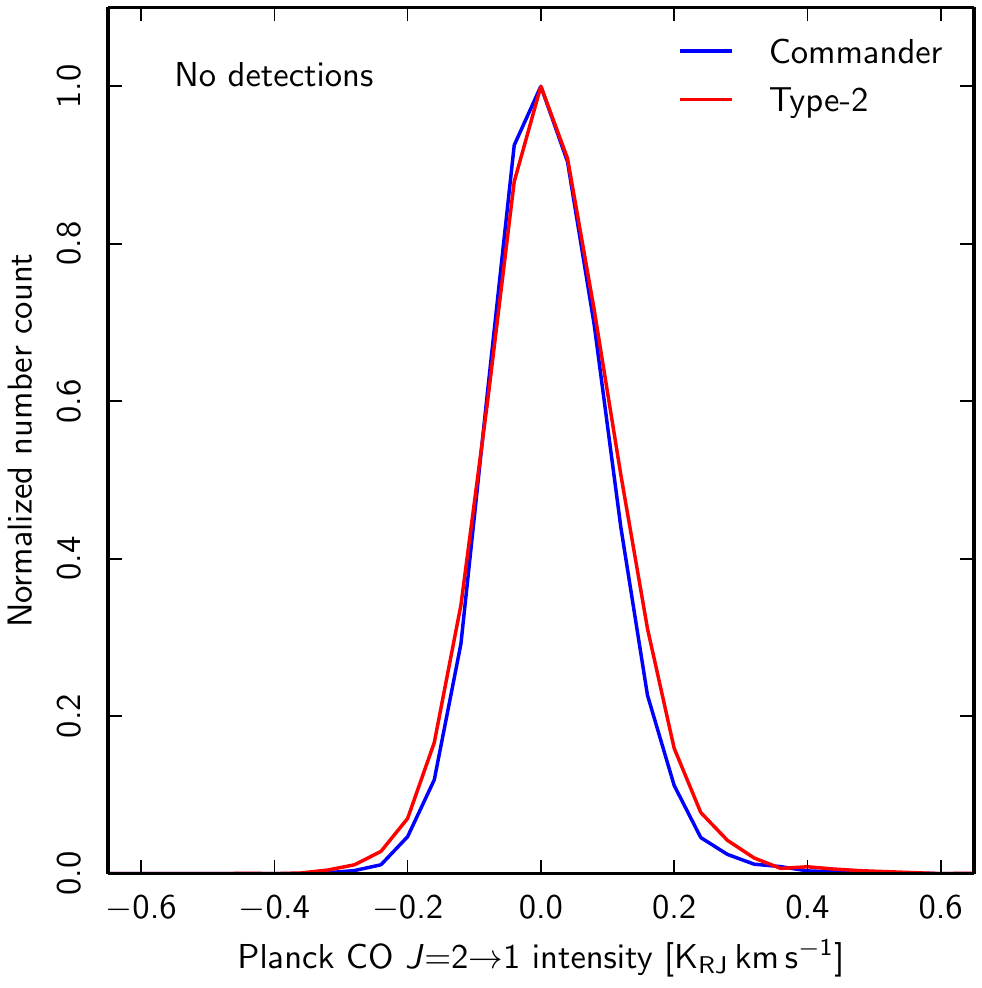,width=\linewidth,clip=}
}
\end{center}
\caption{Comparison of \Planck\ CO $J$=2$\rightarrow$1 maps with the
  high Galactic latitude CO detections published by
  \citet{hartmann1998} and \citet{magnani2000} (\emph{top}), and
  corresponding amplitude histograms including only positions in which
  no detections were found in the same surveys (\emph{bottom}).
}
\label{fig:co_highlat}
\end{figure}

Finally, we compare the CO $J$=1$\rightarrow$0, 2$\rightarrow$1 and
3$\rightarrow$2 maps derived in Sects.~\ref{sec:baseline} and
\ref{sec:temp_highres} with independently derived maps and
products. As described in \citet{planck2013-p03a}, \Planck\ has
implemented a multi-algorithm approach to CO extraction, configuring
the \texttt{MILCA} \citep{hurier2013} and \texttt{Commander}
algorithms for dedicated CO reconstruction. In 2013, this resulted in
three different types of CO maps. In short, the Type-1 CO maps are
built from individual bolometer maps within single frequencies, and as
such are only weakly dependent on foreground extrapolations, but this
insensitivity comes at a high cost in terms of instrumental noise. The
Type-2 maps are built per CO line from a small sub-set of frequencies,
carefully selected to be optimal for CO extraction. Since more than
one frequency is involved, a more elaborate foreground model is
required, such as explicit modelling of CMB, dust and free-free,
although several simplifications are imposed, such as the assumption
of constant dust temperature and spectral indices. Finally, the Type-3
map corresponds to a maximum signal-to-noise CO extraction in which a
complete foreground model is fitted with \texttt{Commander}, as
described in this paper, but with only a single CO amplitude per pixel
and otherwise only spatially fixed line ratios accounting for scaling
between frequencies.

In the present release, the \texttt{MILCA}-based Type-1 and Type-2
maps have been updated with the latest data, while the
\texttt{Commander-Ruler}-based Type-3 map from 2013 has been
superseded by the high-resolution \texttt{Commander}-only
$J$=2$\rightarrow$1 map presented in Sect.~\ref{sec:temp_highres}. In
addition to these high-resolution maps, we of course also provide the
low-resolution line maps discussed in
Sect.~\ref{sec:baseline}. Table~\ref{tab:co_summary} summarizes the
CO-related data products provided in the current release, including
angular resolution, instrumental noise, and analysis assumptions.

We start by comparing the maps derived with \texttt{MILCA} and
\texttt{Commander}, both with each other and with the CO
$J$=1$\rightarrow$0 survey presented by \citet{dame2001}. The full-sky
$J$=1$\rightarrow$0 and 2$\rightarrow$1 maps are shown in
Figs.~\ref{fig:co10_maps} and \ref{fig:co21_maps}, while zoom-ins of
the Orion region are shown in Fig.~\ref{fig:co_highres}. All maps are
smoothed to a common resolution of $15\arcm$ in these plots, except
the Dame et al.\ survey, which has an intrinsic resolution of about
$20\arcm$. For reference, the 2013 Type-3 map is shown in the bottom
panel of Fig.~\ref{fig:co10_maps}.

\begin{figure*}
\begin{center}
\mbox{
\epsfig{figure=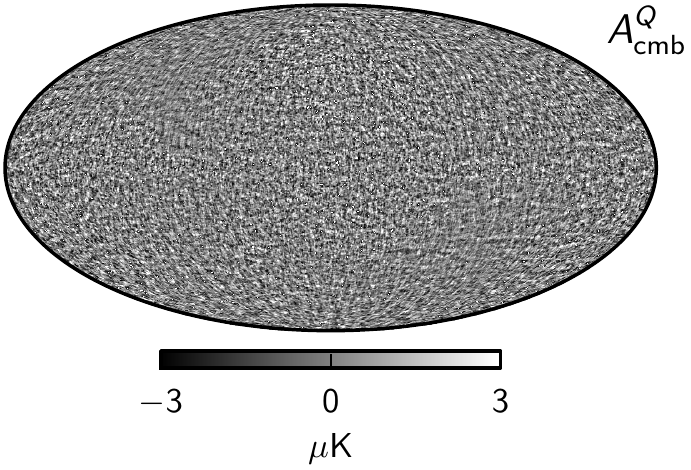,width=88mm,clip=}
\epsfig{figure=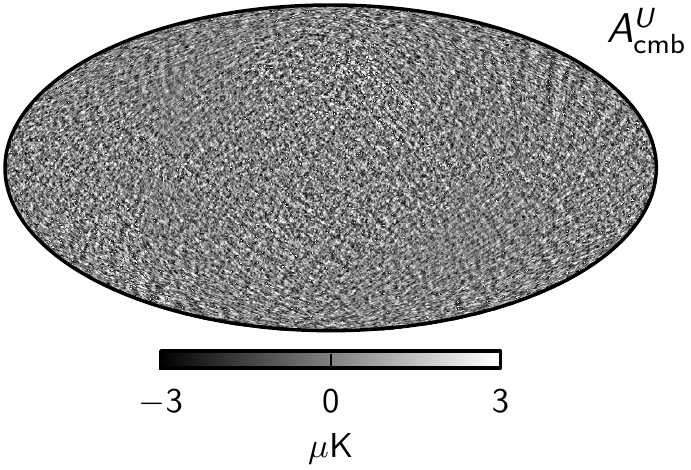,width=88mm,clip=}
}
\mbox{
\epsfig{figure=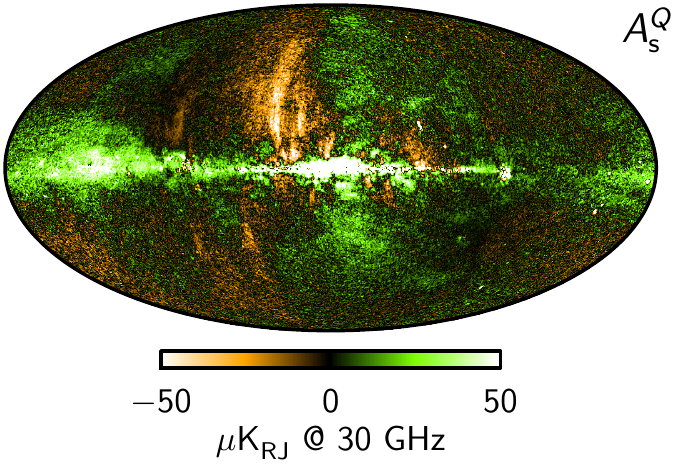,width=88mm,clip=}
\epsfig{figure=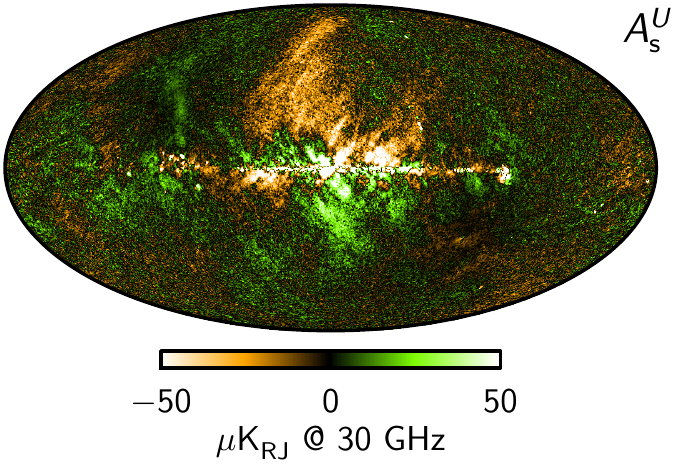,width=88mm,clip=}
}
\mbox{
\epsfig{figure=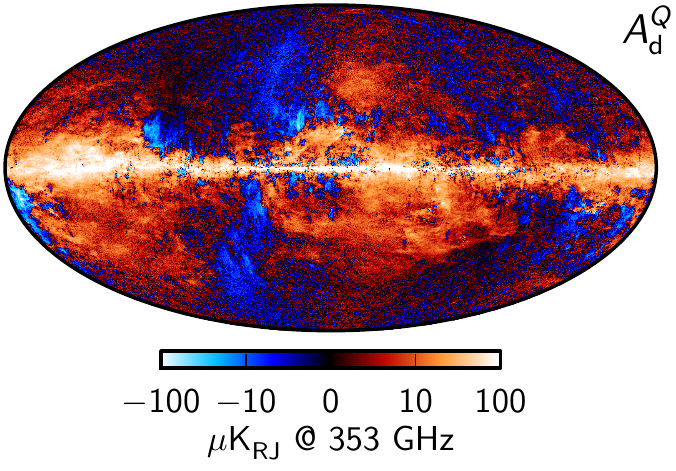,width=88mm,clip=}
\epsfig{figure=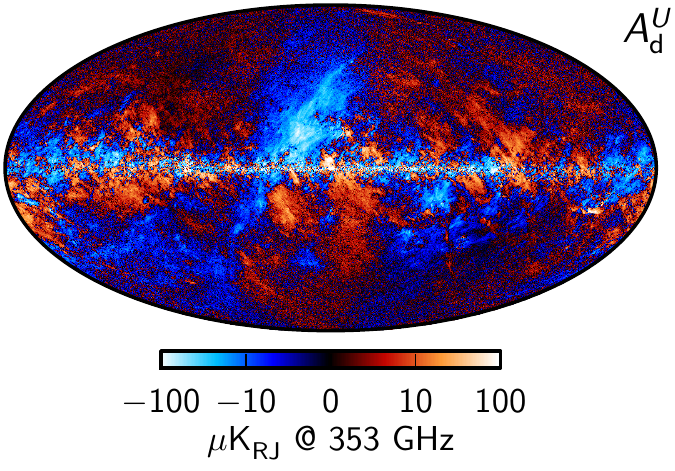,width=88mm,clip=}
}
\end{center}
\caption{Maximum posterior amplitude polarization maps derived from
  the \Planck\ observations between 30 and 353\,GHz. Left and right
  columns show the Stokes $Q$ and $U$ parameters, respectively, while rows show, from
  top to bottom, CMB, synchrotron polarization at 30\,GHz, and
  thermal dust polarization at 353\,GHz. The CMB map has been
  highpass-filtered with a cosine-apodized filter between $\ell=20$
  and 40, and the Galactic plane (defined by the 17\,\% CPM83 mask shown
  in Fig.~\ref{fig:chisq_pol_map}) has been replaced with a
  constrained Gaussian realization \citep{planck2014-a11}.}
\label{fig:pol_amp_map}
\end{figure*}

The first three panels of Fig.~\ref{fig:co_scatter} show $T$--$T$
correlation plots between each of the three CO line maps and the Dame
et al.\ survey, all smoothed to a common resolution of $1\deg$ FWHM;
the fourth panel shows similar correlations between the high- and
low-resolution \Planck\ products. Note that all axes are linear, and
this figure therefore highlights the very brightest CO objects on the
sky. The main points to take away from these scatter plots, and the
maps in Figs.~\ref{fig:co21_maps} and \ref{fig:co_highres}, are the
following.
\begin{enumerate}
\item The Type-2 and low-resolution \texttt{Commander}
  $J$=1$\rightarrow$0 maps agree very well internally, and also
  correlate strongly with the Dame et al.\ survey. However, they both
  show an overall multiplicative scaling factor of about 1.4 relative
  to Dame et al. This level of amplitude difference is similar to
  what was observed in the 2013 release \citep{planck2013-p03a}, and
  is due to a combination of bandpass uncertainties in the
  \Planck\ observations and the overall 10\,\% calibration uncertainty in the
  Dame et al.\ survey. The Type-1 $J$=1$\rightarrow$0 map shows bigger
  differences with respect to the Dame et al.\ survey, both in the
  scatter plot and the Orion zoom-in. Possible explanations include
  the presence of a second significant line emission mechanism, such
  as $^{13}$CO $J$=1$\rightarrow$0 (at 110\,GHz), or, possibly,
  thermal dust leakage.
\item In the CO $J$=2$\rightarrow$1 case, the Type-2 map shows some
  evidence of contamination, both in the form of significant curvature
  in the scatter plot (top right panel of Fig.~\ref{fig:co_scatter}) ,
  and as notable diffuse emission along the Galactic plane in the
  Orion region and full-sky map. The agreement between the Type-1 and
  \texttt{Commander} maps is, however, good, and both show tight
  correlations with the Dame et al.\ survey.
\item For the CO $J$=3$\rightarrow$2 map, the scatter with respect to
  Dame et al.\ is substantial for both Type-1 and \texttt{Commander},
  and in particular the latter shows clear evidence of curvature in
  the $T$--$T$ plot due to contamination from thermal dust.
\item A significant residual dipole aligned with the CMB dipole may be
  seen in the 2013 Type-3 map, and similar residuals are also present
  in the 2013 Type-1 and Type-2 maps. These dipole residuals have been
  greatly suppressed in the new 2015 maps, due to the new estimation of
  the CMB dipole directly from the \Planck\ sky maps, as described in
  \citet{planck2014-a01}.
\end{enumerate}

Figure~\ref{fig:co_noise} shows histograms of the various
high-resolution maps, each map being smoothed to a common resolution
of $15\arcm$ FWHM and re-pixelized at a \healpix\ level of
$N_{\textrm{side}}=512$. As already noted, here we clearly see that
the Type-1 maps have significantly larger instrumental noise (wider
histograms near zero) than the corresponding Type-2 and
\texttt{Commander} maps. However, we also see that the three maps
converge well at intermediate amplitudes between, say, 3 and
200$\,\textrm{K}_{\textrm{RJ}}\,\textrm{km}\,\textrm{s}^{-1}$, for both
$J$=1$\rightarrow$0 and 2$\rightarrow$1.

Next, in Fig.~\ref{fig:co_highlat} we compare the Type-2 and
\texttt{Commander} $J$=2$\rightarrow$1 maps with targeted CO
$J$=1$\rightarrow$0 high-latitude observations published by
\citet{hartmann1998} and \citet{magnani2000}. The same test was
performed for the \Planck\ 2013 CO maps, and full details regarding
methodology and data processing can be found in
\citet{planck2013-p03a}. The top panel shows a $T$--$T$ scatter plot
between each of their 133 detected objects and the corresponding
objects in our maps. The correlation is strong for both Type-2 and
\texttt{Commander}, and with comparable effective line ratios in the
two cases. The bottom panel shows histograms including positions for
which Hartmann/Magnani et al.\ did not find any significant CO
detections, and, correspondingly, our maps also do not exhibit any
detections at these positions; both histograms are consistent with
noise.

Finally, as in the 2013 analysis \citep{planck2013-p03a}, we have also
cross-correlated the Type-1, Type-2, and \texttt{Commander}
$J$=2$\rightarrow$1 maps against the CO $J$=2$\rightarrow$1
AMANOGAWA-2SB survey \citep{yoda2010,handa2012}, and find very good
morphological agreement among all maps. The best-fit slopes are 0.95,
0.82 and 0.86 for the Type-1, Type-2, and \texttt{Commander} maps,
respectively. Thus, while the Type-1 CO $J$=2$\rightarrow$1 map is more
noisy than the other two, it is also less affected by dust and
free-free contamination, and it is a robust estimation of CO
$J$=2$\rightarrow$1 in the Galactic plane. From this test, we also
estimate that the overall re-calibration factors for the Type-2 and
\texttt{Commander} CO $J$=2$\rightarrow$1 maps are 15--20\,\%.

Combining all of these results, we make the following
recommendations. First, the \texttt{Commander} CO $J$=2$\rightarrow$1
map supersedes the 2013 Type-3 map as the primary ``CO detection''
map, and is also our preferred CO $J$=2$\rightarrow$1 map. The main
advantages of this map as compared to the Type-2 map are higher
angular resolution ($7\parcm5$ versus $15\arcm$ FWHM), slightly lower
noise (0.09 versus
0.11$\,\textrm{K}_{\textrm{RJ}}\,\textrm{km}\,\textrm{s}^{-1}$), and a
tighter correlation with respect to the Dame et al.\ survey (see
Fig.~\ref{fig:co_highres}).  However, in specific regions where the
signal is such that the noise level of the Type-1 CO
$J$=2$\rightarrow$1 map becomes less of an issue, i.e., in the
Galactic plane, the Type-1 map is a better alternative in terms of
overall calibration and contamination. For a dedicated
$J$=1$\rightarrow$0 analysis, we recommend the Type-2 map, which has
higher angular resolution than the corresponding \texttt{Commander}
map ($15\arcm$ versus $1\deg$ FWHM), with similar correlation
properties with respect to Dame et al. Finally, for CO
$J$=3$\rightarrow$2 we recommend the Type-1 map, due to a
significantly stronger correlation with respect to Dame et al.\ than
the corresponding \texttt{Commander} map.

\begin{figure*}
\begin{center}
\mbox{
  \epsfig{figure=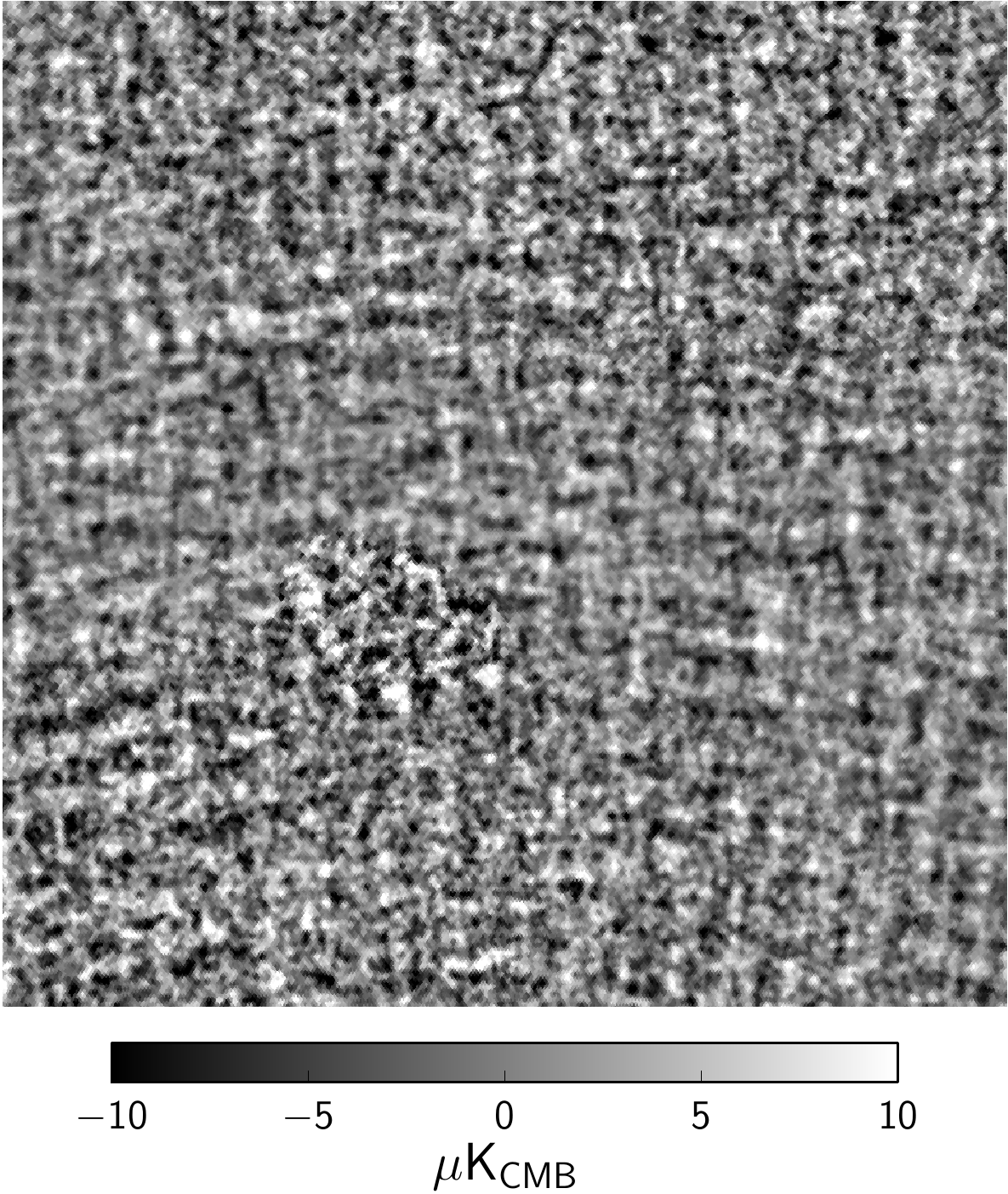,width=0.36\linewidth,clip=}
  \epsfig{figure=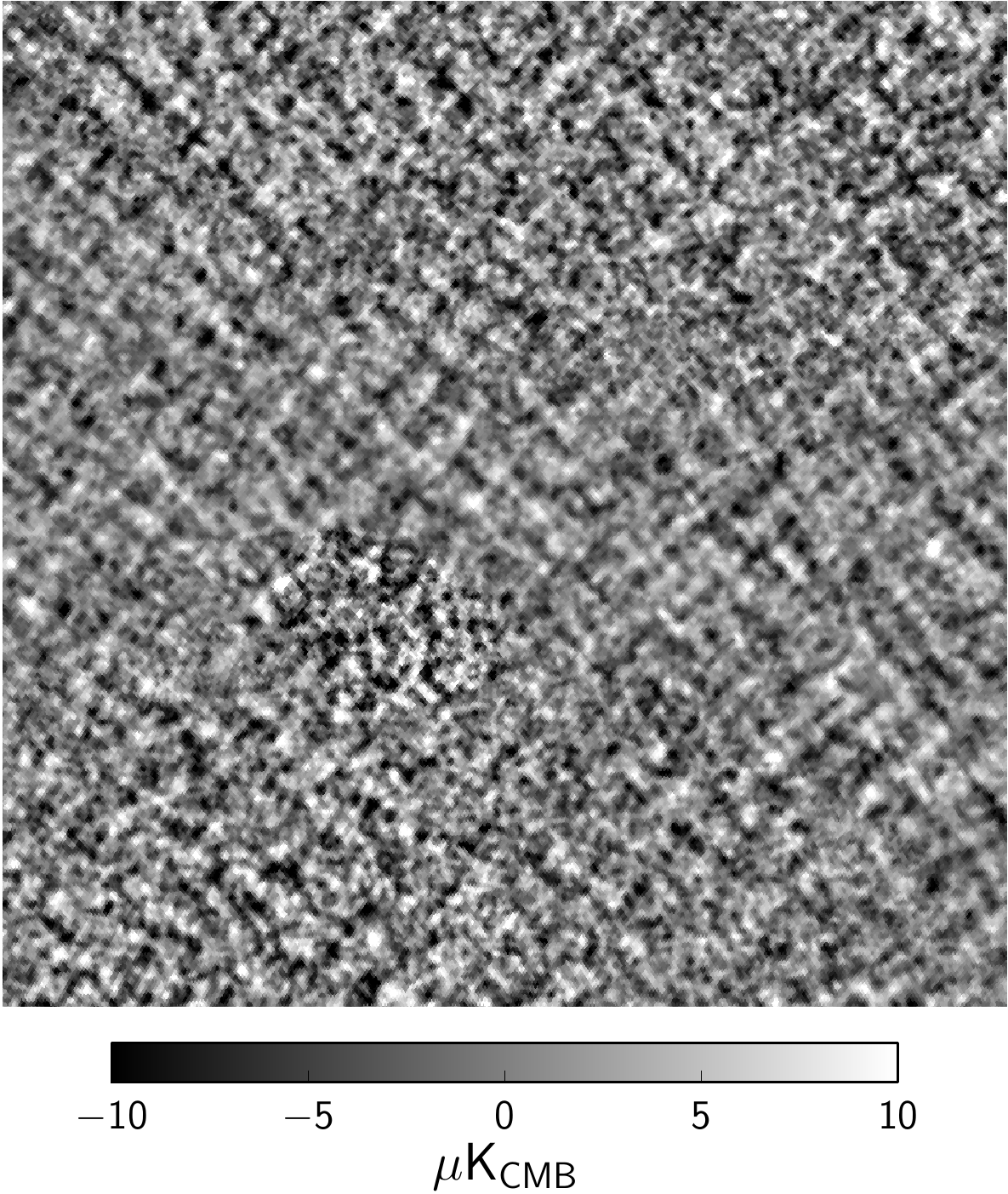,width=0.36\linewidth,clip=}
}
\mbox{
  \epsfig{figure=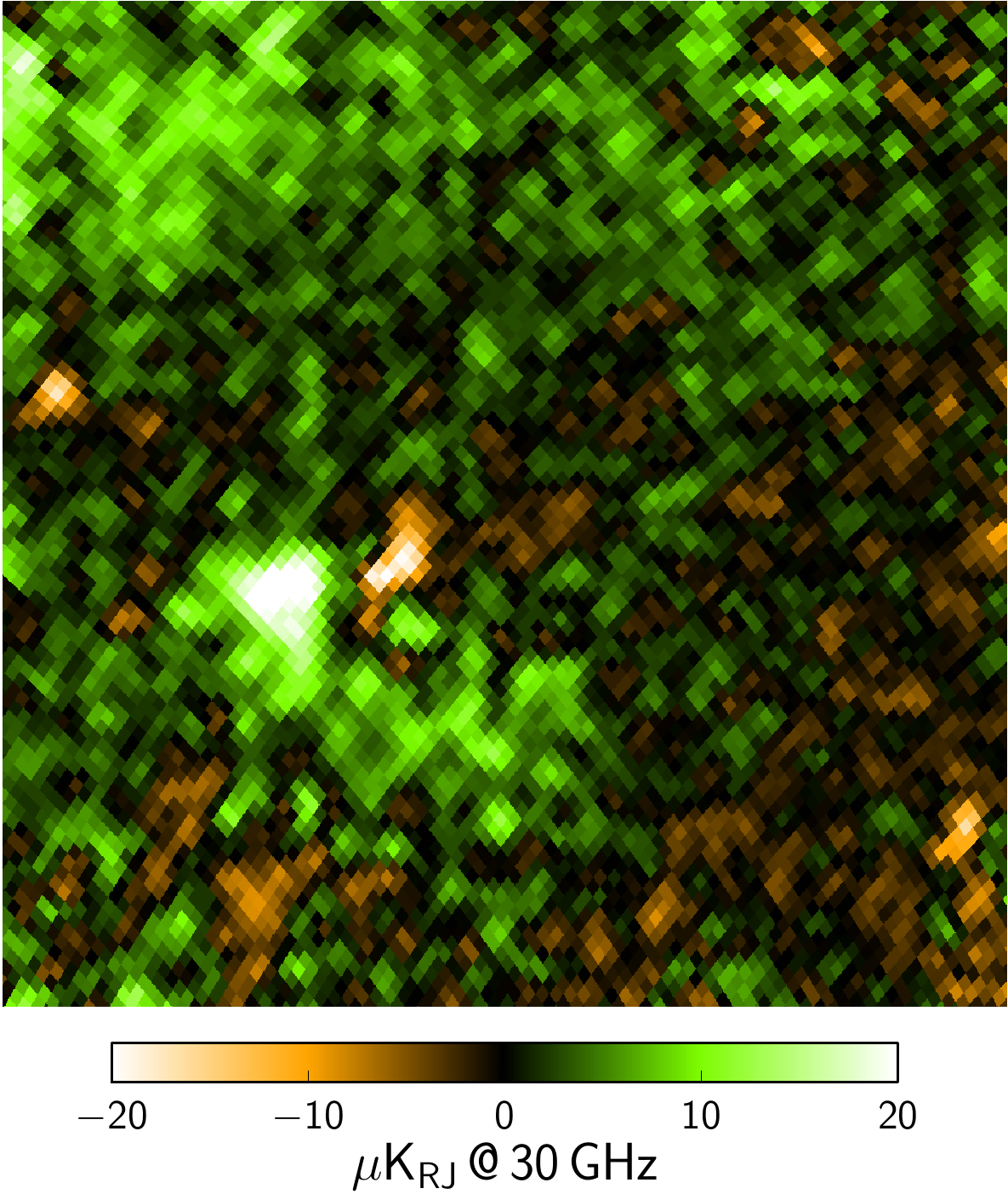,width=0.36\linewidth,clip=}
  \epsfig{figure=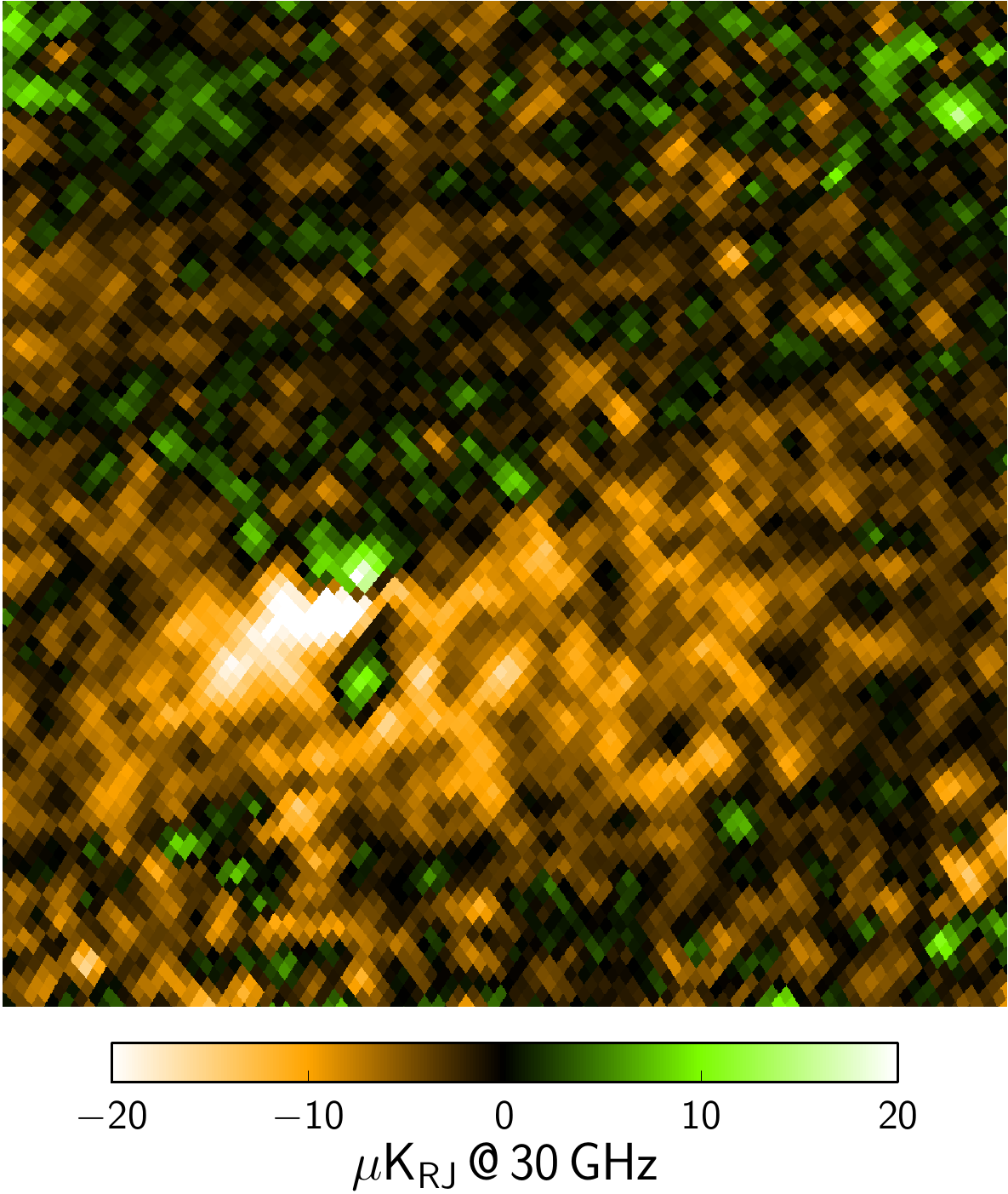,width=0.36\linewidth,clip=}
}
\mbox{
  \epsfig{figure=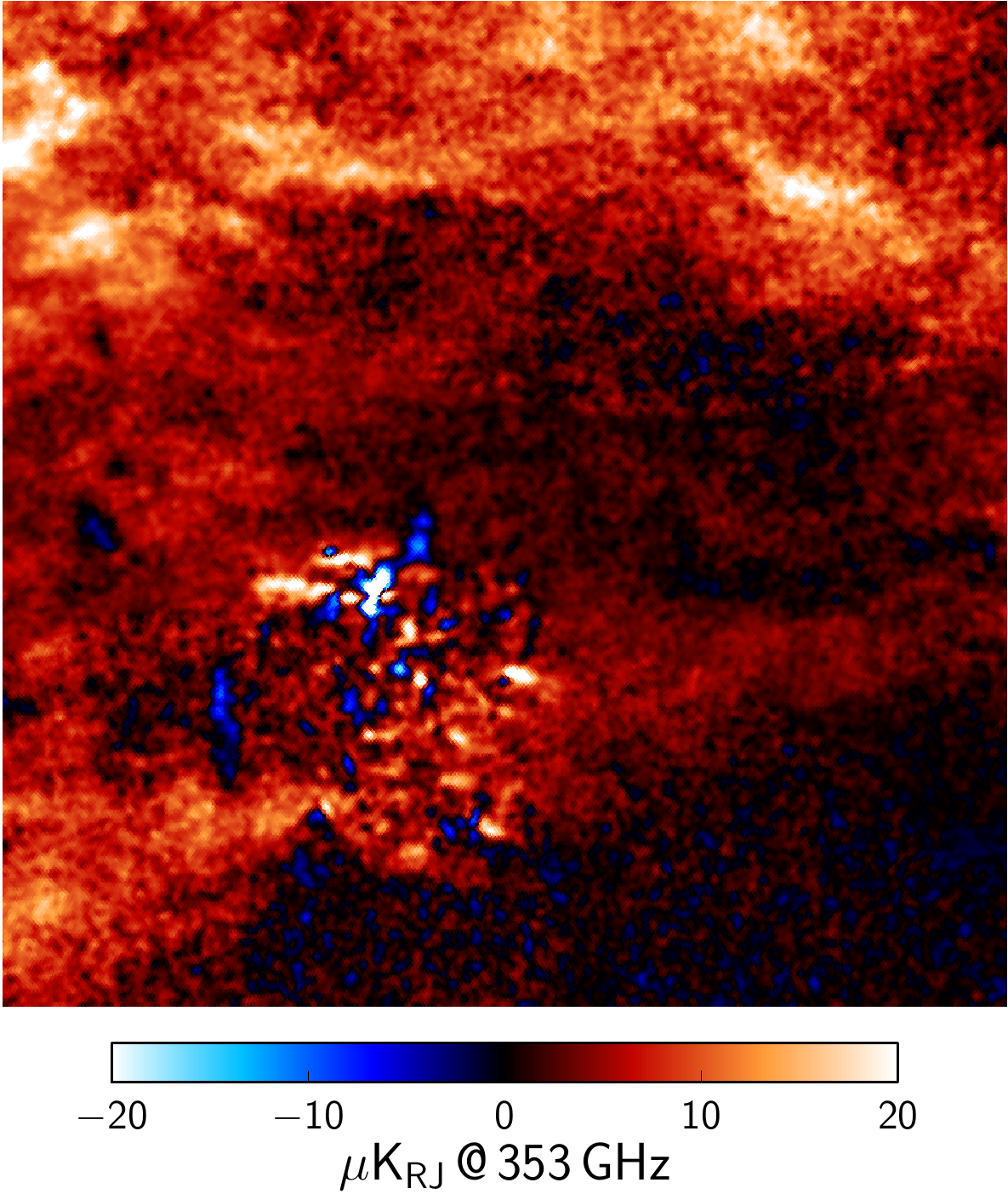,width=0.36\linewidth,clip=}
  \epsfig{figure=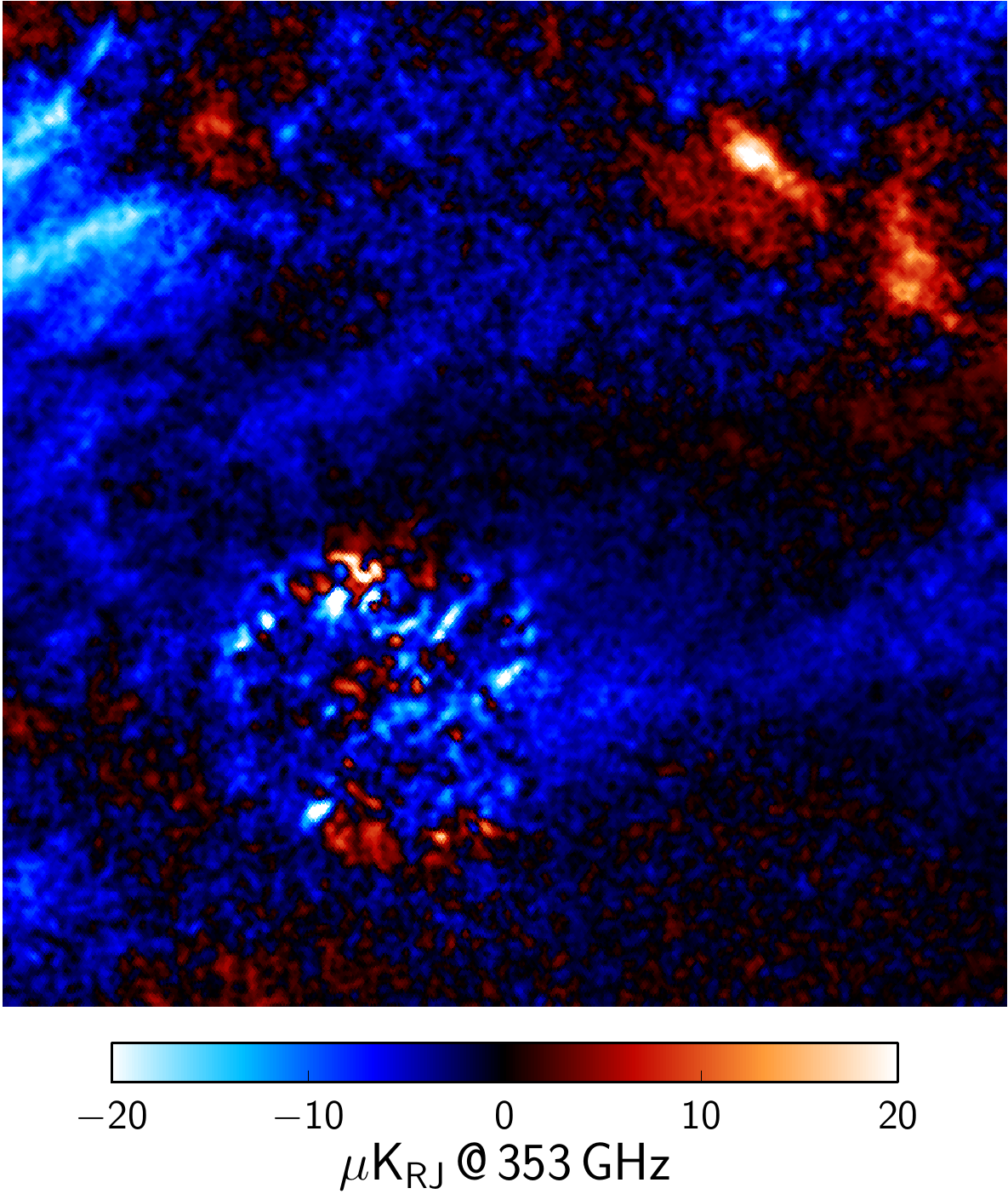,width=0.36\linewidth,clip=}
}
\end{center}
\caption{$20\deg\times20\deg$ polarization zooms centred on the south
  ecliptic pole with Galactic coordinates $(l,b) = (276\deg,-30\deg)$
  of CMB (\emph{top row}), synchrotron (\emph{middle row}), and
  thermal dust emission (\emph{bottom row}). Left and right columns
  show Stokes $Q$ and $U$ parameters, respectively. The object in the
  lower left quadrant is the Large Magellanic Cloud (LMC). }
\label{fig:pol_amp_zoom}
\end{figure*}

\begin{figure*}
\begin{center}
\mbox{
\epsfig{figure=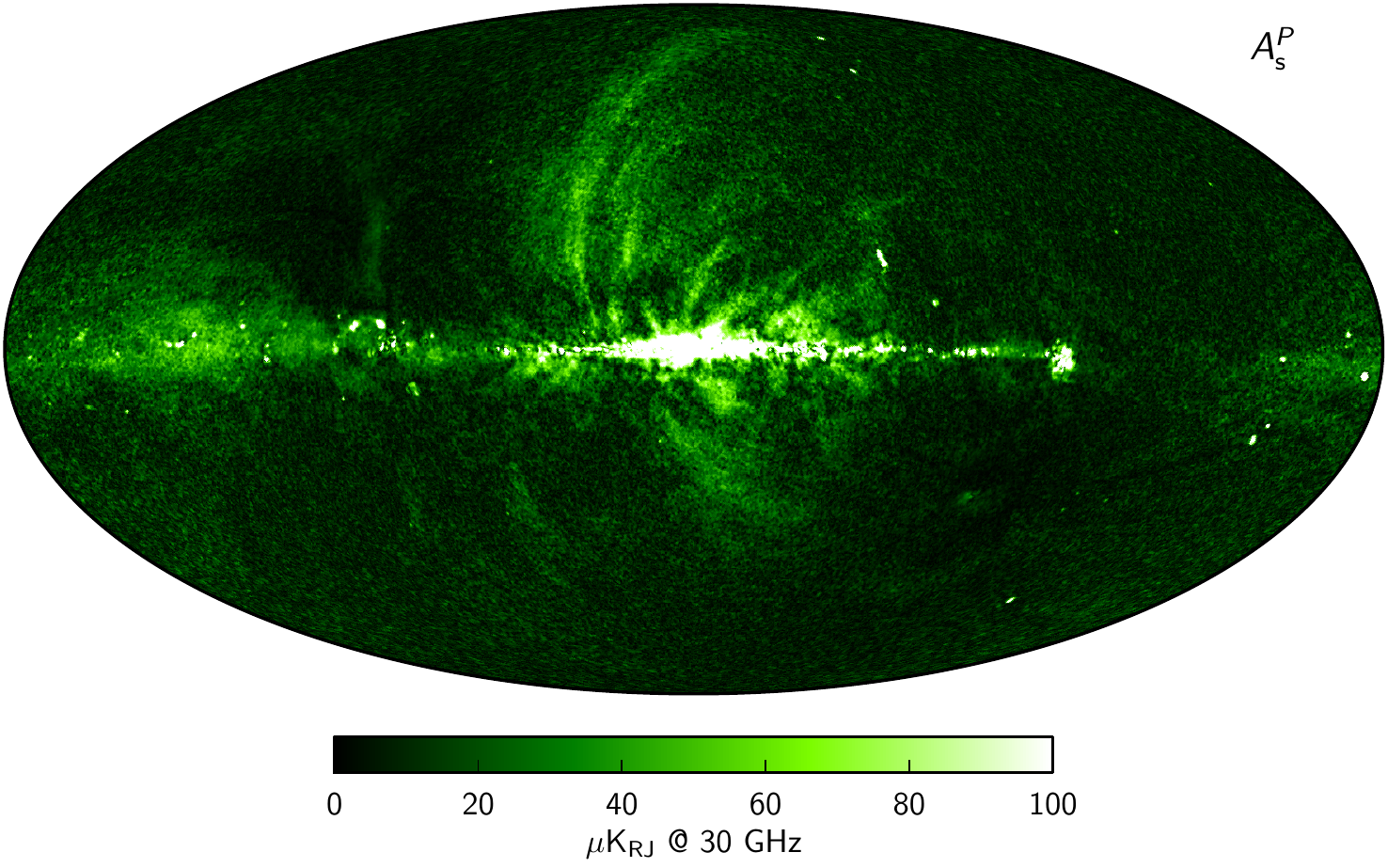,width=\linewidth,clip=}
}
\mbox{
\epsfig{figure=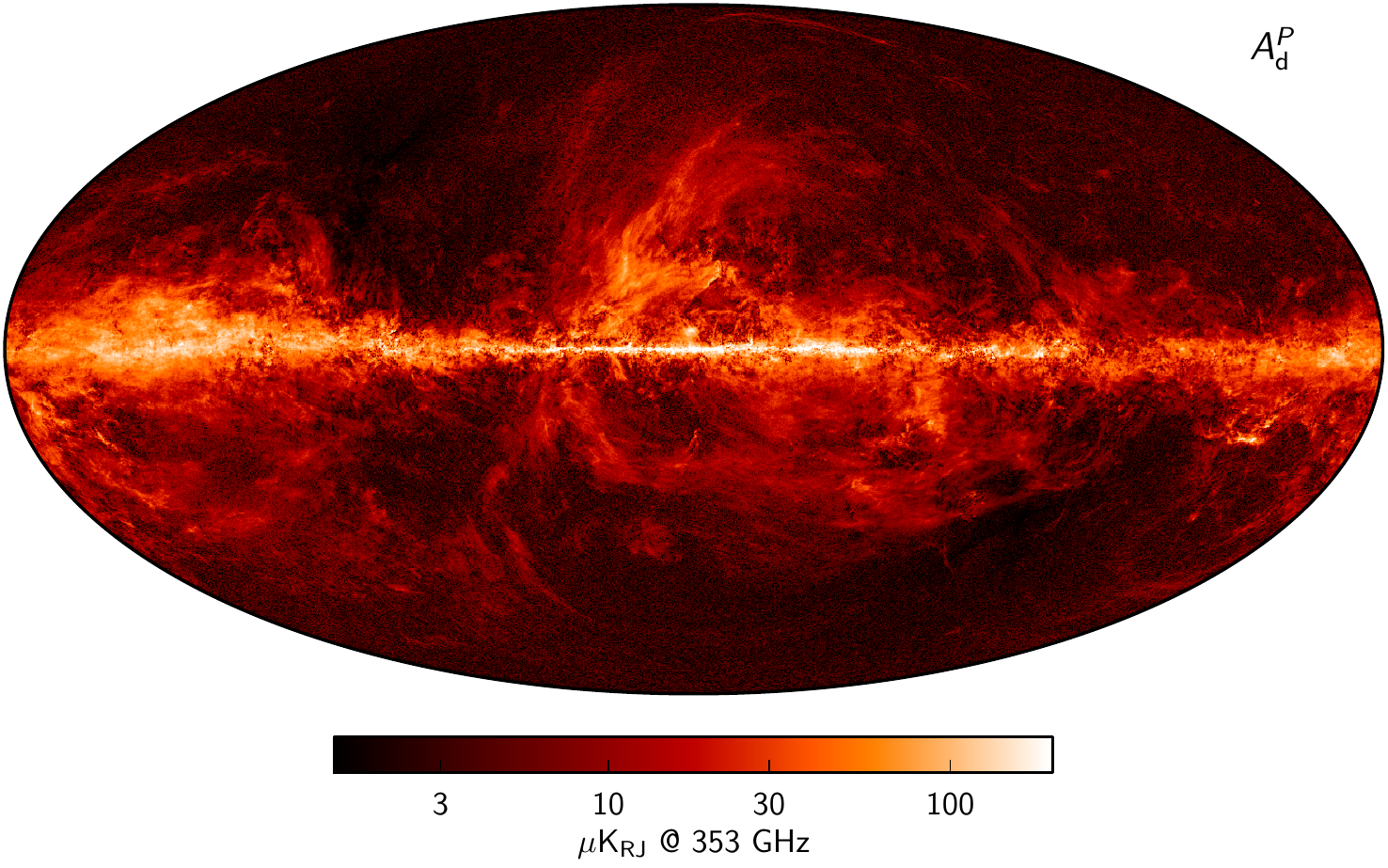,width=\linewidth,clip=}
}
\end{center}
\caption{\Planck\ polarization amplitude maps, $P=\sqrt{Q^2+U^2}$. The
top panel shows synchrotron emission at 30\,GHz, smoothed to an
angular resolution of $40\arcm$, and the bottom panel shows thermal
dust emission at 353\,GHz, smoothed to an angular resolution of
$10\arcm$.}
\label{fig:pol_amp_map2}
\end{figure*}

\begin{figure*}
\begin{center}
  \mbox{
    \epsfig{figure=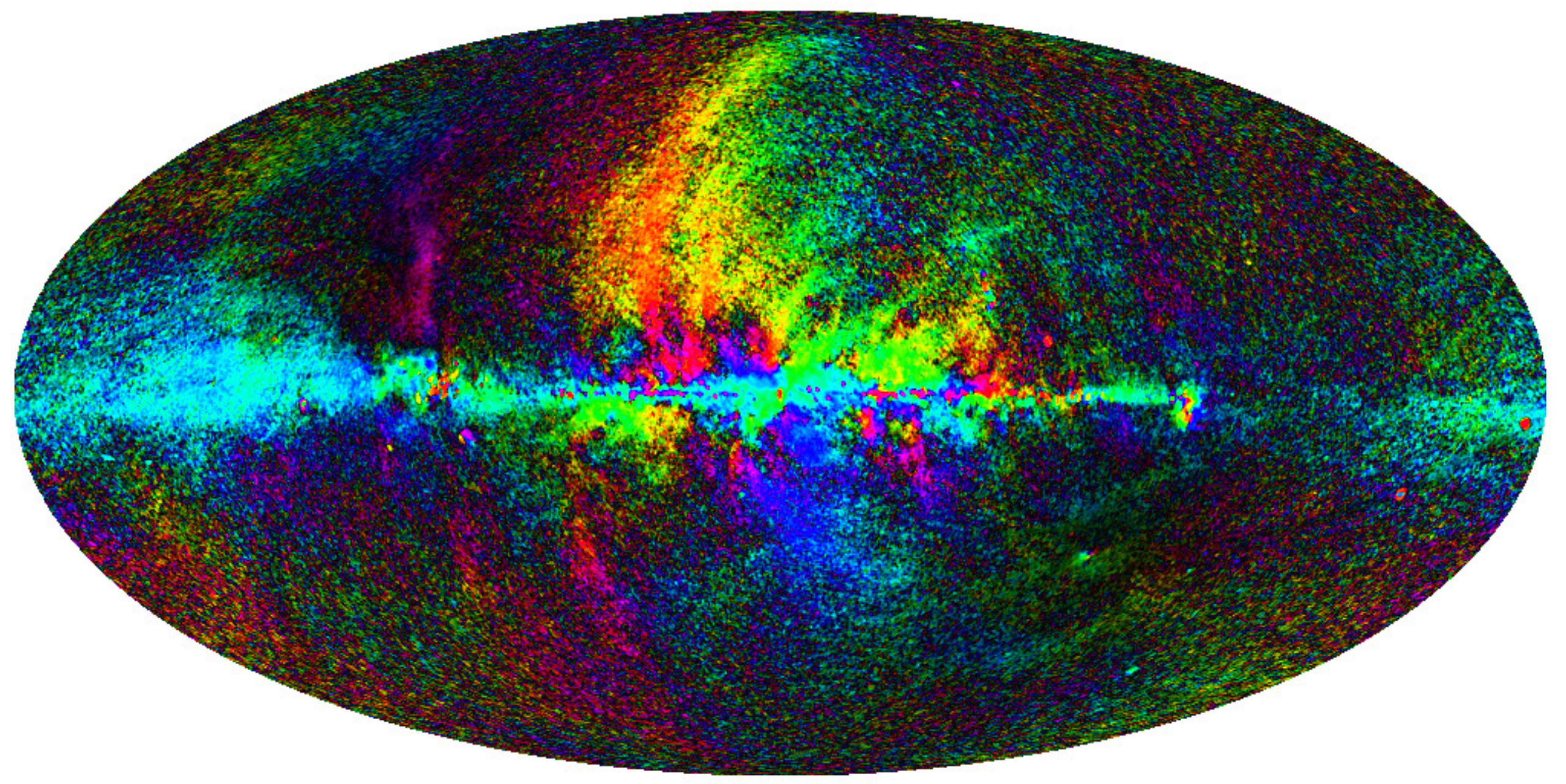,height=\linewidth,clip=,angle=90}
  }
    \mbox{
    \epsfig{figure=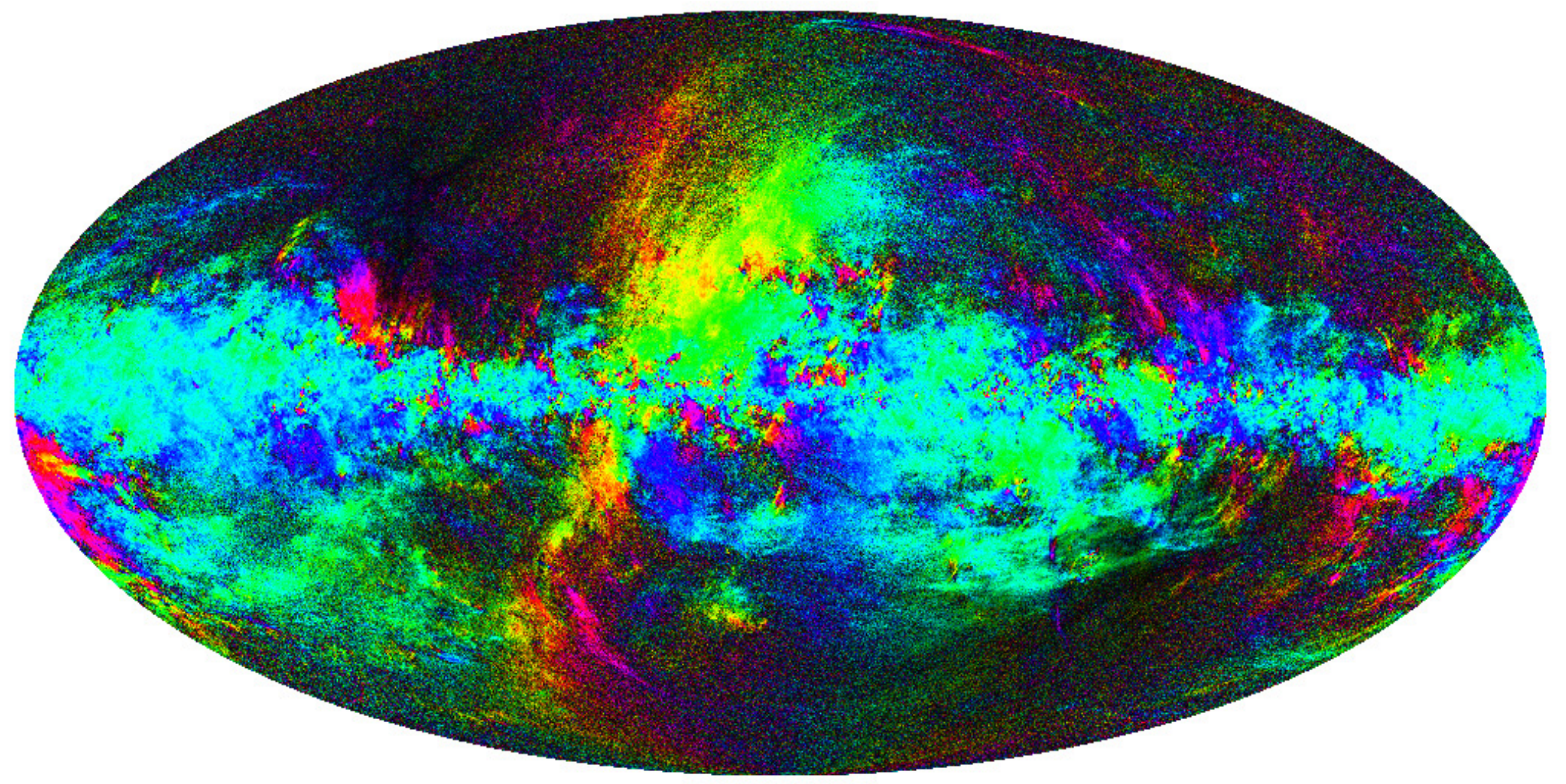,height=\linewidth,clip=,angle=90}
  }
\end{center}
\caption{\Planck\ polarization angle maps for synchrotron emission,
  smoothed to $40\arcm$ FHWM (\emph{top}) and thermal dust emission,
  smoothed to $10\arcm$ FWHM (\emph{bottom}). Light blue and red
  colours indicate polarization angles aligned with meridians
  ($\psi=0\deg$) and parallels ($\psi=90\deg$), respectively, while
  yellow and purple indicate polarization angles rotated by $-45$ and
  $+45\deg$ with respect to the local meridian in the
  \healpix\ polarization angle convention. Colours are saturated at
  10\,$\mu{\textrm K}_{\textrm{RJ}}$. }
\label{fig:pol_angles}
\end{figure*}

\section{Polarization analysis}
\label{sec:polarization}

We now turn our attention to the 2015 \Planck\ measurements of
polarized foregrounds. As described in Sect.~\ref{sec:data}, and
summarized in Table~\ref{tab:data}, we employ full co-added frequency
maps for this analysis, in order to maximize the signal-to-noise
ratio, and we fit only CMB, synchrotron, and thermal dust emission
amplitudes, fixing all calibration and spectral parameters to their
temperature counter-parts (see Sect.~\ref{sec:sky_pol}). The main data
products are summarized in the bottom two sections of
Table~\ref{tab:products}.

From an algorithmic point of view, the polarization analysis is
essentially identical to the temperature analysis, with the same
summary statistics and goodness-of-fit statistics applying equally
well to the Stokes $Q$ and $U$ parameters as to the temperature (i.e.,
Stokes $I$). We therefore proceed in the same manner as for the
temperature case, and first present the main data products, then
discuss internal goodness-of-fit tests. However, in this case we
additionally discuss the $EE$ and $BB$ angular power spectra for
synchrotron and thermal dust emission, recognizing the high importance
of these quantities in current cosmology. Finally, we consider
external validation.

\subsection{Baseline model}

We start by presenting the full-sky Stokes $Q$ and $U$ parameter maps
for polarized CMB, synchrotron, and thermal dust in
Fig.~\ref{fig:pol_amp_map}, and a $20\deg\times20\deg$ zoom-in of the
south ecliptic pole in Fig.~\ref{fig:pol_amp_zoom}. Note that the CMB
map has been high-pass filtered with a cosine-apodized harmonic space
filter, removing all structures below $\ell=20$, in order to suppress
large-scale instrumental systematics. Additionally, the Galactic plane
has been replaced with a constrained Gaussian realization in order to
avoid ringing effects during filtering (see \citealp{planck2014-a11} for
further details).

Figures~\ref{fig:pol_amp_map2} and \ref{fig:pol_angles} show the
polarization amplitude and direction, as defined by direct and naive
estimators,
\begin{align}
  P &= \sqrt{Q^2 + U^2} \\
  \psi &= \frac{1}{2}\,\textrm{atan}(U,Q).
\end{align}
Note that the former of these is a quadratic estimator, and therefore
biased by instrumental noise, and the latter is also a nonlinear
function. More sophisticated approaches are applied and described in
\mbox{\citet{planck2014-XIX}}, which properly accounts for this
bias. However, in this paper the main purpose of these quantities is
visual interpretation, rather than quantitative model comparisons, and
the naive estimators are then useful for providing information
regarding both the underlying physical structures and the instrumental
noise levels. Specifically, the noise level at a given position on the
sky can be inferred from the polarization amplitude maps in
Fig.~\ref{fig:pol_amp_map2} by comparing the intensity of any region
to the the deep fields near the ecliptic poles, where the
\Planck\ signal-to-noise ratio is maximum.

Considering first the CMB maps in Fig.~\ref{fig:pol_amp_zoom}, it is
clear that \Planck\ measures primordial polarized anisotropies with
high signal-to-noise ratio at $10\arcm$ FWHM angular
scales. Furthermore, it is also visually obvious that the CMB signal
is strongly dominated by primordial $E$-modes, in the form of a clear
$+$-type pattern in Stokes $Q$ and a $\times$-type pattern in Stokes
$U$; see Sect.~\ref{sec:sky_pol} for further discussion.

Of course, this particular region of the sky is more deeply observed
than any other region on the sky (except the north ecliptic pole), and
the apparent high signal-to-noise level seen in the middle of the
panel is not representative of the full sky. However, due to very
sharply defined features in the \Planck\ scanning strategy in this
region, the signal seen near the bottom left corner of these plots is
indeed representative, within a factor of a few.

In the same region, we also clearly see the Large Magellanic Cloud
(LMC) in both synchrotron and thermal dust emission (see
\citealp{planck2014-a31} for a detailed analysis of the LMC), but only
very weakly in the CMB maps. Indeed, the primary signature of the LMC
is a slight increase in variance rather than a systematic bias. This
is quite reassuring in terms of CMB reconstruction fidelity, since the
LMC represents one of the richest astrophysical objects on the sky. 

\begin{figure}
\begin{center}
\mbox{
\epsfig{figure=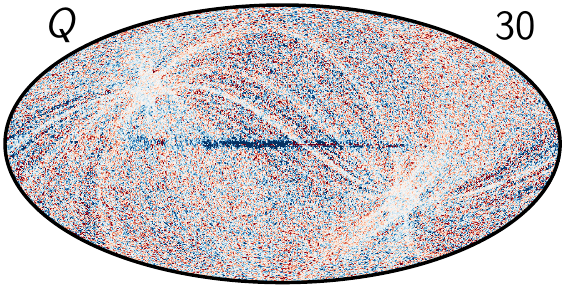,width=0.5\linewidth,clip=}
\epsfig{figure=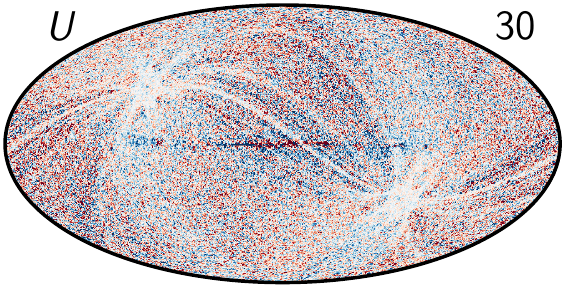,width=0.5\linewidth,clip=}
}
\mbox{
\epsfig{figure=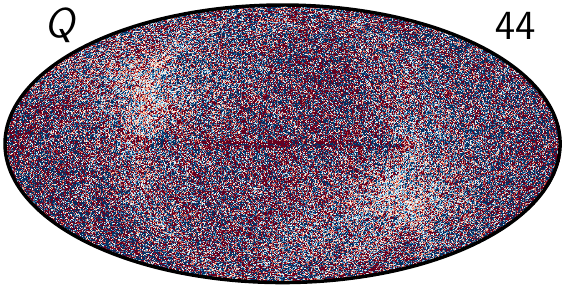,width=0.5\linewidth,clip=}
\epsfig{figure=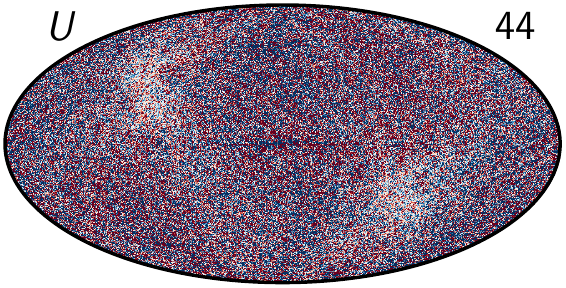,width=0.5\linewidth,clip=}
}
\mbox{
\epsfig{figure=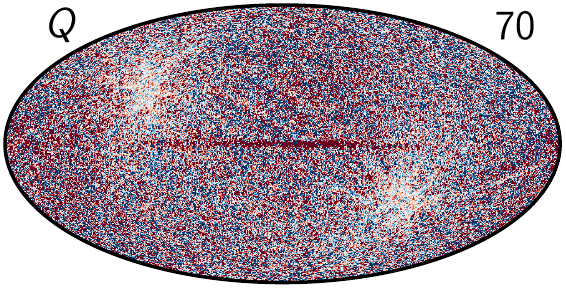,width=0.5\linewidth,clip=}
\epsfig{figure=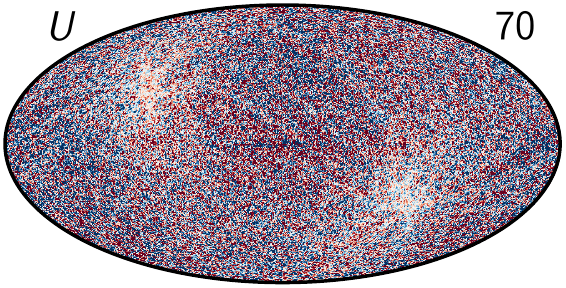,width=0.5\linewidth,clip=}
}
\mbox{
\epsfig{figure=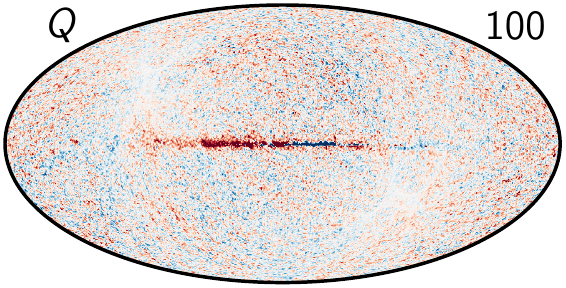,width=0.5\linewidth,clip=}
\epsfig{figure=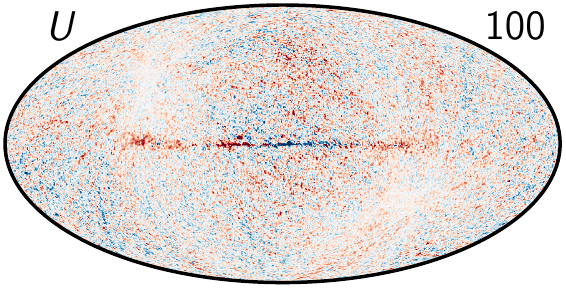,width=0.5\linewidth,clip=}
}
\mbox{
\epsfig{figure=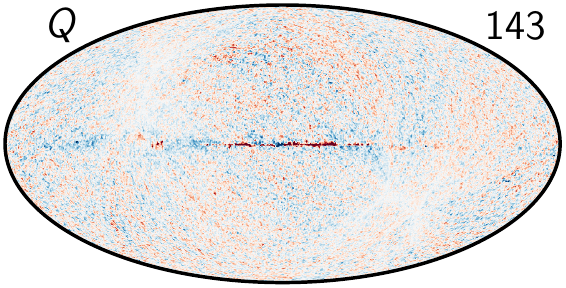,width=0.5\linewidth,clip=}
\epsfig{figure=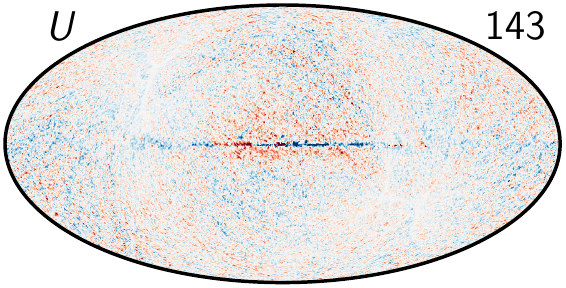,width=0.5\linewidth,clip=}
}
\mbox{
\epsfig{figure=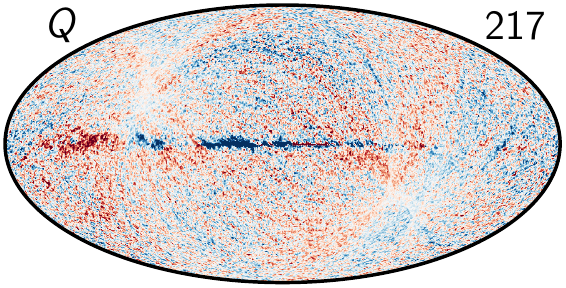,width=0.5\linewidth,clip=}
\epsfig{figure=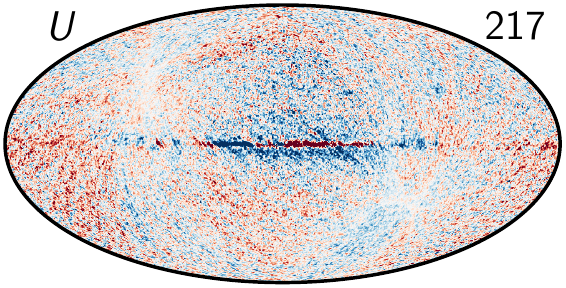,width=0.5\linewidth,clip=}
}
\mbox{
\epsfig{figure=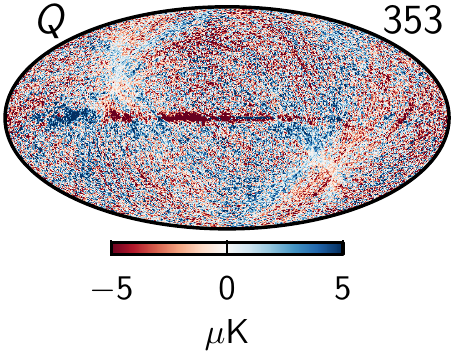,width=0.5\linewidth,clip=}
\epsfig{figure=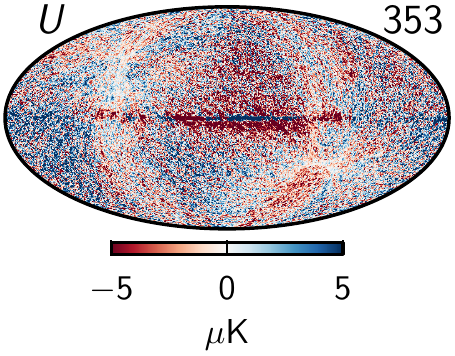,width=0.5\linewidth,clip=}
}
\end{center}
\caption{Polarization residual maps, $\d_{\nu}-\s_{\nu}$. Each row
corresponds to one frequency map, with 30\,GHz in the top row and
353\,GHz in the bottom row; left and right columns show the Stokes
$Q$ and $U$ parameters, respectively. All panels employ the same linear colour
scale. }
\label{fig:pol_residuals}
\end{figure}

\begin{figure}
\begin{center}
\mbox{
\epsfig{figure=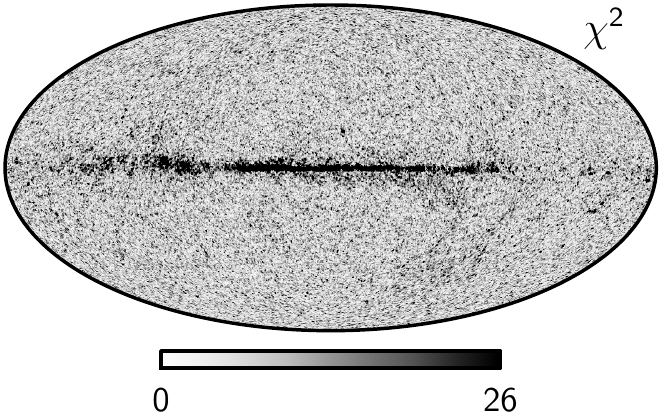,width=88mm,clip=}
}
\mbox{
\epsfig{figure=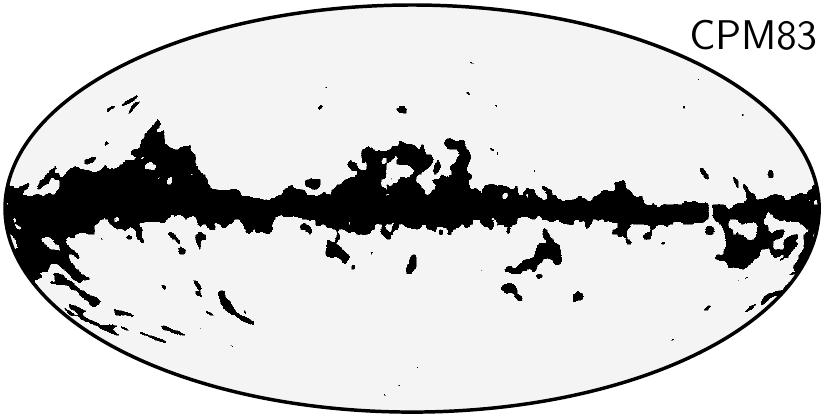,width=88mm,clip=}
}
\mbox{
\epsfig{figure=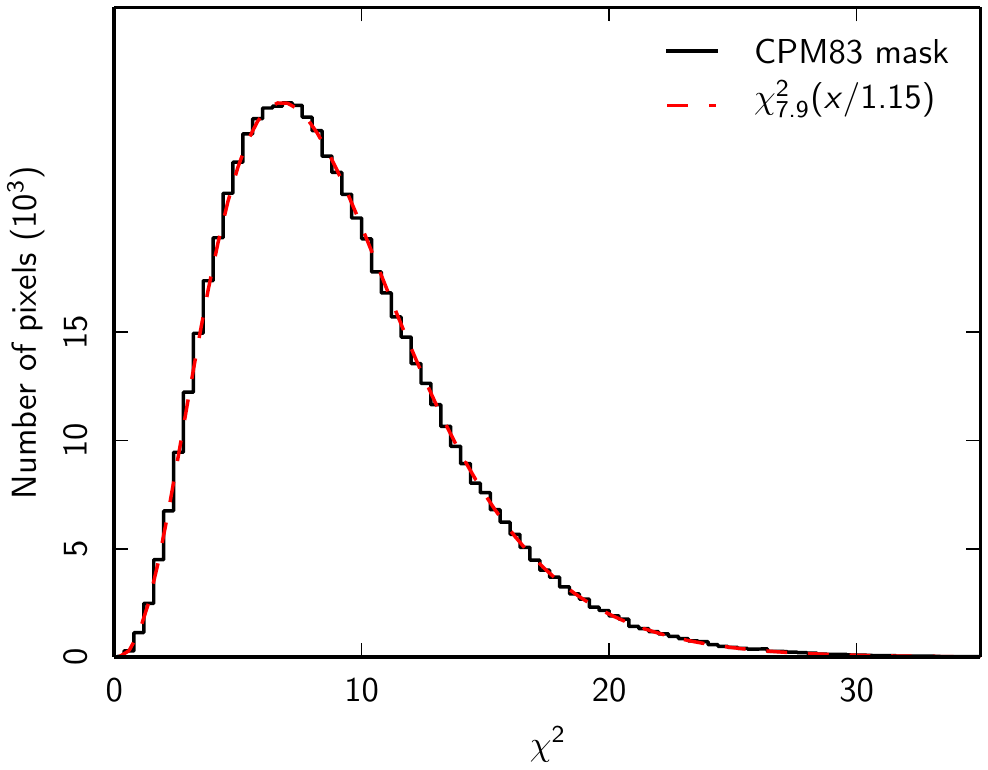,width=88mm,clip=}
}
\end{center}
\caption{\emph{Top}: $\chi^2$ per pixel for the polarization analysis
  of \Planck\ observations between 30 and 353\,GHz, summed over Stokes
  $Q$ and $U$ parameters. \emph{Middle}: \texttt{Commander} polarization mask
  (CPM), defined as the product of the CO $J$=1$\rightarrow$0 emission
  map thresholded at
  $0.5\,\textrm{K}_{\textrm{RJ}}\,\textrm{km}\,\textrm{s}^{-1}$, and the
  smoothed $\chi^2$ map thresholded at a value of 26. This mask
  retains a total of 83\% of the sky. \emph{Bottom}: Histogram of
  $\chi^2$ values outside the conservative CPM83 mask. The grey dashed
  line shows the best-fit $\chi^2$ distribution with a variable degree
  of freedom and scaling, used to account for noise modelling
  effects. }
\label{fig:chisq_pol_map}
\end{figure}

\begin{figure*}
\begin{center}
\mbox{
\epsfig{figure=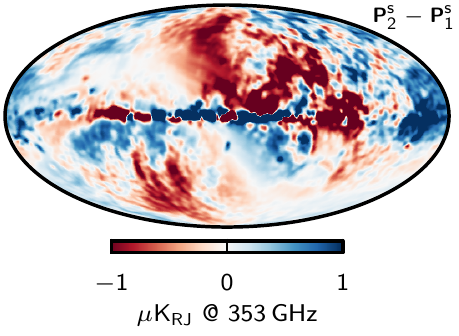,width=0.5\linewidth,clip=}
\epsfig{figure=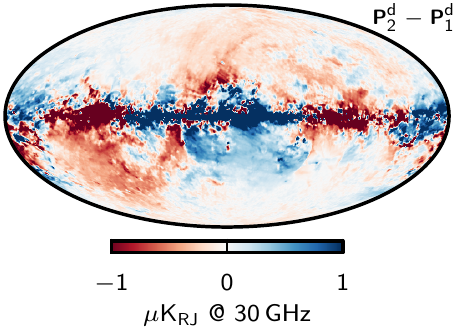,width=0.5\linewidth,clip=}
}
\mbox{
\epsfig{figure=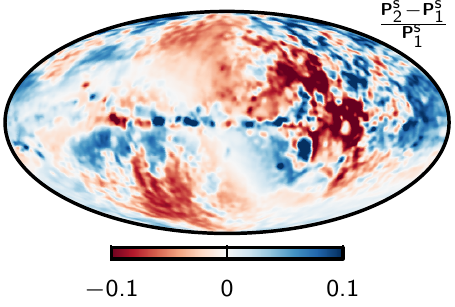,width=0.5\linewidth,clip=}
\epsfig{figure=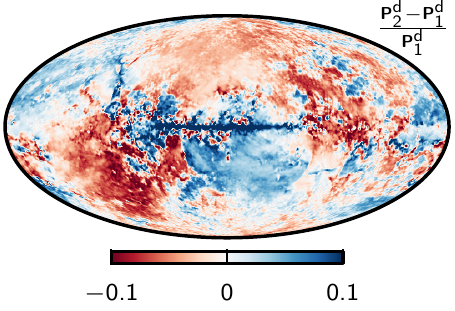,width=0.5\linewidth,clip=}
}
\end{center}
\caption{Difference maps (\emph{top}) and fractional difference maps
  (\emph{bottom}) between the synchrotron (\emph{left}) and thermal
  dust (\emph{right}) polarization solutions derived with two
  different HFI temperature-to-polarization leakage templates. The
  synchrotron polarization amplitude maps are smoothed to $3\deg$ FHWM
  before computing absolute and fractional differences, and the
  thermal dust polarization amplitude maps are smoothed to
  $1\deg$. Maps labelled by a subscript ``1'' correspond to the default leakage
  templates used in the \Planck\ 2015 release, and maps labelled by a subscript ``2''
  correspond to the experimental leakage templates; see
  \citet{planck2014-a09} for further discussion. }
\label{fig:ggf_vs_case1}
\end{figure*}

\begin{figure}
\begin{center}
\mbox{
  \epsfig{figure=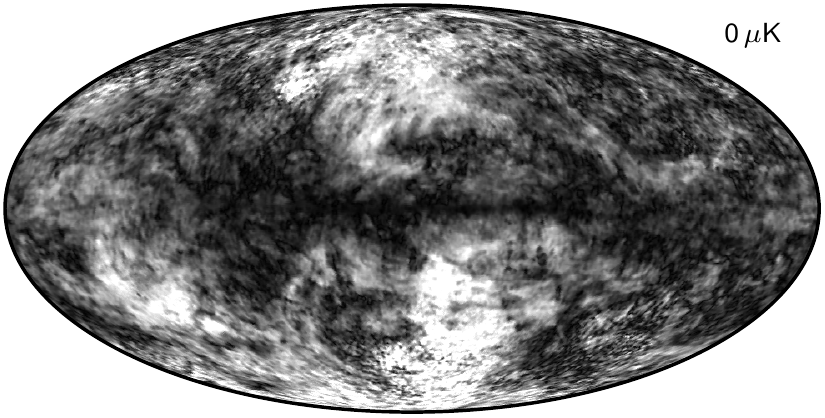,width=\linewidth,clip=}
}
\mbox{
\epsfig{figure=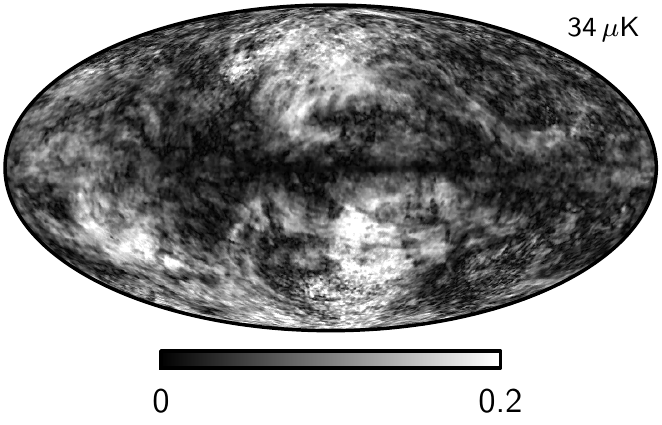,width=\linewidth,clip=}
}
\end{center}
\caption{Thermal dust polarization fraction for Galactic emission
  zero-level corrections of $0\,\mu\textrm{K}$ (\emph{top}) and
  $34\,\mu\textrm{K}$ (\emph{bottom}). A value of $34\,\mu\textrm{K}$
  corresponds to our current best estimate of the residual zodiacal
  light offset in the 353\,GHz channel \citep{planck2014-a09}. The
  statistical uncertainty on the Galactic emission zero-level from
  \ion{H}{i} cross-correlation is 0.0067\,MJy\,sr$^{-1}$ or
  23$\,\mu\textrm{K}_{\textrm{CMB}}$.}
\label{fig:polfrac}
\end{figure}

In this respect it is worth recalling that we fix all calibration and
spectral parameters (thermal dust index and temperature and
synchrotron spectrum) in the polarization analysis to those derived in
the temperature analysis. Several attempts have been made at
re-estimating these parameter independently from the polarization
observations, but we find that the resulting parameters invariably
become significantly biased by the same large-scale systematics that
are responsible for the remaining large-scale residuals in the CMB map
\citep{planck2014-a03,planck2014-a09,planck2014-a11}. However, other
analyses that explicitly exploit spatial correlations (e.g., template
fitting) to suppress such systematics have been able to produce robust
results, and yield only small differences between the temperature and
polarization spectral indices. For instance, \citet{planck2014-XXII}
reports full-sky estimates of $\beta_{\textrm{d}}=1.51\pm0.01$ for
intensity and $\beta_{\textrm{d}}=1.59\pm0.02$ for polarization,
corresponding to a difference of only $3.6\,\sigma$ even after averaging
over most of the sky. Thus, assuming identical temperature and
polarization spectral indicies is a very good approximation at the
precision level of the current data, considering the additional
stability with respect to instrumental systematic errors it provides.

\subsubsection{Goodness-of-fit}

We now consider the statistical goodness-of-fit of this simple
baseline model, following the same procedure as for the temperature
analysis. First, Fig.~\ref{fig:pol_residuals} shows the residual maps
$\d_{\nu}-\s_{\nu}$, for each of the seven \Planck\ frequency maps,
all smoothed to a common resolution of $40\arcm$ FWHM angular
scale. Note that the colour range is linear between
$\pm5\,\mu\textrm{K}$, and the same in all panels. As expected, we see
that 143\,GHz is the most sensitive frequency channel, followed by
the 100 and 217\,GHz channels. In addition to instrumental noise,
these channels also exhibit large-scale patterns tracing the
\Planck\ scanning strategy at the $\lesssim0.5\,\mu\textrm{K}$
level. Although small in an absolute sense, it is important to recall
that the expected peak-to-peak amplitude of a cosmological signal from
reionization corresponding to an optical depth of, say,
$\tau\approx0.07$ is also about $0.5\,\mu\textrm{K}$
\citep{planck2014-a13}. As a result, we do not consider the CMB
polarization map presented here suitable for cosmological parameter
estimation on large angular scales. Instead, the \Planck\ 2015
low-$\ell$ polarization likelihood relies only on the 30, 70, and
353$\,$GHz data, for which instrumental systematics are subdominant
\citep{planck2014-a07, planck2014-a09, planck2014-a13}.

\begin{table}[tmb]                                                                                                                                              
\begingroup                                                                                                                                 
\newdimen\tblskip \tblskip=5pt
\caption{Goodness-of-fit statistics for polarization analysis.\label{tab:Pgof}}
\vskip -4mm
\footnotesize                                                                                                                                       
\setbox\tablebox=\vbox{                                                                                                                                                                            
\newdimen\digitwidth                                                                                                                      
\setbox0=\hbox{\rm 0}
\digitwidth=\wd0
\catcode`*=\active
\def*{\kern\digitwidth}
\newdimen\signwidth
\setbox0=\hbox{+}
\signwidth=\wd0
\catcode`!=\active
\def!{\kern\signwidth}
\newdimen\decimalwidth
\setbox0=\hbox{.}
\decimalwidth=\wd0
\catcode`@=\active
\def@{\kern\signwidth}
\halign{ \hbox to 1in{#\leaderfil}\tabskip=2em&
    \hfil#\hfil\tabskip=2em&
    \hfil#\hfil\tabskip=0pt\cr
\noalign{\doubleline}
\omit&\multispan2\hfil Rms outside CPM83\hfil\cr 
\noalign{\vskip -3pt}
\omit\hfil Frequency\hfil&\multispan2\hrulefill\cr
\noalign{\vskip 3pt}
\omit\hfil[GHz]\hfil&$\sigma_\nu^{\rm res}$ [\muK]&$\sigma^{\rm res}_{\nu}/\sigma^{\rm inst}_{\nu}$\cr
\noalign{\vskip 5pt\hrule\vskip 5pt}
*30&   2.12& 0.28\cr
*44&   7.59& 1.01\cr
*70&   4.85& 1.01\cr
100&   1.18& 0.90\cr
143&   0.90& 0.81\cr
217&   1.52& 0.95\cr
353&   2.93& 0.42\cr
\noalign{\vskip 5pt\hrule\vskip 3pt}
}}
\endPlancktable                                                                                                                                    
\endgroup
\end{table}

Next, the top panel of Fig.~\ref{fig:chisq_pol_map} shows the
corresponding $\chi^2$ map, co-adding over both frequencies and Stokes
parameters. Compared to the temperature case, it is here much
easier to determine the appropriate number of degrees of freedom,
since no spectral parameters are fitted to the data, and no positivity
priors are imposed on the amplitudes. Specifically, there are in total
14 data points (two Stokes parameters in seven frequencies) and 6 free
parameters (two Stokes parameters in three components), resulting in a
net 8 degrees of freedom. The nominal 95\,\% confidence region for
this number of degrees of freedom is $\chi^2 = (2,17)$. 

As usual, the Galactic plane is the most significant feature in the
$\chi^2$ map. Furthermore, when comparing this $\chi^2$ map (and the
individual frequency residual maps) with the various component
amplitude maps derived in the temperature analysis, we find strong
correlations between the $\chi^2$ map and the CO emission maps. This
is expected from the mapmaking analyses presented in
\citet{planck2014-a13}, and, as already noted, a general
recommendation regarding these maps is to reject any pixels with
significant CO intensity contribution, because of
temperature-to-polarization leakage. The \texttt{Commander}
polarization mask (CPM) is accordingly defined as the product of the
(smoothed and thresholded) $\chi^2$ map shown in the top panel of
Fig.~\ref{fig:chisq_pol_map}, and the low-resolution
\textrm{Commander} CO $J$=1$\rightarrow$0 intensity map thresholded at
$0.5\,\textrm{K}_{\small{\textrm{RJ}}}\textrm{km}\,\textrm{s}^{-1}$. The
resulting mask is shown in the middle panel of
Fig.~\ref{fig:chisq_pol_map}, and excludes 17\,\% of the sky.

The bottom panel of Fig.~\ref{fig:chisq_pol_map} shows a histogram of
the $\chi^2$ values outside the CPM83 mask, with the best-fit $\chi^2$
distribution with variable number of degrees of freedom and width,
fully analogous to the temperature case in
Sect.~\ref{sec:baseline}. In this case, the best-fit distribution has
7.9 degrees of freedom, in excellent agreement with the theoretical
expectation of 8, while the width rescaling factor that accounts for
correlated noise and smoothing is 1.15.

Table~\ref{tab:Pgof} lists the rms of the residual maps for each
frequency, analogous to Table~\ref{tab:Tgof} for temperature, averaged
over the two Stokes $Q$ and $U$ parameters. The third column in this
table shows the ratio between these rms values and instrumental noise;
again, we observe good agreement with expectations. As for the
temperature case, the values for the 30 and 353$\,$GHz channels are
significantly lower than unity, because these two frequencies dominate
the synchrotron and thermal dust amplitude parameters, respectively. 

Next, we assess the impact of temperature-to-polarization leakage from
the CMB temperature monopole and dipole and from Galactic temperature
emission by computing the synchrotron and thermal dust amplitude maps
when adopting two different HFI leakage models. The first is simply
the default template set adopted for the main release, corresponding
to the results already discussed, while the second is the experimental
template set described in \citet{planck2014-a09} and
\citet{planck2014-a13}. From the resulting amplitude maps, we perform
the following steps: first compute the polarization amplitude, $P$;
smooth this to $3\deg$ FWHM for synchrotron and $1\deg$ FWHM for
thermal dust; and finally compute the difference, $P_2 - P_1$, and
fractional difference, $(P_2 - P_1)/P_1$, maps. These are shown in
Fig.~\ref{fig:ggf_vs_case1}. Here we see that the absolute
polarization amplitude difference between the two leakage models is
around 1$\,\mu\textrm{K}_{\textrm{RJ}}$ for both synchrotron and
thermal dust at high Galactic latitudes, increasing to a few tens of
$\mu\textrm{K}_{\textrm{RJ}}$ in the Galactic plane. Accordingly, the
fractional residuals are $\lesssim10$\,\% at high Galactic latitudes,
and they increase to about 30\,\% in the central Galactic plane.

Finally, we comment on the polarization fractions that may be derived
from these maps. First of all, we emphasize that the delivered
products are maps of the Stokes $Q$ and $U$ parameters, not of
polarization amplitude and polarization angle and fractions. This
choice is primarily driven by the fact that the Stokes parameters are
linear, and therefore have much simpler noise properties than the
corresponding nonlinear parameters. Second, when computing
polarization fractions, $P/I$, it is of utmost importance to recognize
and account for the considerable uncertainty in this quantity with
respect to the zero-level of the corresponding temperature map.  To
make this point concrete, we show in the top panel of
Fig.~\ref{fig:polfrac} the naive polarization fraction derived
directly from the delivered \texttt{Commander} thermal dust intensity
and polarization maps. This map saturates the colour scale over
extended regions in the southern Galactic hemisphere, nominally
suggesting a polarization fraction well above 20\,\%. However, as
discussed both in Sect.~\ref{sec:data} and \citet{planck2014-a09},
there is an offset in the zero-level of the zodiacal light emission of
$34\,\mu\textrm{K}$ in the current 353\,GHz temperature
data. Correcting for this offset results in the polarization fraction
shown in the lower panel of Fig.~\ref{fig:polfrac}, which shows
significantly smaller values. Further, the raw statistical uncertainty
of the 353\,GHz Galactic emission zero level from \ion{H}{i}
cross-correlation alone is $23\,\mu\textrm{K}$
\citep{planck2014-a09}. The conclusion is therefore that any analysis
that relies directly on the polarization fraction, as opposed to the
much more stable polarization amplitude, needs to account for the
significant uncertainties in the Galactic emission zero-level at
353\,GHz.

\begin{figure*}
\begin{center}
\mbox{
  \epsfig{figure=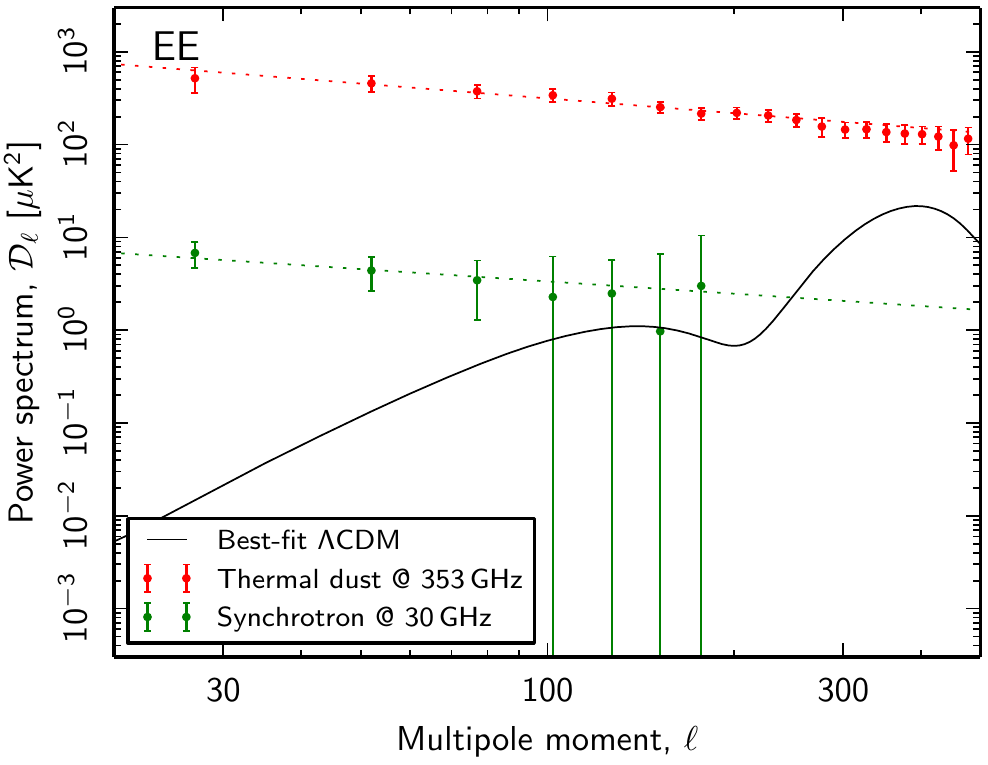,width=0.5\linewidth,clip=}
  \epsfig{figure=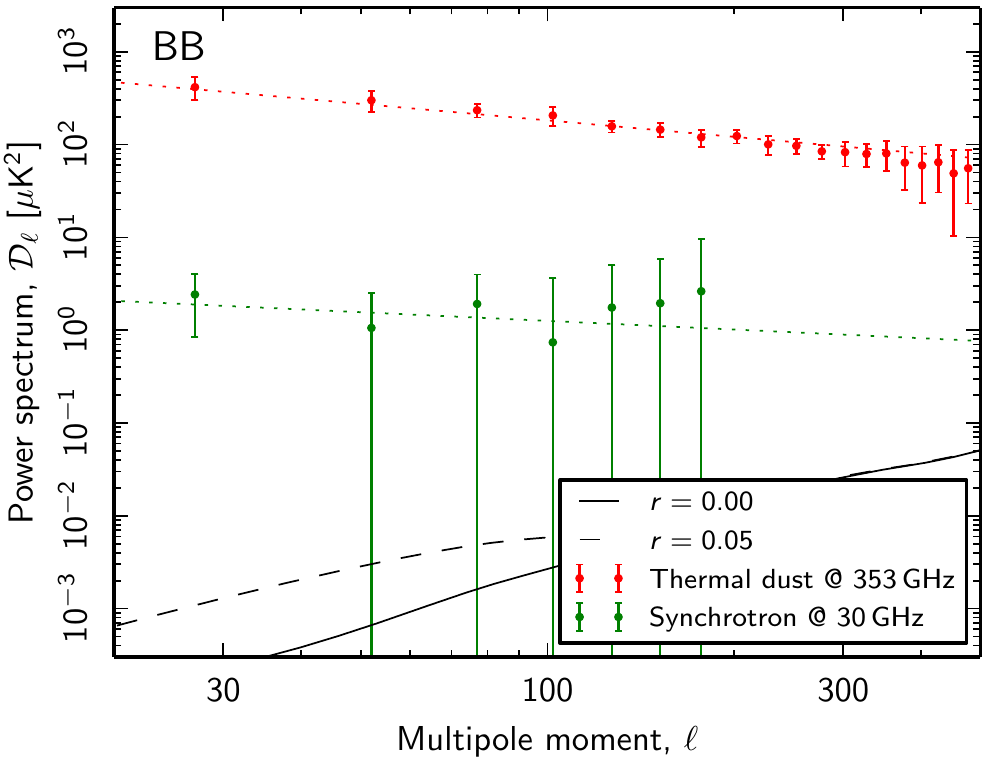,width=0.5\linewidth,clip=}
}
\end{center}
\caption{Angular $EE$ (\emph{left panel}) and $BB$ (\emph{right panel})
  power spectra for polarized synchrotron (at 30\,GHz) and thermal
  dust emission (at 353\,GHz), evaluated with $1\deg$ FWHM apodization
  and including a total effective sky fraction of 73\,\% of the
  sky. The dashed lines show the best-fit power-law models to each
  case, and the solid black lines shows the best-fit $\Lambda$CDM
  power spectrum as fitted to temperature observations only
  \citep{planck2014-a13,planck2014-a15}. The dashed black line in the
  $BB$ panel shows the spectrum for a model with a tensor-to-scalar
  ratio of $r=0.05$.}
\label{fig:cl_pol}
\end{figure*}

\begin{table}[t] 
\begingroup 
\newdimen\tblskip \tblskip=5pt
\caption{Best-fit power-law parameters to the angular power spectra of
  synchtrotron (at 30\,GHz) and thermal dust emission (at 353\,GHz) as
  a function of mask apodization. The parameters are defined by the
  model $D_{\ell} = q\, (\ell/80)^\alpha$, and the fits include
  multipoles between $\ell=10$ and 150 for synchrotron, and between
  $\ell=10$ and 300 for thermal dust emission. All uncertainties are
  statistical, and do not account for systematic or modelling
  uncertainties. The last case is reproduced from Table~1 in
  \citet{planck2014-XXX}. \label{tab:clpar}}
\nointerlineskip                                                                                                                                                                                     
\vskip -6mm
\footnotesize 
\setbox\tablebox=\vbox{ %
\newdimen\digitwidth 
\setbox0=\hbox{\rm 0}
\digitwidth=\wd0
\catcode`*=\active
\def*{\kern\digitwidth}
\newdimen\signwidth
\setbox0=\hbox{+}
\signwidth=\wd0
\catcode`!=\active
\def!{\kern\signwidth}
\newdimen\decimalwidth
\setbox0=\hbox{.}
\decimalwidth=\wd0
\catcode`@=\active
\def@{\kern\signwidth}
\def\s#1{\ifmmode $\rlap{$^{\rm #1}$}$ \else \rlap{$^{\rm #1}$}\fi}
\halign{ \hbox to 0.9in{#\leaderfil}\tabskip=1em&
    \hfil#\hfil\tabskip=0.5em&
    \hfil#\hfil\tabskip=1em&
    \hfil#\hfil\tabskip=0.5em&
    \hfil#\hfil\tabskip=0pt\cr
\noalign{\doubleline}
\omit&\multispan2\hfil Synchrotron\hfil& \multispan2\hfil Thermal dust\hfil\cr
\noalign{\vskip -3pt}
\omit&\multispan2\hrulefill&\multispan2\hrulefill\cr
\noalign{\vskip 2pt}
\omit\hfil \hfil& $q$ [$\mu\textrm{K}_{\textrm{CMB}}^2$]& $\alpha$& $q$ [$\mu\textrm{K}_{\textrm{CMB}}^2$]& $\alpha$\cr
\noalign{\doubleline}
\noalign{\vskip 3pt}
\multispan5 {\bf Common mask; apod = $1\deg$ FWHM; $f_{\textrm{sky}}^{\textrm{eff}}=0.73$}\hfil\cr
\noalign{\vskip 3pt}
\hglue 1em$EE$& $3.7\pm0.2$& $-0.44\pm0.07$& $354\pm6$& $-0.53\pm0.02$\cr
\hglue 1em$BB$& $1.3\pm0.2$& $-0.31\pm0.13$& $208\pm4$& $-0.59\pm0.02$\cr
\noalign{\vskip 2pt}
\hglue 1em$BB$/$EE$& \multispan2 \hfil$0.36\pm0.06$\hfil& \multispan2 \hfil$0.59\pm0.01$\hfil\cr
\noalign{\vskip 3pt}
\multispan5 {\bf Common mask; apod = $2\deg$ FWHM; $f_{\textrm{sky}}^{\textrm{eff}}=0.68$}\hfil\cr
\noalign{\vskip 3pt}
\hglue 1em$EE$& $3.2\pm0.2$& $-0.49\pm0.08$& $285\pm5$& $-0.53\pm0.02$\cr
\hglue 1em$BB$& $1.1\pm0.2$& $-0.02\pm0.17$& $161\pm3$& $-0.62\pm0.02$\cr
\noalign{\vskip 2pt}
\hglue 1em$BB$/$EE$& \multispan2 \hfil$0.34\pm0.07$\hfil& \multispan2 \hfil$0.56\pm0.01$\hfil\cr
\noalign{\vskip 3pt}
\multispan5 {\bf Common mask; apod = $5\deg$ FWHM; $f_{\textrm{sky}}^{\textrm{eff}}=0.55$}\hfil\cr
\noalign{\vskip 3pt}
\hglue 1em$EE$& & & $188\pm3$& $-0.44\pm0.02$\cr
\hglue 1em$BB$& & & $*99\pm2$& $-0.51\pm0.03$\cr
\noalign{\vskip 2pt}
\hglue 1em$BB$/$EE$& \multispan2& \multispan2 \hfil$0.53\pm0.01$\hfil\cr
\noalign{\vskip 3pt}
\multispan5 {\bf CO mask; apod = $5\deg$ FWHM; $f_{\textrm{sky}}^{\textrm{eff}}=0.73$; Planck Int.\ XXX (2014)}\hfil\cr
\noalign{\vskip 3pt}
\hglue 1em$EE$& & & $328\pm3$& $-0.43\pm0.02$\cr
\hglue 1em$BB$& & & $\cdots$& $-0.46\pm0.02$\cr
\noalign{\vskip 2pt}
\hglue 1em$BB$/$EE$& \multispan2& \multispan2 \hfil$0.53\pm0.01$\hfil\cr
\noalign{\vskip 2pt\hrule\vskip 2pt}
}}
\endPlancktable
\endgroup
\end{table}

\subsection{Synchrotron and thermal dust angular power spectra}
\label{sec:pol_cls}

One of the most important goals of modern CMB cosmology is to detect
primordial $B$-mode polarization on large angular scales, a direct
observable signature of inflationary gravitational waves. The main
obstacles in this search are the polarized synchrotron and thermal
emission signals discussed above. In order to quantify the magnitude
of this problem, we compute in this section their angular power
spectra, and compare them to the expected primordial CMB spectrum. A
more comprehensive analysis of the same type, but based only on the
353\,GHz frequency channel, was recently published in
\citet{planck2014-XXX}.

We employ the same cross-spectrum power spectrum estimator as used in
\citet{planck2014-XXX}, but introduce two specific changes. First, we
adopt the so-called common mask from the CMB component separation
analysis presented in \citet{planck2014-a11}, rather than the CO mask
employed in the original paper, and second, we consider three
different mask apodization scales (1, 2, and $5\deg$ FWHM) as opposed
to only $5\deg$ FWHM as in \citet{planck2014-XXX}.

The $EE$ and $BB$ spectra resulting from the evaluation using $1\deg$ FWHM
apodization are shown in the left and right panels of
Fig.~\ref{fig:cl_pol}, respectively, both plotted in terms of
$D_{\ell} = C_{\ell}\,\ell(\ell+1)/2\pi$ in thermodynamic units. Red
data points show the angular power spectra for thermal dust emission
at 353\,GHz, and green points show synchrotron emission at
30\,GHz. Each spectrum is binned with $\Delta\ell=25$, and the plotted
uncertainties are the standard deviation of the single-$\ell$ spectrum
values within each bin. Black solid lines indicate the best-fit
$\Lambda$CDM spectrum \citep{planck2014-a13}, and (in the $BB$ panel
only) the dashed black line shows the spectrum for a tensor-to-scalar
ratio of $r=0.05$. Dotted coloured lines indicate the best-fit
power-law fit, $D_{\ell} = q\, (\ell/80)^{\alpha}$, to each foreground
spectrum, where the pivot scale of $\ell_0=80$ is chosen to match that
used in \citet{planck2014-XXX}. The corresponding best-fit parameters
are tabulated in Table~\ref{tab:clpar} for all three apodization
scales, and including multipoles in the range $\ell=(10,150)$ for
synchrotron and $\ell=(10,300)$ for thermal dust emission. Note that
no synchrotron results are shown for the $5\deg$ FWHM apodization
scale. In this case, the effective sky fraction is too small to allow
a robust estimate of the synchrotron spectrum.

\begin{figure*}
\begin{center}
\mbox{
  \epsfig{figure=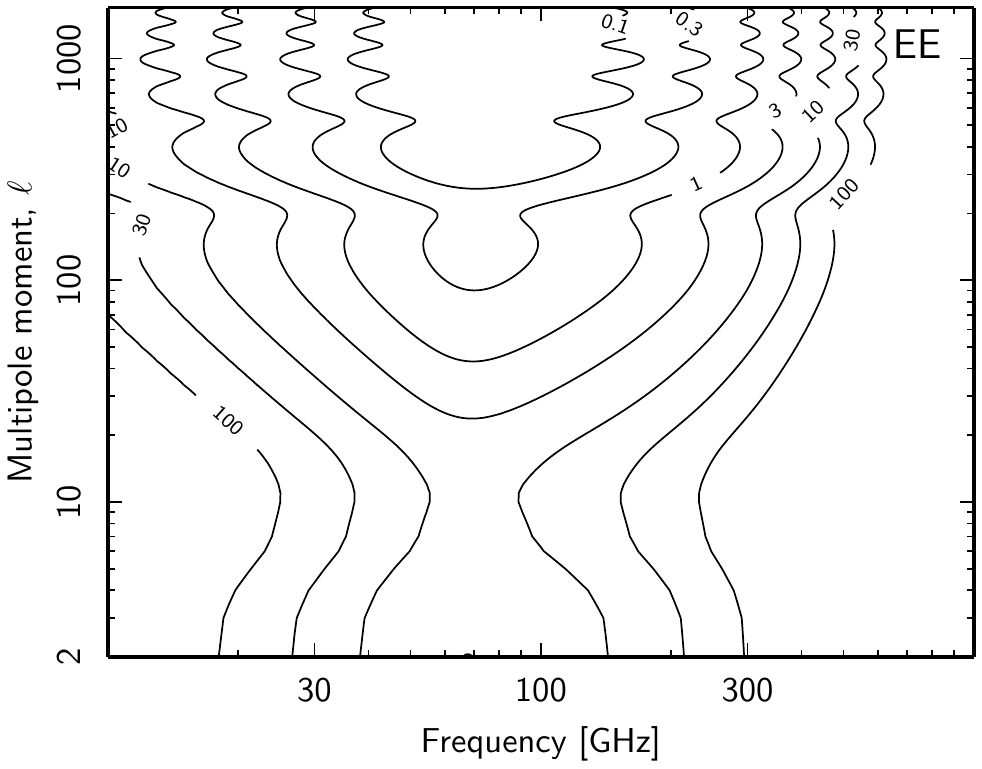,width=0.5\linewidth,clip=}
  \epsfig{figure=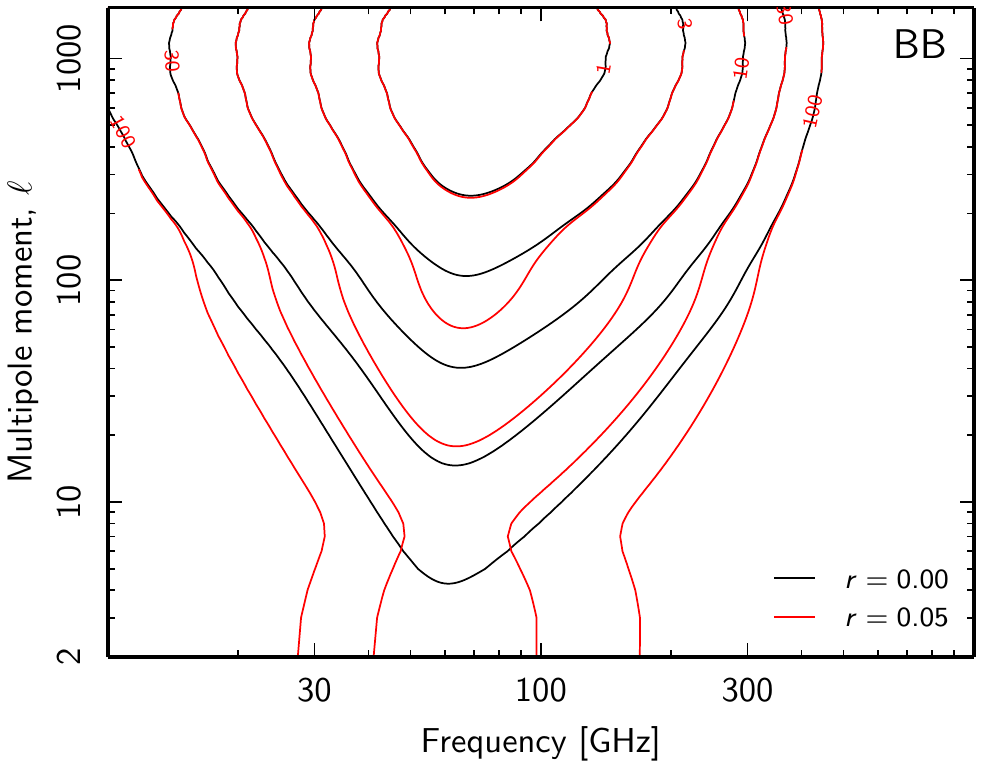,width=0.5\linewidth,clip=}
}
\end{center}
\caption{Amplitude ratio between total polarized foregrounds and CMB
  as a function of both multipole moment and frequency, defined by
  $f(\ell,\nu) =
  [C_{\ell}^{\mathrm{fg}}(\nu)/C_{\ell}^{\mathrm{CMB}}]^{1/2}$, as
  defined Eq.~\ref{eq:pol_fg_model} with parameters derived from 73\,\%
  of the sky. The left and right panels show the $EE$ and $BB$ spectra,
  and the black and red contours in the latter corresponds to
  tensor-to-scalar ratios of $r=0.0$ and $0.05$, respectively.}
\label{fig:cl_pol_ratio}
\end{figure*}

For thermal dust emission these parameters may be compared to the
results presented in \citet{planck2014-XXX}, although a few caveats
are in order. In particular: (1) the masks used in the two analyses are
different, and the mask adopted in this paper effectively removes more
sky around bright point sources after apodization; (2) the map
considered in the present analysis is the \commander\ thermal dust
map, whereas the original analysis considered the raw 353$\,$GHz map;
(3) the multipole ranges adopted for the parameter fits are slightly
different; and (4) we make the fit to the single-$\ell$ power spectrum,
not the binned spectrum.

Nevertheless, we see that the results derived here are in good
agreement with those found in \citet{planck2014-XXX}. In particular,
when considering the same apodization scale of $5\deg$ FWHM, we
recover an identical $BB$/$EE$ ratio of $0.53\pm0.01$, and the $EE$
power-law index agree to $0.5\,\sigma$. For $BB$, the spectral index
difference is slightly larger, but still within
$2\,\sigma$. The power spectrum amplitudes, on the other
hand, are different because of the different effective sky fractions
of the two corresponding apodized masks.

Comparing the different apodization scales, we note both that the
$BB$/$EE$ ratio increases slightly, and that the power-law indices
steepens slightly, as the mask smoothing scale increases. This is due
to thermal dust emission being a highly non-isotropic and non-Gaussian
field, as discussed in \citet{planck2014-XXX}. It is not surprising
that its statistical properties may vary between the Galactic plane
and the high Galactic latitudes. In addition, there is an algorithmic
uncertainty in the form of so-called $E$-to-$B$ leakage, due to
ambiguous polarization modes near the mask edge.  This leakage is far
stronger for foregrounds than for CMB, simply because the foreground
field by construction is at its maximum near the mask boundary. As a
result, it is important to specify the properties of the analysis mask
when summarizing the power spectrum of a foreground field, as
demonstrated in Table~\ref{tab:clpar}.

Overall, however, the mask dependence on the angular power spectrum is
modest, and $D_{\ell}$ provides a useful summary for foreground fields
as well as for the CMB field. Indeed, one of the interesting results
reported by \citet{planck2014-XXX} was the asymmetry between the $B$-
and $E$-mode thermal dust power spectra, with a power ratio of
$BB$/$EE\approx0.5$. This has strong implications for the underlying
astrophysics, and indicates the presence of significant filamentary
structures on intermediate angular scales. In this paper, we find that
the same holds also for synchrotron emission, with an even stronger
asymmetry of $BB$/$EE\approx0.35$. Thus, polarized synchrotron emission
appears to be more strongly aligned along filamentary structures than
thermal dust.

We also find similar power-law indices for synchrotron emission as for
thermal dust, with $\alpha$$\approx$$-0.4$. However, the uncertainties
are relatively larger, because of the lower signal-to-noise ratio of
the 30$\,$GHz channel compared to the 353$\,$GHz channel.

These power-law fits can be used to model the total foreground level
as a function of both multipole moment and frequency. This is
illustrated in Fig.~\ref{fig:cl_pol_ratio} for the $1\deg$ FWHM
apodization case in terms of iso-contour plots of the following
amplitude ratio,
\begin{align}
  f &= \sqrt{\frac{D_{\ell}^{\textrm{s}}(\nu) + D_{\ell}^{\textrm{d}}(\nu)}{D_{\ell}^{\textrm{CMB}}}}\\
  &= \sqrt{\frac{q_{\textrm{s}}\,\left(\frac{\ell}{80}\right)^{\alpha_{\textrm{s}}}\,\frac{s_{\textrm{s}}(\nu)}{s_{\textrm{s}}(30\,\textrm{GHz})}+q_{\textrm{d}}\,\left(\frac{\ell}{80}\right)^{\alpha_{\textrm{d}}}\,\frac{s_{\textrm{d}}(\nu)}{s_{\textrm{d}}(353\,\textrm{GHz})}}{D_{\ell}^{\textrm{CMB}}}},
  \label{eq:pol_fg_model}
\end{align}
where subscripts 's' and 'd' refer to synchrotron and thermal
dust. The frequency spectra, $s_{\textrm{s}}(\nu)$ and
$s_{\textrm{d}}(\nu)$, are the synchrotron (\texttt{GALPROP}) and
thermal dust (one-component greybody) spectra defined in
Table~\ref{tab:model} converted to thermodynamic units, with
parameters defined by the average parameters listed in
Table~\ref{tab:products}. This function is thus simply a model of the
foreground-to-CMB amplitude ratio as a function of multipole and
frequency.

Considering first the $EE$ case shown in the left panel of
Fig.~\ref{fig:cl_pol_ratio}, we note several interesting
features. First, the horizontal ripples seen at $\ell\gtrsim100$
correspond to the CMB acoustic oscillations. Next, we see that the
foregrounds-to-CMB ratio is smaller than unity for all multipoles
above $\ell\gtrsim40$ for frequencies around 70\,GHz, and smaller than
10\,\% for $\ell\gtrsim200$. Also, recall that the corresponding power
spectrum ratio goes as the square of these ratios, and we thus find
that polarized foregrounds have a small effect on the $EE$ spectrum at
multipoles above a few hundred, in agreement with the results pesented
in \citet{planck2014-a13}. However, we also see that the same is by no
means true at low multipoles; the foregrounds-to-CMB ratio is larger
than 3 throughout the reionization peak for $\ell=2$--10.

The right panel of Fig.~\ref{fig:cl_pol_ratio} shows the corresponding
ratio for $BB$, but in this case two different contour sets are plotted; one
for the standard $\Lambda$CDM with a vanishing tensor-to-scalar ratio
(black contours), and one with a tensor-to-scalar ratio of $r=0.05$
(red contours). The peak around $\ell\approx1000$ corresponds to the
signature of weak gravitational lensing, converting E-modes into
B-modes, whereas the ``plateau'' at low multipoles in the red contours
corresponds to additional primordial fluctuations from inflationary
gravitational waves. First of all, we see that foregrounds are
sub-dominant to the lensing signal at multipoles above
$\ell\gtrsim200$ for frequencies around 70\,GHz in this model,
although they never fall below the 10\,\% level. Second, for a vanishing
tensor-to-scalar ratio the foreground-to-CMB around the recombination
peak of $\ell\approx100$ is about 3, and at the reionization peak,
below $\ell\lesssim10$, it is about 100. Increasing the
tensor-to-scalar ratio to $r=0.05$ decreases these numbers to about 2
and 20, respectively.

Before concluding this section, several caveats regarding the above
observations are in order. First of all, it is important to remember
that the angular power spectra reported here are computed over a large
sky fraction including 72\,\% of the sky. For a dedicated B-mode
experiment, it obviously makes sense to consider more conservative
masks. Second, it is also important to bear in mind that the angular
spectra presented here covers only a limited multipole range, and the
extrapolation to small angular scales is therefore associated with
considerable uncertainty. Clearly, extrapolating actual observations
that are made between $\ell\approx10$--100 to $\ell\approx1000$ for
synchrotron emission implies strong assumptions regarding the foreground
composition of both diffuse foregrounds and compact objects.

\subsection{Comparison with independent data products}

\begin{figure*}
\begin{center}
\mbox{
\epsfig{figure=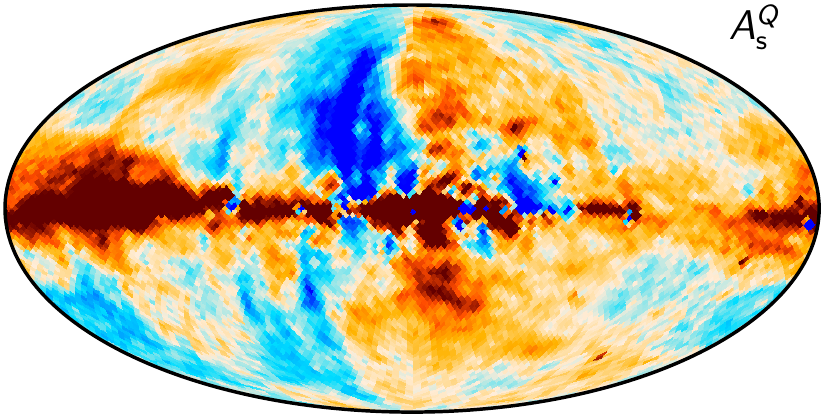,width=0.5\linewidth,clip=}
\epsfig{figure=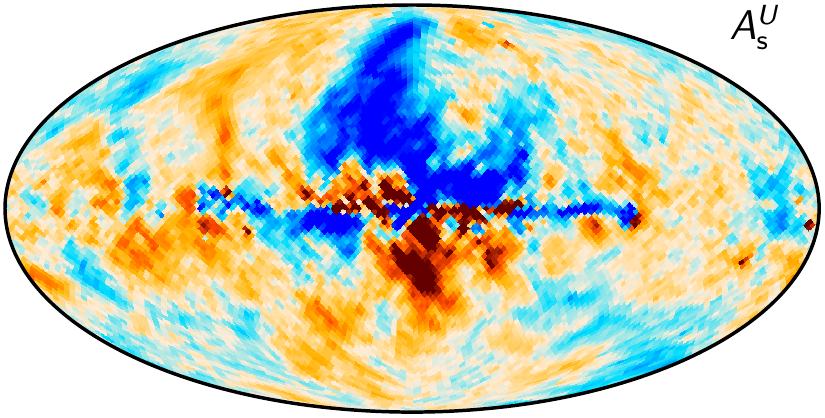,width=0.5\linewidth,clip=}
}
\mbox{
\epsfig{figure=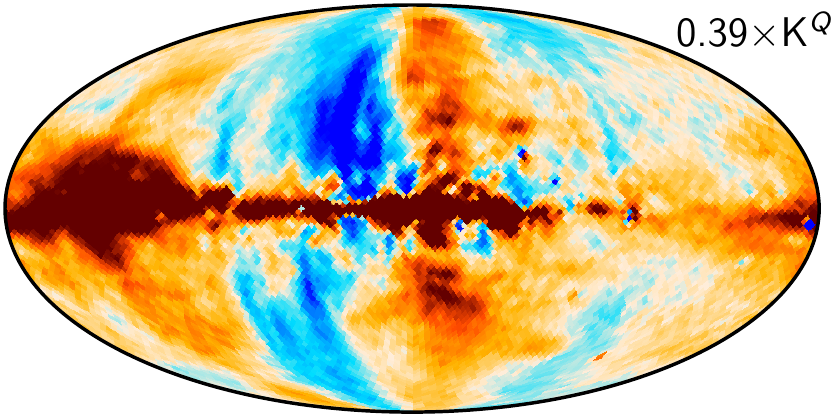,width=0.5\linewidth,clip=}
\epsfig{figure=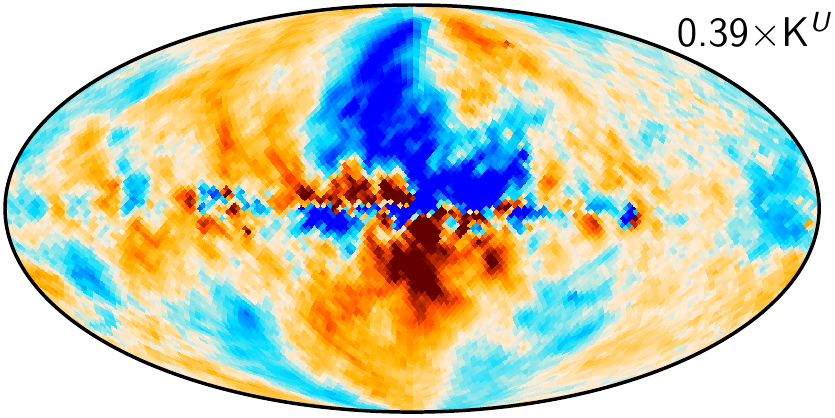,width=0.5\linewidth,clip=}
}
\mbox{
\epsfig{figure=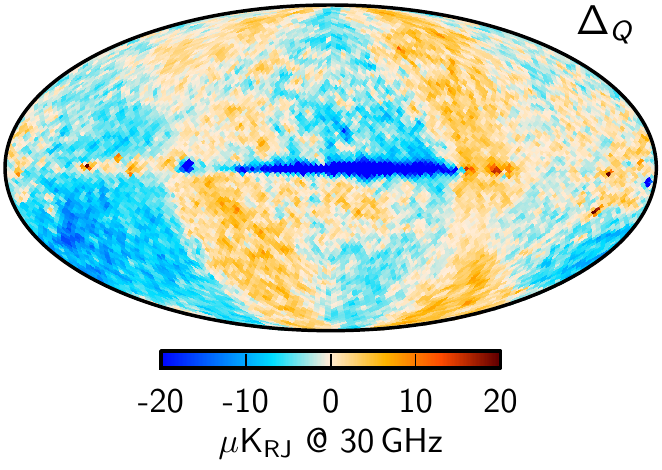,width=0.5\linewidth,clip=}
\epsfig{figure=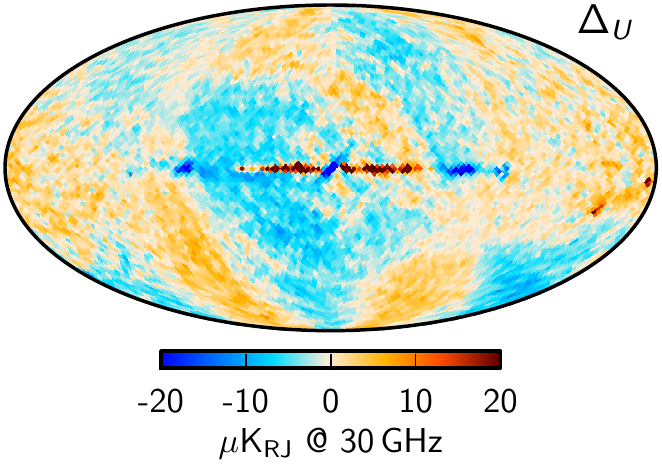,width=0.5\linewidth,clip=}
}
\end{center}
\caption{Comparison of the \Planck\ polarized synchrotron map
(\emph{top}) and the 9-year \WMAP\ K-band map, scaled to 30\,GHz
assuming a spectral index of $\beta_{\textrm{s}}=-3.2$
(\emph{middle}); the bottom row shows the difference between the two
maps. All maps are smoothed to a common resolution of $2\deg$ FWHM. }
\label{fig:wmapK_vs_commander}
\end{figure*}

\begin{figure*}
\begin{center}
\mbox{
\epsfig{figure=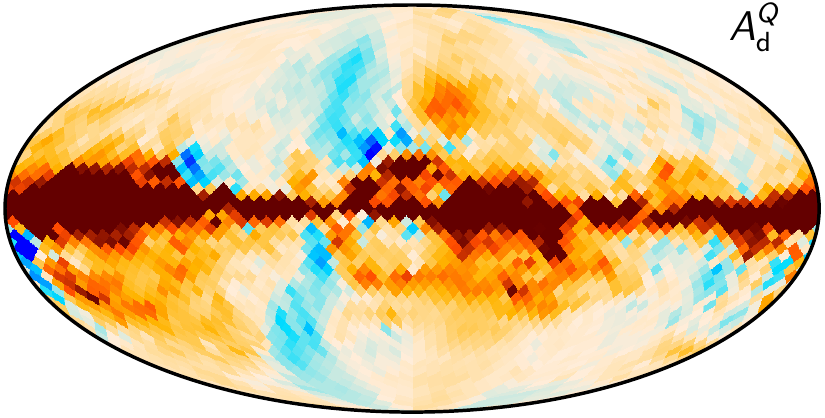,width=0.5\linewidth,clip=}
\epsfig{figure=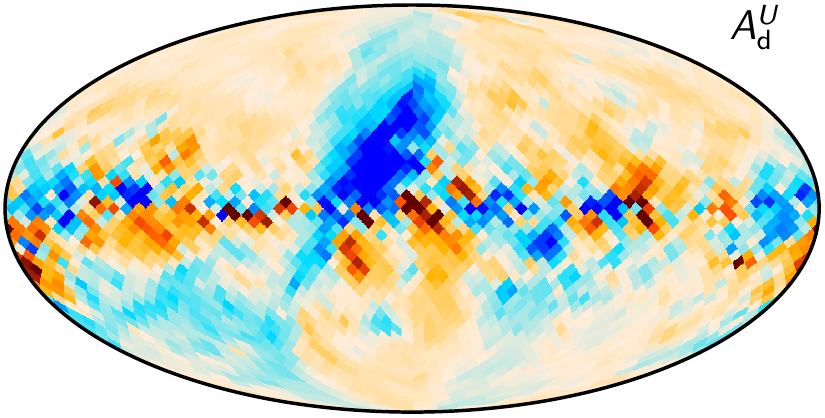,width=0.5\linewidth,clip=}
}
\mbox{
\epsfig{figure=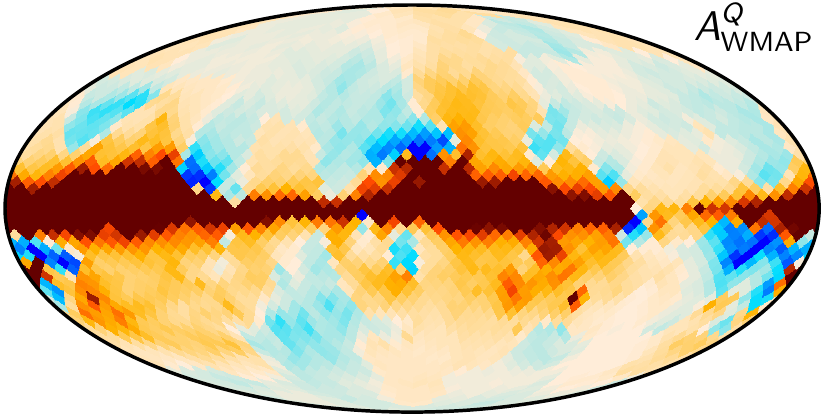,width=0.5\linewidth,clip=}
\epsfig{figure=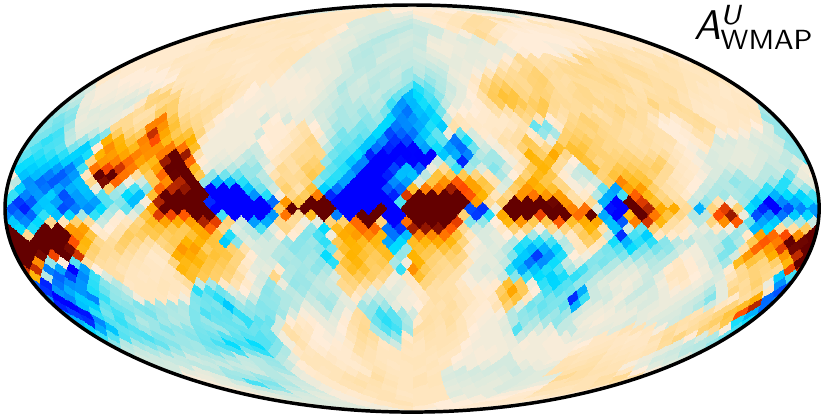,width=0.5\linewidth,clip=}
}
\mbox{
\epsfig{figure=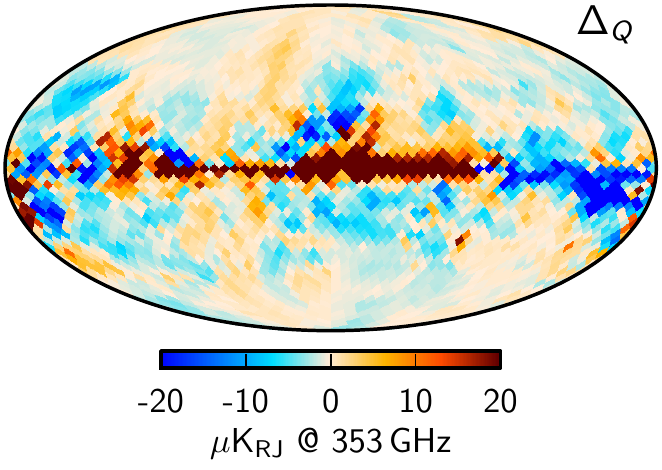,width=0.5\linewidth,clip=}
\epsfig{figure=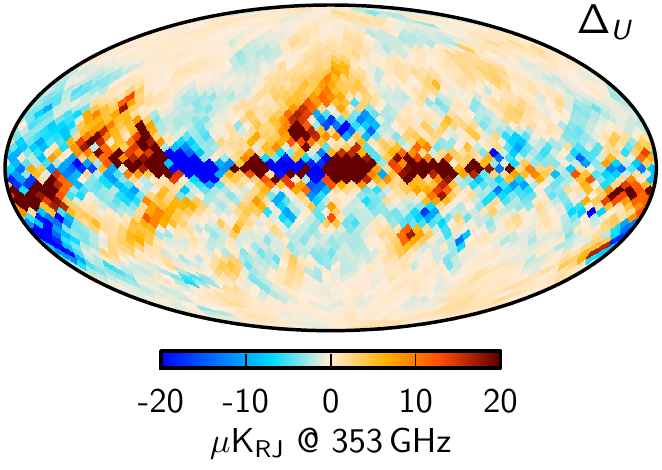,width=0.5\linewidth,clip=}
}
\end{center}
\caption{Comparison of the \Planck\ polarized thermal dust map at
  353\,GHz (\emph{top}) and the \WMAP\ polarized dust template,
  scaled to 353\,GHz assuming a scaling factor of $480\,\mu\textrm{K}$
  (\emph{middle}). The bottom row shows the difference between the two
  maps. All maps are pixelized at a \healpix\ resolution of
  $N_{\textrm{side}}=16$. }
\label{fig:wmap_vs_commander_dust}
\end{figure*}

We now turn to consistency tests based on external (or at least
independently derived) data products. Of course, given the pioneering
nature of the \Planck\ polarization observations, the number of
available external cross-checks is significantly sparser compared to
the temperature case. On the low-frequency side, the \WMAP\ K-band
data represents an excellent comparison for the synchrotron map, while
no products of comparable data quality exists on the high-frequency
side at the moment. This lack of polarized dust measurements has of
course been a major limitation for the entire CMB field for a long
time, and the \WMAP\ solution to this problem was to construct a
polarized dust template by combining the FDS thermal dust intensity
map \citep{finkbeiner1999} with polarization directions from starlight
polarization observations (see \citealt{page2007} for full details).

\begin{figure}
\begin{center}
\mbox{
\epsfig{figure=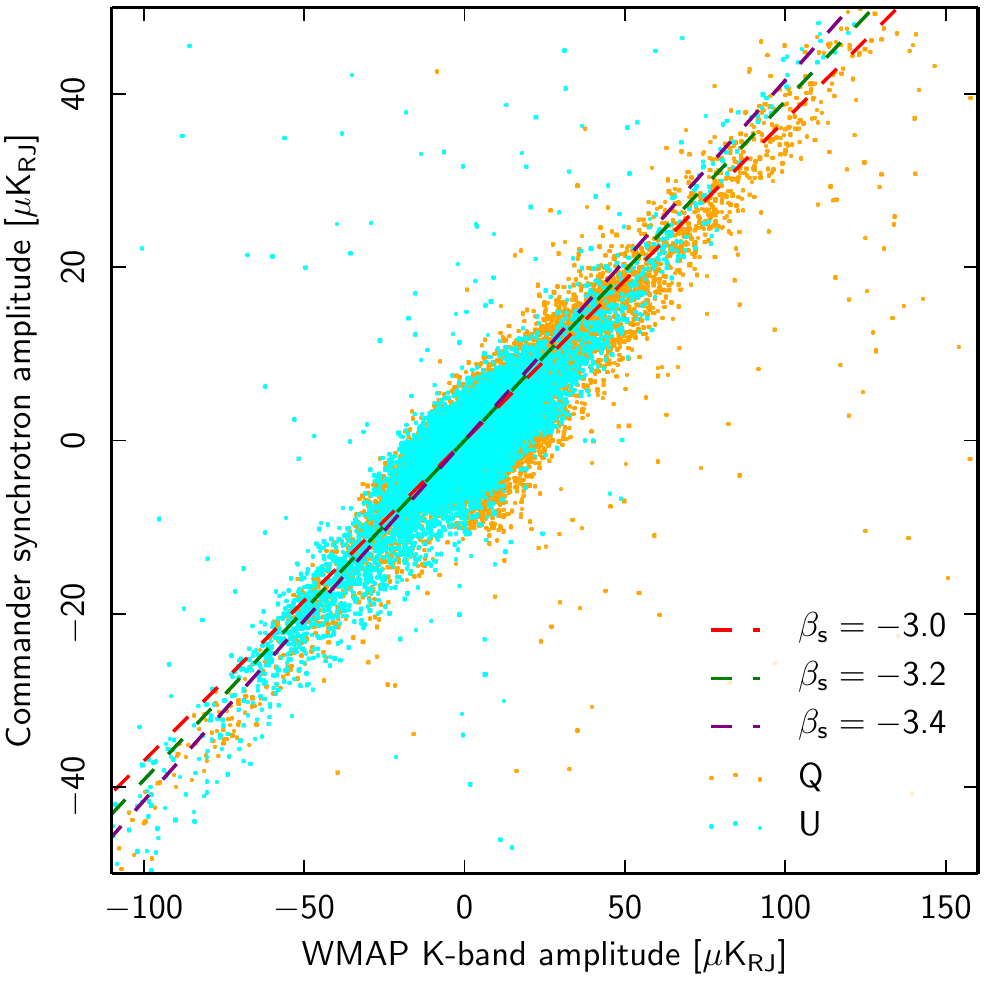,width=88mm,clip=}
}
\end{center}
\caption{$T$--$T$ correlation plot between the \Planck\ polarized
  synchrotron map at 30\,GHz and the \WMAP\ K-band map at 23\,GHz
  for both Stokes $Q$ and $U$ parameters. The dashed coloured lines
  indicate synchrotron spectral indices of $\beta_{\textrm{s}}=-3.0$
  (red), $-3.2$ (green) and $-3.4$ (violet), respectively.}
\label{fig:scatter_wmapK_vs_commander}
\end{figure}

\begin{figure}[!]
\begin{center}
\mbox{
\epsfig{figure=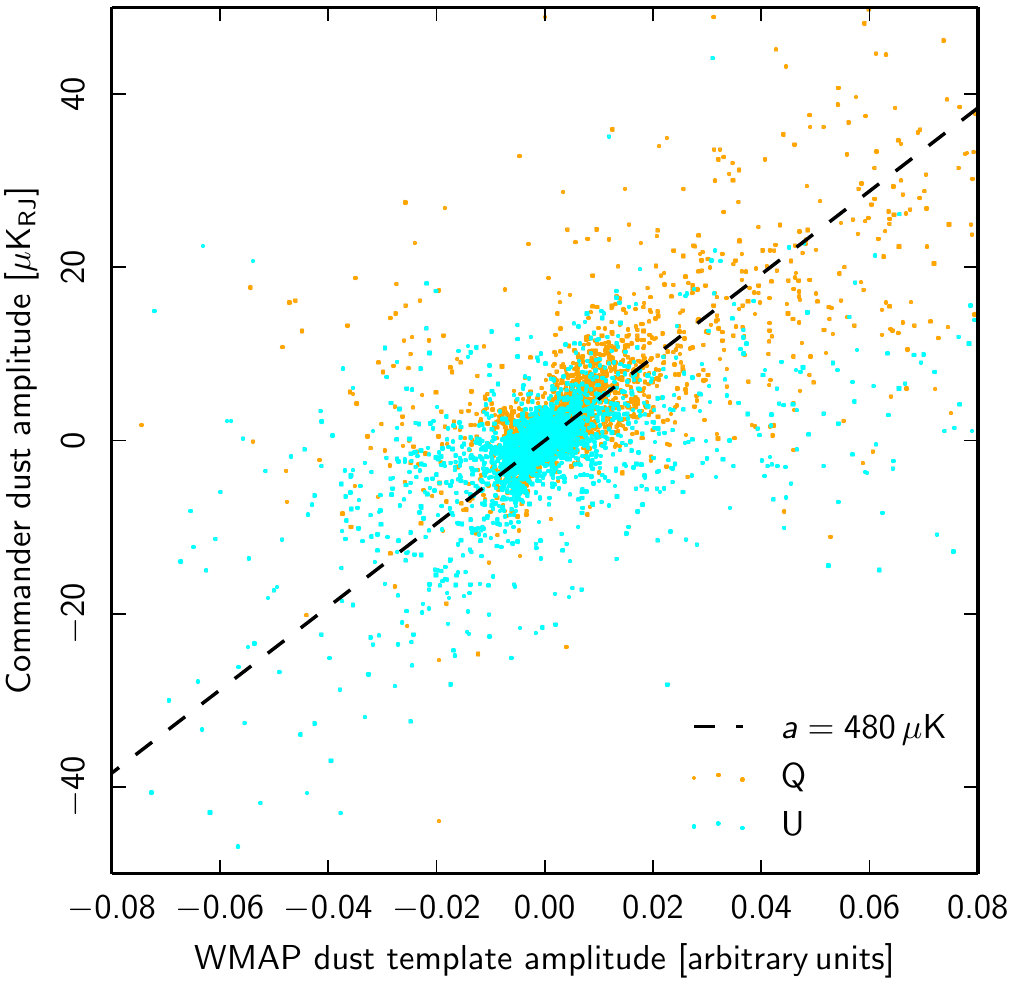,width=88mm,clip=}
}
\end{center}
\caption{$T$--$T$ correlation plot between the \Planck\ polarized
  thermal dust map (at 353\,GHz) and the \WMAP\ polarized dust
  template map (in arbitrary units) for both Stokes $Q$ and $U$
  parameters. The dashed black line corresponds to a relative scaling
  factor of $480\,\mu\textrm{K}_{\textrm{RJ}}$.}
\label{fig:scatter_wmap_vs_commander_dust}
\end{figure}

We compare our new polarized thermal dust and synchrotron maps with
the \WMAP\ maps/templates in Figs.~\ref{fig:wmapK_vs_commander} and
\ref{fig:wmap_vs_commander_dust}, and show corresponding $T$--$T$
scatter plots in Figs.~\ref{fig:scatter_wmapK_vs_commander} and
\ref{fig:scatter_wmap_vs_commander_dust}.

Starting with the synchrotron case, we see first of all in
Fig.~\ref{fig:scatter_wmapK_vs_commander} that the overall
pixel-to-pixel scatter between the \Planck\ synchrotron map and the
\WMAP\ K-band map is substantial. Indeed, based on this full-sky
scatter plot, any synchrotron spectral index between
$\beta_{\textrm{s}}=-3.4$ and $-3.0$ appears consistent with the
observations. Adopting a mean value of $\beta_{\textrm{s}}=-3.2$, and
assuming an effective K-band frequency of $22.6\,\textrm{GHz}$, this
translates into a total scaling factor of 0.39 between K-band and
$30\,\textrm{GHz}$\footnote{Note that \Planck\ 2015 foreground product
  maps are defined at sharp frequencies, and not as bandpass-averaged
  channel maps. The relevant comparison for synchrotron emission is
  therefore indeed 30\,GHz, and not the effective frequency of the
  \Planck\ 30\,GHz band, which is 28.4\,GHz.}. This scaling factor
has been applied to the K-band map shown in
Fig.~\ref{fig:wmapK_vs_commander}, and it also allows us to form a
meaningful residual map, as seen in the bottom row of the same figure.

The relative residuals between the \Planck\ and \WMAP\ synchrotron
maps are clearly substantial, with amplitudes reaching
$5\,\mu\textrm{K}$ at high Galactic latitudes, and with a morphology
clearly associated with the scanning strategy of either \Planck\ or
\WMAP, both of which have symmetries defined by the Ecliptic reference
frame. Furthermore, the residuals are very large-scale in nature, and
obviously dominated by the two lowest multipoles, $\ell=2$ and 3. It
is therefore natural to consider what effects may cause such
large-scale features. Starting with \Planck, a large suite of
null-tests and simulations, specifically targeting large-scale
systematics in the LFI observations, is presented in
\citet{planck2014-a03}. One noteworthy conclusion from that work is a
significant null-test failure in the 44\,GHz polarization frequency
map for $\ell=2$--4, and for two 70$\,$GHz surveys. In addition, the
HFI channels between 100 and 217$\,$GHz are also affected by low level
residual systematics \citep{planck2014-a09}. As a result, these
observations are currently excluded from the \Planck\ 2015 low-$\ell$
likelihood, which instead only relies on the 30, 70 and 353\,GHz
channels \citep{planck2014-a13}. These large-scale 44\,GHz modes are
likely to contribute significantly to the residuals seen in
Fig.~\ref{fig:wmapK_vs_commander}. For WMAP, on the other hand, the
statistical uncertainties in the \WMAP\ $EE$ $\ell=2$ and $BB$ $\ell=3$
modes are very large \citep{bennett2012}, due to the
combination of the differential detectors of \WMAP, and an opening
angle of $141\deg$ between the A and B side reflectors. Although these
uncertainties are appropriately described by the low-resolution
\WMAP\ covariance matrices, it is algorithmically non-trivial to
account for this effect properly in component separation at higher
resolution, and they are also likely to contribute to the residuals in
Fig~\ref{fig:wmapK_vs_commander}.

\begin{figure}
\begin{center}
\mbox{
\epsfig{figure=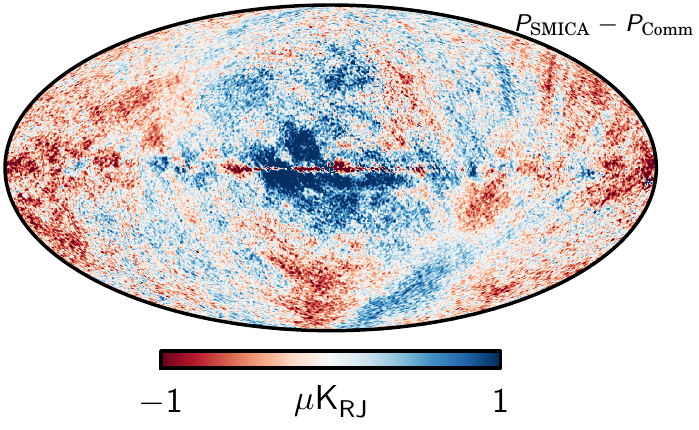,width=0.9\linewidth,clip=}
}
\mbox{
\epsfig{figure=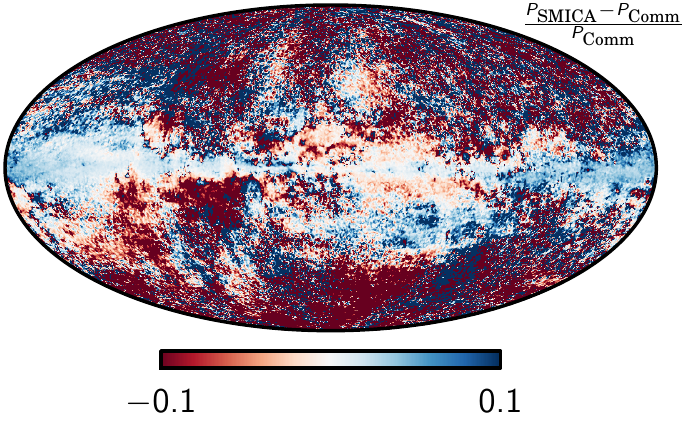,width=0.9\linewidth,clip=}
}
\mbox{
\epsfig{figure=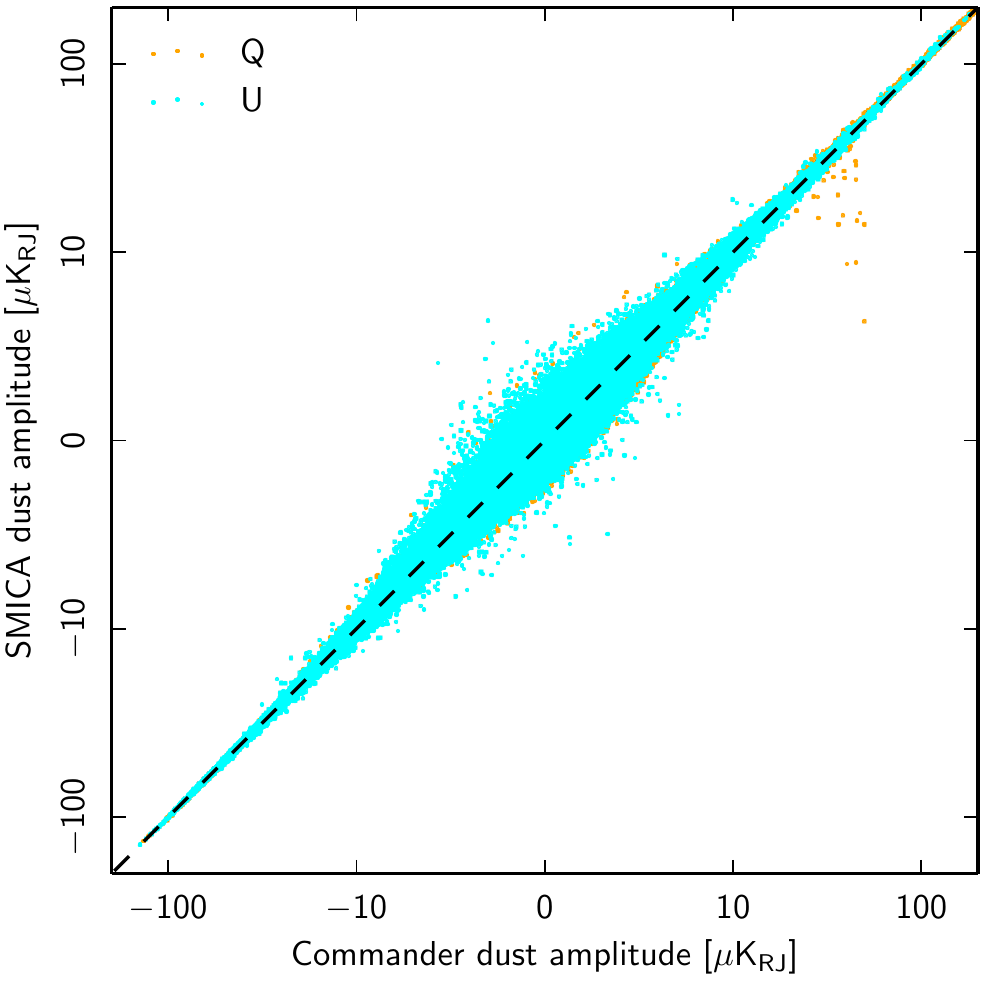,width=0.9\linewidth,clip=}
}
\end{center}
\caption{Comparison between the polarized dust amplitude maps,
  $P=\sqrt{Q^2+U^2}$, derived by \texttt{SMICA} and \texttt{Commander}
  at 353\,GHz. The three panels show the difference map,
  $\d_{\texttt{SMICA}}-\d_{\texttt{Comm}}$ (\emph{top}), the
  fractional difference map,
  $(\d_{\texttt{SMICA}}-\d_{\texttt{Comm}})/\d_{\texttt{Comm}}$
  (\emph{middle}), and a $T$--$T$ correlation plot. Units are
  $\mu\textrm{K}_{\small{\textrm{RJ}}}$. }
\label{fig:smica_vs_commander}
\end{figure}

These differences are also discussed in \citet{planck2014-a31}, with
consistent conclusions. However, that analysis proceeds with
co-adding the \Planck\ and \WMAP\ data sets to produce an alternative
synchrotron map with higher signal-to-noise ratio, and use that map to
identify new polarized synchrotron features and loops. A similar
approach is not straightforward within the framework adopted in the
current paper, which relies sensitively on parametric fits in the
frequency domain. Within such a methodology, even just a single
large-scale mode can severely bias the resulting spectral index map,
rendering the full reconstruction meaningless; for a fully analogous
problem, see the offset discussion in Sect.~\ref{sec:sky} for the
temperature case. The proper way to solve this problem is either to
account for the full covariance matrix in the analysis (as for
instance was done by \citealt{dunkley2009}), but this is only possible at
very low angular resolution. An alternative approach is to marginalize
over just a few modes using the template formalism described in
Sect.~\ref{sec:method}. This requires accurate spatial templates in
the first place, which possibly may be extracted as high-noise
eigenmodes from the full noise covariance matrices. Exploratory work
in this direction is already on-going, but the results have
unfortunately not yet converged.

For now, we recognize that there are significant uncertainties in the
synchrotron model provided, with respect to the very lowest
multipoles. Fortunately, higher-ordered modes are in much better
agreement between \Planck\ and \WMAP. This is important for several
reasons, not least of which is estimation of the optical depth of
reionization, $\tau$, which depends sensitively on the low-$\ell$
modes. However, as shown in \citet{planck2014-a13}, most of the
statistical power for $\tau$ comes from $\ell=4$--6, not $\ell=2$ and
3. Thus, after removing the 44\,GHz channel from the likelihood data
set, all remaining null-tests pass.

A similar comparison between the \Planck\ thermal dust map and the
\WMAP\ dust template is presented in
Figs.~\ref{fig:wmap_vs_commander_dust} and
\ref{fig:scatter_wmap_vs_commander_dust}. Of course, this is not a
validation test of the \Planck\ products in any way, since the two
maps are by no means equivalent data products. Rather, this comparison
provides an interesting reality check on the \WMAP\ model.

The \WMAP\ polarized dust template is provided in arbitrary units, and
must therefore be re-scaled to match the amplitude of the
\Planck\ thermal dust map. The appropriate scale factor is given as
the slope of the $T$--$T$ correlation plot in
Fig.~\ref{fig:scatter_wmap_vs_commander_dust}. Averaging over the two
Stokes parameters, we estimate this to be
$480\,\mu\textrm{K}_{\textrm{RJ}}$ per \WMAP\ unit. This allows us to
form the difference map in the bottom row of
Fig.~\ref{fig:wmap_vs_commander_dust}. With the benefit of hindsight,
we see that the \WMAP\ model is accurate to within 20--30\,\% over
most of the high-amplitude sky, although some regions show deviations
at the 50\,\% level. Unexpectedly, larger scales are reproduced with
greater fidelity than smaller scales.

Given the lack of proper external validation data sets for the
polarized thermal dust emission maps, we instead compare our products
with an independent internal \Planck\ product, similar to what was
done for the CO intensity maps in Sect.~\ref{sec:CO}. Specifically, we
compare the \texttt{Commander} dust map with an equivalent map derived
with the \texttt{SMICA} algorithm
\citep{planck2013-p06,planck2014-a11}.  The results from this
comparison are summarized in Fig.~\ref{fig:smica_vs_commander}, in
terms of: the polarization amplitude difference between the two maps
(top panel); the fractional polarization amplitude difference map
(middle panel); and a $T$--$T$ correlation plot of the individual
Stokes $Q$ and $U$ parameters.\footnote{The only reason for not
  showing the \texttt{SMICA} map itself is that it appears nearly
  identical to the \texttt{Commander} map shown in
  Fig.~\ref{fig:pol_amp_map2}, and the map therefore does not provide
  much new information.} First, we see that the two codes agree to
better than 2\,\% in the Galactic plane, which is very good, considering
the quite different effective bandpass treatments in the two
approaches. In particular, no explicit bandpass integration
corrections are applied in the \texttt{SMICA} analysis, but all
calculations are performed at the bandpass integrated frequency
channel level.

However, the agreement is not equally good at high Galactic latitudes,
with residuals at the roughly $1\,\mu\textrm{K}_{\textrm{RJ}}$
level. Furthermore, these residuals have a morphology that resembles
known instrumental systematics, in particular in the form of monopole
and dipole leakage and ADC corrections \citep{planck2014-a11}. Thus,
the two algorithms clearly respond differently to known systematics,
and we accordingly estimate that the systematic uncertainty in the
large-scale modes of the \Planck\ polarized dust map due to
instrumental effects to be (at least) $1\,\mu\textrm{K}_{\textrm{RJ}}$
at $353\,$GHz. Fortunately, these instrumental effects are expected to
be significantly reduced in the final \Planck\ data release.

\section{Summary and conclusions}
\label{sec:conclusions}

\begin{figure}
\begin{center}
\mbox{
\epsfig{figure=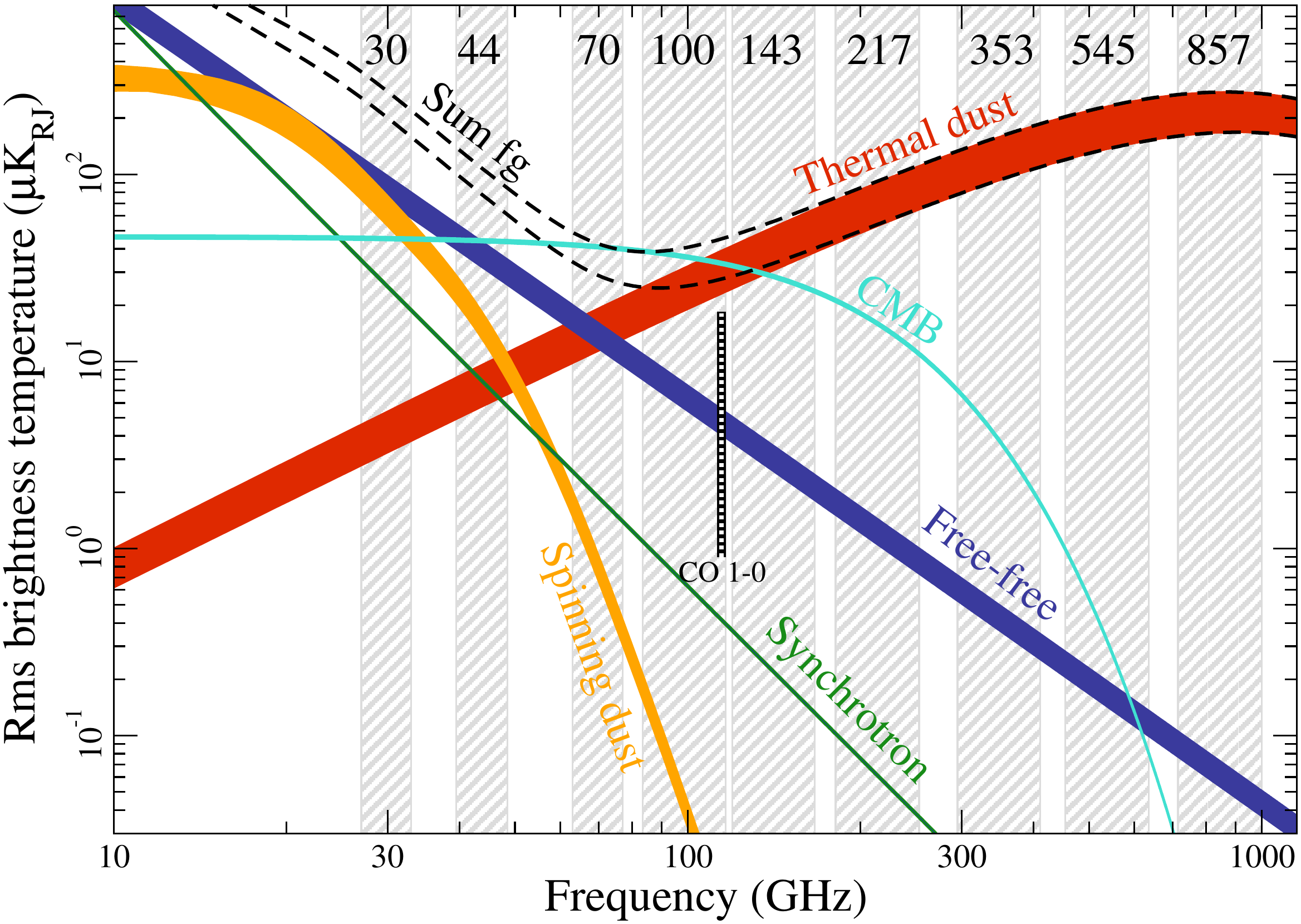,width=88mm,clip=}
}
\vskip 2mm
\mbox{
\epsfig{figure=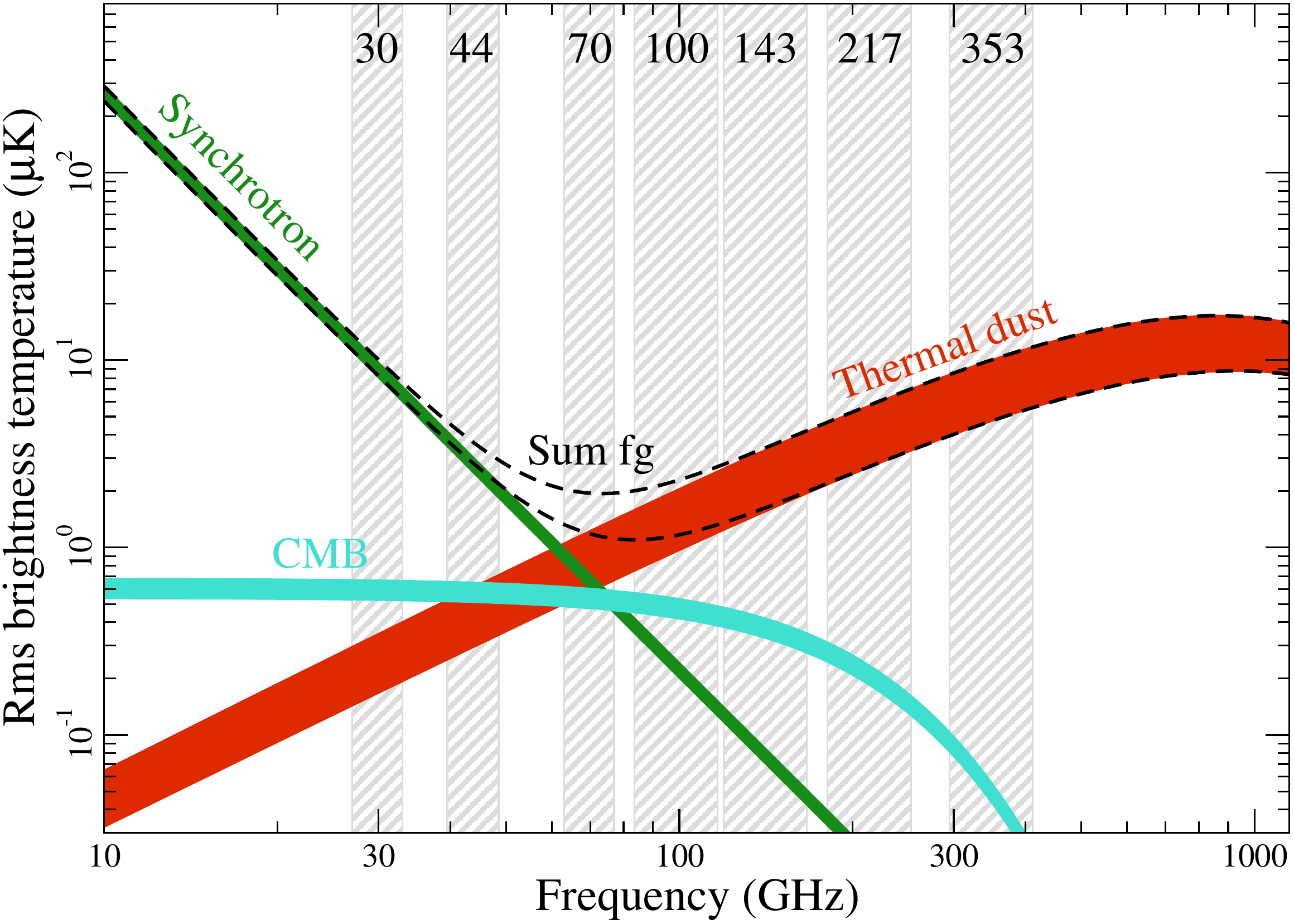,width=88mm,clip=}
}

\end{center}
\caption{Brightness temperature rms as a function of frequency and
  astrophysical component for temperature (\emph{top}) and
  polarization (\emph{bottom}). For temperature, each component is
  smoothed to an angular resolution of $1\deg$ FWHM, and the lower and
  upper edges of each line are defined by masks covering 81 and 93\,\%
  of the sky, respectively. For polarization, the corresponding
  smoothing scale is $40\arcm$, and the sky fractions are 73 and
  93\,\%. Note that foreground rms values decrease nearly monotonically
  with sky fraction, whereas the CMB rms is independent of sky
  fraction, up to random variations. }
\label{fig:overview}
\end{figure}

We have presented the baseline \Planck\ 2015 astrophysical foreground
products in both temperature and polarization, as derived within a
Bayesian parameter estimation framework. Combining the new
\Planck\ sky maps with complementary ancillary data in the form of the
9-year \WMAP\ temperature sky maps and the low-frequency 408\,MHz
Haslam et al.\ survey, we are able to reconstruct a total of six
primary emission temperature mechanisms -- CMB, synchrotron,
free-free, spinning dust, CO, and thermal dust emission -- in addition
to two secondary components, namely thermal SZ emission around the
Coma and Virgo clusters, and line emission between 90 and
100\,GHz. For polarization, we reconstruct three primary emission
mechanisms -- CMB, synchrotron, and thermal dust. In addition to these
astrophysical parameters, we account jointly for calibration and
bandpass measurement errors, as well as monopole and dipole
uncertainties. Statistical uncertainties are propagated from raw sky
maps to final results by means of standard MCMC sampling techniques,
while various model errors are assessed by end-to-end simulations.
All data products are made publicly available, as summarized in
Table~\ref{tab:products}.

Three particularly noteworthy highlights from this analysis is the
following.
\begin{itemize}
\item We have presented the first full-sky polarized thermal dust map,
  which is a direct result of the exquisite sensitivity of the HFI
  instrument. This map will remain a cornerstone of future CMB
  cosmology for the next decade or more, as the search for primordial
  gravitational waves enters the next phase in which foregrounds are
  more important than instrumental noise.
\item We have also presented a full-sky spinning dust intensity
  map. In addition to its obvious scientific value, this map is also
  interesting for algorithmic reasons, as a clear demonstration of
  both the importance and power of joint global analysis. Neither
  \WMAP\ nor \Planck\ have the statistical power to disentangle
  spinning dust from synchrotron, but together beautiful new results
  emerge. We believe that this will be the default approach for
  virtually all future microwave surveys, as no experiment will have
  the power to replace \Planck\ and \WMAP\ by themselves. Rather, each
  new experiment will contribute with a new critical piece of
  information regarding a given phenomenon, frequency coverage, or
  range of angular scales, and thereby help refining the overall
  picture. Global Bayesian analysis provides a very natural framework
  for this work.
\item Another useful illustration of the power of global analysis
  presented in this paper is the identification of important
  instrumental systematic errors. One example is the detection of, and
  correction for, systematic errors in the \Planck\ bandpass
  measurements. More generally, the residual maps shown in
  Figs.~\ref{fig:bad_channels}, \ref{fig:map_residuals}, and
  \ref{fig:pol_residuals} comprise a treasure trove of information on
  instrumental systematics that should prove very valuable for
  improving the raw \Planck\ sky maps before the next data release.
\end{itemize}

All things considered, the sky model presented in this paper provides
an impressive fit to the current data, with temperature residuals at
the few microkelvin level at high latitudes across the CMB-dominated
frequencies, and with median fractional errors below 1\,\% in the
Galactic plane across the \Planck\ frequencies. For polarization, the
residuals are statistically consistent with instrumental noise at high
latitudes, but limited by significant temperature-to-polarization
leakage in the Galactic plane. Overall, this model represents the most
accurate and complete description currently available of the
astrophysical sky between 20 and 857\,GHz.

Figure~\ref{fig:overview} provides an overview of the main components
in both temperature (top panel) and polarization (bottom panel),
summarized in terms of the brightness temperature rms evaluated over
93\,\% and 73\,\% of the sky, respectively. For polarization, this is the
first version of such a plot that is based on observations alone. For
temperature, the most recent previous version is figure~22 of
\citet{bennett2012}, summarizing the \WMAP\ temperature foreground
model. While the two versions agree well in terms of total foreground
power and location of the foreground minimum, there are a few subtle
differences as well. The most important of these is the relative
amplitude of synchrotron and spinning dust. Specifically, synchrotron
dominates over spinning dust at all frequencies in the \WMAP\ model,
whereas in our new model spinning dust dominates over synchrotron
between 15 and 60\,GHz. Such differences are not surprising,
considering the complexity of the astrophysical foregrounds at low
frequencies. As emphasized repeatedly, even when combining the
\Planck\ and \WMAP\ observations, as done in this paper, degeneracies
between synchrotron, free-free and spinning dust remain the leading
source of uncertainty on the low frequency side. Additional
observations between, say, 2 and 20\,GHz are essential to break
these degeneracies. For a more complete analysis of the low-frequency
foreground model presented here, we refer the interested reader to
\citet{planck2014-a31}.

On the high-frequency side, the main outstanding issue are
uncertainties in the net 545 and 857\,GHz calibration, i.e., the
product of calibration and bandpass uncertainties. As of today, the
545\,GHz calibration is uncertain at least at the 1--2\,\% level, and
this translates into an effective 3--6\,\% uncertainty for the 857\,GHz
channel in our fits (in order to maintain a physical thermal dust
frequency scaling). Cross-correlations with \ion{H}{i} observations suggests a
total systematic error on the thermal dust temperature at high
Galactic latitudes of 1--2\,K. Recognizing both calibration and
modelling errors, we emphasize that the thermal dust model presented
here does not represent an accurate model of frequencies beyond
857\,GHz. For instance, naive extrapolation to the DIRBE
$100\,\mu\textrm{m}$ channel results in residuals as large as
40\,\%. Both more physical models and better calibration are needed to
extend into this regime.  In addition, it is important to note that
the current model makes no attempt at separating Galactic
thermal dust emission from CIB fluctuations, and these therefore
constitute a significant contaminant in our thermal dust model on
small angular scales.

Finally, for polarization the main limitations are instrumental
systematics, primarily in the form of temperature-to-polarization
leakage, uncertainties in the analogue-to-digital conversion, and very
long time constants
\citep{planck2014-a01,planck2014-a07,planck2014-a09}. Thus, although
the new \Planck\ 2015 observations already have opened up a completely
new window on the physics of our own Galaxy, through its deep
observations of polarized thermal dust, more work is required in order
to fully realize the science potential of the
\Planck\ measurements. This will be the main focus of the
\Planck\ analysis efforts in the coming months.

\begin{acknowledgements}
\label{sec:acknowledgements}

The Planck Collaboration acknowledges the support of: ESA; CNES, and
CNRS/INSU-IN2P3-INP (France); ASI, CNR, and INAF (Italy); NASA and DoE
(USA); STFC and UKSA (UK); CSIC, MICINN, and JA (Spain); Tekes, AoF,
and CSC (Finland); DLR and MPG (Germany); CSA (Canada); DTU Space
(Denmark); SER/SSO (Switzerland); RCN (Norway); SFI (Ireland);
FCT/MCTES (Portugal); ERC and PRACE (EU).  A description of the Planck
Collaboration and a list of its members, indicating which technical or
scientific activities they have been involved in, can be found at
\url{http://www.cosmos.esa.int/web/planck/planck-collaboration}.

\end{acknowledgements}
\bibliographystyle{aat}
\bibliography{Planck_bib,A12_bib}

\raggedright
\end{document}